\documentclass[12pt,a4paper,notitlepage]{report}
\usepackage[T1]{fontenc}
\usepackage{amsfonts}
\usepackage{latexsym}
\usepackage{dsfont}
\usepackage{amsmath}
\usepackage{amssymb}
\usepackage{amssymb}
\usepackage{amsthm}
\usepackage{mathrsfs}
\usepackage[nosort]{cite}
\newcommand{\tr}{\mathrm{tr}}
\usepackage{setspace}
\usepackage{color}
\usepackage{graphicx}
\usepackage{accents}
\usepackage[font=footnotesize,labelfont=bf]{caption}
\usepackage[utf8]{inputenc}
\usepackage{geometry}
\usepackage[toc,page]{appendix}
\usepackage{braket}
\usepackage{enumitem}

\makeatletter
\newcommand{\labeltext}[2]{%
  \@bsphack
  \MakeLinkTarget*{#1}%
  \def\@currentlabel{#1}{\label{#2}}%
  \@esphack
}
\makeatother

\usepackage{tikz}
\usetikzlibrary{arrows.meta,backgrounds,calc,fit,positioning,scopes,shadows}

\makeatletter
\def\tikzsavelastnodename#1{\let#1=\tikz@last@fig@name}
\makeatother

\usepackage[hidelinks]{hyperref}
\hypersetup{
colorlinks=false,
citecolor= blue,
linkcolor= blue,
urlcolor= blue,
breaklinks=true
}
\usepackage{cleveref}
\crefformat{section}{\S#2#1#3} % see manual of cleveref, section 8.2.1
\crefformat{subsection}{\S#2#1#3}
\crefformat{subsubsection}{\S#2#1#3}

\renewcommand{\theequation}{\thesection.\arabic{equation}}

\def\appendix#1{\addtocounter{section}{1}\setcounter{equation}{0}
\renewcommand{\thesection}{\Alph{section}}
\section*{Appendix \thesection\protect\indent \parbox[t]{11.15cm}{#1}}
\addcontentsline{toc}{section}{Appendix \thesection\ \ \ #1}}

\numberwithin{equation}{section}

\def\dd{\text{d}}
\def\lp{\lambda_+}
\def\lm{\lambda_-}

\newcommand{\inv}[1]{\frac{1}{#1}}
\newcommand{\brac}[1]{ \left( #1 \right)}
\renewcommand{\cal}[1]{\mathcal{#1}}
\newcommand{\calz}{\cal{Z}}
\newcommand{\bcalz}{\bar{\cal{Z}}}

\newcommand{\mn}{\mu\nu}
\newcommand{\ch}{\cal{H}}

\makeatletter
\def\user@resume{resume}
\def\user@intermezzo{intermezzo}
\newcounter{previousequation}
\newcounter{lastsubequation}
\newcounter{savedparentequation}
\renewenvironment{subequations}[1][]{%
      \def\user@decides{#1}%
      \setcounter{previousequation}{\value{equation}}%
      \ifx\user@decides\user@resume 
           \setcounter{equation}{\value{savedparentequation}}%
      \else  
      \ifx\user@decides\user@intermezzo
           \refstepcounter{equation}%
      \else
           \setcounter{lastsubequation}{0}%
           \refstepcounter{equation}%
      \fi\fi
      \protected@edef\theHparentequation{%
          \@ifundefined {theHequation}\theequation \theHequation}%
      \protected@edef\theparentequation{\theequation}%
      \setcounter{parentequation}{\value{equation}}%
      \ifx\user@decides\user@resume 
           \setcounter{equation}{\value{lastsubequation}}%
         \else
           \setcounter{equation}{0}%
      \fi
      \def\theequation  {\theparentequation  \alph{equation}}%
      \def\theHequation {\theHparentequation \alph{equation}}%
      \ignorespaces
}{%
  \ifx\user@decides\user@resume
       \setcounter{lastsubequation}{\value{equation}}%
       \setcounter{equation}{\value{previousequation}}%
  \else
  \ifx\user@decides\user@intermezzo
       \setcounter{equation}{\value{parentequation}}%
  \else
       \setcounter{lastsubequation}{\value{equation}}%
       \setcounter{savedparentequation}{\value{parentequation}}%
       \setcounter{equation}{\value{parentequation}}%
  \fi\fi
  \ignorespacesafterend
}
\makeatother

\begin{document}

%%%%%%%%%%%%%%%%%%%%%%%%%%%%%%%%%%%%%%%%%%%%%%%%%%%%%%%%%%%%%%%%%%%%%%%%%%%%%%%%%%%%%%%%%%%%%%%%%%%%%%%%%%%%%%%%%%%%%%%%%%%%%%%%%%%%%%%%%%%%%%%%%%%%%%%%%%%%%%%%%%%%%%%%%%%%%%%%%%%%%%%%%%%% TITLE PAGE %%%%%%%%%%%%%%%%%%%%%%%%%%%%%%%%%%%%%%%%%%%%%%%%%%%%%%%%%%%%%%%%%%%%%%%%%%%%%%%%%%%%%%%%%%%%%%%%%%%%%%%%%%%%%%%%%%%%%%%%%%%%%%%%%%%%%%%%%%%%%%%%%%%%%%%%%%%%%%%%%%%%%%%%%%%%%%%%%%%%%%%%%%%%%%%%%%%%%%%%

\begin{titlepage}
\vspace*{-1.0cm}

\begin{center}
\vspace{2.0cm}

{\LARGE  {\fontfamily{lmodern}\selectfont \bf An Introduction to 
String Newton-Cartan  \\
\vspace{7mm}
Holography and Integrability}} \\[.2cm]

\normalsize

\vskip 1.5cm
\textsc{Andrea Fontanella \footnotesize and \normalsize Juan Miguel Nieto Garc\'ia}\\
\vskip 1.2cm

\begin{small}
\textit{School of Mathematics $\&$ Hamilton Mathematics Institute, \\
Trinity College Dublin, Ireland}\\
\vspace{1mm}
\href{mailto:andrea.fontanella[at]tcd.ie}{\texttt{andrea.fontanella[at]tcd.ie}}, 
\href{mailto:juanmiguel.nietogarcia[at]upm.es}{\texttt{juanmiguel.nietogarcia[at]upm.es}}

\end{small}

\vskip 1 cm
\begin{abstract}
\vskip1cm

\noindent \emph{String Newton-Cartan holography} is a new example of gauge/gravity duality relating non-relativistic string theory and gauge theories. We review how to construct a family of string and $p$-brane Newton-Cartan holographic dualities by consistently taking the non-relativistic limit of the AdS/CFT correspondence. 
We also review classical string solutions, quantisation, string coset action and integrability of the non-relativistic string theory appearing in the String Newton-Cartan limit of the AdS$_5$/CFT$_4$ correspondence.

\end{abstract}

\vskip2cm

\emph{Invited review for J. Phys. A}

\end{center}

\end{titlepage}

\tableofcontents
\vspace{5mm}
\hrule

%%%%%%%%%%%%%%%%%%%%%%%%%%%%%%%%%%%%%%%%%%%%%%%%%%%%%%%%%%%%%%%%%%%%%%%%%%%%%%%%%%%%%%%%%%%%%%%%%%%%%%%%%%%%%%%%%%%%%%%%%%%%%%%%%%%%% END OF TITLE PAGE %%%%%%%%%%%%%%%%%%%%%%%%%%%%%%%%%%%%%%%%%%%%%%%%%%%%%%%%%%%%%%%%%%%%%%%%%%%%%%%%%%%%%%%%%%%%%%%%%%%%%%%%%%%%%%%%%%%%%%%%%%%%%%%%%%%%%%%%%%%%%%%%%%%%%%%

\setcounter{section}{0}
\setcounter{footnote}{0}

%%%%%%%%%%%%%%%%%%%%%%%%%%%%%%%%%%%%%%%%%%%%%%%%%%%%%%%%%%%%%%%%%%%%%%%%%%%%%%%%%%%%%%%%%%%%%%%%%%%%%%%%%%%%%%%%%%%%%%%%%%%%%%%%%%%%%%%%%%%%   MAIN BODY   
%%%%%%%%%%%%%%%%%%%%%%%%%%%%%%%%%%%%%%%%%%%%%%%%%%%%%%%%%%%%%%%%%%%%%%%%%%%%%%%%%%%%%%%%%%%%%%%%%%%%%%%%%%%%%%%%%%%%%%%%%%%%%%%%%%%%%%%%%%%%%%%%%%%%%%%%%%%%%%%%
\chapter{Introduction}

Holography is a beautiful paradigm that establishes the equivalence between a gravity theory in a given spacetime and a gauge theory defined at the boundary of such spacetime. The first concrete proposal was due to Maldacena \cite{Maldacena:1997re} (see also \cite{Gubser:1998bc, Witten:1998qj, Maldacena:1998im}) establishing a duality between type IIB string theory in AdS$_5\times$S$^5$ and $\mathcal{N}=4$ super Yang-Mills (SYM) theory. Although the generalisation of holography to Anti-de Sitter spacetimes with dimension different than five was pointed out in the original work by Maldacena, it was not fully consolidated until much later, see e.g. \cite{Aharony:2008ug, Gukov:2004ym, Tong:2014yna, Eberhardt:2017pty}. However, at this point it is very natural to ask whether holography is just an artifact of the Anti-de Sitter spacetime, or whether it is a more general principle of physics. 

A possible direction where to explore such question is by considering holography in spacetimes with \emph{non-relativistic} symmetries. The topic of non-relativistic holography is not new, but it started already twenty years ago with the study of Schr\"odinger and Lifshitz holography \cite{Dobrev:2013kha, Aizawa:2009yc, Christensen:2013lma, Christensen:2013rfa, Hartong:2013cba, Hartong:2014pma, Hartong:2014oma, Bergshoeff:2014uea, Son:2008ye, Maldacena:2008wh, Guica:2010sw, CaldeiraCosta:2010ztk, Taylor:2008tg, Taylor:2015glc}. 
In these non-relativistic holographic dualities, the bulk geometry is a Schr\"odinger or Lifshitz spacetime.\footnote{The difference between Schr\"odinger and Lifshitz spacetimes is that the former has more symmetries that the latter, namely, the Galilean boosts and particle number symmetry. See \cite{Taylor:2008tg, Taylor:2015glc} for reviews of the topic.} They are characterised by an anisotropic dilatation isometry that scales time and space with different weights, namely $x \to \lambda x$, $t\to \lambda^z t$, with $z \neq 1$. In the holographic paradigm, the global symmetries of the bulk are realised as spacetime and internal symmetries of the dual gauge theory. This means that, in the context of Schr\"odinger and Lifshitz spacetimes, the dual gauge theory must have non-relativistic spacetime symmetries.

A very important point to remark, is that the geometry of Schr\"odinger and Lifshitz spacetimes is \emph{Lorentzian}, in the sense that it is defined via a single non-degenerate metric tensor that is invariant under local Lorentz transformations acting on its vielbein. It is only at the level of its Killing vectors that one discovers the anisotropic dilatation symmetry. A very different question is whether holography can be formulated in spacetimes that \emph{do not} have local Lorentz invariance. These spacetimes belong to the class of Non-Lorentzian geometries \cite{Bergshoeff:2022eog}, which will be of main focus in this review.    

Non-Lorentzian geometries emerge when taking a singular limit of Lorentzian metrics. They are of two types, one is a Newton-Cartan geometry that appears when taking the non-relativistic limit. The second is a Carroll geometry, appearing when taking the ultra-relativistic limit. In both cases, the spacetime geometry is described by a pair of degenerate metric tensors invariant under, respectively, Galilean and Carrollian local symmetries, corresponding to the appropriate \.In\"on\"u-Wigner contractions of the Poincar\'e algebra \cite{Bergshoeff:2022eog, Bagchi:2025vri}. An interesting context where these geometries appear is when taking the non- or ultra-relativistic limits of string and gauge theories. In this review we focus on the non-relativistic limit, which is precisely what we discuss next.

\section{Non-relativistic string theory}

Since Einstein's relativity, we are used to the fact that physical theories are demanded to be Lorentz invariant. Even more, as string theory is one of the best candidate for the quantisation of General Relativity, which is based on demanding local relativistic invariance, it might seem pointless and even contradictory with our aim to consider non-relativistic string theory. Actually, this is not the case. The action can be shown to be free of Weyl anomalies provided the background satisfies a non-relativistic version of Einstein equations, therefore providing simple toy models for ultraviolet-finite string theories. 

The non-relativistic limit of string theory in flat spacetime was first studied in \cite{Gomis:2000bd, Danielsson:2000gi}, hence called Gomis-Ooguri action\footnote{Sometimes this action is also called wound string theory.}, and later generalised in the AdS$_5\times$S$^5$ background \cite{Gomis:2005pg}, also referred to as GGK theory. When taking the non-relativistic limit, the time-like component of the spacetime metric diverges, making the Polyakov action ill-defined. To avoid such inconsistency, it was pointed out that one has to introduce a 
closed B-field, that needs to be fine-tuned to a critical value in order to remove such divergent term. This sort of ``renormalisation'' procedure does not alter the equations of motion (as the B-field is closed), but it will alter the global conserved charges. The string energy would diverge in the non-relativistic limit, but it becomes finite after taking into account the B-field contribution. This is the analogue of taking the non-relativistic limit of a particle's energy: one always gets the $mc^2$ term that diverge as $c\to \infty$, which is eliminated by a critical 1-form. The B-field plays precisely the same role as the critical 1-form of removing the string analogue of the particle rest mass. The non-relativistic limit has therefore the effect of zooming into a corner of the string energy spectrum (sometimes referred to in the literature as BPS bound).  

After this, it was understood that the spacetime probed by a non-relativistic string is described by the \emph{String Newton-Cartan} geometry \cite{Andringa:2012uz, Bergshoeff:2015uaa, Bergshoeff:2018yvt, Bergshoeff:2019pij}. This geometry is a co-dimension two singular foliation, characterised by a pair of degenerate metric tensors: one called longitudinal metric $\tau_{\mu\nu}$ and one called transverse tensor $h_{\mu\nu}$. It is important to mention that at this point the non-relativistic limit acts only at the level of the target space geometry. Therefore, the worldsheet remains Lorentz invariant (i.e. relativistic).\footnote{It is possible to take an additional limit that makes the worldsheet non-Lorentzian as well. This was studied in \cite{Harmark:2017rpg, Harmark:2018cdl, Harmark:2019upf, Harmark:2020vll, Baiguera:2020jgy, Baiguera:2020mgk, Baiguera:2021hky}.} The action has a well-defined beta function \cite{Gomis:2019zyu, Gallegos:2019icg}, which gives us a set of non-relativistic Einstein equations. The action is free of Weyl anomalies provided the String Newton–Cartan target space satisfies such set of equations. 

In the past years, non-relativistic string theory has been extensively studied in many different formal facets. For instance, it was found that there are different, but equivalent, ways to derive a non-relativistic string action - all revisited in this review. They can be summarised as the limit procedure \cite{Gomis:2000bd, Gomis:2005pg, Bergshoeff:2018yvt, Bergshoeff:2019pij}, the null-reduction method \cite{Harmark:2017rpg, Harmark:2018cdl, Harmark:2019upf} and the expansion approach \cite{Hartong:2021ekg, Hartong:2022dsx}.\footnote{\samepage There is also a more recent approach, based on gauging the string Galilei global symmetry of non-relativistic strings in flat spacetime in order to derive the action in a generic curved background \cite{Banerjee:2020xwa, Moinuddin:2021oft, Nandi:2025fmk}. Although very promising, this method is still lacking in few aspects. For example, it is unclear why the gauging involves only the transverse embedding fields, and therefore how a curved longitudinal metric may appear.} 
Other formal aspects of non-relativistic string theory concern T-duality \cite{Bergshoeff:2018yvt, Kluson:2018vfd, Kluson:2019xuo, Kluson:2019avy}, symmetries of the action \cite{Bergshoeff:2019pij, Harmark:2019upf, Bidussi:2021ujm}, double field theory \cite{Ko:2015rha, Morand:2017fnv, Blair:2019qwi, Blair:2020gng, Blair:2020ops}, Hamiltonian formalism \cite{Kluson:2017abm,Kluson:2018egd,Kluson:2018grx}, integrability \cite{Fontanella:2020eje, Fontanella:2022fjd, Fontanella:2022pbm, Fontanella:2022wfj}, classical string solutions \cite{Fontanella:2021btt, Fontanella:2023men, Gomis:2004ht}, quantisation \cite{Sakaguchi:2007ba, Fontanella:2021hcb, deLeeuw:2024uaq}, open strings \cite{Gomis:2020fui,Gomis:2020izd, Hartong:2024ydv}, non-relativistic supergravity \cite{Bergshoeff:2021bmc, Bergshoeff:2021tfn, Bergshoeff:2023ogz, Bergshoeff:2024nin, Bergshoeff:2025grj, Lescano:2026dhv} and heterotic string theory \cite{Bergshoeff:2023fcf, Bergshoeff:2025uut, Lescano:2024url, Lescano:2025ixp, Lescano:2025yio, Lescano:2025asm, Lescano:2025xzs}. For a review on some of these topics,  we refer the reader to \cite{Bergshoeff:2022eog, Oling:2022fft, Bergshoeff:2022iyb,  Hartong:2022lsy}. 

String and M-theory do not contain only strings, but a family of extended objects called $p$-branes. Therefore, it is natural to generalise the particle and string non-relativistic limits to the case of $p$-branes. It turns out that the spacetime probed by a non-relativistic $p$-brane is a singular foliation of codimension $p+1$. Such geometry is called $p$-brane Newton-Cartan. As for the string case, the longitudinal metric will generate a divergent term in the Dirac-Born-Infeld (DBI) action describing the dynamics of the $p$-brane. To remove such divergence, one has to introduce a critical $p+1$ Ramond-Ramond field via a Wess-Zumino term. All these examples of decoupling limits turns out to be related to each others via a web of dualities inherited from String and M-theory \cite{Blair:2023noj}.

\section{Non-relativistic gauge theories}

Quantum field theories with broken Lorentz invariance are particularly relevant for describing condensed matter systems, where Lorentz invariance of the microscopic description is no longer necessary.  
Non-relativistic quantum field theories find various applications, among which the fractional quantum Hall effect \cite{Geracie:2014nka}, cold atoms \cite{Son:2008ye} unitary Fermi gas \cite{Nishida:2010tm}, strange metals \cite{Hartnoll:2009ns}, superconductivity \cite{PhysRev.108.1175}. Focusing on String and M-theory context, one of the simplest example of non-relativistic gauge theory is Galilean Electrodynamics \cite{Santos:2004pq}. It was derived from a five-dimensional spacetime framework, as suggested in \cite{OKTO}, and proposed as a Lagrangian theory that captures the so-called electric and magnetic non-relativistic limits of the relativistic Maxwell equations, identified by Le Bellac and L'evy-Leblond in \cite{LeBellac}.  Later, various aspects of Galilean Electrodynamics (GED) were studied. Its action was found to appear as a non-relativistic limit of Maxwell theory coupled to a massless scalar field \cite{Bergshoeff:2015sic}, and also via a null-reduction\footnote{The null-reduction procedure is based on the fact that the Bargmann algebra in $d$ dimensions, i.e. the central extension of the Galilei algebra, can be consistently embedded into the Lorentz algebra in $d+1$ dimensions, see e.g. \cite{OKTO}.} \cite{Duval:1990hj, Duval:1984cj, Julia:1994bs, Festuccia:2016caf, Bagchi:2022twx}. The symmetries of its action are given by a twisted version of the conformal Milne algebra, where the scaling generator of spacetime coordinates is replaced by two independent generators that scale time and space separately \cite{Fontanella:2024hgv} - see also \cite{Festuccia:2016caf, Bagchi:2022twx, Bagchi:2014ysa}. A preliminary work on quantisation of GED theory was done in \cite{Chapman:2020vtn, Banerjee:2022uqj}, and a supersymmetric version of GED in 2+1 dimensions was found in \cite{Baiguera:2022cbp}.

Another non-relativistic gauge theory of great importance in the String and M-theory context is the non-abelian version of GED, called Galilean Yang-Mills (GYM) \cite{Bagchi:2015qcw}. This theory can also be derived via null reduction \cite{Bagchi:2022twx}. The symmetries of the Galilean Yang-Mills action are given by the conformal Milne algebra \cite{Fontanella:2024hgv}. The BRST symmetry was studied in \cite{Islam:2023iju}, whereas coupling GYM with fermions was discussed in \cite{Bagchi:2017yvj} - see also \cite{Lambert:2024yjk}.

Both GED and GYM theories can also be derived from a non-relativistic limit of the DBI action coupled to a critical B-field. The DBI action \cite{Tseytlin:1997csa,Myers:1999ps} describes the dynamics of a stack of D$p$-branes. In the case of a single spacetime-filling brane, the non-relativistic abelian DBI action precisely gives the GED theory, once the low-energy limit $\alpha'\to 0$ is taken \cite{Gomis:2020fui, Hartong:2024ydv}. A similar situation happens for a stack of spacetime-filling branes, where in such case the non-relativistic non-abelian DBI action reproduces the GYM theory. In the case where the brane under consideration is not spacetime-filling, the situation is slightly different. For a stack of D3-branes in type IIB string theory, it was shown that the non-relativistic non-abelian DBI action reproduces at low-energy the GYM theory with five interacting scalar fields \cite{Fontanella:2024kyl, Fontanella:2024rvn}. This theory is now understood as the non-relativistic limit of the bosonic sector of $\mathcal{N}=4$ super Yang-Mills in four dimensions (for its supersymmetric completion, see \cite{Lambert:2024yjk}).

Another example of a gauge theory relevant in the context of String and M-theory is the BLG theory \cite{Bagger:2006sk, Bagger:2007jr, Bagger:2007vi, Gustavsson:2007vu}, or its generalisation ABJM theory \cite{Aharony:2008ug}, which underlies the AdS$_4$/CFT$_3$ duality. The non-relativistic limit of the ABJM theory was proposed in \cite{Lambert:2024uue} and will be discussed in this review.
There are several other topics regarding non-relativistic gauge theories which we will not discuss in this review, such as the state-operator correspondence \cite{Lambert:2021nol, Baiguera:2024vlj}, deformations \cite{Lambert:2020zdc, Lambert:2020jjm, Lambert:2019jwi}, supersymmetry \cite{Lambert:2020scy, Lambert:2019nti, Lambert:2018lgt, Meyer:2017zfg, Harmark:2019zkn, Bergman:1995zr}.
Since the topic of non-relativistic gauge theories is quite extensive, we refer the reader to existing reviews on non-relativistic gauge theories \cite{Baiguera:2023fus, Bergshoeff:2022eog}.

\section{And now, relate them: holography}

At this point, having introduced non-relativistic string and gauge theories, it is natural to ask: \emph{is there a non-relativistic gauge/gravity duality that relates them?} A first attempt based on a contraction of the Anti-de Sitter spacetime symmetries was suggested in \cite{Bagchi:2009my}. A second approach, suggested in \cite{Sakaguchi:2007ba}, is based on the observation that GGK string theory in static gauge coincides with the leading-order term in the expansion of the relativistic string in AdS$_5\times$S$^5$ around minimal AdS$_2$ surfaces \cite{Giombi:2017cqn}.
These ideas were proposed well before the modern understanding of String Newton–Cartan geometry and its role in non-relativistic string theory. A different approach has been put forward more recently, making use of earlier work on non-relativistic strings and gauge theories. In \cite{Fontanella:2024kyl, Fontanella:2024rvn}, it was suggested to take the non-relativistic limit directly at the level of the brane construction of AdS$_5$/CFT$_4$, while imposing a set of consistency conditions. This construction led to propose that non-relativistic string theory in the String Newton–Cartan version of AdS$_5\times$S$^5$ (known as GGK theory \cite{Gomis:2005pg}) is holographically dual to Galilean Yang–Mills theory with five interacting scalar fields in $3+1$ dimensions. This provides the first example of \emph{String Newton–Cartan holography}, i.e. a holographic duality between a Gomis-Ooguri type of non-relativistic string theory and a non-relativistic gauge theory. 
It is amusing to see that the B-field required in order to have a well-defined non-relativistic string action is also required in the dual gauge theory. Furthermore, this duality is supported by a match between  the bulk global symmetries and the spacetime and internal symmetries of the dual gauge theory.

The GGK/GYM duality was also studied from a different point of view, namely by considering the S-dual non-relativistic limit \cite{Lambert:2024yjk}. In that case, the string theory is not in the form of a Gomis-Ooguri action, as there is no critical B-field involved. Instead, it involves critical Ramond-Ramond two and four forms, and the spacetime is a singular codimension two foliation describing the non-relativistic limit of intersecting D1-D3 branes. On the gauge theory side, taming the divergence produces a constraint, such that the action describes a quantum mechanics on monopole moduli space. 

In the same spirit, a limit producing a codimension four singular foliation of the AdS$_5$/CFT$_4$ was considered in \cite{Lambert:2024yjk}, where it was suggested to describe the non-relativistic limit of intersecting D3-D3$'$ branes. Again, the limit is not of the Gomis-Ooguri form, but requires turning on some critical Ramond-Ramond four forms. As before, taming the divergence in the dual gauge theory introduces a constraint, such that the theory describes a 2d sigma model on Hitchin moduli space.

A non-relativistic limit of the AdS$_4$/CFT$_3$ from the M-theory perspective was then proposed in \cite{Lambert:2024uue}. The limit produces a codimension three foliation on the bulk spacetime (i.e. an M2-brane Newton-Cartan limit), whereas on the dual theory it gives a non-relativistic ABJM theory, which describes a quantum mechanics on Hitchin moduli space.  

On a more general ground, in \cite{Lambert:2024ncn} it was pointed out that if a codimension $p+1$ foliation of a D$q$-brane gives the non-relativistic limit of an intersecting D$p$-D$q$ brane system, then the same result is obtained by inverting the order of $p$ and $q$, namely by taking the codimension $q+1$ foliation of a D$p$-brane. This implies that the two limits in the two configurations lead to the same result. Therefore, not every non-relativistic limit produces a genuine novel result. This reciprocity result can nevertheless turn complicated configurations into simpler ones, as illustrated by the M2–M5 example discussed in \cite{Lambert:2024ncn}.

As it is clear by now, String Newton-Cartan holography can be generalised to what may be called \emph{$p$-brane Newton-Cartan holography}, where the idea of having a singular codimension two foliation is now extended to generic geometries with singular codimension $p+1$ foliations. This generalisation started with the results of \cite{Lambert:2024yjk, Lambert:2024uue, Lambert:2024ncn, Guijosa:2023qym, Guijosa:2025mwh}, later extended for generic $p$ in \cite{Blair:2024aqz}, and also analysed from the point of view of string dualities \cite{Blair:2025prd}. 

A particularly interesting example of $p$-brane Newton-Cartan holography appears when considering the codimension $p+1$ foliation that aligns with the world-volume of a $p$-brane. In this specific case, the non-relativistic limit acts precisely as the Maldacena's near-horizon limit $\alpha'\to 0$ \cite{Guijosa:2025mwh}. At the level of the dual gauge theory, the $\mathcal{N} = 4$ super Yang-Mills theory (which is a relativistic field theory in four dimensions) can be seen as a non-relativistic gauge theory from the ten dimensions perspective.    
This interesting observation makes the usual relativistic AdS/CFT a particular example of $p$-brane Newton-Cartan holography.

\paragraph{Reasons of interest.} Interest in String Newton–Cartan holography, and its generalisations, stems from two fundamental features. First, they provide a novel setting, distinct from the usual Anti-de Sitter spacetime, in which to study and test holography. In other words, they offer a quantum-consistent setting, free of Weyl anomalies, in which to explore non-Lorentzian holography. Second, non-relativistic limits provide a way to zoom into \emph{corners} of relativistic AdS/CFT (the so-called BPS bounds). In these limits the theory simplifies and there is already enough evidence that it becomes exactly solvable (see the next section for further discussion in the context of integrability). Observables can then be computed exactly, making AdS/CFT more tractable. Testing AdS/CFT in these corners also addresses a conceptual question: if AdS/CFT is regarded as a black box, does it remain valid when we zoom into its corners?

On more speculative grounds, the Gomis–Ooguri action appears in the recently suggested string-dual description of 2d Yang–Mills \cite{Komatsu:2025dqv}, a toy model that shares many interesting features with 4d QCD, such as colour confinement and an infinite tower of resonances. Since the flat space limit of the GGK theory reproduces the Gomis–Ooguri action, this provides an interesting motivation to explore whether the GGK/GYM holographic duality may be related to the proposal of \cite{Komatsu:2025dqv}, and possibly shed light on the stringy mechanisms underlying confinement in 4d QCD. 

In addition, holographic superconductors have been discussed in literature \cite{Hartnoll:2008kx}, also in the context of Lifshitz holography \cite{Bu:2012zzb}. Given that field theories with non-relativistic symmetries capture condensed-matter systems more closely, it is an interesting avenue of future research to investigate whether String Newton–Cartan holography could help provide a holographic string description of high-temperature superconductivity.

\section{Integrability and the power of symmetries}

Symmetries are a powerful tool to understand a physical theory. A clear example relevant for the AdS$_5$/CFT$_4$ correspondence is how supersymmetry heavily constrains $\mathcal{N}=4$ SYM. Supersymmetry constrains quantities to depend holomorphically on quantum fields and coupling constants. For theories with $\mathcal{N}=2$ supersymmetry, this implies that the $\beta$ function does not gain perturbative corrections beyond one-loop \cite{Gates:1983nr}. For theories with $\mathcal{N}=4$ supersymmetry, it is even more restrictive, fixing the $\beta$ function to vanish \cite{Sohnius:1981sn, Mandelstam:1982cb, Seiberg:1988ur}.

The most important point for us is that both theories appearing in the AdS$_5$/CFT$_4$ correspondence have more symmetries than just supersymmetry and conformal symmetry: they are \emph{integrable theories}. We say that a theory is classically integrable, in the sense of Liouville, if it has at least as many independent conserved quantities as degrees of freedom. If that is the case, we can fully determine the dynamics of the system.\footnote{In this review we focus only on classical integrability. We will not discuss integrability of $\mathcal{N}=4$ SYM, as it inherently involves quantum integrability. Compared to classical integrability, quantum integrability is more difficult to define. We may think that a requirement equivalent to Liouville integrability, like ``having enough linearly independent operators constituting a complete set of commuting operators'', would be enough, but this does not have any special implication for a quantum theory \cite{WEIGERT1992107}. In fact, for a quantum Hamiltonian $H$, we can always construct a complete set of commuting operators, for instance, by considering powers of $H$. Usually, quantum integrability is defined as the existence of a Bethe Ansatz.} Classical integrability of type IIB string theory in AdS$_5\times$S$^5$ action was first described in \cite{Bena:2003wd} and later refined in \cite{Alday:2005gi}. Since then, it has been the main driving force in understanding and testing the AdS$_5$/CFT$_4$ correspondence. One of the most celebrated tests of the correspondence, namely the fact that the dispersion relation of some classical strings with large angular momenta can be mapped to the one-loop anomalous dimensions of single trace operators of $\mathcal{N}=4$ SYM, can actually be extended to a matching between the eigenvalues of an infinite number of hidden higher commuting charges from both sides of the correspondence \cite{Arutyunov:2003rg}.

As we are considering a field theory, Liouville integrability would require us to find an infinite number of conserved quantities. There exists a method to systematically construct these conserved charges. One can combine the equations of motion obtained from the Metsaev-Tseytlin action of the string into the flatness condition of a current called \emph{Lax connection} that depends on a parameter called \emph{spectral parameter} \cite{Bena:2003wd}.
The path ordered exponential of this current over a loop that winds the worldsheet cylinder once is called \emph{monodromy matrix}. Flatness of the Lax connection implies that the eigenvalues of the monodromy matrix are independent of the worldsheet time, and thus can be used as generating functions for conserved quantities of the action when we expand them on the spectral parameter. 

A very convenient method to collect and access the information encoded in the eigenvalues of the monodromy matrix is to consider its characteristic equation. The algebraic curve defined by this characteristic equation is called \emph{classical spectral curve}. As it contains all the information about the conserved quantities, it can be used to identify and study classical string solutions \cite{Kazakov:2004qf, Kazakov:2004nh, Beisert:2004ag, Beisert:2005bm}. For example, rigid rotating string solutions appear as algebraic curves with one square-root cut. A more interesting and powerful use of the spectral curve is to compute quantum corrections to a given classical solution. As cuts on the algebraic curve correspond to bosonic classical solutions, adding small enough cuts (i.e. poles) would correspond to adding quantum corrections. This idea was formalised in \cite{Gromov:2007aq}, where it was applied to circular strings in S$^3$ and AdS$_3$.

\paragraph{Integrability in GGK theory.} Similarly to how we asked before if holography appears in the non-relativistic limit, we can also ask if integrability appears in this corner of the theory. Integrability of the AdS$_5 \times$S$^5$ string action is more evident when the theory is written as coset model, so the first step to answer that question is to rewrite the non-relativistic AdS$_5 \times$S$^5$ string action in this form. A natural way to construct the non-relativistic limit of the coset action is to perform a Taylor expansion of the algebras involved in the coset construction, with the condition that the leading order gives us the \.In\"on\"u-Wigner contraction. This process is called Lie Algebra Expansion \cite{deAzcarraga:2002xi,deAzcarraga:2007et, Hatsuda:2001pp}. The Lie Algebra Expansion method was applied to the coset construction at the level of the Maurer-Cartan currents in \cite{Fontanella:2020eje}, while a modified coset construction was proposed in \cite{Fontanella:2022fjd, Fontanella:2022pbm}. These two methods can be used to rewrite GGK theory as a coset model. Furthermore, following similar steps as for the relativistic case, a Lax connection for GGK theory was constructed.

Once we have shown that GGK theory is classically integrable, in the sense of having a Lax connection, the next step is to take advantage of it and increase our understanding of the theory. The classical spectral curve of GGK theory was constructed in \cite{Fontanella:2022wfj}. It was proven that, if the Lax connection evaluated on a given solution is independent of the $\sigma$ variable, then the eigenvalues of the monodromy matrix are independent of the spectral parameter. Therefore, from this analysis, the classical spectral curve turns out ot be trivial. As the expansion in the spectral parameter is what encodes the conserved charges into the monodromy matrix, Liouville integrability of the non-relativistic action is not straightforward. Nevertheless, the monodromy matrix is \emph{not} completely independent of the spectral parameter, but its dependency is confined to non-diagonalisable contributions. This happens because the algebra behind the GGK theory is non-semisimple, leading to non-diagonalisable matrices. This is not special of just non-relativistic systems. In fact, it was shown in \cite{Fontanella:2022wfj} that this issue also appears for relativistic strings in Minkowski space.
From this point of view, it is natural that integrability is not realised in the standard form and that a characteristic equation would fail to capture all the physical properties of the system.

\newpage

\section{What we will \emph{not} discuss in this review}

The main focus of this review is on String Newton–Cartan holography and its integrability structure. There are some topics which are connected to it, but which will not be treated here. These topics have already been discussed in past reviews, and they are:
\begin{itemize}
    \item Lifshitz and Schr\"odinger holography (review \cite{Taylor:2008tg, Taylor:2015glc});
    \item holographic duality between strings with a non-relativistic worldsheet and Spin Matrix Theory (review \cite{Oling:2022fft});
    \item non-relativistic gauge theories (review \cite{Baiguera:2023fus}); 
    \item integrability in the relativistic AdS/CFT (review \cite{Beisert:2010jr}).
\end{itemize}

%%%%%%%%%%%%%%%%%%%%%%%%%%%%%%%%%%%%%%%%%%%%%%%%%%%%%%%%%%%%%%
%%%%%%%%%%%%%%%%%%%%%%%%%%%%%%%%%%%%%%%%%%%%%%%%%%%%%%%%%%%%%%%%%%%%%%%%%%%%%%%%%%%%%%%%%%%%%%%%%%%%%%%%%%%%%%%%%%%%%%%%%%%%%%%%%%%%%%%%%%%%%%%%%%%%%%%%%%%%%%%%%%%%%%%%%%%%%%%%%%%%%%%%%%%%%%%%%%%%%%%%%%%%%%%%%%%%%%%%%%%%%%%%%%%%%%%%%%%%%%%%%%%%%%%%%%%%%%%%%%%%%%%%%%%%%%%%%%%%%%%%%%%%%%%%%%%%%%%%%%%%%%%%%%%%%%%%%%%%%%%%%%%%%%%%%%%%%%%%%%%%%%%%%%%%%%%%%%%%%%%%%%%%%%%%%%%%%%%%%%%%%%%%%%%%%%%%%%%%%%%%%%%%%%%%%%%%%%%%%%%%%%%%%%%%%%%%%%

\chapter{A brief review of non-relativistic string theory}

In this chapter we review aspects of non-relativistic string theory that will be instrumental for our subsequent discussion of non-relativistic holography and integrability. 
We shall review various methods for deriving the non-relativistic action, along with its local and global symmetries, quantum consistency, the landscape, and the coordinate systems used to describe certain background geometries. For a more dedicated review on some of the topics treated in this chapter, we refer the reader to \cite{Oling:2022fft}.

%Nowadays altenatives formulation of the NR limit in the AdS5xS5 have been developed but, paraphasing Sidney Coleman, they consist of treating the GGK limit in ever-increasing levels of abstraction (emphasise that this is not a bad thing).

\section{Constructing non-relativistic string actions}

\subsection{The limit procedure}
\label{subsec:limit}

\subsubsection{Polyakov formalism} \label{sec:PolyakovFormalism}

We start with the closed string Polyakov action coupled to a Kalb-Ramond B-field, 
\begin{equation}
\label{rel_Pol_action}
    S=- \frac{T}{2}  \int{\dd^2 \sigma \bigg( \gamma^{\alpha \beta} \partial_\alpha X^\mu \partial_\beta X^\nu g_{\mu \nu} +  \varepsilon^{\alpha\beta} \partial_\alpha X^\mu \partial_\beta X^\nu b_{\mu \nu} \bigg)} \,,
\end{equation}
where $T$ is the string tension; the string worldsheet coordinates are collected as $\sigma^{\alpha} = (\tau, \sigma)$, with the periodicity condition $\sigma \equiv \sigma + 2 \pi$; $\gamma^{\alpha\beta} \equiv \sqrt{-\mathsf{h}} \mathsf{h}^{\alpha\beta}$ is the Weyl invariant combination of the inverse worldsheet metric $\mathsf{h}^{\alpha\beta}$ and $\mathsf{h} =$ det$(\mathsf{h}_{\alpha\beta})$; $g_{\mu\nu}$ is the target space metric; $b_{\mu \nu}$ is the antisymmetric Kalb-Ramond B-field; $X^{\mu}$ are the embedded coordinates, with $\mu \in \{0, ..., d-1\}$. We will often use the notation $g_{\alpha\beta} \equiv g_{\mu\nu} \partial_{\alpha}X^{\mu} \partial_{\beta} X^{\nu}$ for the pullback of the target space metric into the worldsheet. For the Levi-Civita symbol, we choose the convention $\varepsilon^{01} = - \varepsilon_{01} = +1$.

The first intuitive attempt of taking a non-relativistic limit is using dimensional analysis to reinstate the speed of light $c$ in the target space part of the action \eqref{rel_Pol_action} while keeping the worldsheet relativistic, and take the limit $c\to \infty$.\footnote{Since $c$ is a dimensionful quantity, such limit cannot be taken on it directly. What we are implicitly doing is to rescale $c\to \omega c$, where $\omega$ is dimensionless, and then take $\omega \to \infty$. We will keep this notation throughout the review.} This naive attempt is too simple, since the resulting non-relativistic string is non-vibrating, as shown in \cite{Batlle:2016iel} for the case of strings in flat spacetime.  

\begin{figure*}[t!]
    \centering
    \includegraphics[keepaspectratio,width=\textwidth]{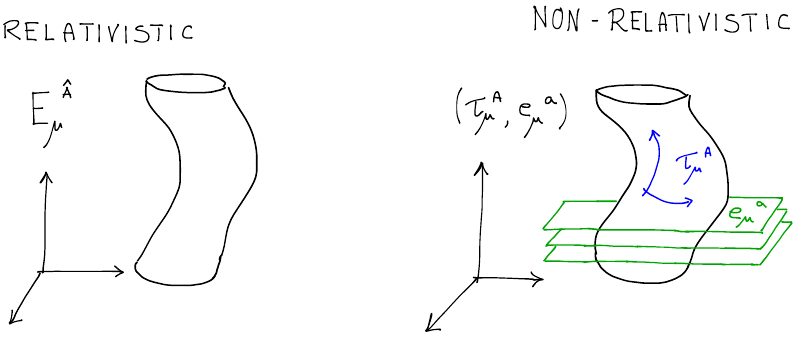}

  \caption{The non-relativistic limit produces a codimension two singular foliation of the target space, where the relativistic vielbein $E_{\mu}{}^{\hat{A}}$ is replaced by the degenerate SNC vielbeine $(\tau_{\mu}{}^A, e_{\mu}{}^a)$, with $A=0,1$, $a=2, ..., 9$.}
    \label{fig:SNC_spacetime}
\end{figure*}
    
In order to get a more interesting action, we have two options: either we also make the worldsheet non-relativistic, leading us to actions that might have Weyl anomalies, or we also scale a spatial direction with $c$. In this review, we take the second choice. 

This rescaling of time and space will produce a singular co-dimension 2 foliation of the spacetime, see fig. \ref{fig:SNC_spacetime}. In particular, we assume that the vielbein $ E_{\mu}{}^{\hat{A}}$ associated with $g_{\mu \nu}$ will expand in $c$ at $c\to \infty$ as follows,
\begin{align}\label{E_scaling}
 E_{\mu}{}^A &= c \tau_{\mu}{}^A + \frac{1}{c} m_{\mu}{}^A + \mathcal{O}(c^{-3})\, , &
 E_{\mu}{}^a &= e_{\mu}{}^a + \mathcal{O}(c^{-2}) \, ,
\end{align}
where the flat index $\hat{A}\in \{0, ..., d-1\}$ has been divided into a \emph{longitudinal} part $A\in\{ 0 , 1\}$ and a \emph{transverse} part  $a \in\{ 2, ..., d-1 \}$. This makes the metric expand as\footnote{One may also include odd powers of $c$ in $g_{\mu \nu}$ but, under a weak field assumption and some physical constraints on energy-momentum, the first odd power corrections are shadow fields, which must be eliminated anyway by using the field equations, relegating odd powers to subleading orders \cite{Dautcourt:1996pm}. A physical way of understanding it is the fact that gravitational potentials of even power in $c$ cannot source potentials of odd power in $c$. If one drops the weak field assumption, odd terms in the expansion reappear \cite{Ergen:2020yop}.}
\begin{gather} \label{gscaling}
    g_{\mu\nu} = c^2 \tau_{\mu\nu} + H_{\mu\nu} + \mathcal{O}(c^{-2}) \, ,\\
    \tau_{\mu\nu} = \tau_{\mu}{}^A \tau_{\nu}{}^B \eta_{AB} \, , \qquad    h_{\mu\nu} = e_{\mu}{}^a e_{\nu}{}^b \delta_{ab}  \, , \qquad    H_{\mu\nu} = h_{\mu\nu} + \left(\tau_{\mu}{}^A m_{\nu}{}^B + \tau_{\nu}{}^A m_{\mu}{}^B \right) \eta_{AB} \, . \notag 
\end{gather}
To fully define a String Newton-Cartan geometry, one has to introduce also the inverse vielbeine $(\tau^{\mu}{}_A, e^{\mu}{}_a)$, although they will not directly enter in the Polyakov action. The SNC vielbeine, together with their inverses, satisfy the following set of orthogonality and completeness relations, 
\begin{subequations}
\begin{align}
\tau_{\mu}{}^A \tau^{\mu}{}_B &= \delta^A_B \, , & \tau_{\mu}{}^A e^{\mu}{}_b = e_{\mu}{}^a \tau^{\mu}{}_B = 0 \, , \\
e_{\mu}{}^a e^{\mu}{}_b &= \delta^a_b \, ,   & \tau_{\mu}{}^A \tau^{\nu}{}_A + e_{\mu}{}^a e^{\nu}{}_a = \delta^{\mu}_{\nu} \, . 
\end{align}
\end{subequations}

At this point, if we naively substitute these expansions in \eqref{rel_Pol_action} and take the $c\to \infty$ limit, we will encounter that the $\tau_{\alpha\beta}$ term diverges. To tame this divergence, the Kalb-Ramond field needs to have a critical component \cite{Gomis:2000bd, Gomis:2005pg}, namely
\begin{eqnarray}\label{critical_B}
    b_{\mu\nu} = c^2 b^{\text{crit.}}_{\mu\nu} + b^{\text{NR}}_{\mu\nu} \, , \qquad\qquad
    b^{\text{crit.}}_{\mu\nu} \equiv \tau_{\mu}{}^A \tau_{\nu}{}^B \varepsilon_{AB} \, .
\end{eqnarray}
Notice that $b^{\text{crit.}} = \tau^A \wedge \tau^B\varepsilon_{AB}$ is closed if $\tau_\mu{}^A$ satisfies the torsionless condition \cite{Bergshoeff:2018yvt, Bergshoeff:2019pij},
\begin{eqnarray}\label{torsionless_constraint}
    \partial_{[\mu} \tau_{\nu]}{}^A + \varepsilon^A{}_B \tau_{[\mu}{}^B \Omega_{\nu]} = 0 \, ,
\end{eqnarray}
where $\Omega_{\mu}$ is the $\mathfrak{so}(1,1)$ spin connection associated with longitudinal Lorentz rotation. Therefore, for any background where this condition is fulfilled, the critical B-field has no effect on the equations of motion, although it can alter the value of \emph{globally} defined physical quantities, such as Noether charges. In fact, we will discuss in section \ref{flowing} that the energy contribution from the critical B-field is the analogue of the leading $mc^2$ in the expansion of the relativistic energy for the particle.

At this stage the metric and B-field can be combined into a Lorentz square. There are two options, one of which requires us to write the longitudinal internal space in light-cone coordinates. The option that does not involve light-cone indices was found in \cite{Gomis:2005pg}. It consists in rewriting
\begin{equation}
\label{action_F2}
c^2 \partial_{\alpha} X^{\mu} \partial_{\beta} X^{\nu} \left(\gamma^{\alpha\beta}\tau_{\mu\nu} + \varepsilon^{\alpha\beta} b^{\text{crit.}}_{\mu\nu} \right) = c^2 \gamma^{00} \mathcal{F}^A \mathcal{F}^B \eta_{AB}\, ,
\end{equation}
where
\begin{equation}
\label{F}
 \mathcal{F}^A = \tau_{\mu}{}^A \partial_0 X^{\mu} - \frac{1}{\gamma_{11}} \varepsilon^{AB} \eta_{BC} \tau_{\mu}{}^C \partial_1 X^{\mu} - \frac{\gamma_{01}}{\gamma_{11}} \tau_{\mu}{}^A \partial_1 X^{\mu} \, . 
\end{equation}
We can now implement a Hubbard-Stratonovich transformation, which consists in rewriting any divergent quadratic action $c^2 \mathcal{A}^2$ in terms of a finite action, by introducing Lagrange multiplier fields. In our case, this amounts to:  
\begin{equation}
\label{rewriting}
S_{\text{div}} = \int \dd^2 \sigma  \, c^2 \gamma^{00} \mathcal{F}^A \mathcal{F}^B \eta_{AB} =\int \dd^2 \sigma \bigg(  \lambda^A \mathcal{F}^B \eta_{AB} - \frac{1}{4 c^2 \gamma^{00}} \lambda^A \lambda^B \eta_{AB}\bigg) \, .
\end{equation}
Notice that this equivalence only holds on-shell, i.e. the two forms of writing (\ref{rewriting}) are the same only after eliminating $\lambda_A$ via their equations of motion. The advantage of the rewriting (\ref{rewriting}) is to trade a divergent term for a finite one, at the price of introducing extra non-dynamical degrees of freedom $\lambda_A$. 
After having performed the Hubbard-Stratonovich transformation \eqref{rewriting}, we are finally allowed to take $c \rightarrow \infty$, obtaining the NR string action:
\begin{equation}
\label{NR_action}
S^{NR}_{\text{Poly}} = - \frac{T}{2} \int \dd^2 \sigma \, \bigg( \gamma^{\alpha\beta}\partial_{\alpha} X^{\mu} \partial_{\beta} X^{\nu} H_{\mu\nu} +  \varepsilon^{\alpha\beta} \partial_\alpha X^\mu \partial_\beta X^\nu b^{\text{NR}}_{\mu\nu}  + \lambda_A \mathcal{F}^A \bigg) \, . 
\end{equation}

The second option for treating the divergent term was proposed in \cite{Bergshoeff:2018yvt}, and it consists in introducing the zweibein for the worldsheet metric $\mathsf{h}_{\alpha\beta} = \theta_{\alpha}{}^{i}\theta_{\beta}{}^{j} \eta_{ij}$, where $i,j \in \{0,1\}$, such that the NR string action (\ref{NR_action}) is equivalent to
\begin{align}
\label{NR_action_Eric}
\notag
S^{NR}_{\text{Poly}} = - \frac{T}{2} \int \dd^2 \sigma \, \bigg( \gamma^{\alpha\beta}\partial_{\alpha} X^{\mu} \partial_{\beta} X^{\nu} H_{\mu\nu} +  \varepsilon^{\alpha\beta} \partial_\alpha X^\mu \partial_\beta X^\nu b^{\text{NR}}_{\mu\nu}& \\
+ \varepsilon^{\alpha\beta} (\lambda_+ \theta_{\alpha}{}^+ \tau_{\mu}{}^+ + \lambda_- \theta_{\alpha}{}^- \tau_{\mu}{}^- )\partial_{\beta}X^{\mu}  \bigg)& \, ,
\end{align}
where we introduced the light-cone basis $\theta_{\alpha}{}^{\pm} \equiv \theta_{\alpha}^0 \pm \theta_{\alpha}^1$, $\tau_{\mu}{}^{\pm} \equiv \tau_{\mu}{}^0 \pm \tau_{\mu}{}^1$ and $\lambda_{\pm} \equiv \frac{1}{2} \left( \lambda_0 \pm \lambda_1 \right)$. As it was remarked in \cite{Fontanella:2021btt}, such equivalence holds provided the following positivity condition is satisfied\footnote{One can always choose the worldsheet zweibein such that \eqref{positivity_Zweibein} is fulfilled. If instead one had chosen $\theta_1{}^+ \theta_0{}^- - \theta_0{}^+\theta_1{}^-$ to be non-positive, then \eqref{NR_action_Eric} would still be the same, but with $\theta_{\alpha}{}^+$ and $\theta_{\alpha}{}^-$ swapped. }
\begin{equation}
\label{positivity_Zweibein}
\theta_1{}^+ \theta_0{}^- - \theta_0{}^+\theta_1{}^- \geq 0 \, .
\end{equation}  
The advantage of writing the action in the form (\ref{NR_action_Eric}) is to make the following $\mathbb{Z}_2$ symmetry manifest,
\begin{eqnarray}\label{Z2_symmetry}
    \theta_{\alpha}{}^\pm \rightarrow \theta_{\alpha}{}^\mp \, , \qquad \tau_{\beta}{}^\pm \rightarrow \tau_{\beta}{}^\mp  \, , \qquad \lambda_\pm \rightarrow \lambda_\mp \, ,
\end{eqnarray}
where $\tau_{\beta}{}^A$ is the pullback of $\tau_{\mu}{}^A$, i.e. $\tau_{\beta}{}^A \equiv \tau_{\mu}{}^A \partial_{\beta} X^{\mu}$. This symmetry is important, for example, when introducing polar coordinates. For Lorentzian strings in AdS$_5\times$S$^5$ in polar coordinates, one can solve for the radial coordinate in the full $\mathbb{R}$ domain, and only at the very end, use the $\mathbb{Z}_2$ symmetry to restrict to the positive domain. In this way, one can bypass the problem of having to impose a domain restriction when solving the PDEs. This desirable feature is guaranteed for the NR string action, thanks to the $\mathbb{Z}_2$ symmetry \eqref{Z2_symmetry}. For an in-depth discussion, see \cite{Fontanella:2023men}.

\subsubsection{Nambu-Goto formalism}

So far, we have derived the NR string action in the Polyakov formalism. However, the NR limit could also have been taken for the string action in the Nambu–Goto formalism. The starting point is the Lorentzian Nambu-Goto action coupled to a Kalb-Ramond B-field,
\begin{eqnarray}
    S_{\footnotesize\text{NG}} = - T \int \dd^2 \sigma \,  \left( \sqrt{- \det \left(\partial_{\alpha} X^{\mu}\partial_{\beta} X^{\nu}  g_{\mu\nu} \right)} 
    + \frac{1}{2} \varepsilon^{\alpha\beta} \partial_\alpha X^\mu \partial_\beta X^\nu b_{\mu \nu} \right)\, . 
\end{eqnarray}
We perform the non-relativistic $c$-expansion of the target space vielbein as in \eqref{E_scaling}, and we arrive at  
\begin{eqnarray}\label{NG_div}
    S_{\footnotesize\text{NG}} = - T \int \dd^2 \sigma \, \left( 
    c^2 \sqrt{-\tau} 
    + \frac{1}{2} \sqrt{-\tau} \tau^{\alpha\beta} H_{\alpha\beta}  
    + \frac{1}{2} \varepsilon^{\alpha\beta} b_{\alpha\beta} \right)\, , 
\end{eqnarray}
where the pullback of $H$ and $b$ are defined as the one of $g$. Now, by using Binet's formula, and assuming that $\tau_{\mu\nu}$ is a $2\times 2$ matrix (e.g. in the $X^0$ and $X^1$ directions), we get
\begin{eqnarray}
    \det \tau_{\alpha\beta} = \det \tau_{\mu\nu} \left( \det \partial_{\alpha} X^{\mu} \right)^2 \, , 
\end{eqnarray}
where $\tau_{\alpha\beta}$, $\tau_{\mu\nu}$ and $\partial_{\alpha} X^{\mu}$ are all $2\times 2$ matrices. Then, 
\begin{eqnarray}
   \det \partial_{\alpha} X^{\mu} = \partial_{\tau} X^0 \partial_{\sigma} X^1 -  \partial_{\sigma} X^0 \partial_{\tau} X^1 = \varepsilon^{\alpha\beta} \partial_{\alpha} X^0 \partial_{\beta} X^1 \, . 
\end{eqnarray}
By plugging this result inside \eqref{NG_div}, we obtain
\begin{eqnarray}\label{NG_div_2}
    S_{\footnotesize\text{NG}} = - T \int \dd^2 \sigma \, \left( 
    c^2 \sqrt{-\det \tau_{\mu\nu}} \varepsilon^{\alpha\beta} \partial_{\alpha} X^0 \partial_{\beta} X^1 
    + \frac{1}{2} \sqrt{-\tau} \tau^{\alpha\beta} H_{\alpha\beta}  
    + \frac{1}{2} \varepsilon^{\alpha\beta} b_{\alpha\beta} \right)\, .
\end{eqnarray}
From here, it is clear that the divergent term neatly disappears if we fine tune the B-field with a critical component only in the $X^0$ and $X^1$ directions, i.e.  
\begin{eqnarray}\label{critical_B_NG}
    b_{\mu\nu} = c^2 b^{\text{crit.}}_{\mu\nu} + b^{\text{NR}}_{\mu\nu} \, , \qquad
    b^{\text{crit.}}_{01} \equiv - \sqrt{-\det \tau_{\mu\nu}} \, .
\end{eqnarray}
Then, the final form of the NR Nambu-Goto action is
\begin{eqnarray}\label{NR_NG} S^{NR}_{\footnotesize\text{NG}} 
    = - \frac{T}{2} \int \dd^2 \sigma \, \left(  \sqrt{-\tau} \tau^{\alpha\beta} H_{\alpha\beta}  
    +  \varepsilon^{\alpha\beta} b^{\text{NR}}_{\alpha\beta} \right)\, .
\end{eqnarray}
Some remarks are in order. The critical B-field \eqref{critical_B_NG} is consistent with the one we found in the Polakov formulation \eqref{critical_B}, as it can be shown by using the identity $\sqrt{- \det \tau_{\mu\nu}} = - \frac12 \varepsilon^{\mu\nu} \varepsilon_{AB} \tau_{\mu}{}^A \tau_{\nu}{}^B$, when $\mu, \nu$ are restricted to the longitudinal two-dimensional subspace. As we have seen, in the Nambu-Goto formulation there is no need to perform a Hubbard-Stratonovich transformation, since the divergent term is exactly cancelled. As a consequence, there are no Lagrange multipliers appearing in the action. Because of that, the $\mathbb{Z}_2$ symmetry \eqref{Z2_symmetry} is trivially realised. Finally, it is important to remark that the cancellation of the divergence with the critical B-field is a mechanism that is only possible because $\tau_{\mu\nu}$ is a $2\times 2$ matrix.

Another way to arrive to the NR Nambu-Goto action is by starting with the NR Polyakov action \eqref{NR_action_Eric}, computing the equation of motion associated with the Lagrange multipliers, and solving them in terms of the worldsheet zweibein as follows: 
\begin{eqnarray}
\varepsilon^{\alpha\beta} \theta_{\alpha}{}^{\pm} \tau_{\beta}{}^{\pm} = 0 \, ,  \qquad\Longrightarrow \qquad
\theta_{\alpha}{}^{\pm} = f_{\pm}(\tau, \sigma)\, \tau_{\alpha}{}^{\pm} \, , \label{PolytoNG}
\end{eqnarray}
where $f_{\pm}(\tau, \sigma)$ are generic functions of the worldsheet coordinates. An intuitive way to arrive at the solution of these equations of motion is by interpreting them as the vanishing of a vector product in two dimensions, which implies that the involved vectors are parallel. By plugging this solution inside the NR Polyakov action \eqref{NR_action_Eric}, one obtains, 
\begin{equation}
S^{NR}_{\footnotesize\text{NG}} = - \frac{T}{2} \int \dd^2 \sigma \,  \left( \sqrt{-\tau}\tau^{\alpha\beta}\partial_{\alpha} X^{\mu} \partial_{\beta} X^{\nu} H_{\mu\nu}  +  \varepsilon^{\alpha\beta} \partial_\alpha X^\mu \partial_\beta X^\nu b^{\text{NR}}_{\mu\nu} \right)\, , 
\end{equation}
which is precisely the NR Nambu-Goto action that we found in \eqref{NR_NG}.

\subsection{The null reduction method}
\label{sec:null_red}

The NR string action can also be derived by reducing a relativistic string action along a light-like isometry at fixed momentum. This is called the null reduction approach, developed for the particle action in \cite{PhysRevD.31.1841,Duval:1990hj} and for the string action in \cite{Harmark:2017rpg,Harmark:2018cdl,Harmark:2019upf}.

As in the previous section, we start with the Polyakov action coupled to a Kalb-Ramond B-field. We assume that the target space has a light-like isometry, and we adapt our coordinate system to it. We choose the coordinates as $x^{\mu} = (u, x^i)$, such that $u$ is not compact and $\partial/\partial u$ is the light-like isometry. Then, the metric and B-field read as    
\begin{subequations}
\begin{align}
    g_{\mu \nu} \dd x^\mu \dd x^\nu &=2(\tau_i \dd x^i) (\dd u - m_i \dd x^i) + h_{ij} \dd x^i \dd x^j \, , \\
b_{\mu \nu} \dd x^\mu \wedge \dd x^\nu &= b_{i j} \dd x^i \wedge \dd x^j + 2 b_{i} \dd x^i \wedge \dd u \, .
\end{align}
\end{subequations}
The main step now is to force conservation of the momentum in the null direction via a Lagrange multiplier. The momentum is given by
\begin{eqnarray}
   P^{\alpha}_u = \frac{\partial \mathcal{L}}{\partial (\partial_{\alpha} X^u)} = - T \gamma^{\alpha\beta} \tau_{\beta} - T \varepsilon^{\alpha\beta} b_{\beta} \, , 
\end{eqnarray}
where $X^u$ is the string embedding coordinate of $u$. 
This leads us to the following action
\begin{align}
\notag
    S_{\text{null}}=-\frac{T}{2} \int{\dd^2 \sigma}\Bigg[ &\gamma^{\alpha \beta} \partial_\alpha X^i \partial_\beta X^j h_{ij} + \varepsilon^{\alpha \beta} \partial_\alpha X^i \partial_\beta X^j b_{i j} \\
    &+2\left( \gamma^{\alpha \beta}\tau_i \partial_\alpha X^i + \varepsilon^{\alpha \beta}b_i \partial_\alpha X^i +\varepsilon^{\alpha \beta} \partial_\alpha \eta\right) A_\beta  \Bigg] \, .
\end{align}
It is immediate to check that the equation of motion for $\eta$, i.e. $\varepsilon^{\alpha\beta} \partial_{\alpha} A_{\beta} = 0$,  can be solved by setting $A_\alpha=\partial_\alpha U$, with $U$ a scalar field. By identifying $U = X^u$, we get back to the Polyakov action. On the other hand, the equation of motion associated with $A_\alpha$ becomes $P^{\alpha}_u=T \varepsilon^{\alpha \beta} \partial_\beta \eta$, and the conservation of the momentum $P_u$ is guaranteed off-shell.

As the field $\eta = X^v$ can be regarded as a formal T-dual of $X^u$, a fixed momentum condition on $P_u$ on the coordinates $(u,x^i)$ can be exchanged for a fixed winding condition along $v$ on the coordinates $(v,x^i)$, implying that $\eta$ has to be $2\pi$-periodic in $\sigma$ and $v$ has to be a compact direction.

Despite the lack of an explicit non-relativistic limit in this approach, the final action is equivalent to the one obtained via a limit in the previous section. For the proof, see \cite{Harmark:2018cdl,Harmark:2019upf}.

\subsection{The expansion approach} \label{sec.expansionapproach}

Instead of a limit $c\to \infty$, the non-relativistic action can also be obtained by expanding in $1/c^2$ the relativistic action. This has been proposed for closed strings in \cite{Hartong:2021ekg, Hartong:2022dsx}, and also for open strings ending on D$p$-branes in \cite{Hartong:2024ydv}. The idea is to expand fields (i.e. the embedding string coordinates and worldsheet metric) in $1/c^2$, where each order of the expansion is treated as a new independent field. This creates more degrees of freedom than necessary, but the idea is that one can remove the excess by imposing constraints. The question is at which order one should truncate the equations of motion. This is answered by the method of Lie algebra expansion \cite{deAzcarraga:2002xi,deAzcarraga:2007et, Hatsuda:2001pp} applied to the local Lorentz symmetry of the initial relativistic string action. The Lie algebra expansion method comes equipped with the so-called \emph{truncation rules}, which allow stopping the infinite series at a finite value, keeping consistency with the Maurer-Cartan equations. An equivalent procedure has been applied to string theory in homogeneous spaces in \cite{Fontanella:2020eje}, where the Lie algebra expansion acts directly on the coset algebras. We will further discuss this in section \ref{sec:NR_action_as_coset}. 

Here, we review the expansion method for the closed string action. The starting point is the Polyakov action \eqref{rel_Pol_action} where the B-field is now set to zero, $b_{\mu\nu} = 0$. 
We expand the embedding fields $X^{\mu}$ as,
\begin{equation}\label{field_exp}
    X^{\mu} = x_{(0)}^{\mu} + c^{-2}x_{(2)}^{\mu} + c^{-4}x_{(4)}^{\mu} + \mathcal{O}(c^{-6})\, ,
\end{equation}
and the worldsheet metric $\mathsf{h}_{\alpha\beta}$ as,
\begin{equation}
    \mathsf{h}_{\alpha\beta} =  \mathsf{h}_{(0)\alpha\beta} +c^{-2} \mathsf{h}_{(2)\alpha\beta} + c^{-4}\mathsf{h}_{(4)\alpha\beta} + \mathcal{O}(c^{-6}) \, .
\end{equation}
Compatible with the field expansion \eqref{field_exp}, the metric $g_{\mu \nu}$ needs to expand as,
\begin{equation}
    g_{\mu \nu} = c^2\tau_{\mu \nu} + H_{\mu \nu} + c^{-2}\phi_{\mu \nu} + \mathcal{O}(c^{-4})\, .
\end{equation}
In turns, these imply that the Polyakov action expands as 
\begin{subequations}
\begin{align}
    \mathcal{L} &= c^2 \mathcal{L}_{(0)} + \mathcal{L}_{(2)} + \mathcal{O}(c^{-2})\, , \\
    \mathcal{L}_{(0)} &= -\frac{T}{2}\sqrt{-\mathsf{h}_{(0)}} \,\mathsf{h}_{(0)}^{\alpha\beta} \tau_{\alpha\beta}(x_{(0)})\, , \\
    \mathcal{L}_{(2)} &= -\frac{T}{2}\sqrt{-\mathsf{h}_{(0)}} \,\mathsf{h}_{(0)}^{\alpha\beta} \mathsf{H}_{\alpha\beta}(x_{(0)},x_{(2)}) + \frac{T}{2}\sqrt{-\mathsf{h}_{(0)}} \, \tau_{\alpha\beta}(x_{(0)}) \mathsf{h}_{(2)\gamma\delta}\left[\mathsf{h}_{(0)}^{\alpha\gamma}\mathsf{h}_{(0)}^{\delta\beta} - \frac{1}{2}\mathsf{h}_{(0)}^{\alpha\beta}\mathsf{h}_{(0)}^{\gamma\delta} \right] \notag \\
    &=\hat{\mathcal{L}}_{(2)} (x_{(0)})+ x^\mu_{(2)} \frac{\delta \mathcal{L}_{(0)}}{\delta x^\mu_{(0)}} \, ,
\end{align}
\end{subequations}
where we have introduced the following pullbacks of the components of the metric
\begin{subequations}
\begin{align}
    \tau_{\alpha\beta}(x_{(0)}) &= \tau_{\mu\nu} \partial_\alpha x_{(0)}^\mu \partial_\beta x_{(0)}^\nu\,,\\
\mathsf{H}_{\alpha\beta}(x_{(0)},x_{(2)}) &= H_{\mu\nu} \partial_\alpha x_{(0)}^\mu \partial_\beta x_{(0)}^\nu + 2\tau_{\mu\nu}\partial_{(\alpha} x_{(0)}^\mu \partial_{\beta)}x_{(2)}^\nu + \partial_\alpha x_{(0)}^\mu\partial_\beta x_{(0)}^\nu x_{(2)}^{\rho} \partial_{\rho} \tau_{\mu\nu}\, , 
\end{align}
\end{subequations}
and $\hat{\mathcal{L}}_{(2)} (x_{(0)})$ is the part of $\mathcal{L}_{(2)}$ that is independent of $x_{(2)}$.

The equations of motion for $\mathcal{L}_{(0)}$ with respect to $\mathsf{h}_{(0)}$ give us exactly the leading order of the Virasoro constraints. Similarly, the equations of motion for $\mathcal{L}_{(2)}$ with respect to $\mathsf{h}_{(0)}$ give us the next-to-leading order of the Virasoro constraints, while the equation of motion with respect to $\mathsf{h}_{(2)}$ are redundant because they give us again the leading order of the Virasoro constraints. A similar situation happens to subsequent terms in the expansion.

This situation is mirrored for the embedding fields. The variation of the action is given by
\begin{equation}
    \delta S=\sum_{m=0}^\infty c^{2-2m} \delta S_{(2m)}= \sum_{i,j=0}^\infty c^{2-2(i+j)} \mathcal{E}_\mu^{(2i)} \delta x^\mu_{(2j)} \,,
\end{equation}
where $\mathcal{E}_\mu^{(m)}$ is the expansion of the relativistic equations of motion at $m$-th order. Similar to the case of the equations of motion for $\mathsf{h}_{(2n)}$, the equations of motion for $x_{(2n)}$ fulfil
\begin{equation}
    \frac{\delta S_{(m)}}{\delta x^\mu_{(n)}}= \frac{\delta S_{(m-n)}}{\delta x^\mu_{(0)}}=\mathcal{E}_\mu^{(m-n)} \,,
\end{equation}
as a consequence, only the variations with respect to $x_{(0)}$ carry physical meaning. 

Notice that $\mathcal{E}_\mu^{(2m)}$ only depend on fields up to $x^\mu_{(2m)}$, so we can solve them in an echelon form: the equations of motion at leading order only contain the fields $x^\mu_{(0)}$. After finding a solution $x^{\text{sol}}_{(0)}$, one moves the next order $S_{(2)}$. Here, the equations of motion are 
\begin{equation}
    \delta S_{(2)}= \mathcal{E}_\mu^{(0)} \delta x^\mu_{(2)}+\mathcal{E}_\mu^{(2)} \delta x^\mu_{(0)}= \left( \frac{\delta \mathcal{L}_{(0)}}{\delta x^\mu_{(0)}} \right) \delta x^\mu_{(2)}+\left( \frac{\delta \mathcal{L}_{(2)}}{\delta x^\mu_{(0)}} + x^\nu_{(2)}\frac{\delta^2 \mathcal{L}_{(0)}}{\delta x^\mu_{(0)} \delta x^\nu_{(0)}} \right) \delta x^\mu_{(0)} \,.
\end{equation}
where $\mathcal{E}_\mu^{(0)}$ are the equations of motion we solved at leading order, so they automatically vanish, while $\mathcal{E}_\mu^{(2)}$ give us algebraic equations for $x^\mu_{(2)}$ where $x^{\text{sol}}_{(0)}$ plays the role of a source. Higher orders are solved similarly.

A different approach to solve these equations was proposed in \cite{Hartong:2021ekg}. It was suggested that imposing the condition
\begin{equation}
    \dd \tau^A = \alpha^A{_B}\wedge\tau^B \, , \qquad\qquad
    \alpha^A{}_A = 0 \, , \label{tracelessFrob}
\end{equation}
ensures that the differential equations $\mathcal{E}_\mu^{(0)}=0$ hold for any $x^\mu_{(0)}$. As a consequence, $\mathcal{L}_{(2)}$ is independent of $x^\mu_{(2)}$.\footnote{Although the authors of \cite{Hartong:2021ekg} believe that this is automatic from substituting $\frac{\delta \mathcal{L}_{(0)}}{\delta x^\mu_{(0)}}=0$ at the level of the Lagrangian, one has to be careful as the substitution must be done at the level of the equations of motion, and not at the level of the Lagrangian. Vanishing of the derivative does not imply vanishing of the curvature. As an example, consider the Klein-Gordon action where, regardless of the solution we consider, the mass term will always contribute to the second functional derivative of the action. Thus, we believe that the condition \eqref{tracelessFrob} may have to be supplemented with \begin{displaymath}
    \frac{\delta^2 \mathcal{L}_{(0)}}{\delta x^\mu_{(0)} \delta x^\nu_{(0)}}=0 \,.
\end{displaymath}} After rewriting $\hat{\mathcal{L}}_{(2)}$ by trading $\mathsf{h}_{(2)}$ for Lagrange multipliers, it takes the form of the non-relativistic action \eqref{NR_action_Eric}.

A justification for the condition \eqref{tracelessFrob} comes from requiring the vanishing of the beta-function \cite{Hartong:2022dsx}. The leading order of the expanded Einstein equations implies the Frobenius integrability condition
\begin{equation}
    \dd \tau^A = \alpha^A{_B}\wedge\tau^B \, .
\end{equation}
Notice that \eqref{tracelessFrob} is nothing more than the Frobenius condition with traceless $\alpha^A{_B}$.

\section{Structures of the non-relativistic string action} \label{sec:MathStructures}

\subsection{Local symmetries} \label{sec:localsym}

The NR string action has a relativistic worldsheet, and therefore the non-relativistic limit does not affect the invariance under worldsheet diffeomorphisms and Weyl transformations of the worldsheet metric, which are realised in the usual sense \cite{Bergshoeff:2018yvt}.  
The difference with respect to the relativistic string action concerns the local symmetries of the target space. The local target space symmetries of the NR string action have been identified as the \emph{String Newton-Cartan algebra} \cite{Bergshoeff:2019pij}, whose generators are 
\begin{subequations} \label{eq:generatorslistbulk}
\begin{align}
    	\text{longitudinal translations} \qquad & H_A \\[2pt]
    	\text{transverse translations} \qquad & P_{a} \\[2pt]
	\text{longitudinal Lorentz rotation} \qquad & M \\[2pt]
    	\text{string-Galilean boosts} \qquad & G_{Aa} \\[2pt]
    	\text{transverse rotations} \qquad & J_{ab} \\[2pt]
	\text{non-central extensions} \qquad & Z_A \,, Z_{AB} \ \mathrm{with} \ Z^A{}_A = 0 \,.
\end{align}
\end{subequations}
with commutation relations,
\begin{subequations}
	\begin{align}
	[H_A,M] & = \varepsilon_A{}^B H_B \, , &
	[G_{Aa}, M]  & = \varepsilon_A{}^B G_{Ba} \, , \\
	[H_A, G_{Ba}]  & = \eta^{}_{AB} P_{a} \, , &
	[G_{Aa}, J_{bc}] & = 2\delta_{a[b} G_{|A|c]} \, , \\
    [P_{a}, J_{bc}] & = 2\delta_{a[b} P_{c]} \, , &
	 [J_{ab}, J_{cd}] & = 4\delta_{[b[c} J_{a]d]} \, , \\
    [P_a, G_{Ab}] & = - \delta_{ab} Z_A \, , &
    [G_{Aa}, G_{Bb}] & = \delta_{ab} Z_{[AB]} \, ,\\
    [Z_A, M] & = \varepsilon_A{}^B Z_B \, , &
    [Z_{AB}, M] & = \varepsilon_A{}^C Z_{CB} + \varepsilon_B{}^C Z_{AC} \, , \\
    [H_A, Z_{BC}] & = 2 \eta_{AC} Z_B - \eta_{BC} Z_A \, , & & 
	\end{align}
\end{subequations}
where $Z_{[AB]} \equiv \frac12 (Z_{AB} - Z_{BA})$. 
The String Newton-Cartan geometry transforms under these symmetries as follows,
\begin{subequations} \label{eq:local_trasf_SNC}
\begin{align}
    \delta \tau_\mu{}^A  &= -\Lambda \, \varepsilon^A{}_B \, \tau_\mu{}^B , \\
	\delta e_\mu{}^{a} &= \Lambda_A{}^{a} \tau_\mu{}^A - \Lambda^{a}{}_{b} e_\mu{}^{b}\, , \\
    \delta m_\mu{}^A  &= \partial_\mu \sigma^A + \varepsilon^A{}_B \sigma^B \Omega_\mu - \Lambda \, \varepsilon^A{}_B m_\mu{}^B - \Lambda^A{}_a e_{\mu}{}^a + \tau_\mu{}^B \sigma^A{}_B \, ,
    \end{align}
\end{subequations}
where $\Omega_{\mu}$ is the spin connection associated with longitudinal Lorentz rotation $M$.
The parameters $\Lambda$, $\Lambda^{ab}$ and $\Lambda^{Aa}$ are associated with the longitudinal Lorentz rotation $M$, transverse rotations $J_{ab}$ and string-Galilean boosts $G_{Aa}$. The traceless two-tensor $\sigma^{AB}$ (i.e. $\sigma^A{}_A=0$) and $\sigma^A$ parameterise instead the transformations associated with the non-central extensions $Z_{AB}$ and $Z_A$. 
From these local symmetries, it follows that $\delta \tau_{\mu\nu}= 0$, but $\delta h_{\mu\nu} = \Lambda_{Aa} (\tau_{\mu}{}^A e_{\nu}{}^{a} + \tau_{\nu}{}^A e_{\mu}{}^{a} )$. Therefore, $\tau_{\mu\nu}$ is a metric, but $h_{\mu\nu}$ is not.   

The transformations \eqref{eq:local_trasf_SNC} are local symmetries of the NR string action, provided the Lagrange multipliers transform as follows,
\begin{subequations}\label{l_lb_gauge}
\begin{eqnarray}
\delta\lp  =& - \Lambda \lp + 2 \theta^{\alpha}{}_+  \left(\partial_\alpha \sigma^- - \sigma^- \Omega_{\alpha} - \frac{1}{2}\tau_{\alpha}{}^+ \sigma_{++} \right)\, , \\
\delta \lm =& \Lambda \lm - 2 \theta^{\alpha}{}_-  \left(\partial_\alpha \sigma^+ + \sigma^+ \Omega_{\alpha} - \frac{1}{2} \tau_{\alpha}{}^- \sigma_{--} \right) \, .
\end{eqnarray}
\end{subequations}
Moreover, the transformation associated with $Z_A$ is a symmetry only if the torsionless condition \eqref{torsionless_constraint} holds.
This condition is fulfilled in the case of NR strings in SNC AdS$_5\times$S$^5$ that will be of interest in this review. However, it was noticed from the analysis of the supergravity \cite{Bergshoeff:2021bmc}, the limit approach \cite{Bergshoeff:2019pij}, the null-reduction approach \cite{Harmark:2018cdl}, and from beta function computations \cite{Gomis:2019zyu, Gallegos:2019icg, Yan:2021lbe, Yan:2019xsf}, that this condition is too restrictive. It turns out that the torsionless condition can be removed by fixing a St\"uckelberg-type symmetry of the NR action.

The NR action has a St\"uckelberg-type symmetry \cite{Bergshoeff:2019pij, Harmark:2019upf, Bidussi:2021ujm}. This is due to the fact that there is an ambiguity in defining a $c$-expansion of the relativistic vielbein. In the string action \eqref{NR_NG}, this ambiguity translates into the invariance under the following transformation,
\begin{eqnarray}
    \delta H_{\mu\nu} = 2 C_{(\mu}{}^A \tau_{\nu)}{}^B \eta_{AB} \, , \qquad
    \delta b^{\text{NR}}_{\mu\nu} = - 2 C_{[\mu}{}^A \tau_{\nu]}{}^B \varepsilon_{AB} \, , 
\end{eqnarray}
where $C_{\mu}{}^A$ are arbitrary parameters. By fixing these parameters as $C_{\mu}{}^A = - m_{\mu}{}^A$,
we can move the term involving $m_{\mu}{}^A$ from $H_{\mu\nu}$ to the B-field. Specifically, after this transformation the Nambu-Goto NR action takes the form 
\begin{equation}\label{NR_NG_fix_Stuckelberg}
S^{NR}_{\footnotesize\text{NG}} = - \frac{T}{2} \int \dd^2 \sigma \,  \left( \sqrt{-\tau}\tau^{\alpha\beta}\partial_{\alpha} X^{\mu} \partial_{\beta} X^{\nu} h_{\mu\nu}  +  \varepsilon^{\alpha\beta} \partial_\alpha X^\mu \partial_\beta X^\nu \mathfrak{B}_{\mu\nu} \right)\, , 
\end{equation}
where $h_{\mu\nu}$ is defined as in \eqref{gscaling} and $\mathfrak{B}_{\mu\nu}$ is the following combination of the NR B-field and the contribution from the longitudinal SNC metric data,
\begin{equation}
    \mathfrak{B}_{\mu\nu} \equiv b^{\text{NR}}_{\mu\nu} + 2 m_{[\mu}{}^A \tau_{\nu]}{}^B \varepsilon_{AB} \, . 
\end{equation}
The advantage of fixing the St\"uckelberg symmetry is that the non-central extension generators $Z_A$ and $Z_{AB}$ in \eqref{eq:generatorslistbulk} are no longer part of the local symmetry algebra of the NR action \eqref{NR_NG_fix_Stuckelberg}, and therefore there is no torsionless constraint to be imposed. For a detailed discussion, we refer the reader to \cite{Bergshoeff:2019pij, Bidussi:2021ujm}.

We remark that the same procedure could have been applied to the NR string action in the Polyakov formalism. In that case, one has to demand that also the Lagrange multipliers transform with the St\"uckelberg symmetry, see \cite{Bergshoeff:2019pij} for the detail. The final Polyakov NR action reads 
\begin{eqnarray}\label{NR_Polyakov_fix_Stuckelberg}
\notag
S^{NR}_{\text{Poly}} &=& - \frac{T}{2} \int \dd^2 \sigma \, \bigg( \gamma^{\alpha\beta}\partial_{\alpha} X^{\mu} \partial_{\beta} X^{\nu} h_{\mu\nu} +  \varepsilon^{\alpha\beta} \partial_\alpha X^\mu \partial_\beta X^\nu \mathfrak{B}_{\mu\nu} \\
&& \hspace{3.5cm}+ \varepsilon^{\alpha\beta} (\lambda_+ \theta_{\alpha}{}^+ \tau_{\mu}{}^+ + \lambda_- \theta_{\alpha}{}^- \tau_{\mu}{}^- )\partial_{\beta}X^{\mu}  \bigg) \, .
\end{eqnarray}
where $h_{\mu\nu}$ and $\mathfrak{B}_{\mu\nu}$ are defined as before.

In addition, the String Newton-Cartan algebra spanned by \eqref{eq:generatorslistbulk} admits an extension by a dilatation generator $D$, which acts as a St\"uckelberg symmetry on the NR string action. The SNC metric data and Lagrange multipliers transform under its action as, 
\begin{eqnarray}
    \delta_D \tau_{\mu}{}^{\pm} = \omega_{\pm} \tau_{\mu}{}^{\pm} \, , \qquad
    \delta_D m_{\mu}{}^{\pm} = \frac{1}{\omega_{\pm}} m_{\mu}{}^{\pm} \, , \qquad
    \delta_D \lambda_{\pm} = \frac{1}{\omega_{\pm}} \lambda_{\pm} \, .
\end{eqnarray}

\subsection{Global symmetries}
\label{sec:global_symmetries}

The global symmetries of the NR string action are transformations of the string embedding fields $\delta X^{\mu} = \xi^{\mu}$ that leave the action invariant \cite{Batlle:2016iel, Bidussi:2023rfs, Fontanella:2024rvn}. It is immediate to check that demanding $\xi^{\mu}$ to be a global symmetry is equivalent to demanding $\xi^{\mu}$ to be a Killing vector that preserves the SNC metric tensors. To find such Killing vectors, one has to demand that their Lie derivative acting on the background geometry is zero up to the local symmetries. For the Lorentzian string action, this boils down to demanding the Lie derivative of the metric $g_{\mu\nu}$ to be zero up to Lorentz transformations. However, since the metric is Lorentz-invariant, this effectively amounts to demand that the Lie derivative of $g_{\mu\nu}$ vanishes, namely:
\begin{eqnarray}
    (\mathsterling_{\xi} \, g)_{\mu\nu} = \delta g_{\mu\nu} = \left( \Lambda_{\hat{A}\hat{B}} + \Lambda_{\hat{B}\hat{A}}\right) E_{\mu}{}^{\hat{A}} E_{\mu}{}^{\hat{B}} = 0 \, .
\end{eqnarray}
Exactly the same logic can be applied to the NR string action. The difference is that the SNC metric tensors transform non-trivially under local symmetries, and therefore their Lie derivative will not be set to zero. For example, the transverse metric tensor $h_{\mu\nu}$ is not invariant under local string Galilei boosts. By taking into account all local symmetries that we discussed before, the global symmetries of the NR string action \eqref{NR_NG_fix_Stuckelberg} are given by solving the following system of equations:
\begin{subequations}\label{SNC_Killing}
\begin{align}
\label{Lie_tau}
(\mathsterling_{\xi} \,\tau )_{\mu\nu}&= \delta \tau_{\mu\nu} = 2 \omega \tau_{\mu\nu} \, , \\
\label{Lie_h}
(\mathsterling_{\xi}\, h )_{\mu\nu} &= \delta h_{\mu\nu} =\Lambda_{Aa} (\tau_{\mu}{}^A e_{\nu}{}^{a} + \tau_{\nu}{}^A e_{\mu}{}^{a} ) \, , \\
\label{Lie_m}
(\mathsterling_{\xi}\, \mathfrak{B} )_{\mu\nu} &= \delta \mathfrak{B}_{\mu\nu} = \varepsilon_{AB} \Lambda^B{}_{a} (\tau_{\mu}{}^A e_{\nu}{}^{a} - \tau_{\nu}{}^A e_{\mu}{}^{a} ) + \partial_{\mu} \Sigma_{\nu} -  \partial_{\nu} \Sigma_{\mu} \, ,
\end{align}
\end{subequations}
where $\omega$ parameterises the dilatation symmetry, $\Lambda_{Aa}$ the string Galilei boost symmetry and $\Sigma_{\mu}$ the gauge symmetry of the 2-form $\mathfrak{B}_{\mu\nu}$.
Solving this set of equations is enough to determine the global symmetries of the NR string action, viewed as a 2d bosonic sigma model. The situation is more involved if instead we are considering a superstring action. In that case, the global symmetries are given by the Killing vectors that preserve the full NR supergravity solution, up to local symmetries. This will be discussed in detail in section \ref{sec:matching_symmetries} for NR type IIB superstring with target space SNC AdS$_5\times$S$^5$.    

Notice that due to the metric being degenerate and the additional symmetries that introduce non-vanishing terms in the right-hand-side of the (conformal) Killing equations, we cannot recover all the non-relativistic Killing vectors by a limit of the relativistic Killing vectors. In particular, the fact the String Newton-Cartan geometry $(\tau, h)$ is degenerate is responsible for generating an infinite amount of Killing vectors. This is in striking contrast with a $d$-dimensional pseudo-Riemannian geometry, where the maximal amount of Killing vectors is $d(d+1)/2$.

\subsection{Weyl anomalies and the beta function} \label{subsec:weyl}

As in the case of the relativistic string, it is important for the NR string path integral to be free of Weyl anomalies. Because the worldsheet is relativistic, and the action is invariant under Weyl transformations in the usual sense, the computation of the beta function closely resemble the one for the relativistic string. This was performed in \cite{Gomis:2019zyu, Gallegos:2019icg}, where it has been shown that setting to zero the beta functional associated with the various SNC couplings imposes a set of NR Einstein equations, together with a critical condition on the dimension of the spacetime, which turns out to be $d=26$ for the bosonic string, likewise the relativistic case. 

The derivation of the beta function has been extensively reviewed in \cite{Oling:2022fft}. Here, we limit ourselves to list the NR Einstein equations for the background geometry implied by the vanishing of the beta function:
\begin{subequations}
\label{beta=0_eqn}
    \begin{align}
    \label{SNC_EOM1}
     P_{ab}=Q_{ab}=P_{A a}+\epsilon_{A}{}^{B}Q_{Ba}=\eta^{AB}P_{AB}-\epsilon^{AB}Q_{AB}&=0 \, ,
    \\
    \label{SNC_DilatonEOM}
     \nabla_{a}\nabla^{a}\Phi-\nabla^{a}\Phi\nabla_{a}\Phi+\frac{1}{4}R_{a}{}^{a}-\frac{1}{48}\mathcal{H}_{abc}\mathcal{H}^{abc}&=0 \, ,\\
     \epsilon_{C}{}^{\left(A \right.}\tau_{\left[\mu\right.}^{\left.B\right)}\partial_{\nu}\tau_{\left.\rho \right]}^{C}&=0 \, .
    \end{align}
\end{subequations}
where 
\begin{subequations}
\label{PQ_Def}
    \begin{align}
    \label{def_H_P_Q}
    \mathcal{H}_{\mu\nu\rho}&\equiv \partial_{\mu}B_{\nu\rho}+\partial_{\rho}B_{\mu\nu}+\partial_{\nu}B_{\rho\mu} \, ,
    \\
    \label{P_Def}
    P_{\mu\nu}&\equiv R_{\mu\nu}+2\nabla_{\mu}\nabla_ {\nu}\Phi-\frac{1}{4}\mathcal{H}_{\mu a b}\mathcal{H}_{\nu}{}^{ab} \, ,
    \\
    \label{Q_Def}
    Q_{\mu\nu}&\equiv -\frac{1}{2}\nabla^{a}\mathcal{H}_{a\mu\nu}+\mathcal{H}_{a\mu\nu}\nabla^{a}\Phi \, .
    \end{align}
\end{subequations}
In this notation, the flat indices stand for curved indices contracted with either the longitudinal or transverse vielbein, e.g. $\partial_A B_{aB} \equiv \tau^{\mu}{}_A e^{\nu}{}_a \tau^{\rho}{}_B \partial_{\mu} B_{\nu\rho}$. 
As noticed in \cite{Bergshoeff:2019pij}, these equations of motion can also be derived by taking the $c\to \infty$ of the beta function equations associated with the relativistic string action.

\subsection{Landscape}

Although NR string theory has been extensively explored at the level of the formalism, less effort  has been devoted regarding exploring its landscape, namely the set of all possible target space geometries that set the beta function to zero. So far, we know two examples of them:
\begin{itemize}
\item \underline{SNC flat spacetime} (Gomis, Ooguri \cite{Gomis:2000bd}) 
\begin{eqnarray}  
\tau_{\mu}{}^A = \delta_{\mu}^A \, , \qquad
m_{\mu}{}^A = 0 \, , \qquad
e_{\mu}{}^a = \delta_{\mu}^a \, . 
\end{eqnarray}
The $\tau_{\mu\nu}$ metric describes Mink$_2$ and the $h_{\mu\nu}$ tensor describes $\mathbb{R}^8$. 
\item \underline{SNC AdS$_5\times$S$^5$} (Gomis, Gomis, Kamimura - in short, GGK - \cite{Gomis:2005pg}) 
\begin{eqnarray}  \label{GGK_geom_Poinc}
\tau_{\mu}{}^A = \frac{R}{z} \delta_{\mu}^A \, , \qquad
m_{\mu}{}^A = 0 \, , \qquad
e_{\mu}{}^a = ( \frac{R}{z} \delta_{\mu}^i \, , \, R \,  \delta_{\mu}^{i'}) \, ,
\end{eqnarray}
where we split $e_{\mu}{}^a$ into the AdS part, $i\in \{1,2,3\}$, and the 5-sphere part, $i' \in\{ 1,..., 5\}$. $R$ is the radius, and $z$ is the holographic coordinate in the Poincar\'e patch (see section \ref{sec:coordinates} for an exhaustive discussion of coordinates).    

The $\tau_{\mu\nu}$ metric describes AdS$_2$ and the $h_{\mu\nu}$ tensor describes $w(z) \mathbb{R}^3\times \mathbb{R}^5$. The transverse AdS directions gain a warped factor $w(z)$.\footnote{The condition $m_{\mu}{}^A = 0$ is not a coordinate-invariant statement. This holds for Poincar\'e coordinates used in \eqref{GGK_geom_Poinc} but, as we will see in section \ref{sec:coordinates}, it can be non-zero in other set of coordinates. Moreover, it turns out that the warped factor disappears in some set of coordinates, e.g. in the GGK coordinates. The only coordinate independent statement is that $\tau_{\mu\nu}$ describes AdS$_2$. }   
\end{itemize}
It is important to remark that only $\tau_{\mu\nu}$ has meaning of an actual metric, whereas $h_{\mu\nu}$ is just regarded as a tensor. This is because $h_{\mu\nu}$ is not invariant under string Galilei boosts, due to the local SNC transformations \eqref{eq:local_trasf_SNC}. Instead, $\tau_{\mu\nu}$ does not transform under string Galilei boosts, and is invariant under longitudinal Lorentz transformations. This explains why $h_{\mu\nu}$ can depend on the longitudinal coordinates as well, whereas $\tau_{\mu\nu}$ strictly describes a 2d Lorentzian manifold. 

Other examples of target space geometries have recently been derived via a limit procedure \cite{Lambert:2024yjk, Blair:2024aqz}. In these geometries, the metric $\tau_{\mu\nu}$ cannot be written in the form of a strictly $2 \times 2$ block with vanishing components elsewhere, and therefore divergences cannot be cancelled via the critical B-field mechanism reviewed in section \ref{subsec:limit}. These geometries fall into a more exotic class of geometries, still to be understood from the string action point of view, and they will be introduced in section \ref{sec:Holo_other_NR_limits}.

\section{Non-relativistic limit via coordinate rescaling}
\label{sec:coordinates}

So far, we discussed the NR limit from a formal point of view. Now we want to take a more hands-on approach. Consider the relativistic string action on a concrete target space geometry, e.g. AdS$_5\times$S$^5$, equipped with a set of coordinates. As described in section \ref{sec:null_red}, a possible way to derive the corresponding NR action is by identifying a non-compact light-like isometry direction of the target space, and perform a null reduction.

Alternatively, suppose that we want to obtain the NR action via the limit procedure introduced in section \ref{subsec:limit}. Then, we need to engineer a $c$-expansion of the vielbein as in \eqref{E_scaling}. This is achieved via rescaling the coordinates by a dimensionless parameter, which will be called $c$. For the case of flat spacetime equipped with Cartesian coordinates, such rescaling is immediate. It suffices to rescale the time coordinate $X^0$ and one of the space coordinates, say $X^1$. Due to isotropy of flat spacetime, rescaling any other spatial coordinate instead of $X^1$ will lead to the same final result. This implies that the relativistic vielbein will rescale as in \eqref{E_scaling}, generating the appropriate String Newton-Cartan geometry. The coordinate rescaling is less obvious when the target space is \emph{curved}. Intuitively, the time coordinate should still be rescaled, but then it is not obvious at all which spatial coordinate should be rescaled. This raises the natural question: \emph{In a curved manifold, how should the coordinates be rescaled in order to take the NR limit?}

Answering this question for a fully general curved manifold is still an open problem. However, it is possible to give an answer for manifolds which are \emph{coset spaces} $G/H$, with $G$ and $H$ Lie groups. For this class of manifolds, there exist a method to write their metric in a coordinate-free language, via Maurer-Cartan 1-forms. Here we will quickly review the important details, see \cite{Metsaev:1998it,Arutyunov:2009ga, Zarembo:2017muf} and section \ref{sec:symmetricspaces} for a more extensive explanation. Let us denote $\mathfrak{g}$ and $\mathfrak{h}$ the Lie algebras associated with $G$ and $H$ respectively, and denote $\mathfrak{p} = \mathfrak{g}\setminus \mathfrak{h}$, with
generators $\mathfrak{p} = \{ P_{\hat{A}}\}$ (not necessarily abelian), where $\hat{A} \in\{ 0, ..., d-1\}$, with $d$ being the dimension of the manifold. To construct the metric, one needs to choose a coset representative $g \in G$, and consider the associated Maurer-Cartan (MC) 1-form $A \equiv g^{-1} \dd g$. Then the metric is given by 
\begin{eqnarray}\label{metric_from_MC}
    g_{\mu\nu} = \langle \mathcal{P} A_{\mu} , \mathcal{P} A_{\nu} \rangle \, , 
\end{eqnarray}
where $\mathcal{P} : \mathfrak{g} \to \mathfrak{p}$ is a projector onto $\mathfrak{p}$, and $\langle \cdot , \cdot \rangle$ is an inner product. The formula \eqref{metric_from_MC} is justified if we recall that the $\mathfrak{p}$-projection of the MC 1-form is the vielbein, i.e. $\mathcal{P} A_{\mu} = E_{\mu}{}^{\hat{A}} P_{\hat{A}}$. If the inner product is chosen diagonally as $\langle P_{\hat{A}}, P_{\hat{B}} \rangle = \eta_{\hat{A}\hat{B}}$, then formula \eqref{metric_from_MC} follows immediately. The metric $g_{\mu\nu}$ will be described in a set of coordinates that is dictated by the specific choice of coset representative $g$. A very simple coset representative, but in general producing a complicated metric, is given by:
\begin{eqnarray}\label{g_exp(xP)}
    g = e^{x^{\hat{A}} P_{\hat{A}}} \, . 
\end{eqnarray}
Thanks to this coordinate-free language of writing the metric, we are now able to define the NR limit for homogeneous spaces $G/H$. The steps are in order:
\begin{enumerate}
    \item At the pure kinematical level, one needs first to identify the NR \.In\"on\"u-Wigner contraction of $\mathfrak{g}$. This is typically done by scaling $P_A$ and $P_a$ differently, together with a suitable rescaling of the generators in $\mathfrak{h}$. This study is fully coordinate-free. 
    \item From the specific coset representative $g$, the rescaling of generators demanded by the NR \.In\"on\"u-Wigner contraction induces a ``dual'' rescaling of coordinates, where coordinates rescale but generators do not. For example, for the coset representative \eqref{g_exp(xP)}, the generator rescaling $P_A = \frac{1}{c} H_A$ is equivalent to the coordinate rescaling $x^A = c \,y^A$. 
    \item The coordinate rescaling is then plugged into the metric $g_{\mu\nu}$ and expanded in large $c$. By comparing the expansion coefficients with the SNC expansion \eqref{E_scaling}, one reads off the SNC metric tensors.  
\end{enumerate}

This procedure may also be reconstructed by working backward. 
First, we recall that the vielbein needs to expand in the NR limit as in \eqref{E_scaling}, which we write here for convenience, 
\begin{align}
 E_{\mu}{}^A &= c \tau_{\mu}{}^A + \frac{1}{c} m_{\mu}{}^A + \mathcal{O}(c^{-3})\, , &
 E_{\mu}{}^a &= e_{\mu}{}^a + \mathcal{O}(c^{-2}) \, .
\end{align}
Then, thanks to the fact the $\mathfrak{p}$-projection of the MC 1-form is the vielbein, the leading order of this expansion is equivalently captured by the following rescaling of generators
\begin{eqnarray}\label{gen_resc}
  \mathcal{P} A_{\mu} = E_{\mu}{}^{\hat{A}} P_{\hat{A}} \, , \qquad \Longrightarrow \qquad
  P_A = \frac{1}{c} H_A \, , \quad 
  P_a \ \text{not rescaled} \, .
\end{eqnarray}
This operation of rescaling the generators identifies the \.In\"on\"u-Wigner contraction of the algebra $\mathfrak{p}$, and more generically of $\mathfrak{g}$. The next-to-leading order in the longitudinal vielbein expansion also has an algebraic interpretation: it corresponds to the subleading generator $Z_A$ in the Lie algebra expansion of $P_A$ \cite{Hansen:2019vqf, Hansen:2019pkl, Bergshoeff:2019ctr}. 
The identification of the coordinate rescaling goes as before: the choice of coset representative induces a correspondence between the rescaling of generators and coordinates.

\subsection{SNC flat spacetime} \label{sec:SNCflatspacetime}

As a trivial yet instrumental example, we consider the case of flat spacetime. Using the Maurer-Cartan formalism, we review how the NR limit is implemented at the level of coordinate rescaling. We consider two types of coordinates: Cartesian and Polar.   

First, we recall that the $d$-dimensional flat spacetime is defined as a coset space,
\begin{eqnarray}
    \text{Mink}_d = \frac{ISO(1,d-1)}{SO(1,d-1)} \, .
\end{eqnarray}
The associated Lie algebras are spanned by the generators $\mathfrak{iso}(1,d-1) = \{ P_{\hat{A}}, J_{\hat{A}\hat{B}} \}$ and $\mathfrak{so}(1,d-1) = \{ J_{\hat{A}\hat{B}} \}$, with non-vanishing commutation relations
\begin{eqnarray}\label{Poincare}
    [P_{\hat{A}}, J_{\hat{B}\hat{C}}] = 2 \eta_{\hat{A}[\hat{B}} P_{\hat{C}]} \, , \qquad\qquad
    [J_{\hat{A}\hat{B}}, J_{\hat{C}\hat{D}}] = 4\eta_{[\hat{B}[\hat{C}} J_{\hat{A}]\hat{D}]} \, .
\end{eqnarray}
The string NR contraction of the Poincar\'e algebra gives rise to the \emph{string Galilei} algebra. This is obtained via the rescaling \eqref{gen_resc}, supplemented by a rescaling of the angular momentum generators. The full rescaling reads \cite{Barducci:2019jhj}, 
\begin{eqnarray}\label{Galieli_rescaling}
    P_A = \frac{1}{c} H_A \, , \qquad 
  J_{Aa} = c \,G_{Aa} \, .
\end{eqnarray}
Since the Poincar\'e algebra is invariant under an overall rescaling of the $P_{\hat{A}}$ generators, an equivalent NR rescaling is,  
\begin{eqnarray}\label{Galieli_rescaling_2}
    P_a = c \tilde{P}_a \, , \qquad 
  J_{Aa} = c \,G_{Aa} \, .
\end{eqnarray}
By implementing this rescaling in the algebra \eqref{Poincare} and by taking $c\to \infty$, we get the string Galilei algebra:
\begin{subequations}
	\begin{align}
	[H_A,M] & = \varepsilon_A{}^B H_B \, , &
	[G_{Aa}, M]  & = \varepsilon_A{}^B G_{Ba} \, , \\
	[H_A, G_{Ba}]  & = \eta^{}_{AB} P_{a} \, , &
	[G_{Aa}, J_{bc}] & = 2\delta_{a[b} G_{|A|c]} \, , \\
    [P_{a}, J_{bc}] & = 2\delta_{a[b} P_{c]} \, , &
	 [J_{ab}, J_{cd}] & = 4\delta_{[b[c} J_{a]d]} \, , 
	\end{align}
\end{subequations}
where we used the notation $J_{AB} = - \varepsilon_{AB} M$. 

\subsubsection{Cartesian coordinates}

The Cartesian coordinates are given in terms of the coset representative
\begin{eqnarray}
    g = e^{x^{\hat{A}} P_{\hat{A}}} \, ,  
\end{eqnarray}
and the inner product $\langle P_{\hat{A}}, P_{\hat{B}} \rangle = \eta_{\hat{A}\hat{B}}$, such that the associated MC 1-form produces the following metric,
\begin{eqnarray}\label{metric_Cartesian_flat}
    \dd s^2 = \eta_{\hat{A}\hat{B}} \dd x^{\hat{A}} \dd x^{\hat{B}} \, . 
\end{eqnarray}
From the generator rescaling \eqref{Galieli_rescaling}, and from the specific coset representative that relates generators with coordinates, we identify the dual rescaling of coordinates as
\begin{eqnarray}\label{resc_Cart_flat}
     x^A = c \, y^A \, , \qquad x^a = y^a\, . 
\end{eqnarray}
By plugging \eqref{resc_Cart_flat} inside the metric \eqref{metric_Cartesian_flat}, and by taking the large $c$ expansion, we get the SNC metric tensors 
\begin{eqnarray}
    \tau =  \eta_{AB} \dd y^A \dd y^B \, , \qquad
    H = \delta_{ab} \dd y^a \dd y^b \, .  
\end{eqnarray}
Alternatively, one could have plugged the rescaling \eqref{resc_Cart_flat} inside the vielbein, and expand in large $c$. In this way, one reads off the SNC vielbeine, 
\begin{eqnarray}  
\tau_{\mu}{}^A = \delta_{\mu}^A \, , \qquad
m_{\mu}{}^A = 0 \, , \qquad
e_{\mu}{}^a = \delta_{\mu}^a \, . 
\end{eqnarray}

\subsubsection{Polar coordinates}
In Mink$_d$, the coset representative is
\begin{eqnarray}
    g =  e^{t P_0} e^{\theta_{d-2} J_{d-2,d-1}} e^{\theta_{d-3} J_{1,d-2}} \cdots e^{\theta_{1} J_{12}} e^{r P_1} \, .  
\end{eqnarray}
The associated MC 1-form produces the following metric,
\begin{eqnarray}\label{metric_polar_flat}
    \dd s^2 =  - \dd t^2 + \dd r^2 + r^2 \dd \Omega_{d-2} \, . 
\end{eqnarray}
where $\dd \Omega_{d-2}$ is the line element of the $(d-2)$-sphere, 
\begin{eqnarray}
  \hspace{-5mm} \dd \Omega_{d-2} = \dd \theta_1^2 + \cos^2\theta_1 \dd\theta_2^2 + \cos^2\theta_1 \cos^2\theta_2  \dd\theta_3^2 + ... 
   + \left(\prod_{i=1}^{d-4} \cos^2\theta_i\right) \sin^2\theta_{d-3}  \dd\theta_{d-2}^2 \, .
\end{eqnarray}
For example, when $d=5$ this reads
\begin{eqnarray}\label{metric_polar_flat_d=5}
    \dd \Omega_{3} =  \dd \theta_1^2 + \cos^2 \theta_1 \left( \dd \theta_2^2 + \sin^2 \theta_2 \dd \theta_3^2 \right) \, . 
\end{eqnarray}
From the algebra rescaling, the dual rescaling of coordinates is\footnote{The cases of $d\leq 3$ are special, as all $\theta_i$ variables need to be rescaled. } 
\begin{equation}\label{resc_polar_flat}
  \theta_i = \frac{\tilde{\theta}_i}{c} \, , \qquad
  \forall\ i \in\{ 1, ..., d-3\} \, , 
\end{equation}
with $t$ and $r$ not rescaled. By plugging \eqref{resc_polar_flat} inside the metric \eqref{metric_polar_flat}, and by taking the large $c$ expansion, we get the SNC metric tensors\footnote{Notice that with the coordinate rescaling \eqref{resc_polar_flat} the SNC metric data is not in its ``canonical'' form, namely $\tau$ appears at order 1 and $h$ at order $c^{-2}$. In the string action, it is possible to introduce an overall $c^2$ via rescaling the string tension, therefore getting the canonical form. } 
\begin{eqnarray}
    \tau =  - \dd t^2 + \dd r^2 \, , \qquad
    H = r^2 \left( \tilde{\theta}_{d-3}^2 \dd \tilde{\theta}_{d-2}^2 + \sum_{i=1}^{d-3} \dd \tilde{\theta}_i^2 \right)   \, .  
\end{eqnarray}
The metric $\tau$ describes Mink$_2$ and the tensor $h$ describes $w(r) \mathbb{R}^{d-2}$ in polar coordinates, where $w(r) = r^2$ is the warped factor, which was trivial in Cartesian coordinates. 

The associated SNC vielbeine reads,
\begin{eqnarray}
\notag
\tau_{\mu}{}^A &=& \text{diag} \left(1, 1, 0, ..., 0 \right) \, , \\
m_{\mu}{}^A &=& 0 \, , \\
\notag
e_{\mu}{}^a &=& r \, \text{diag} \left(0, 0, 1, ..., 1, \tilde{\theta}_{d-3} \right) \, . 
\end{eqnarray}

\subsection{SNC AdS$_5\times$S$^5$} \label{subsec:SNCAdscoords}

The second SNC landscape that has been studied is the one arising from the relativistic AdS$_5\times$S$^5$ geometry. This geometry is of particular interest because the AdS$_5$/CFT$_4$ correspondence is the most extensively studied example of holography. Accordingly, the remainder of this review is devoted to understanding holography, quantisation, and integrability in the non-relativistic setting derived from this background.

Here we shall review how to rescale the coordinates and how the final SNC AdS$_5\times$S$^5$ geometry looks like in a number of set of coordinates: Cartesian, polar, GGK and Poincar\'e. Most of the results that will be presented are based on Appendix A of \cite{Fontanella:2021hcb}. 

We recall that AdS$_5\times$S$^5$ is described in terms of the coset,
\begin{eqnarray}
    \text{AdS}_5\times\text{S}^5 = \frac{SO(2,4)\times SO(6)}{SO(1,4) \times SO(5)} \, . 
\end{eqnarray}
We denote the generators in AdS$_5$ as $\mathfrak{so}(2,4) = \{ P_{\hat{A}}, J_{\hat{A}\hat{B}} \}$, $\mathfrak{so}(1,4) = \{ J_{\hat{A}\hat{B}} \}$ and in S$^5$ as $\mathfrak{so}(6) = \{ P_{a'}, J_{a'b'} \}$, $\mathfrak{so}(5) = \{ J_{a'b'} \}$. The indices run as follows: $\hat{A} \in\{ 0, ..., 4\}$, $A \in\{0,1\}$, $a\in\{2,3,4\}$, $a' \in\{ 5, ..., 9\}$. The reader should be careful with the indices: here, $\hat{A}$ denotes indices in AdS$_5$, which split as $(A,a)$, whereas $a'$ denotes indices in S$^5$. The non-vanishing commutation relations are\footnote{Here we adopt the convention in which the $P$ generators have units of inverse length because this gives a physical meaning to the algebra contraction of \cite{Gomis:2005pg}. This convention implies that the coefficients of the algebra parameterisation have unit of length, so they can be interpreted as coordinates. Nevertheless, $\hat{R}$ can always be set to 1 by redefining the algebra generators and parameterisation.}
\begin{subequations}\label{so(4,2)+so(6)}
	\begin{align}
	[P_{\hat{A}}, P_{\hat{B}}] &= \frac{1}{\hat{R}^2} J_{\hat{A}\hat{B}} \, , &
	[P_{a'}, P_{b'}] &= - \frac{1}{\hat{R}^2} J_{a'b'} \, , \\
	[P_{\hat{A}}, J_{\hat{B}\hat{C}}] &= 2 \eta_{\hat{A}[\hat{B}} P_{\hat{C}]} \, , &
	[P_{a'}, J_{b'c'}] &= 2 \delta_{a'[b'} P_{c']} \, , \\
	[J_{\hat{A}\hat{B}}, J_{\hat{C}\hat{D}}] &= 4\eta_{[\hat{B}[\hat{C}} J_{\hat{A}]\hat{D}]} \, , &
	[J_{a'b'}, J_{c'd'}] &= 4\delta_{[b'[c'} J_{a']d']} \, ,
	\end{align}
\end{subequations}
where $\hat{R}$ denotes the radius of AdS$_5$ and S$^5$, which must be equal in order to satisfy the type IIB supergravity equations of motion.

The stringy non-relativistic \.In\"on\"u-Wigner contraction of the conformal algebra gives the \emph{string Newton-Hooke} algebra. This contraction rule is defined for AdS$_5$, but it can be extended also to S$^5$ by treating the sphere directions as transverse spatial direction. 
The rescaling was given in \cite{Gomis:2005pg}:
\begin{eqnarray}\label{Newton_Hooke_rescaling}
    P_A = \frac{1}{c} H_A \, , \qquad 
  J_{Aa} = c \,G_{Aa} \, , \qquad
  \hat{R} = c R \, ,  
\end{eqnarray}
while the remaining generators do not rescale. Notice that $\hat{R}$ is required to rescale as well. A different rescaling, which does \emph{not} involve the radius $\hat{R}$, but gives the same contraction, is the following \cite{Fontanella:2021hcb,Fontanella:2023men}:
\begin{eqnarray}
\label{Newton_Hooke_rescaling_2}
    P_a = c\, \tilde{P}_a \, , \qquad
    P_{a'} = c \, \tilde{P}_{a'} \, , \qquad
    J_{Aa} = c \, G_{Aa} \, .
\end{eqnarray}
By implementing this rescaling in the algebra \eqref{so(4,2)+so(6)} and by taking $c\to \infty$, we get the string Newton-Hooke algebra (for the AdS$_5$ part) plus the Euclidean algebra (for the S$^5$ part):
\begin{subequations} \label{Newton-Hooke-algebra}
	\begin{align}
	[H_A, H_B]& = -\varepsilon_{AB} M \, , &
	[P_{a'}, J_{b'c'}] &= 2 \delta_{a'[b'} P_{c']} \, , \\
	[H_A, P_b] &= G_{Ab} \, , &
	[J_{a'b'}, J_{c'd'}] &= 4\delta_{[b'[c'} J_{a']d']} \, , \\
    [H_A,M] & = \varepsilon_A{}^B H_B \, , &
	&  \\
    [H_A, G_{Ba}]  & = \eta^{}_{AB} P_{a} \, , &
	&  \\
    [P_{a}, J_{bc}] & = 2\delta_{a[b} P_{c]} \, , &
	&  \\
    [G_{Aa}, M]  & = \varepsilon_A{}^B G_{Ba} \, , &
	&  \\
    [G_{Aa}, J_{bc}] & = 2\delta_{a[b} G_{|A|c]} \, , &
	&  \\
    [J_{ab}, J_{cd}] & = 4\delta_{[b[c} J_{a]d]} \, , &
	& 
	\end{align}
\end{subequations}
where, again, we used the notation $J_{AB} = - \varepsilon_{AB} M$.

\subsubsection{Cartesian coordinates}

This is a set of \emph{global} coordinates, and their name stems from the fact that in the flat space limit they reproduce precisely the Cartesian coordinates for Minkowski. 
The coset representative is 
\begin{eqnarray}
    g = g_a g_s \, , \qquad
    g_a = \Lambda(t) G(z) \, ,\qquad
    g_s = \Lambda(\phi) G(y) \, , 
\end{eqnarray}
with
\begin{eqnarray}
\notag
  &\Lambda(t) = \exp \left(t P_0 \right) \, , \qquad
  G(z) = \exp \left(z^i P_i\right) \, , \\
  &\Lambda(\phi) = \exp \left(\phi P_5 \right) \, , \qquad
  G(y) = \exp \left(y^{i'} P_{i'}\right) \, ,
\end{eqnarray}
where $i \in\{ 1, ..., 4\}$ and $i' \in\{ 6, ..., 9\}$. The associated MC 1-form produces the metric, 
\begin{equation}
\label{metric_cartesian}
\dd s^2 = - \bigg(\frac{1+ \frac{z^2}{4 \hat{R}}}{1-\frac{z^2}{4 \hat{R}}}\bigg)^2 \dd t^2 + \frac{1}{(1-\frac{z^2}{4 \hat{R}})^2} \dd z^2  + \bigg(\frac{1 - \frac{y^2}{4 \hat{R}}}{1 + \frac{y^2}{4 \hat{R}}}\bigg)^2\dd \phi^2 + \frac{1}{(1+\frac{y^2}{4 \hat{R}})^2} \dd y^2 \, ,
\end{equation}
where we denoted $z^2 \equiv z^i z^j \delta_{ij}$, and the same for $y^2$.  The fact the generators are rescaled as in \eqref{Newton_Hooke_rescaling} implies the dual coordinate rescaling
\begin{eqnarray}\label{rescaling_Cartesian_AdSxS}
    t = c \tilde{t} \, , \qquad
    z_1 = c \tilde{z}_1 \, , \qquad
    \hat{R} = c R \, .
\end{eqnarray}
By plugging this rescaling inside the metric, and expanding in large $c$, one reads off the SNC metric tensors\footnote{Looking beyond the \.In\"on\"u-Wigner contraction to Newton-Hooke, one might be tempted to see what happens when we rescale the time direction and an S$^5$ direction. If we substitute $z_1$ by $\phi$ in the rescaling \eqref{rescaling_Cartesian_AdSxS}, we arrive to an SNC flat spacetime.}
\begin{subequations}\label{All_cartesian}
\begin{align}
\label{Tau_cartesian}
    \tau &=  - \left[\frac{1+ (\frac{z_1}{2R})^2}{1- (\frac{z_1}{2R})^2}\right]^2 \dd t^2 + \frac{1}{\left[1 - (\frac{z_1}{2R})^2\right]^2} \dd z_1^2  \, , \\
    \label{Hcartesian}
   H &= - \frac{1+ (\frac{z_1}{2R})^2}{\left[1- (\frac{z_1}{2R})^2\right]^3} \frac{\zeta^2}{R^2}\, \dd t^2 
+ \frac{\zeta^2}{2R^2 \left[1- (\frac{z_1}{2R})^2\right]^3}\,  \dd z_1^2 + \frac{1}{\left[1- (\frac{z_1}{2R})^2\right]^2} \, \dd \zeta^2  + \dd \phi^2
+ \dd y^2  \, ,   
\end{align}
\end{subequations}
where $\zeta \equiv (z_2, z_3, z_4)$ are the transverse coordinates in AdS$_5$, and we removed the tildes in $t$ and $z_1$ to avoid cluttering the notation. The associated SNC vielbeine are:    
\begin{subequations}
\begin{align}
\tau_{\mu}{}^A &= \text{diag}\bigg(-\frac{1+ (\frac{z_1}{2R})^2}{1- (\frac{z_1}{2R})^2},   \frac{1}{1 - (\frac{z_1}{2R})^2}, 0, ... , 0 \bigg) \, , \\
m_{\mu}{}^A &= \text{diag}\bigg(-\frac{\zeta^2}{2R^2(1-(\frac{z_1}{2R})^2)},  \frac{\zeta^2}{4R^2(1-(\frac{z_1}{2R})^2)}, 0, ... , 0 \bigg) \, , \label{NRdatacartesian}\\
e_{\mu}{}^a &=\text{diag} \left(0, 0, \frac{1}{1 - (\frac{z_1}{2R})^2}, \frac{1}{1 - (\frac{z_1}{2R})^2}, \frac{1}{1 - (\frac{z_1}{2R})^2}, 1,..., 1 \right) \, .
\end{align}
\end{subequations}

\subsubsection{Polar coordinates}

This is a set of \emph{global} coordinates, where $t$ is the global AdS time, $\rho$ is the radial coordinate in AdS$_5$ and $\beta_i$ and $\varphi_j$ are angles in AdS$_5$ and S$_5$ respectively. The coset representative is given by, 
\begin{eqnarray} \label{PolarCoset}
    g = g_a g_s \, , \qquad
    g_a = \Lambda (t, \beta_3) \Theta (\beta_1, \beta_2) G(\rho) \, ,\qquad
    g_s = \Lambda (\varphi_3, \varphi_4, \varphi_5 ) \Theta (\varphi_2) G(\varphi_1) \, , 
\end{eqnarray}
with
\begin{subequations}
	\begin{align}
	\Lambda (t, \beta_3) &= e^{t P_0} e^{\beta_3 J_{34}}\, , &
	\Lambda (\varphi_3, \varphi_4, \varphi_5 ) &= e^{\varphi_5 J_{78}} e^{(\varphi_4 + \frac{\pi}{2}) J_{57}} e^{\varphi_3 J_{56}}\, , \\
	\Theta (\beta_1, \beta_2) &= e^{(\beta_2 + \frac{\pi}{2})J_{13}} e^{\beta_1 J_{12}} \, , &
	\Theta (\varphi_2) &= e^{(\varphi_2 + \frac{\pi}{2}) P_5} \, , \\
	G(\rho) &= e^{\rho P_1} \, , &
	G(\varphi_1) &= e^{\varphi_1 P_9}\, .
	\end{align}
\end{subequations}
Notice that, with this parameterisation, the angles $\varphi_1$ and $\varphi_2$ have units of length. Here we set $\hat{R}=1$ and completely ignore this issue. The MC 1-form associated to this coset representative produces the metric, 
\begin{align}
    \dd s^2 &=  \dd s^2_{\text{AdS}}+\dd s^2_{\text{S}}  \, , \label{coordpolar}\\
    \notag
    \dd s^2_{\text{AdS}} &= -\cosh^2 \rho \, \dd t^2 +\dd\rho^2 + \sinh^2 \rho \, \dd\beta_1^2 + \sinh^2 \rho \, \cos^2 \beta_1 \, \dd\beta_2^2 \\
    \notag
    &\hspace{5mm} + \sinh^2 \rho \, \cos^2 \beta_1 \, \cos^2 \beta_2 \, \dd\beta_3^2 \, , \notag \\
    \notag
    \dd s^2_{\text{S}} &= \dd \varphi_1^2 + \cos^2 \varphi_1 \, \dd \varphi_2^2 + \cos^2 \varphi_1 \, \cos^2 \varphi_2 \, \dd\varphi_3^2 + \cos^2 \varphi_1 \, \cos^2 \varphi_2 \, \cos^2 \varphi_3 \, \dd\varphi_4^2   \\
    &\hspace{5mm}  +\cos^2 \varphi_1 \, \cos^2 \varphi_2 \, \cos^2 \varphi_3 \, \cos^2 \varphi_4 \, \dd\varphi_5^2  \, ,  \notag 
\end{align}
From the rescaling of generators as in \eqref{Newton_Hooke_rescaling_2} and from the specific coset representative, the coordinate rescaling in the dual picture is 
\begin{subequations}\label{rescaling_polar_AdSxS}
    \begin{align}
    \beta_1 &= \frac{\tilde{\beta}_1}{c} \, , &
    \beta_2 + \frac{\pi}{2} &= \frac{1}{c} \left(\tilde{\beta}_2 + \frac{\pi}{2}\right) \, , \\
    \varphi_1 &= \frac{\tilde{\varphi}_1}{c} \, , &
    \varphi_2 + \frac{\pi}{2} &= \frac{1}{c} \left(\tilde{\varphi}_2 + \frac{\pi}{2}\right) \, .
\end{align}
\end{subequations}

By plugging this rescaling inside the metric and expanding in large $c$, one reads off the SNC metric tensors\footnote{The SNC metric data does not appear in the ``canonical'' form, as discussed for the polar coordinates in flat spacetime. The same trick of rescaling the string tension to bring it to the canonical form applies here. Typically, the string tension in the AdS$_5\times$S$^5$ string action is normalised by the radius, and therefore rescaling the string tension is a ``hidden'' effect of rescaling the radius $\hat{R}$.}
\begin{subequations}\label{All_polar}
\begin{align}
\label{Tau_polar}
    \tau&=  -\cosh^2 \rho\,  \dd t^2  + \dd \rho^2 \, , \\
    \notag
    H&= \sinh^2 \rho \, \dd\beta_1^2 + \sinh^2 \rho \, \dd\beta_2^2 + \sinh^2 \rho \, (\beta_2 + \pi/2)^2 \dd\beta_3^2 + \dd \varphi_1^2 + \dd \varphi_2^2 + (\varphi_2 - \pi/2)^2 \dd \varphi_3^2 \\
    &\hspace{4mm}+ (\varphi_2 - \pi/2)^2 \, \cos^2 \varphi_3 \dd \varphi_4^2 + (\varphi_2 - \pi/2)^2 \, \cos^2 \varphi_3 \, \cos^2 \varphi_4 \dd \varphi_5^2  \, , \label{Hpolar}
\end{align} 
\end{subequations}
where we removed the tildes on the transformed coordinates to avoid cluttering the notation. 
The associated SNC vielbeine are:
\begin{align}\label{NRdatapolar}
    \notag
    \tau_\mu{}^A &=\text{diag} \left( \cosh \rho , 1 , 0, ... , 0 \right) \, , \\
    m_\mu{}^A &= 0 \, ,  \\
    \notag
    e_\mu{}^a &=\text{diag} \left( 0 , 0 , -\sinh \rho , -\sinh \rho , \sinh \rho \, (\beta_2 + \pi/2) , 1 , 1 , \right.  \\
    \notag
    &\hspace{14mm}\left.-(\varphi_2 - \pi/2) , -(\varphi_2 - \pi/2) \, \cos \varphi_3 , (\varphi_2 - \pi/2) \, \cos \varphi_3 \, \cos \varphi_4 \right) \, .
\end{align}

\subsubsection{GGK coordinates}

This is a set of \emph{global} coordinates, used in \cite{Gomis:2005pg} to make an embedding of AdS$_2$ inside AdS$_5$ manifest. In particular, they make a space-like isometry inside AdS$_2$ manifest, where the NR string can wrap. The AdS coordinates are $(\hat{x}^0, \hat{x}^1, x^a)$, where $(\hat{x}^0, \hat{x}^1)$ describe the embedded AdS$_2$, whereas the sphere coordinates are $x^{a'}$. The coset representative is,   
\begin{eqnarray}\label{coset_GGK}
    g = g_a g_s \, , \qquad\qquad
    g_a = e^{\hat{x}^1 P_1} e^{\hat{x}^0 P_0} e^{x^a P_a} \, ,\qquad
    g_s = e^{x^{a'} P_{a'}} \, . 
\end{eqnarray}
The metric generated by the MC 1-form is quite cumbersome before taking the NR limit, and given by 
\begin{align}
    \dd s^2 &=  \dd s^2_{\text{AdS}}+\dd s^2_{\text{S}}  \, , \label{coordGGK}\\
    \notag
    \dd s^2_{\text{AdS}} &= \cosh^2\theta     \left[    -(\dd \hat{x}^0)^2+\cos^2\left({\hat{x}^0\over
    \hat{R}}\right)(\dd \hat{x}^1)^2\right] \\
    & \hspace{5mm}+\left[ \left({\sinh \theta\over \theta}\right)^2(\dd x^a)^2-\left({\sinh^2
    \theta\over \theta^2}-1\right){(x_a \dd x^a)^2\over {\theta}^2 \hat{R}^2}\right] \, , \notag \\
    \notag
    \dd s^2_{\text{S}} &= \left({\sin \zeta\over \zeta}\right)^2(\dd x^{a'})^2-\left({\sin^2
    \zeta\over \zeta^2}-1\right){(x_{a'}\dd x^{a'})^2\over {\zeta}^2 \hat{R}^2} \, ,  \notag 
\end{align}
where $\theta \equiv \sqrt{x_a x^a}/\hat{R}$ and $\zeta \equiv \sqrt{x_{a'} x^{a'}}/\hat{R}$. From the rescaling \eqref{Newton_Hooke_rescaling}, and from the GGK coset representative, we read off the dual coordinate rescaling, 
\begin{eqnarray}
    \hat{x}^A = c x^A \, , \qquad
    \hat{R} = c R \, .
\end{eqnarray}
By plugging this rescaling inside the metric and expanding in large $c$, one extracts the following SNC metric tensors:
\begin{subequations}\label{all_GGK}
    \begin{align}
\label{Tau_GGK}
    \tau&=  - (\dd x^0)^2 + \cos^2 \left(\frac{x^0}{R} \right)\, (\dd x^1)^2 \, , \\
\label{H_GGK}
    H&= \frac{x^a x_a}{R^2} \left[-(\dd x^0)^2 + \cos^2 \left(\frac{x^0}{R} \right)\, (\dd x^1)^2 \right]+ (\dd x^a )^2 + (\dd x^{a'})^2  \, .
    \end{align}
\end{subequations}
The associated SNC vielbeine are: 
\begin{subequations}
    \begin{align}
     \tau_{\mu}{}^A &= \text{diag} \left( -1,  \cos \frac{x^0}{R} , 0, ..., 0\right) \, ,  \\
m_{\mu}{}^A &= \frac{x^a x_a}{2R^2} \ \text{diag} \left(-1,  \cos \frac{x^0}{R} , 0, ...., 0 \right) \, , \label{SNC_data_GGK} \\
e_{\mu}{}^a &=\text{diag} (0,  0, 1, ...., 1) \, .   
    \end{align}
\end{subequations}

\subsubsection{Poincar\'e coordinates}

These are \emph{local} coordinates for AdS, as they patch only half of the spacetime. They are spanned by time $t$, the holographic radial coordinate $z$ and the spatial coordinates $x^a$. The sphere can be patched by any of the previous set of coordinates. The coset representative is:
\begin{eqnarray}
    g_{\text{AdS}} = e^{t \mathfrak{P}_0 + x^a \mathfrak{P}_a}e^{-z \mathfrak{P}_0}e^{-\frac{\hat{R}^2}{2z} \mathfrak{K}_0}\, , 
\end{eqnarray}
where we introduced the generators in the ``conformal basis'',
\begin{subequations}
	\begin{align}
	\mathfrak{P}_0 &\equiv P_0 - P_1\, , &
	 \mathfrak{P}_a &\equiv J_{a0} - J_{a1}\, , \\
	\mathfrak{K}_0 &\equiv P_0 + P_1 \, , &
	\mathfrak{K}_a &\equiv J_{a0} + J_{a1} \, , \\
	\mathfrak{J}_{0a} &\equiv P_{a} \, , &
    \mathfrak{J}_{ab} &\equiv J_{ab}\, , \\
	\mathfrak{D} &\equiv J_{10}\, . &
	\end{align}
\end{subequations}
The AdS$_5$ metric generated by the associated MC 1-form is,
\begin{eqnarray}
    \dd s^2_{\text{AdS}} = \frac{-\dd t^2 + \dd z^2 + (\dd x^a)^2}{z^2} \, ,  
\end{eqnarray}
which is conformally flat. From the rescaling \eqref{Newton_Hooke_rescaling_2} and from the specific coset representative, we read off the dual coordinate rescaling,
\begin{eqnarray}
    x^a = \frac{\tilde{x}^a}{c}  \, . 
\end{eqnarray}
By plugging this rescaling inside the metric, removing the tilde on $x^a$ and expanding in large $c$, one reads off the SNC metric tensors\footnote{Similarly to what happened in polar coordinates, the SNC metric data does not appear in the ``canonical'' form. The trick of rescaling the string tension to bring it to the canonical form still applies here. }
\begin{align}
\label{Tau_H_Poincare}
    \tau=  \frac{-\dd t^2 + \dd z^2}{z^2} \, , \qquad\qquad
    H= \frac{(\dd x^a)^2}{z^2}  \, .
\end{align}
The corresponding SNC vielbeine are:
\begin{eqnarray}
\tau_{\mu}{}^A = \frac{1}{z}\delta_{\mu}^A \, , \qquad
m_{\mu}{}^A = 0 \, , \qquad
e_{\mu}{}^a = \frac{1}{z}\delta_{\mu}^a \, . 
\end{eqnarray}

%%%%%%%%%%%%%%%%%%%%%%%%%%%%%%%%%%%%%%%%%%%%%%%%%%%%%%%%%%%%%%%%%%%%%%%%%%%%%%%%%%%%%%%%%%%%%%%%%%%%%%%%%%%%%%%%%%%%%%%%%%%%%%%%%%%%%%%%%%%%%%%%%%%%%%%%%%%%%%%%%%%%%%%%%%%%%%%%%%%%%%%%%%%%%%%%%%%%%%%%%%%%%%%%%%%%%%%%%%%%%%%%%%%%%%%%%%%%%%%%%%%%%%%%%%%%%%%%%%%%%%%%%%%%%%%%%%%%%%%%%%%%%%%%%%%%%%%%%%%%
\chapter{Non-relativistic gauge/gravity correspondence}
\label{chap:holography}

The AdS/CFT correspondence is one of the most active lines of work in the fields of string theory and quantum field theory. This success is easily understandable because, starting from the assumption that the duality is true, we can map very complicated strong-coupling computations to more manageable weak-coupling computations. It is not a surprise that many people work on answering the question: \emph{how much can we modify the setting without breaking the correspondence?} For example, we can take the field theory and add exactly marginal deformations and study how those appear in the geometrical framework. This would lead us, for example, to the duality between Leight-Strassler deformations of $\mathcal{N}=4$ SYM and the Lunin-Maldacena deformations of AdS$_5\times$S$^5$ \cite{Lunin:2005jy}. In our case, we are interested in the case of singular limits of the geometric framework. Here we have to be careful, because nothing guarantees that the singular limit of the geometry and the near-horizon/decoupling limit in Maldacena’s construction of the holographic correspondence commute. In this chapter, we present five consistency conditions that the limit of the brane construction must satisfy in order to establish that holography survives in such a limit. We will use these points as a guiding principle to construct the GGK/GYM duality. We will also discuss other non-relativistic holographic dualities derived from the AdS$_{n+1}$/CFT$_n$ correspondence from this point of view.

\section{Maldacena's AdS$_5$/CFT$_4$ correspondence}
\label{sec:Maldacena_AdS/CFT}

In this section, we shall shortly review Maldacena's construction of the AdS$_5$/CFT$_4$. For reviews of the topic, one can refer to the classical references \cite{Aharony:1999ti, Maldacena:2003nj, Nastase:2007kj,Polchinski:2010hw} or to more modern books \cite{Ammon_Erdmenger_2015}. 

The basic idea laid down in \cite{Maldacena:1997re} is to consider type IIB superstring theory in 10d flat spacetime with a stack of $N$ D3-branes. We can study two different regimes of this setting depending on the value of $g_s N$. On the one hand, we can study the limit $g_s N \ll 1$, where the D3-branes can be treated as hypersurfaces where open strings are forced to end. On the other hand, for $g_s N \gg 1$ the interaction among the D3-branes is so strong that they are more suitably described as heavy objects that modify the spacetime background. In this setting, the system is described by closed strings propagating on a black D3-brane spacetime.

\subsection{The gravity perspective: $g_s N \gg 1$} \label{MaldacenaGravity}

In the $g_s N \gg 1$ regime, the system is well approximated by type IIB closed strings, propagating on a curved target space sourced by the D3-branes as heavy objects. Such spacetime is described in terms of the following metric
\begin{eqnarray}
\label{D3_metric_z}
\notag
    \dd s^2_{\text{D3-brane}} &=& \frac{4 \pi  g_s N}{\sqrt{f(z)}} \left( -\dd t^2 + \dd x^i \dd x_i   \right) + \alpha'^2 \sqrt{f(z)} \left(\frac{\dd z^2}{z^4} + \frac{1}{z^2} \dd \Omega^2_5 \right) \, , \\
f (z) &=& 1 + \frac{4 \pi  g_s N}{\alpha'^2} z^4 \, , \\
    \notag
        \dd \Omega_5^2 &=& \left(\frac{4-y^2}{4+y^2}\right) \dd \phi^2 + \frac{16\ \dd y^2 }{\left( 4+y^2\right)^2} \, ,
\end{eqnarray}
where $g_s$ is the string coupling, $N$ is the number of D3-branes, $(t, x^i)$, with $i\in\{1,2,3\}$, are coordinates along the world-volume of the D3-brane, and $\dd \Omega_5^2$ is the metric of the unit 5-sphere. Here, we choose to describe the 5-sphere metric $\dd \Omega_5^2$ in terms of Cartesian coordinates $(\phi, y^m)$, with $m\in\{1,..., 4\}$ and $y^2 \equiv y^m y^n \delta_{mn}$.

Then, the crucial idea is that this metric admits two different regions: one asymptotic, and one ``near-horizon''. The asymptotic region, that is, the region $0\ll z \ll \frac{(\alpha')^{1/2}}{(4 \pi  g_s N)^{1/4}}$, is described by a 10d flat spacetime.\footnote{Technically speaking, the asymptotic region looks like a 3-brane NC geometry. This is because one has to rescale $z\to \varepsilon z$, with $\varepsilon$ a dimensionless parameter, and then take the limit $\varepsilon \to 0$.} The near-horizon region, namely the region $\frac{(\alpha')^{1/2}}{(4 \pi  g_s N)^{1/4}} \ll z \ll \infty$, is described by the AdS$_5\times$ S$^5$ metric. This can be seen by taking the limit $\alpha' \to 0$ of the metric \eqref{D3_metric_z}, leading to
\begin{eqnarray}\label{AdS5xS5_metric}
\dd s^2_{\text{AdS}_5\times\text{S}^5} &=& R^2 \bigg(\frac{-\dd t^2 +  \dd z^2 + \dd x^i \dd x_i}{z^2} + \dd \Omega_5^2 \bigg)\, , 
\end{eqnarray}
where we defined $R^2 \equiv \sqrt{4 \pi  g_s N} \alpha'$, i.e. the common AdS$_5$ and S$^5$ radius.

When considering low energy excitations, the spectrum of closed strings propagating in each of these two regions completely decouples, and their dynamics can be treated independently.

\subsection{The gauge perspective: $g_s N \ll 1$}

In the $g_s N \ll 1$ regime, the system is described by closed and open strings propagating in 10d Minkowski, and the interaction between them cannot be neglected at this stage. Here, the D3-branes play the role of boundary conditions for the open strings. 
The action describing this system is: 
\begin{eqnarray}
\label{S_initial}
    S = S_{\text{closed}} + S_{\text{open}} + S_{\text{interaction}} \, .
\end{eqnarray}
The part of the action $S_{\text{closed}}$ is given by the closed string action in flat 10d spacetime, while the part $S_{\text{open}} + S_{\text{interaction}}$ is described by the non-abelian Dirac-Born-Infeld (DBI) action \cite{Tseytlin:1997csa,Myers:1999ps}. 

For simplicity, let us consider the case of the abelian DBI, namely the case of $N=1$. The action is given by
\begin{equation}\label{abelian_DBI}
    S_{\text{DBI}} = - \frac{1}{(2 \pi)^3 \alpha'^2 g_s} \int \dd\sigma^4 e^{-\Phi} \sqrt{- \det \left(g_{\alpha\beta} + 2 \pi \alpha' F_{\alpha \beta}\right)} \, ,  
\end{equation}
where $\sigma^{\alpha} = (t, x^i)$ are the coordinates of the D3-brane world-volume, with $i\in\{1,...,3\}$;  $g_{\alpha\beta} \equiv \partial_{\alpha} X^{\mu} \partial_{\beta} X^{\nu} g_{\mu\nu}$ is the induced metric, $g_{\mu\nu}$ is the 10d target space metric, $F_{\alpha\beta}$ is the field strength on the D3-brane world-volume and $\Phi$ is the dilaton field. Let us consider static gauge, which amounts to fix the first four spacetime coordinates equal to the world-volume coordinates, i.e. $X^\alpha = \sigma^\alpha$. Additionally, since we want to zoom into the D3-brane world-volume, we rescale the remaining transverse coordinates as\footnote{This rescaling, as we will comment in section \ref{sec:NRasNH}, can actually be interpreted as the 3-brane NC limit of the DBI action.} 
\begin{eqnarray}\label{D3_brane_transverse_resc}
    X^{\hat{I}+3}=2\pi \alpha' S^{\hat{I}} \, , 
\end{eqnarray}
where $\hat{I} \in \{ 1, ..., 6\}$.
Finally, by expanding in small $\alpha'$ around the flat spacetime background $g_{\mu\nu} = \eta_{\mu\nu}$ and $\Phi = 0$, we get
\begin{equation}
    S_{\text{DBI}}\approx - \frac{1}{2\pi g_s} \int \dd^4\sigma \left[ \frac{1}{4} F_{\alpha\beta}F^{\alpha\beta} + \frac{1}{2} \partial_\alpha S^{\hat{I}} \partial^\alpha S^{\hat{I}} \right]  + \mathcal{O} (\alpha ') \, ,
\end{equation}
where we see that no interaction between open and closed strings survives.

If one repeats these steps with the full non-abelian DBI action (i.e. $N >1$), one would find the bosonic part of the $\mathcal{N}=4$ supersymmetric Yang-Mills theory in 4d, 
\begin{equation}\label{N=4_SYM}
    S^{\text{bos.}}_{\text{SYM}} = - \frac{1}{2\pi g_s} \int \dd^4\sigma \left[ \frac{1}{4} F_{\alpha\beta}^\mathtt{a} F^{\alpha\beta \, \mathtt{a}} + \frac{1}{2} D_\alpha S^{\hat{I}\, \mathtt{a}} D^\alpha S^{\hat{I}\, \mathtt{a}} - \frac{1}{4} (f_{\mathtt{bc}}{}^\mathtt{a} S^{\hat{I}\, \mathtt{b}} S^{\hat{J}\, \mathtt{c}}) (f_{\mathtt{de}}{}^\mathtt{a} S^{\hat{I}\, \mathtt{d}} S^{\hat{J}\, \mathtt{e}})\right] \, .
\end{equation} 
where $\hat{I} \in \{ 1, ..., 6\}$; $\mathtt{a, b}, ...$ are $U(N)$ gauge indices arising from Chan-Paton indices; $F_{\alpha\beta}^\mathtt{a}$ is the non-abelian field strength of a 1-form potential $A_{\alpha}^\mathtt{a}$, given by 
\begin{equation}
    F_{\alpha\beta}^\mathtt{a} \equiv \partial_{\alpha} A_{\beta}^\mathtt{a} - \partial_{\beta} A_{\alpha}^\mathtt{a} + f_{\mathtt{bc}}{}^\mathtt{a} A_{\alpha}^\mathtt{b}A_{\beta}^\mathtt{c} \, ,
\end{equation}
and the covariant derivative $D$ is defined as, 
\begin{eqnarray}
    D_\alpha S^{\hat{I}\, \mathtt{a}} \equiv \partial_{\alpha} S^{\hat{I}\, \mathtt{a}} + f_{\mathtt{bc}}{}^\mathtt{a} A_{\alpha}^\mathtt{b} S^{\hat{I}\, \mathtt{c}} \, , 
\end{eqnarray}
where $f_{\mathtt{ab}}\null^\mathtt{c}$ are the structure constants of $U(N)$.

\begin{figure*}[t!]
    \centering
    \includegraphics[keepaspectratio,width=\textwidth]{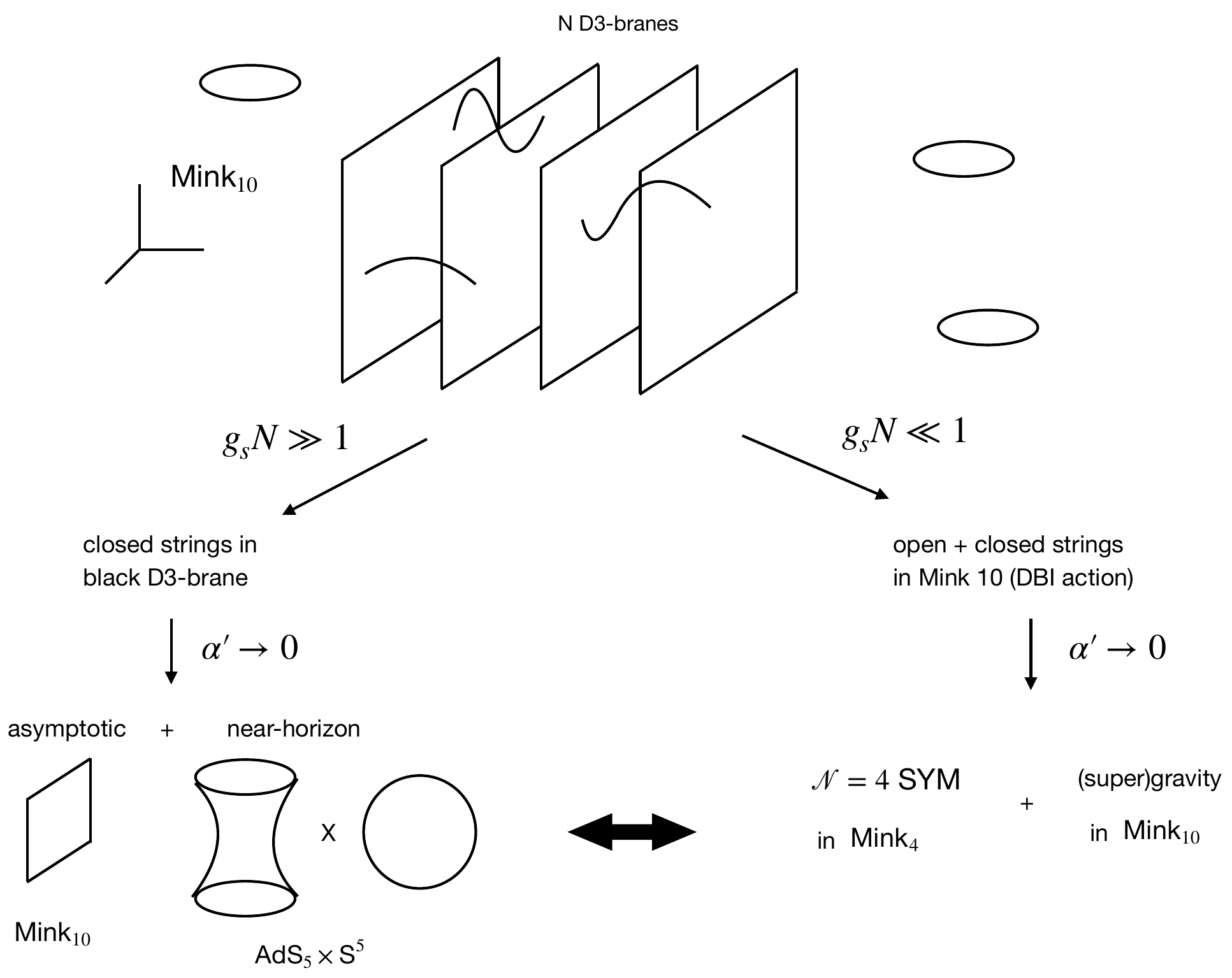}

  \caption{Maldacena's construction of the AdS$_5$/CFT$_4$.}
    \label{fig:Maldacena_holography}
\end{figure*}

\subsection{The holographic duality}

To better illustrate the main idea of Maldacena, let us consider a completely different physical situation: a viscous fluid. On the one hand, we can approximate it as a perfect fluid (no viscosity), and compute corrections to the Euler equations due to the non-vanishing viscosity. On the other hand, if the fluid is highly viscous, it is better described as small elastic deformations of a solid. Nevertheless, these are two descriptions of a fluid and, provided we compute corrections up to the appropriate accuracy, they have to match. In particular, a \emph{small elasticity} in the ``solid picture'' corresponds to a \emph{high viscosity} in the ``fluid picture'', and vice versa.

Coming back to the topic of gauge theories, it was observed by G. 't Hooft \cite{HOOFT1974461} that, for a large number of colours $N$, a gauge theory start displaying a geometrisation of its perturbative expansion. This is because the expansion at large number of colours is controlled by the Euler characteristic of the corresponding Feynman diagrams. To him, this was heavily reminiscent of the sum over topologies that appears in String Theory. This intuition is what Maldacena's holography formalises.

The idea of Maldacena is that the two \emph{perspectives} that we have presented are two descriptions of the same physics, a D3-brane, from different points of view. On the one hand, the $g_s N \gg 1$ limit gives us supergravity in asymptotically 10d flat spacetime together with type IIB supergravity in AdS$_5\times$S$^5$ near the horizon of the black D3-brane. On the other hand, the $g_s N \ll 1$ limit gives us supergravity in 10d flat spacetime together with $\mathcal{N}=4$ super Yang-Mills. If we ignore the supergravity in flat space that appears in both sides, we are led to believe that type IIB supergravity (i.e. the $\alpha'\to 0$ limit of string theory) in AdS$_5\times$S$^5$ and $\mathcal{N}=4$ super Yang-Mills should be identified as the same physical theory.

\section{Constructing a new holography from a limit of the AdS$_5$/CFT$_4$}
\label{sec:consistency_conditions}

The AdS/CFT duality is a beautiful example of holographic duality that is derived from examining different regimes of a given brane setting. A relevant question to ask is whether one could take a limit on a physical parameter (e.g. the AdS radius $R$, or the speed of light $c$) and still be able to  derive a new gauge/gravity correspondence from such brane construction. 
When taking a limit on a physical parameter entering in the brane construction of the AdS/CFT, there are consistency conditions, like axioms, that one has to demand:
\begin{enumerate}
    \item \textbf{Horizon condition.} \labeltext{`Horizon condition'}{Horizon condition} We demand that after taking the limit on the physical parameter, the D3-brane geometry still retains a near-horizon and an asymptotic region. This condition is necessary due to the fact that, in the weak string coupling regime, the DBI action at low energy decouples into a ``near-horizon'' gauge theory plus an ``asymptotic'' gravity theory, see fig. \ref{fig:Maldacena_holography}.
    \item \textbf{Uniqueness.} \labeltext{`Uniqueness'}{Uniqueness} The near-horizon (or low-energy) limit $\alpha'\to 0$ must commute with the limit on the physical parameter that we are interested in. This condition is necessary in order to guarantee a unique final theory.   
    \item \textbf{Holographic realisation.} \labeltext{`Holographic realisation'}{Holographic realisation} Since the duality must be \emph{holographic}, one has to demand that the gauge theory is defined on a spacetime that is precisely the (conformal) boundary of the spacetime where the gravity theory lives on. 
    \item \textbf{Symmetry matching.} \labeltext{`Symmetry matching'}{Symmetry matching} The symmetries of the gauge theory must be in a one-to-one correspondence with the symmetries acting on the boundary of spacetime of the gravity theory.\footnote{It is important to stress that only the symmetries that act at the \emph{boundary} of the gravitational theory needs to be matched holographically. There are coincidences, like for string theory in AdS$_5\times$S$^5$, where the symmetries at the boundary coincide with the symmetries in the bulk.}  
    \item \textbf{Quantitative test.} \labeltext{`Quantitative test'}{Quantitative test} There must exist a dictionary that relates  observables and their Hilbert spaces in the gauge theory to the ones in the gravity theory. The expectation values of these observables related by the dictionary need to match numerically. 
\end{enumerate}
At this point, this is a set of five consistency conditions that can be applied to any limit. The purpose of this review is to focus on the non-relativistic limit.  
In this section, we begin by considering the non-relativistic limit which, on the string theory side, leads to the Gomis–Gomis–Kamimura (GGK) theory \cite{Gomis:2005pg}. This is also called SNC limit, and it gives rise to String Newton-Cartan holography. In section \ref{sec:other_NR_limits}, we will consider a higher-dimensional generalisation to $p$-brane Newton-Cartan holographies. Finally, in section \ref{sec:Holo_other_NR_limits} we will apply this methodology to other types of limits, such as the Carroll and flat space limits. Throughout the discussion, we shall keep in mind the set of five consistency conditions listed above, and check whether they are fulfilled.

\subsection{The non-relativistic limit in the gravity side}
\label{sec:NR_limit_gravity}

The idea is to incorporate inside the AdS/CFT brane picture the non-relativistic limit that leads from type IIB string theory in AdS$_5\times$S$^5$ to the GGK non-relativistic string action \cite{Gomis:2005pg}. To do that, we shall demand the above five consistency conditions.

\vspace{-2mm}
\paragraph{The commuting limits.} The starting point is the metric for a stack of $N$ D3-branes given in \eqref{D3_metric_z}, 
\begin{subequations} \label{D3_metric_z_2}
\begin{align}
    \dd s^2_{\text{D3-brane}} &= \frac{4 \pi  g_s N}{\sqrt{f(z)}} \left( -\dd t^2 + \dd x^i \dd x_i   \right) + \alpha'^2 \sqrt{f(z)} \left(\frac{\dd z^2}{z^4} + \frac{1}{z^2} \dd \Omega^2_5 \right) \, , \\
    \dd \Omega_5^2 &= \left(\frac{4-y^2}{4+y^2}\right) \dd \phi^2 + \frac{16\ \dd y^2 }{\left( 4+y^2\right)^2} \, , \\
f (z) &= 1 + \frac{4 \pi  g_s N}{\alpha'^2} z^4 \, ,
\end{align}
\end{subequations}
where $g_s$ is the string coupling, $(t, x^i)$, with $i\in \{1,2,3\}$, are coordinates along the world-volume of the D3-brane, and $\dd \Omega_5^2$ is the metric of the unit 5-sphere. The 5-sphere metric $\dd \Omega_5^2$ is described in terms of Cartesian coordinates $(\phi, y^m)$, with $m\in\{1,..., 4\}$ and $y^2 \equiv y^m y^n \delta_{mn}$.

\vspace{4mm}
\noindent \emph{Near-horizon first, NR second.} Maldacena's ``near-horizon'' (or ``near-throat'') limit $\alpha'\to 0$ produces the famous AdS$_5\times$S$^5$ metric, 
\begin{eqnarray}
\dd s^2_{\text{AdS}_5\times\text{S}^5} &=& R^2 \bigg(\frac{-\dd t^2 +  \dd z^2 + \dd x^i \dd x_i}{z^2} + \dd \Omega_5^2 \bigg)\, . 
\end{eqnarray}
where $R^2 \equiv \sqrt{4 \pi  g_s N} \alpha'$ is the common AdS$_5$ and S$^5$ radius. As described in section \ref{sec:Maldacena_AdS/CFT}, the near-horizon limit has the effect of decoupling the AdS$_5\times$S$^5$ region inside the stack of D3-brane geometry from the Mink$_{10}$ asymptotic spacetime. 

Then, the GGK non-relativistic limit consists in rescaling the coordinates as\footnote{This rescaling is equivalent of the ones discussed in section \ref{subsec:SNCAdscoords}. The only difference is that, here, the Cartesian coordinates $(\phi, y^m)$ are dimensionless, whereas in section \ref{subsec:SNCAdscoords}, they are dimensionful. They differ by a factor of $R$. } 
\begin{eqnarray}
\label{NR_rescaling}
    R \to c \, R \, , \qquad
    x^i \to \frac{x^i}{c} \, , \qquad 
    \phi \to \frac{\phi}{c} \, , \qquad
    y^m \to \frac{y^m}{c} \,  .
\end{eqnarray}
By applying this rescaling to the AdS$_5\times$S$^5$ metric, and by taking the limit $c\to \infty$, one obtains the String Newton Cartan (SNC) geometry of AdS$_5\times$S$^5$, given by 
\begin{subequations}
\label{SNC_AdS5xS5_metric}
\begin{align}
    \dd s^2_{\text{SNC AdS}_5\times\text{S}^5} &= (c^2 \tau_{\mu\nu} + h_{\mu\nu}) \dd X^{\mu} \dd X^{\nu}\,  , \\
    \tau_{\mu\nu} \dd X^{\mu} \dd X^{\nu} &= \frac{R^2}{z^2} \left( -\dd t^2 + \dd z^2 \right) \, , \\
    h_{\mu\nu} \dd X^{\mu} \dd X^{\nu} &= \frac{R^2}{z^2} \dd x^i \dd x^j \delta_{ij} + R^2 \dd x^{i'} \dd x^{j'} \delta_{i' j'} \, , 
\end{align}
\end{subequations}
where we denoted collectively the flat coordinates originating from the 5-sphere as $x^{i'} = ( \phi, y^m )$, $i'\in\{5,...,9\}$.
The longitudinal metric $\tau$ describes AdS$_2$, whereas the transverse tensor $h$ describes $w(z) \mathbb{R}^3\times \mathbb{R}^5$, where $w(z) = R^2 / z^2$ is the warped factor for the transverse AdS directions.  As explained in section \ref{subsec:limit}, when taking this limit from the point of view of the string action, one has to introduce a critical closed B-field $b^{\text{crit.}}_{\mu\nu} = \tau_{\mu}{}^A \tau_{\nu}{}^B \varepsilon_{AB}$, which in this case corresponds to the volume form of AdS$_2$.  

The NR rescaling is also acting on the decoupled Mink$_{10}$ asymptotic spacetime, leading to SNC Mink$_{10}$. Here the dynamics is described by the low-energy limit of the NR closed string action in flat spacetime.

\vspace{4mm}
\noindent \emph{NR first, near-horizon second.} For consistency, the same GGK string theory should appear by inverting the order of the limits. If we implement the NR rescaling \eqref{NR_rescaling} first in the D3-brane metric \eqref{D3_metric_z_2}, and then take $c\to\infty$, the resulting geometry is immediately the one given in \eqref{SNC_AdS5xS5_metric}. Taking the limit in this straightforward way destroys the asymptotic flat region that is required in the brane construction of the AdS/CFT. In other words, the limit does not fulfil the \ref{Horizon condition} given at the beginning of section \ref{sec:consistency_conditions}. Instead, the correct non-relativistic rescaling turns out to be 
\begin{eqnarray}
\label{NR_rescaling_alpha_p}
    \alpha' \to c \, \alpha' \, , \qquad
    R \to c \, R \, , \qquad
    x^i \to \frac{x^i}{c} \, , \qquad 
    \phi \to \frac{\phi}{c} \, , \qquad
    y^m \to \frac{y^m}{c} \,  .
\end{eqnarray}
where $\alpha'$ needs to scale as well. Notice that we could instead use this NR rescaling in place of \eqref{NR_rescaling} for the ``near-horizon first, NR second'' ordering. In that case, the parameter $\alpha'$ drops out when taking the near-horizon limit first, so the metric becomes sensitive only to the scaling of $R$, and not to the independent scalings of $\alpha'$ and $g_s N$. Therefore, after the near-horizon limit is taken, \eqref{NR_rescaling_alpha_p} effectively acts in the same way as \eqref{NR_rescaling}. Hence, $\alpha' \to 0$ and the rescaling \eqref{NR_rescaling_alpha_p} are the commuting limits.

There is still a subtlety regarding the behaviour of $\alpha'$ in these limits. From the rescaling \eqref{NR_rescaling_alpha_p} it looks like $\alpha'$ becomes large, which is in conflict with the subsequent near-horizon limit where $\alpha'$ is small. This is resolved by the fact that the ``old'' $\alpha'$ needs to go to zero, therefore the ``new'' $\alpha'$ needs to go to zero faster than $1/c$.  

Since $R$ are $\alpha'$ are related, the rescaling \eqref{NR_rescaling_alpha_p} implies that $g_s N \to c^2 g_s N$. As we will explain in a moment, the dilaton also needs to rescale as $\Phi \to \ln c + \Phi$. This fixes the rescaling of $g_s = e^{\Phi}$ and $N$ as follows
\begin{eqnarray}
    g_s \to c g_s \, , \qquad
    N \to c N \, .
\end{eqnarray}
When the NR rescaling \eqref{NR_rescaling_alpha_p} is applied to the metric of a stack of D3-branes, it produces in the limit $c\to \infty$ the following String Newton-Cartan geometry
\begin{subequations}
\label{NR_D3_metric}
\begin{align}
    \dd s^2_{\text{SNC D3-brane}} &= \left(c^2 \tau_{\mu\nu} + h_{\mu\nu} \right) \dd X^{\mu} \dd X^{\nu}  \,  , \\
    \tau_{\mu\nu} \dd X^{\mu} \dd X^{\nu} &= -\frac{R^4}{\alpha^{\prime 2}} \frac{1}{\sqrt{f(z)}} \dd t^2 +  \alpha^{\prime 2}\sqrt{f(z)} \, \frac{\dd z^2}{z^4} \, , \\
    h_{\mu\nu} \dd X^{\mu} \dd X^{\nu} &= \frac{R^4}{\alpha^{\prime 2}} \frac{1}{\sqrt{f(z)}} \dd x^i \dd x^j \delta_{ij} + \alpha^{\prime 2}\sqrt{f(z)} \frac{1}{z^2} \dd x^{i'} \dd x^{j'} \delta_{i' j'}  \, , \\
    f(z)&= 1 + \frac{R^4}{\alpha^{\prime 4}} z^4 \, , 
\end{align}
\end{subequations}
which needs to be supplemented by the critical closed B-field $b^{\text{crit.}}_{\mu\nu} = \tau_{\mu}{}^A \tau_{\nu}{}^B \varepsilon_{AB}$ in order to ensure a well-defined limit at the level of the string action. Here again, we denoted collectively the (now warped) flat coordinates originating from the 5-sphere as $x^{i'} = ( \phi, y^m )$, $i'\in\{5,...,9\}$. 

This stack of SNC D3-branes geometry preserves the \ref{Horizon condition}, namely it has an asymptotic SNC flat spacetime region at $0\ll z \ll \frac{(\alpha')^{1/2}}{(4 \pi  g_s N)^{1/4}}$ and a ``near-throat'' region at $\frac{(\alpha')^{1/2}}{(4 \pi  g_s N)^{1/4}} \ll z \ll \infty$, which is captured by the limit $\alpha'\to 0$.

Finally, the near-horizon limit $\alpha'\to 0$ of \eqref{NR_D3_metric} precisely leads to the SNC AdS$_5\times$S$^5$ geometry \eqref{SNC_AdS5xS5_metric}. This is exactly the same geometry we obtain if we swap the order of the limits. Therefore, the \ref{Uniqueness} condition is fulfilled.

\paragraph{The Penrose boundary.} The conformal boundary of the SNC AdS$_5\times$S$^5$ metric \eqref{SNC_AdS5xS5_metric} can be studied using the Penrose procedure \cite{Penrose}. The idea is that points at infinite distance on a manifold $\mathcal{M}$ equipped with a metric $g$ can be mapped, via a conformal transformation, to points at \emph{finite} distance on a new manifold $\tilde{\mathcal{M}}$ with metric $\tilde{g}$. The two metrics $g$ and $\tilde{g}$ are related by the conformal transformation,   
\begin{eqnarray}
    \tilde{g}_{\mu \nu} = \Omega^2 g_{\mu \nu} \, . 
\end{eqnarray}
where $\Omega$ is a generic function of the spacetime coordinates. 
The points at infinite distance in $\mathcal{M}$ corresponds to the points where $g_{\mu \nu}$ becomes infinite. Since we demand $\tilde{g}_{\mu \nu}$ to remain finite, these points are solution of the equation $\Omega = 0$. The metric $\tilde{g}$ evaluated at the points where $\Omega = 0$ gives rise to a geometry that defines the \emph{conformal boundary} of the manifold $\mathcal{M}$.\footnote{In Penrose's formalism, one should demand an additional condition, namely that the gradient of $\Omega$ is non-vanishing on the conformal boundary, i.e. $\dd \Omega |_{\Omega = 0} \neq 0$. However, as pointed out by \cite{Norman} (see also the second footnote in page 182 of \cite{Penrose}), this condition is necessarily violated for some cosmological models and therefore can somewhat be dropped.}     

The same logic can be applied to a String Newton-Cartan geometry. In particular, the points at infinite distance of SNC AdS$_5\times$S$^5$ equipped with metric tensors \eqref{SNC_AdS5xS5_metric} are located at $z=0$. By performing the conformal rescaling, 
\begin{eqnarray}
    \tilde{\tau}_{\mu\nu} = \Omega^2 \tau_{\mu\nu} \, , \qquad\qquad
    \tilde{h}_{\mu\nu} = \Omega^2 h_{\mu\nu} \, ,\qquad\qquad \Omega= z^2\, , 
\end{eqnarray}
we obtain a new set of SNC metric tensors $(\tilde{\tau}, \tilde{h})$ where the divergence at $z=0$ has been removed. The Penrose conformal boundary of SNC AdS$_5\times$S$^5$ is then defined by $(\tilde{\tau}, \tilde{h})$ evaluated at $z=0$, namely
\begin{eqnarray}\label{Penrose_boundary_SNCAdS5xS5}
  \left. \tilde{\tau}_{\mu\nu} \dd X^{\mu}\dd X^{\nu} \right|_{z=0} = - \dd t^2 \, , \qquad\qquad
  \left. \tilde{h}_{\mu\nu} \dd X^{\mu}\dd X^{\nu} \right|_{z=0} = \dd x^i \dd x^j \delta_{ij} \, ,
\end{eqnarray}
which describes a 3+1d Newton-Cartan flat spacetime. It is interesting to observe that although the bulk is a String Newton-Cartan manifold, the conformal boundary is just a Particle Newton-Cartan geometry. The symmetries are expected to appropriately reduce at the boundary.

\paragraph{The full supergravity field content.} After having discussed the near-horizon and NR limits in the metric sector, we now move to the remaining supergravity fields. The AdS$_5\times$S$^5$ type IIB supergravity solution has a self-dual RR 5-form flux and a constant dilaton. All remaining supergravity fields are vanishing. The self-dual RR 5-form flux is proportional to the sum of the AdS$_5$ and S$^5$ volume forms \cite{Ortin:2015hya, Blau:2002dy}, 
\begin{align}\label{RR_5_form}
    F^{(5)} &= (g_s R)^{-1} \left[ \text{dvol}(\text{AdS}_5) + \text{dvol}(\text{S}^5) \right]  \\
    &= (g_s R)^{-1} \left[ \frac{R^5}{z^5} \dd t\wedge \dd z\wedge \dd x^1 \wedge\dd x^2 \wedge\dd x^3 + R^5 \sqrt{\frac{1-\frac{y^2}{4}}{\left( 1+\frac{y^2}{4} \right)^9}} \dd \phi \wedge \dd y_1 \wedge \dd y_2\wedge \dd y_3\wedge \dd y_4  \right]\, , \notag
\end{align} 
where $g_s = e^{\Phi}$, with $\Phi$ a constant dilaton. In order to get a RR 5-form flux that fullfils the NR type IIB supergravity equations of motion \cite{Bergshoeff:2023ogz}, the NR rescaling \eqref{NR_rescaling_alpha_p} needs to be supplemented by a rescaling of the dilaton $\Phi \to \ln c + \Phi$, or equivalently $g_s\to c g_s$. Then, in the limit $c\to \infty$, the NR 5-form flux becomes 
\begin{eqnarray}
\label{NR_RR_5_form}
    F^{(5)}_{\text{\scriptsize NR\normalsize}} = (g_s R)^{-1} \text{dvol}(\text{AdS}_5) \, . 
\end{eqnarray}
The fact the volume form of the 5-sphere dropped out has a geometrical interpretation: since the 5-sphere has flattened out in the NR limit, there is no need to support a flat geometry via a flux. Moreover, the NR RR 5-form flux \eqref{NR_RR_5_form} is no longer self-dual. However, the self-duality constraint from the Lorentzian theory still imposes a pair of conditions \cite{Bergshoeff:2023ogz}, which are fulfilled by the solution \eqref{NR_RR_5_form} and by the fact the remaining RR fluxes are vanishing. 

The NR type IIB supergravity requires the presence of a 5-form Lagrange multiplier, $\mathcal{A}^{(5)}$. For the solution here presented, $\mathcal{A}^{(5)}$ is proportional to $F^{(5)}_{\text{\scriptsize NR\normalsize}}$, i.e.  
\begin{eqnarray}
  \mathcal{A}^{(5)} = \kappa F^{(5)}_{\text{\scriptsize NR\normalsize}} \, ,   \qquad\qquad
  \kappa = \text{const.}
\end{eqnarray}

\subsection{The non-relativistic limit in the gauge theory side}
\label{sec:NR_gauge_side}

The idea now is to consider the system of D3-branes at weak coupling regime and implement a NR limit that is the corresponding analogue of the NR limit taken in the strong coupling regime (i.e. the bulk theory). The parameter `$c$' entering in the bulk theory is not necessarily the same `$c$' entering in the boundary theory. To find such limit, we will demand the consistency conditions outlined at the beginning of section \ref{sec:consistency_conditions}.

The system of D3-branes at weak coupling is described by interacting open and closed strings in an ambient geometry that is, at first order approximation, 10d flat spacetime. The closed string action is captured by the usual Polyakov action. The dynamics of the open strings together with the interaction with the closed strings, instead, is given by the non-abelian DBI action \cite{Tseytlin:1997csa,Myers:1999ps}. The formula of the abelian DBI action (i.e. $N=1$) was given in \eqref{abelian_DBI}.        

The idea is that the NR limit will act only on the spacetime coordinates $X^{\mu}$, and not on the world-volume coordinates $\sigma^{\alpha}$. Static gauge, i.e. $X^{\alpha} = \sigma^{\alpha}$, will be imposed only at the very end, so it does not interfere with any decoupling limit, and will have the role of selecting
the D3-brane world-volume as the spacetime for the dual gauge theory. 

At strong coupling (gravity side) we applied a \emph{stringy} NR limit (SNC limit), i.e. after taking the limit the spacetime is a co-dimension two singular foliation. The same SNC limit should be implemented at weak coupling. Here, there are two obvious possibilities to implement an SNC rescaling in 10d flat spacetime, either
\begin{eqnarray}\label{NR_resc_DBI_1}
   \text{limit 1:}\hspace{1cm} X^A \to c X^A \, , \qquad 
    X^a \to  X^a \, , 
\end{eqnarray}
or 
\begin{eqnarray}\label{NR_resc_DBI_2}
    \text{limit 2:}\hspace{1cm} X^A \to X^A \, , \qquad 
    X^a \to \frac{X^a}{c}  \, , 
\end{eqnarray}
where $A\in\{0,9\}$ and $a\in\{1, ..., 8\}$. Notice that there is no conformal symmetry acting on the 10d flat coordinates, therefore these two limits are inequivalent.

Demanding the \ref{Holographic realisation} condition fixes the spatial longitudinal direction of the SNC limit (i.e. $X^9$) to not belong to the longitudinal D3-brane coordinates, which instead will be taken as $\{X^0, ..., X^3\}$. This prevents the gauge theory from living on a \emph{String} Newton-Cartan flat spacetime, and instead places it on the appropriate Penrose boundary. Schematically, the SNC limit and D3-brane coordinates are taken as:
\begin{align}\label{SNC_D3}
\begin{array}{rrrrrrrrr}
D3: & 0 & 1 &2&3& & & & \\
SNC:& 0 &  & & & & & & 9 \  \, . \\
\end{array}
\end{align}

\paragraph{The commuting limits.} The starting point is the non-abelian DBI theory in the presence of a B-field. The reader can refer to \cite{Tseytlin:1997csa,Myers:1999ps} for the action's formula. 

\vspace{4mm}
\noindent \emph{Low-energy first, NR second.} First, one needs to perform the rescaling that zooms into the D3-brane world-volume, 
\begin{eqnarray}\label{low_energy_DBI}
   X^{I+3}=2\pi \alpha' S^I \, , \qquad    
   X^{9}=2\pi \alpha' \zeta  \, ,
\end{eqnarray}
where $I, J, ...$ run from 1 to 5 only. The sixth scalar $S^6 \equiv \zeta$ has been written with a different letter, to emphasise that it represents the longitudinal spatial direction for the SNC limit \eqref{SNC_D3}. In terms of the hatted indices, $\hat{I} = (I, 6)$. Moreover, in this limit one needs to turn on a B-field along the $X^0$ and $X^9$ target space directions, $B_{90}$, which is left unfixed for the moment.

By taking the low energy limit $\alpha'\to 0$, one arrives to an action that, upon setting the B-field to zero and fixing static gauge, is the bosonic sector of the $\mathcal{N}=4$ super Yang-Mills theory. Such Lagrangian was given in \cite{Fontanella:2024rvn}, and we report it here:
\begin{align}\label{N=4_SYM_Bfield}
    \mathcal{L}_{\text{SYM + B}}&=-\frac{ \left(\prod_{\alpha=0}^3 \partial_{\alpha} X^{\alpha}\right)}{2\pi g_s} \sum_{\mathtt{a}=1}^{N^2} \left[ \frac{1}{2} \sum_{I=1}^5 \sum_{i=1}^3 \left(\frac{(D_i S^{I \mathtt{a}})^2}{(\partial_i X^i)^2} - \frac{(D_0 S^{I \mathtt{a}})^2}{(\partial_0 X^0)^2}  \right)   \right.  \notag \\
    &+\frac{1}{2} \sum_{i=1}^3 \left(\frac{(D_i \zeta^\mathtt{a})^2}{(\partial_i X^i)^2} - \frac{(D_0 \zeta^\mathtt{a})^2}{(\partial_0 X^0)^2} \right)+ \frac{(B_{90} D_i \zeta^\mathtt{a} (\partial_0 X^0) + F_{i0})^2}{2(\partial_0 X^0\partial_i X^i)^2} +  \frac{1}{4} \frac{F_{ij} F^{ij}}{(\partial_i X^i \partial_j X^j)^2} \notag \\
    &\left. +\sum_{\mathtt{b},\mathtt{c}=1}^{N^2} \left(2 B_{90} f_\mathtt{bc}\null^\mathtt{a} \zeta^\mathtt{b}S^{I \mathtt{c}} \frac{D_0 S^{I \mathtt{a}}}{\partial_0 X^0}  + \sum_{I=1}^5 \left( f_{\mathtt{bc}}\null^\mathtt{a} \zeta^\mathtt{b} S^{I \mathtt{c}} \right)^2 - \sum_{\substack{I,J=1\\I< J}}^5 \left( f_{\mathtt{bc}}\null^\mathtt{a} S^{I \mathtt{b}} S^{J \mathtt{c}} \right)^2 \right)  \right] \, , 
\end{align}
where $i, j, ...$ run from 1 to 3; $\alpha = (0, i)$; $\mathtt{a, b}, ...$ are $U(N)$ gauge indices.

The next step is to take the SNC limit of \eqref{N=4_SYM_Bfield} at the level of the target space coordinates. There are two possibilities for doing that, given by the rescalings \eqref{NR_resc_DBI_1} and \eqref{NR_resc_DBI_2}. Here we describe the two limit scenarios. 

\vspace{2mm}
\noindent {\bf Limit 1}: The limit in \eqref{NR_resc_DBI_1} translates into the following rescaling:
\begin{eqnarray}
    X^0 \to c X^0\, , \qquad
    \zeta \to c \,\zeta \, ,\qquad 
    X^i \to X^i \, , \qquad
    S^I \to S^I \, . 
\end{eqnarray}
By taking this rescaling inside the action \eqref{N=4_SYM_Bfield}, one discovers that the action expands with a leading term of order $c^3$, which does not contain all field content, and a next-to-leading term of order $c$, which does retain all fields. At this stage, one has the freedom of rescaling the dilaton, or equivalently, $g_s$. This can be used to make the next-to-leading term finite by rescaling $\Phi \to \ln c + \Phi$, or equivalently, $g_s \to c g_s$. In this way, the leading term is of order $c^2$ and given by:      
\begin{equation}\label{GYM_div}
    \mathcal{L}_{\text{div.}}=-\frac{c^2}{4\pi g_s} (1-B_{90}^2 ) \sum_{\mathtt{a}=1}^{N^2} \left[ (D_i \zeta^\mathtt{a})^2 + \sum_{\mathtt{b},\mathtt{c}=1}^{N^2}\sum_{I=1}^5 \left( f_{\mathtt{bc}}\null^\mathtt{a} \zeta^\mathtt{b} S^{I \mathtt{c}} \right)^2\right] \, .
\end{equation}
Such term is eliminated by fixing the B-field,\footnote{Another choice of B-field that leads to a finite action would be $B_{90} = \pm 1 + \frac{\text{const.}}{c^2}$. However, this is not a good choice for two reasons. First, the bulk analogue of the B-field does not have this form. Second, the term \eqref{GYM_div} breaks some of the symmetries of \eqref{GYM}. This can be seen by turning off the scalar fields $S^I$ and the gauge field $A_{\alpha}$. In this setting, the spacetime symmetries are expected to not change. However, the finite action will contain a Klein-Gordon term for $\zeta$, which is Lorentz invariant, hence breaking the holographic realisation of the symmetries, as we will comment later.  } 
\begin{eqnarray}\label{B_fix}
    B_{90} = \pm 1 \, . 
\end{eqnarray}
Then we are left just with the finite next-to-leading term, given by 
\begin{align}\label{GYM}
    \mathcal{L}_{\text{GYM + scal.}} &= -\frac{1}{2\pi g_s } \sum_{\mathtt{a}=1}^{N^2} \left( \frac{1}{2} \sum_{I=1}^5 \bigg[ (D_i S^{I \mathtt{a}})^2 \mp \sum_{\mathtt{b},\mathtt{c}=1}^{N^2} 2 f_{\mathtt{bc}}\null^\mathtt{a} \zeta^\mathtt{b} S^{I \mathtt{c}} D_0 S^{I \mathtt{a}}\bigg]  \right. \notag  \\ 
    &\left. - \sum_{\mathtt{b},\mathtt{c}=1}^{N^2}\sum_{I<J} \left( f_{\mathtt{bc}}\null^\mathtt{a} S^{I \mathtt{b}} S^{J \mathtt{c}} \right)^2 + \frac{1}{4} (F^\mathtt{a}_{ij})^2 \pm F^\mathtt{a}_{i0} D_i \zeta^\mathtt{a} - (D_0 \zeta^\mathtt{a})^2 \right) \, .
\end{align}
where the sign $\pm$ reflects the choice of the B-field \eqref{B_fix}, and can be reabsorbed into a sign redefinition of $\zeta$. This action, when the five scalar fields $S^I$ are set to zero, is known as \emph{Galilean Yang-Mills} (GYM) \cite{Bagchi:2015qcw}.

\vspace{2mm}
\noindent {\bf Limit 2}: The second limit given in \eqref{NR_resc_DBI_2} consists in the following rescaling:
 \begin{eqnarray}
    X^0 \to X^0\, , \qquad
    \zeta \to \,\zeta \, ,\qquad 
    X^i \to \frac{X^i}{c} \, , \qquad
    S^I \to \frac{S^I}{c} \, . 
\end{eqnarray}
In order to make this limit commuting with $\alpha'\to 0$, one is required to supplement it with a rescaling of the 1-form gauge field as
\begin{eqnarray}
    A_{\alpha} \to \frac{A_{\alpha}}{c^2} \, . 
\end{eqnarray}
Rescaling the 1-form has the effect of abelianising the original action \eqref{N=4_SYM_Bfield}, since the commutators introduce higher powers of $A_{\alpha}$, which disappear as sub-leading terms in the limit $c\to \infty$. 
By taking this rescaling into the action \eqref{N=4_SYM_Bfield}, one finds a leading term of order $c^{-1}$, which does not contain the full field content, and a next-to-leading term of order $c^{-3}$, which instead does contain all fields. Again, one can use the freedom to rescale the dilaton as $\Phi \to -3 \ln c + \Phi$ or equivalently, $g_s \to c^{-3} g_s$, such that the next-to-leading term is finite. 
The leading term now becomes of order $c^2$, given by,
\begin{equation}
    \mathcal{L}_{\text{div.}}=-\frac{c^2}{4\pi g_s} (1-B_{90}^2 ) \sum_{\mathtt{a}=1}^{N^2} \partial_i \zeta^\mathtt{a}  \partial^i \zeta^\mathtt{a}  \, ,
\end{equation}
which is set to zero by fixing the B-field as in \eqref{B_fix}. After taking $c\to \infty$, one remains with the finite next-to-leading action, 
\begin{equation}
\label{GED_action}
    \mathcal{L}_{\text{GED + scal.}}= -\frac{1}{2\pi g_s} \sum_{\mathtt{a}=1}^{N^2}\left( \frac{1}{2} \sum_{I=1}^5 (\partial_i S^{I \mathtt{a}})^2 + \frac{1}{4} (F^\mathtt{a}_{ij})^2 \mp F^\mathtt{a}_{i0} \partial_i \zeta^\mathtt{a} -\frac{1}{2} (\partial_0 \zeta^\mathtt{a})^2 \right) \, ,
\end{equation}
where, again, the sign $\pm$ reflects the choice of the B-field \eqref{B_fix}, and can be reabsorbed into a sign redefinition of either $\zeta$ or $A_{\alpha}$. This action, when the five scalar fields $S^I$ are set to zero and when $N=1$, is known as \emph{Galilean Electrodynamics} (GED) \cite{Santos:2004pq}. One should note that the action \eqref{GED_action} describes $N^2$ copies of GED supplemented by 5 scalar fields. The index `$\mathtt{a}$' is no longer carrying the non-abelian structure, instead it has the role of counting the number of copies.

\vspace{4mm}
\noindent \emph{NR first, low-energy second.} As demanded by the \ref{Uniqueness} condition of section \ref{sec:consistency_conditions}, reversing the order of taking the limits must produce the same final theory. Let us first take the SNC limit and discuss the two scenarios separately.

\vspace{2mm}
\noindent {\bf Limit 1}: The SNC rescaling is the following, 
\begin{eqnarray}
    X^0 \to c X^0\, , \qquad
    X^9 \to c X^9 \, ,\qquad 
   \Phi \to \ln c + \Phi \, , 
\end{eqnarray}
with the remaining fields invariant. To cancel the divergence from the DBI, one has to set $B_{90} = 1$. After applying this rescaling to the non-abelian DBI, one arrives to a SNC non-abelian DBI, which we refrain from writing down. The next step is to take the low-energy limit, given by rescaling the fields as in \eqref{low_energy_DBI}, and then take $\alpha' \to 0$. This leads precisely to the GYM action with 5 scalar fields given in \eqref{GYM}.

\vspace{2mm}
\noindent {\bf Limit 2}: In this case, the NR rescaling is given by,
\begin{eqnarray}
    X^i \to \frac{X^i}{c} \, , \qquad
    X^{I+3} \to \frac{X^{I+3}}{c} \, , \qquad
    \Phi \to -3 \ln c + \Phi \, , \qquad
    A_{\alpha} \to \frac{A_{\alpha}}{c^2} \, ,
\end{eqnarray}
with the remaining fields invariant. Again, to cancel the divergence from the DBI, one has to set $B_{90} = 1$. After applying this rescaling to the non-abelian DBI, one arrives to an action that was given in \cite{Fontanella:2024rvn}, see also \cite{Gomis:2020fui} for the abelian action obtained from a space-filling brane. The next step consists in taking the low-energy limit, again given by rescaling the fields as in \eqref{low_energy_DBI}, and then take $\alpha' \to 0$. This leads precisely to $N^2$ copies of the GED action with 5 abelian scalar fields given in \eqref{GED_action}. It is interesting to point out that the anisotropic rescaling of $A_{0}$ and $ A_{i}$ with different powers of $c$ does not lead to a consistent commuting limit with $\alpha'\to 0$.      

\vspace{2mm}
The above SNC rescalings apply not only to the DBI action, but also to the closed string action appearing in \eqref{S_initial} when describing the system at weak coupling. When taking $c\to \infty$ and $\alpha'\to 0$, this matches the low-energy limit of the NR closed string action in flat spacetime expected from the gravity side (i.e. the same term appears on both sides of the duality).
At this stage, both limits 1 and 2 are consistent limits on the gauge theory side. Moreover, both limits produce theories that are defined on the Penrose conformal boundary of the bulk SNC AdS$_5 \times$S$^5$ geometry. However, only one of them can be the right holographic dual of the GGK string theory.\footnote{A first indication that the GED theory with scalar fields cannot be the holographic dual of GGK theory comes from the rescaling of the dilaton, as it scales differently in the gauge and gravity pictures. Instead, the dilaton scales in the correct way in the GYM theory with scalar fields. }  The answer comes precisely by demanding the \ref{Symmetry matching} of section \ref{sec:consistency_conditions}, which we are going to discuss next.

\subsection{Matching the symmetries}
\label{sec:matching_symmetries}

Accordingly to the holographic paradigm, the global symmetries of the bulk theory need to be in a one-to-one correspondence with the global (i.e. spacetime and internal, to distinguish from gauge) symmetries of the dual gauge theory. In the bulk theory, typically, there are two types of symmetries that need to be conceptually distinguished: the symmetries that act inside the bulk, $G_{bulk}$, and the symmetries that act asymptotically at the boundary, $G_{asym}$. On the other hand, this distinction is not present at the level of the dual gauge theory, which is expected to be invariant under a unique set of symmetries, called $G_{gauge}$. The natural question is which of the bulk symmetries need to holographically match the gauge theory symmetries. Since the bulk observables organise themselves into representations of the asymptotic symmetries, then one has to demand that $G_{asym} = G_{gauge}$, in a holographic sense. For example, this can be seen from the S-matrix perspective, as the ingoing and outgoing states are prepared at the boundary. The distinction between $G_{bulk}$ and $G_{asym}$ is clear in the context of holography in AdS$_3$ \cite{Brown:1986nw, Giveon:1998ns, deBoer:1998gyt}, and flat spacetime \cite{Bondi:1962px, Sachs:1962zza, Sachs:1962wk}, whereas AdS$_5$/CFT$_4$ is a special case where $G_{bulk}=G_{asym}$.  
As we shall discuss below, holography in SNC AdS$_5\times$S$^5$ is of the first type, namely where the bulk and the asymptotic symmetries are different.

\begin{figure*}[t!]
    \centering
    \includegraphics[keepaspectratio,width=\textwidth]{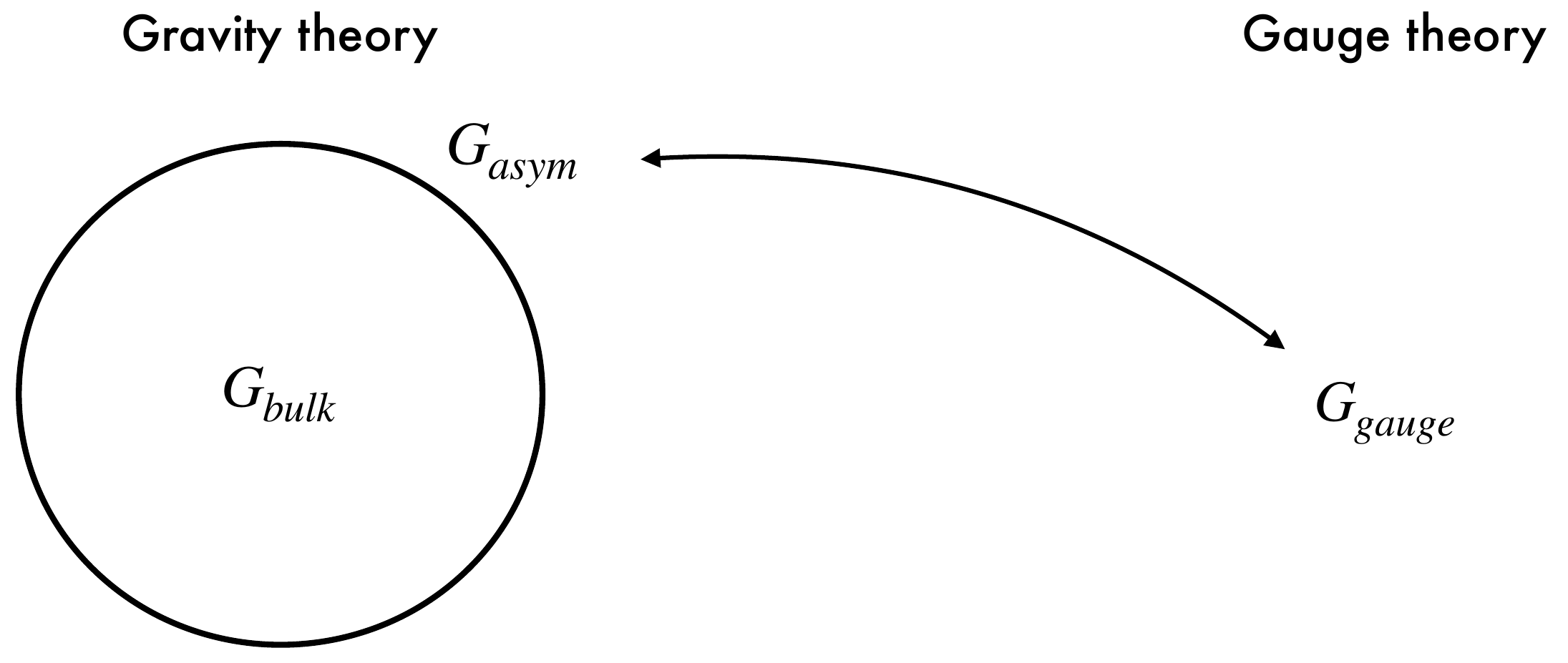}

  \caption{In the gravity theory, there are two conceptually different symmetries: the symmetries acting in the bulk and the asymptotic symmetries acting at the boundary. Only the asymptotic symmetries need to holographically match the symmetries of the dual gauge theory. }
    \label{fig:symmetries}
\end{figure*}

\paragraph{Asymptotic symmetries of the gravity theory.} To determine the asymptotic symmetries, we need first to understand the bulk global symmetries of the NR type IIB superstring action in SNC AdS$_5\times$S$^5$. They are given by the set of the Killing vectors that preserve the full SNC AdS$_5\times$S$^5$ solution of NR type IIB supergravity found in \cite{Bergshoeff:2023ogz}, see also section \ref{sec:global_symmetries}. They can be found by solving a set of equations of the type:
\begin{eqnarray}\label{KV_full_IIB_NR_sugra}
     \mathsterling_{\xi} \, \Theta = \delta \Theta \, , 
\end{eqnarray}
where $\Theta \in \{\tau_{\mu\nu}, h_{\mu\nu}, \mathfrak{B}_{\mu\nu}, \Phi, F^{(5)}, F^{(3)}, F^{(1)}, \mathcal{A}^{(5)}\}$ stands for all field content of NR type IIB supergravity, and $\delta \Theta$ correspond to the local symmetries acting on the field $\Theta$ that leave the NR type IIB supergravity action invariant. 

Solving \eqref{KV_full_IIB_NR_sugra} for the full SNC AdS$_5\times$S$^5$ solution presented in section \ref{sec:NR_limit_gravity} gives the following bulk symmetries $G_{bulk}$, found in \cite{Fontanella:2024rvn}:\footnote{The differential notation used here can be translated into infinitesimal variation of the fields as follows. Any generator of the form $A\frac{\partial \phantom{B}}{\partial B}$ is equivalent to $\delta B=\alpha A$ for an infinitesimal real parameter $\alpha$. For example, the family of symmetries $x^i \partial_j - x^j \partial_i$ is equivalent to the family of variations $\delta x^i=\omega_{ij} x^j$ with $\omega_{ij}$ an antisymmetric rank-2 tensor of infinitesimal parameters.}
\begin{subequations}\label{KV_bulk}
\begin{align}
    H &= \partial_t \, , & &\\
    D &= t\, \partial_t + z \, \partial_z + x^i \partial_i \, , & &\\
    K &= (t^2 + z^2) \partial_t +2 \,z\, t\, \partial_z +2 \, t \, x^i \partial_i\, , & &\\
    P^{(n, \pm)}_i &= (t\pm z)^n (t\mp nz) \partial_i \, ,  &P^{(n, \pm)}_{i'} &=  (t\pm z)^n  \partial_{i'} \, ,  \\
    J_{ij} &= x^i \partial_j - x^j \partial_i \, ,   &J_{i'j'} &= x^{i'} \partial_{j'} - x^{j'} \partial_{i'} \, .
\end{align}
\end{subequations}
The set of Killing vectors $\{ H, D, K \}$ generates an $\mathfrak{sl}(2, \mathbb{R})$ subalgebra, corresponding to the isometries of the AdS$_2$ $\tau$-metric. The Killing vectors $\{ P^{(n, \pm)}_i, P^{(n, \pm)}_{i'} \}$ correspond to translations of the transverse fields that depend on the longitudinal coordinates $x^{\pm} \equiv t\pm z$ of the metric $\tau$, e.g. $\delta x^{i'} = f^{i'}_{\pm}(x^{\pm})$. Finally, the set $\{ J_{ij}, J_{i'j'} \}$ generates the $\mathfrak{so}(3) \oplus \mathfrak{so}(5)$ rotation of the transverse spatial coordinates.

The asymptotic symmetries $G_{asym}$ consist of the set of Killing vectors that leave invariant the metric structure and the supergravity forms at the boundary $z=0$. The rigorous analysis would require to perform a near-boundary expansion of the bulk metric structure and supergravity forms, together with fixing suitable boundary conditions (\`a la Brown-Hennaux), which has not been done yet. A first initial consideration \cite{Fontanella:2024rvn} consists in noticing that $G_{asym}$ consists at least of $G_{bulk}$ evaluated at the boundary $z=0$. There are some indications why there should be no further enhancement. First, the boundary geometry is four dimensional (split into 3+1), so it does not share the special features leading to the symmetry enhancement occurring in two dimensions, e.g. at the 2d boundary of AdS$_3$ \cite{Brown:1986nw, Giveon:1998ns, deBoer:1998gyt}. Second, both bulk and boundary geometries are degenerate, as opposite to the case of flat spacetime, where the Lorentzian bulk metric transitions to a Carrollian geometry at the null boundary \cite{Raclariu:2021zjz, Donnay:2022aba, Donnay:2022wvx, Bagchi:2016bcd, Bagchi:2023cen, Bagchi:2025vri, Nguyen:2025zhg, Ruzziconi:2026bix}. Degenerate metrics have more symmetries than non-degenerate ones, motivating the enhancement of symmetries in flat spacetime when reaching the boundary, which is not expected in SNC AdS$_5\times$S$^5$. Another type of enhancement could come from symmetries of the metric structure that are not symmetries of the supergravity fields in the bulk, but they become only asymptotically. However, this cannot be the case neither, as all bulk symmetries of the metric structure have been shown to also be symmetries of the supergravity fields in \cite{Fontanella:2024rvn}.   

Given these reasonings, the expected $G_{asym}$ have been proposed \cite{Fontanella:2024rvn}:
\begin{subequations}\label{KV_asym}
\begin{align}
    H &= \partial_t \, , & &\\
    D &= t\, \partial_t + x^i \partial_i \, , & &\\
    K &= t^2 \partial_t +2 \, t \, x^i \partial_i\, , & &\\
    P^{(n)}_i &= t^{n+1} \partial_i \, ,  & P^{(n)}_{i'} &=  t^n  \partial_{i'} \, ,  \\
    J_{ij} &= x^i \partial_j - x^j \partial_i \, ,   & J_{i'j'} &= x^{i'} \partial_{j'} - x^{j'} \partial_{i'} \, .
\end{align}
\end{subequations}
It is interesting to note that $G_{\text{asym}} \subset G_{\text{bulk}}$, which is the opposite of what occurs in the AdS$_3$ and flat spacetime cases. There is an elegant geometric explanation for this behaviour. Since both the bulk and boundary geometries are degenerate, they each possess as a symmetry the translations of the transverse fields that depend on the $\tau$-metric coordinates. In the bulk, $\tau$ is two-dimensional, and the two towers of ``non-relativistic supertranslations'' $\{ P^{(n,\pm)}_i,\, P^{(n,\pm)}_{i'} \}$ are labelled by the sign $\pm$, which denotes the possibility of performing shifts that depend on $x^{\pm}$. In contrast, the boundary geometry is characterised by a one-dimensional $\tau$-metric. In this case, there is no ambiguity in the choice of longitudinal direction involved in the shift, as it can only be $t$. Consequently, the two towers of supertranslations collapse into a single one. This is a direct consequence of the fact that the geometry changes from being String Newton–Cartan (a co-dimension two foliation) in the bulk to being simply Newton–Cartan (a co-dimension one foliation) at the boundary.

Another remark is that the generators $\{P^{(n)}_{i'},J_{i'j'}\}$ act at the ``9d boundary'' of SNC AdS$_5\times$S$^5$, understood as the submanifold located at $z=0$. This should not be confused with the Penrose \emph{conformal} boundary of SNC AdS$_5\times$S$^5$, which is a 3+1d NC flat spacetime, and it is the spacetime where the dual gauge theory needs to be defined. From the dual gauge theory perspective, $\{P^{(n)}_{i'},J_{i'j'}\}$ gain meaning of internal symmetries, as we are going to discuss next.

\paragraph{Symmetries of the gauge theory.} In section \ref{sec:NR_gauge_side}, we have seen that there are two candidates of dual gauge theories: one is GYM with five interacting scalar fields, and the other one is GED with five free scalar fields. Both of them fulfil the consistency conditions \ref{Uniqueness} and \ref{Holographic realisation} of section \ref{sec:consistency_conditions}. However, only one of them satisfies the condition \ref{Symmetry matching}, and it is the GYM theory with five interacting scalar fields.

\vspace{4mm}
\noindent \emph{GYM with five interacting scalar fields.} The symmetries $G_{gauge}$ are \cite{Fontanella:2024hgv}:
\begin{subequations}\label{GYM_sym}
\begin{align}
D&=t \partial_t +x^i\partial_i\, , & H&=\partial_t\, , &  K&=t^2\partial_t+2 \, t\, x^i\partial_i\, ,\\
M^{(n)}_i &= t^{n+1}\partial_i\, ,  & & &J_{ij}&=x^i\partial_j-x^j\partial_i\, ,  \\
L^{(n)}_{0I} &= t^n \zeta \frac{\partial\phantom{S^I}}{\partial S^I} - t^n S^I \frac{\partial\phantom{S^I}}{\partial A_0} \, , & & & L_{IJ} &= S^I \frac{\partial\phantom{S^I}}{\partial S^J} - S^J \frac{\partial\phantom{S^I}}{\partial S^I} \, , 
 \end{align}
\end{subequations}
where the gauge indices of the fields are left implicit. The set $\{ H, M^{(n)}_i, J_{ij}\}$ forms the Milne algebra first introduced in \cite{Carter:1993aq}, see also \cite{Duval:1993pe}. On the other hand, $\{ H, D, K\}$ form an $\mathfrak{sl}(2, \mathbb{R})$ algebra, namely the finite part of the conformal algebra in two dimensions. 
For this reason, the set $\{ H, D, K, M^{(n)}_i, J_{ij}\}$ was named as \emph{Conformal Milne Algebra} (CMA), as first introduced in \cite{Fontanella:2024hgv}.\footnote{A different type of ``conformal Milne algebra'' was introduced in \cite{Duval:2009vt}, which however it does not contain the Milne subalgebra (in particular, it does not contain the full infinite tower $M_i^{(n)}$). Such algebra should not be confused with what we call here conformal Milne algebra, which does contain the Milne subalgebra.} They act as spacetime symmetries in the action \eqref{GYM}. The set $\{L^{(n)}_{0I}, L_{IJ}\}$ generates an infinite dimensional algebra, and they act instead as internal symmetries.   
There is a one-to-one correspondence between $G_{gauge}$ and $G_{asym}$, with dictionary given in Table \ref{tab:match_sym}. 

This theory was proposed in \cite{Fontanella:2024rvn, Fontanella:2024kyl} as the holographic dual of GGK string theory, based on fulfilling the condition \ref{Symmetry matching} of section \ref{sec:consistency_conditions}.
\renewcommand{\arraystretch}{1.5}
\begin{table}
\begin{center}
        \begin{tabular}{|c||c|c|c|c|c|c|c|}
        \hline
         $G_{gauge}$ & $H$ & $D$ & $K$ & $M_i^{(n)}$ & $J_{ij}$ & $L^{(n)}_{0I}$ & $L_{IJ}$ \\
         \hline
         $G_{asym}$ & $H$ & $D$ & $K$ & $P_i^{(n)}$ & $J_{ij}$ & $P^{(n)}_{i'}$ & $J_{i'j'}$ \\
         \hline
    \end{tabular}
    \caption{One-to-one correspondence between gauge theory spacetime and internal symmetries, $G_{gauge}$, and asymptotic SNC AdS$_5\times$S$^5$ Killing vectors, $G_{asym}$.}
      \label{tab:match_sym}
    \end{center}
\end{table}

\vspace{4mm}
\noindent \emph{GED with five free scalar fields.} The symmetries of this theory have also been found in \cite{Fontanella:2024hgv}, and they are:
\begin{subequations}\label{GED_sym}
\begin{align}
H&=\partial_t\, , & D_t&=t \partial_t \, , & D_x&= x^i\partial_i\, , \\
K&=t^2\partial_t+2 \, t\, x^i\partial_i\, ,
&J_{ij}&=x^i\partial_j-x^j\partial_i\, , & M^{(n)}_i &= t^{n+1}\partial_i\, , \\
L^{(n)}_{0I} &= t^n \zeta \frac{\partial\phantom{S^I}}{\partial S^I} - t^n S^I \frac{\partial\phantom{S^I}}{\partial A_0} \, , & L_{IJ}^{(n)} &= t^n S^I \frac{\partial\phantom{S^I}}{\partial S^J} - t^n S^J \frac{\partial\phantom{S^I}}{\partial S^I} \, . 
 \end{align}
\end{subequations}
This set of $G_{gauge}$ symmetries is larger than the expected asymptotic symmetries $G_{asym}$, for two reasons: 
\begin{enumerate}
    \item Due to the abelian nature of the GED theory, the \emph{separate} scaling of time and space is a symmetry of the full theory. Only when the structure constants are non-trivial, i.e. GED becomes GYM, the separate scaling of time and space  is no longer a symmetry. This is evidenced by the fact the generators $D_t$ and $D_x$ of GED need to combine into the single $D$ generator in GYM.
    \item The five free scalars, which appear in the action \eqref{GED_action} with no time derivatives, can be rotated with arbitrary functions that depend on time. This generates an infinite tower of rotations $L_{IJ}^{(n)}$ which do not have a counterpart in $G_{asym}$, or $G_{bulk}$.
\end{enumerate}    

This analysis was performed in detail in \cite{Fontanella:2024hgv}, leading to propose in \cite{Fontanella:2024rvn, Fontanella:2024kyl, Fontanella:2024hgv} that GYM theory with five interacting scalar fields is the holographic dual of GGK string theory. From the list of the consistency criteria given at the beginning of section \ref{sec:consistency_conditions}, only the \ref{Quantitative test} is missing in the current state of the art. Although work is still in progress, in chapter \ref{chap:quantisation} we will discuss the results obtained so far regarding the holographic matching of the string spectrum.

\subsection{The S-dual non-relativistic limit}
\label{sec:D1NC_limit}

A different approach to what we discussed so far was given in \cite{Lambert:2024yjk,Lambert:2024ncn,3096754}. Instead of taking the GGK limit (also known as SNC limit), one could take its S-dual limit, called D1NC.   

The setting is similar to the SNC configuration discussed above, in the sense that the NR longitudinal and D3-brane world-volume directions are taken as follows,
\begin{align}
\begin{array}{rrrrrrrrr}
D3: & 0 & 1 &2&3& & & & \\
D1NC:& 0 &  & & & & & & 9 \  \, . \\
\end{array}
\end{align}
The difference with the SNC limit arises at the level of scaling the gauge and scalar fields of $\mathcal{N}=4$ SYM. The natural interpretation of this limit is that it captures a system of intersecting D1 and D3 branes in the non-relativistic limit.  

\paragraph{The gauge theory.} The D1NC limit of \cite{Lambert:2024yjk} was implemented directly at the level of the $\mathcal{N}=4$ SYM, whose bosonic sector is given in \eqref{N=4_SYM}. Supersymmetry was discussed as well, but for simplicity, here we focus on the bosonic sector only. We shall follow the notation already introduced in section \ref{sec:consistency_conditions}, where $(t, x^i)$, with $i \in \{ 1,2,3\}$, are the D3-brane longitudinal coordinates; the scalar field index splits as $\hat{I} = (I, 6)$, with $I\in\{1,..., 5\}$, and $\zeta \equiv S^6$. 

The D1NC limit consists in rescaling the spacetime coordinates as\footnote{Due to the theory's conformal symmetry, this rescaling could be brought into a more standard form where $t\to c t$ and $x^i$ remains invariant.}
\begin{eqnarray}
    t \to \sqrt{c} t \, , \qquad\qquad
    x^i \to \frac{x^i}{\sqrt{c}}\, , 
\end{eqnarray}
together with a rescaling of the scalar and 1-form fields as,
\begin{equation}
    \zeta \to \sqrt{c} \,\zeta \, , \qquad
    S^I \to \frac{S^I}{\sqrt{c}}\, , \qquad
    A_0 \to \frac{A_0}{\sqrt{c}} \, , \qquad
    A_i \to \sqrt{c} \,A_i \, , 
\end{equation}
and the string coupling rescaling, 
\begin{eqnarray}
    g_s \to \frac{g_s}{c} \, . 
\end{eqnarray} 
By inserting the above rescalings in the bosonic $\mathcal{N}=4$ SYM \eqref{N=4_SYM}, one finds that the action expands at large $c$ as
\begin{eqnarray}
    S^{\text{bos.}}_{\text{SYM}} = c^2 S^{(2)} + S^{(0)} + \mathcal{O}(c^{-2})\, .
\end{eqnarray}
The divergent term of order $c^2$ is:
\begin{eqnarray}
    S^{(2)} = - \frac{1}{2\pi g_s} \int \dd t \dd x^3 \left( \frac{1}{4} F_{ij}^\mathtt{a} F^{ij \, \mathtt{a}} + \frac{1}{2} D_i \zeta^{\mathtt{a}}D^i \zeta^{\mathtt{a}}  \right) \, , 
\end{eqnarray}
and it can be rewritten as,
\begin{eqnarray}\label{identity_D1NC}
    \frac{1}{2} F_{ij}^\mathtt{a} F^{ij \, \mathtt{a}} +  D_i \zeta^{\mathtt{a}}D^i \zeta^{\mathtt{a}} = \frac{1}{2} (F_{ij}^\mathtt{a} \mp \varepsilon_{ijk}  D_k \zeta^{\mathtt{a}})^2 \pm \varepsilon_{ijk}F_{ij}^\mathtt{a}D_k \zeta^{\mathtt{a}} \, . 
\end{eqnarray}
By using the Bianchi identity, the second term in \eqref{identity_D1NC} is a total derivative. Although it does not affect the dynamics, this term contributes to the global charges. Since the gravity dual will have a 2-form RR field $C_2$, it was sugggested in \cite{Lambert:2024yjk} that the $C_2$ contribution to the DBI via a Wess-Zumino term will cancel the total derivative in \eqref{identity_D1NC}.  

The first term in \eqref{identity_D1NC} is a square, so it can be made finite via a Hubbard-Stratonovich transformation by introducing an auxiliary field $G_{ij}^\mathtt{a}$. By taking the $c\to \infty$ limit, one arrives to the D1NC $\mathcal{N}=4$ super Yang-Mills action,
\begin{eqnarray}\label{D1NC_gauge}
\notag
    && S_{\text{D1NC}} = - \frac{1}{4\pi g_s} \int \dd t \dd x^3 \left[ 
     F_{0i}^\mathtt{a} F^{0i \, \mathtt{a}} 
    +  D_0 \zeta^{\mathtt{a}}D^0 \zeta^{\mathtt{a}} 
    +  D_i S^{I\, \mathtt{a}} D^i S^{I\, \mathtt{a}}
     \right. \\
    &&\hspace{4.5cm}
    \left.  
    - (f_{\mathtt{bc}}{}^\mathtt{a} \zeta^{\mathtt{b}} S^{I\, \mathtt{c}})^2
    + G^{ij \, \mathtt{a}} (F_{ij}^\mathtt{a} - \varepsilon_{ijk}  D_k \zeta^{\mathtt{a}})
    \right] \, . 
\end{eqnarray}
This action describes a quantum mechanics on monopole moduli space \cite{Lambert:2024yjk}.
The precise relation between the actions obtained by taking the D1NC and the SNC limits of the $\mathcal{N}=4$ SYM theory was found in \cite{Kim:2026hpe}. With our normalisation, the D1NC theory at coupling $g_s$ is mapped into the SNC theory at coupling $1/g_s$.  

It is important to remark that the action \eqref{D1NC_gauge} has been derived directly from the D1NC limit of the $\mathcal{N}=4$ SYM action. It would be interesting to check if the same result appears by first taking the D1NC limit of the DBI action with Wess-Zumino term, followed by the low-energy limit $\alpha'\to 0$, as done for the SNC limit and required by the \ref{Uniqueness} condition outlined at the beginning of section \ref{sec:consistency_conditions}. As reported in \cite{Lambert:2024yjk}, there are higher-order divergences that appear when taking the D1NC limit of the DBI action with Wess-Zumino term. These higher terms should vanish once the constraint is imposed. As this is not guaranteed to be the case, it represents a strong consistency test of the D1NC limit. However, an indication that this might be the case comes from the fact that its S-dual SNC limit commutes with $\alpha'\to 0$.

\vspace{3mm}
\noindent\emph{Symmetries.} The symmetries of the action \eqref{D1NC_gauge}, and their associated Noether currents, have been computed in \cite{Lambert:2024yjk}. These symmetries have been distinguished between \emph{unphysical} and \emph{physical}, depending on whether their Noether charges are vanishing or not, respectively. They are the following:
\begin{subequations}
    \begin{align}
        H &= \partial_t \, , \qquad \qquad
        D = t \partial_t + x^i \partial_i \, , 
        &K &= t^2 \partial_t + 2 t x^i \partial_i \, , \\
        \hat{M}_i^{(n)} &= t^n \partial_i \, , &
        J_{ij} &= x^i \partial_j - x^j \partial_i \, , \\
        L_{IJ} &= S^I \frac{\partial \phantom{S^I}}{\partial S^J} - S^J \frac{\partial \phantom{S^I}}{\partial S^I}\, , \\
        \hat{R}^{(n)}_{I} &= t^n \zeta \frac{\partial \phantom{S^I}}{\partial S^I} - t^n \varepsilon_{ijk} D_k S^I \frac{\partial \phantom{G_{ij}}}{\partial G_{ij}}\, , &
        U &= \zeta \frac{\partial \phantom{A_0}}{\partial A_0} -  \varepsilon_{ijk} F_{0k}\frac{\partial \phantom{G_{ij}}}{\partial G_{ij}} \, ,
    \end{align}
\end{subequations}
where hatted generators indicate unphysical symmetries. 

It was pointed out in \cite{Lambert:2024ncn,3096754} that the non-trivial dynamics occurs when a scalar field possesses a vacuum expectation value (VEV). Classically, this may be interpreted as a boundary condition of the scalar fields at infinity. One can ask whether the above symmetries preserve the boundary conditions. It turns out that $D$ and $K$ do not, and therefore must be disregarded. Additionally, the first two levels of the tower of generators $\hat{M}_i^{(n)}$ gain a non-zero Noether charge due to the non-zero VEV of the scalar fields, and therefore they become physical symmetries. In summary, the physical and VEV preserving symmetries are 
\begin{eqnarray}
    G_{\text{VEV}} = \mathfrak{gal}(3)\oplus \mathfrak{so}(5) \oplus \mathfrak{u}(1) \, . 
\end{eqnarray}

\paragraph{The gravity theory.}  The gauge theory in the D1NC limit was proposed in \cite{Lambert:2024yjk} to be dual to a gravitational system of intersecting D1-D3 branes, where the D1-brane is smeared. The relativistic geometry for this setting is ($i\in\{1,2,3\}, I\in\{1,...,5\}$), 
\begin{subequations}\label{D1_D3_branes}
\begin{align} \nonumber
    \dd s^2_{\text{D1-D3}} &= - H_1^{-1/2} H_3^{-1/2} \dd t^2  + H_1^{1/2} H_3^{-1/2} \delta_{ij} \dd x^i\dd x^j  \\
    & \hspace{1.0cm} + H_1^{-1/2} H_3^{1/2} \dd z^2 + H_1^{1/2} H_3^{1/2} \delta_{IJ} \dd y^I\dd y^J \, , \\
    C_2 &=  H_1^{-1}   \dd t \wedge \dd z \, , \\
    C_4 &=  \brac{H_3^{-1} -1}   \dd t \wedge \dd x^1 \wedge \dd x^2 \wedge \dd x^3 \, , \\
    e^{\Phi} &= g_sH_1^{1/2} \, ,
\end{align}
\end{subequations}
where $H_1$ and $H_3$ satisfy the equations
\begin{subequations}
\begin{align}
    0 &= \partial^I \partial_I H_1 \, , \\
    0 &= H_1 \partial_{z}^2 H_3 + \partial^I \partial_I H_3 \, .\label{H3eq}
\end{align}
\end{subequations}
The corresponding D1NC limit is engineered in such a way that the spacetime acquires a codimension two foliation. This is achieved by taking $H_1$ completely ``smeared'' along the transverse ($x^i$ and $y^i$) directions, and $H_3$ as a ``localised'' solution of \eqref{H3eq} along the transverse ($z$ and $y^i$) directions,
\begin{eqnarray}
    H_1 = c^{-2} \, , \qquad\qquad
    H_3 = 1 + \frac{R^4}{\brac{z^2 + c^{-2} y^I y^I}^2} \, , 
\end{eqnarray}
where $R$ is a constant of integration.
By plugging this choice in \eqref{D1_D3_branes}, and by taking $c\to \infty$, one arrives to the D1NC limit of the intersecting D1-D3 brane system,  
\begin{subequations} \label{eq: metric c expansion D1}
\begin{align}
    \dd s^2_{\text{D1NC D1-D3}} &= c \tau_{\mn} \dd X^{\mu} \dd X^{\nu} + c^{-1} h_{\mn} \dd X^{\mu} \dd X^{\nu} \, , \\
    \tau_{\mn} \dd X^{\mu} \dd X^{\nu} &= -H^{-1/2} \dd t^2  + H^{1/2} \dd z^2  \, , \\
    h_{\mn} \dd X^{\mu} \dd X^{\nu} &= H^{-1/2} \delta_{ij} \dd x^i\dd x^j + H^{1/2} \delta_{IJ} \dd y^I\dd y^J   \, , \\
    C_2 &= c^2 \dd t\wedge\dd z\, ,\\
    C_4 &= (H^{-1}-1) \dd t\wedge \dd x^1\wedge\dd x^2\wedge\dd x^3\wedge\dd x^4 \, ,\\
    e^{\Phi} &= c^{-1} g_s \, ,
\end{align}
\end{subequations}
where $H$ is the leading order of $H_3$, 
\begin{eqnarray}
     H = 1 + \frac{R^4}{z^4} \, .
\end{eqnarray}
Finally, by taking the near-horizon limit $R\to \infty$, one arrives to a spacetime geometry that is the same as in \eqref{eq: metric c expansion D1}, except for replacing $H$ by $(R/z)^4$, namely keeping only the dominant term.
It is important to remark that to arrive to this geometry one does not rescale the coordinates. Instead, the factors of $c$ enter purely by assuming that $H_1$ is $c^{-2}$, which then propagate to $H_3$ upon solving \eqref{H3eq}.

A different D1NC limit was proposed in \cite{Lambert:2024ncn,3096754}, where $H_3$ is taken as a ``smeared'' solution (in contrast with the ``localised'' solution considered above),
\begin{eqnarray}\label{H3_smeared}
   H_3 = 1 + \frac{R^4}{\brac{y^I y^I}^2} 
\end{eqnarray}
Notice that this alternative harmonic function also fulfils the equation \eqref{H3eq}. The steps we have to follow in this case are the same as before mutatis mutandis, and the metric we obtain at the end of the process is the same as \eqref{eq: metric c expansion D1} but with $H$ replaced by \eqref{H3_smeared}.

The \emph{smeared} limit is usually preferred because it does not require the rescaling of the number of branes implicitly required by the non-relativistic limit. Additionally, it has nice T-duality properties \cite{Lambert:2024ncn,3096754}. In this alternative limit, the longitudinal metric $\tau_{\mu\nu}$ depends on a transverse coordinate through the harmonic function, and therefore it does not satisfy the torsionless constraint \eqref{torsionless_constraint}. 
However, since this is not a standard SNC limit, the cancellation of the divergence is not performed by adding a critical closed B-field. Hence, the torsionless constraint that guarantees closeness of the critical B-field might be superfluous in this case.

\vspace{3mm}
\emph{Symmetries}. This solution is expected to arise from the S-dual of the SNC solution \cite{Blair:2023noj}, in particular the supergravity fields have the expected form. The D1NC limit at the level of the type IIB supergravity has been partially developed in \cite{Lambert:2024ncn, 3096754}, and the corresponding Killing vectors preserving the supergravity solution with \emph{smeared} harmonic function is 
\begin{eqnarray}
    G_{bulk - smeared} = \mathfrak{gal}(3)\oplus \mathfrak{so}(5) \oplus \mathfrak{u}(1) \, . 
\end{eqnarray}
which matches the physical symmetries preserving the boundary conditions found in the gauge theory side.

\section{$p$-brane Newton-Cartan holographies}
\label{sec:other_NR_limits}

It is a legitimate question to ask whether the GGK/GYM duality, and its ``S-dual'', presented in section \ref{sec:consistency_conditions} is the only  non-relativistic holography that can be consistently derived from the AdS$_{n+1}$/CFT$_n$ duality, or whether there exists other examples. We believe this question is still open, although some proposals have been suggested by considering more general non-relativistic limits with more than two longitudinal directions. In this section, we shall review the membrane limit of AdS$_4$/ABJM proposed in \cite{Lambert:2024uue}, the 3-brane NC limits of the D3-brane discussed in \cite{Lambert:2024yjk} and \cite{Guijosa:2025mwh}, and a generalisation to $p$-NC limits of D$q$-branes presented in \cite{Blair:2024aqz}.

\subsection{The membrane limit of AdS$_4$/ABJM}

The AdS$_4$/ABJM duality is a holographic duality between M-theory on AdS$_4\times$S$^7/\mathbb{Z}_k$ and the ABJM theory \cite{Aharony:2008ug}. 
Also in this context, one can ask whether holography continues to hold after taking a non-relativistic limit on both sides of the duality. This question has been explored in \cite{Lambert:2024uue}, with a positive conclusion. Below, we shall review their analysis. 

\paragraph{NR limit of ABJM.} For the ABJM theory there is no derivation in terms of a low-energy limit of a DBI-like action. Instead, the ABJM theory appears as the low-energy dynamics of a stack of M2-branes on a $\mathbb{C}^4/\mathbb{Z}_k$ background, via string/M-theory dualities. For this reason, the logic of demanding the commutation of the NR limit with the low-energy limit $\alpha'\to 0$ in a DBI-like action, as demanded for the AdS$_5$/CFT$_4$ in \cite{Fontanella:2024kyl, Fontanella:2024rvn}, ceases to make sense.

The logic that guided \cite{Lambert:2024uue} to craft the NR limit of the ABJM theory was based on the approach \cite{Mouland:2019zjr}, and it first appeared in \cite{Lambert:2019nti}. The idea is to find a scaling limit of the ABJM theory that zooms into solutions preserving half of the supercharges, i.e. $\frac12$-BPS, such that the dynamics of the theory reduces to the Manton approximation of geodesic motion on the moduli space of solutions \cite{Manton:1981mp}.  

The ABJM theory \cite{Aharony:2008ug} is a 3d $\cal{N}=6$ superconformal $U(N) \times U(N)$ Chern-Simons matter theory with two $U(N)$ gauge fields $A^{L/R}$ and four complex scalar fields $\hat{\calz}^M$ in the bifundamental of $U(N) \times U(N)$, where $M\in\{1,2,3,4\}$ is the $SU(4)$ R-symmetry index. The spacetime coordinates are denoted by $\hat{x}^{\mu} = (\hat{t},\hat{x}^i)$, with $i\in \{1,2\}$. In what follows, it will be convenient to consider the complex coordinate $z \equiv x^1 + i x^2$, together with $D\equiv D_z, \bar{D}\equiv D_{\bar{z}}$.  
In the notation of \cite{Lambert:2024uue}, the action is given by  
\begin{align} \label{ABJM}\nonumber
    \hat{S}_{\text{ABJM}} = \tr \int \dd^3\hat{x} \Bigg[& - \hat{D}_{\mu} \hat{\bcalz}_M \hat{D}^{\mu} \hat{\calz}^M + \frac{k \varepsilon^{\mn\rho}}{4\pi} \bigg( \hat{A}_{\mu}^L \hat{\partial}_{\nu} \hat{A}_{\rho}^L - \frac{2i}{3} \hat{A}_{\mu}^L \hat{A}_{\nu}^L \hat{A}_{\rho}^L  \\
    &- \hat{A}_{\mu}^R \hat{\partial}_{\nu} \hat{A}_{\rho}^R + \frac{2i}{3} \hat{A}_{\mu}^R \hat{A}_{\nu}^R \hat{A}_{\rho}^R \bigg) - \frac{8\pi^2}{3k^2 } \hat{\bar{\Upsilon}}^{K}_{IJ} \hat{\Upsilon}^{IJ}_{K} \Bigg] \, ,
\end{align}
where
\begin{equation}
    \hat{D}_{\mu} \hat{\calz}^M = \hat{\partial}_{\mu} \hat{\calz}^M - i \hat{A}_{\mu}^L \hat{\calz}^M + i \hat{\calz}^M \hat{A}_{\mu}^R \, ,
\end{equation}
and
\begin{equation}
    \hat{\Upsilon}^{MN}_{P} = [\hat{\calz}^M , \hat{\calz}^N ; \hat{\bcalz}_P] - \inv{2} \delta^M_P [ \hat{\calz}^Q , \hat{\calz}^N; \hat{\bcalz}_Q] + \inv{2} \delta^N_P [ \hat{\calz}^Q , \hat{\calz}^M ; \hat{\bcalz}_Q]\, ,
\end{equation}
with notation
\begin{equation}
    [\hat{\calz}^M , \hat{\calz}^N ; \hat{\bcalz}_P] \equiv \hat{\calz}^M \hat{\bcalz}_P \hat{\calz}^N - \hat{\calz}^N \hat{\bcalz}_P \hat{\calz}^M \, .
\end{equation}
Demanding that the field configurations preserve half of the supercharges, i.e. $\frac12$-BPS, leads to the following set of equations \cite{Kim:2009ny}:
\begin{subequations} \label{eq: undeformed BPS equations}
\begin{align}
    \hat{D}_i \hat{\calz}^A &= 0 \, , \\ \label{eq: holomorphic BPS equation}
    \hat{\bar{D}} \hat{\calz}^1 &= 0 \, , \\
    [\hat{\calz}^1 , \hat{\calz}^2 ; \hat{\bcalz}_2] &= [\hat{\calz}^1 , \hat{\calz}^3 ; \hat{\bcalz}_3] = [\hat{\calz}^1 , \hat{\calz}^4 ; \hat{\bcalz}_4] \, , \\
    \hat{D}_0 \hat{\calz}^1 &= \frac{2\pi i}{3k} [\hat{\calz}^1 , \hat{\calz}^A ; \hat{\bcalz}_A] \, , \\ \label{eq: undeformed ZA eq}
    \hat{D}_0 \hat{\calz}^A &= \frac{2\pi i}{k} [\hat{\calz}^1, \hat{\calz}^A ; \hat{\bcalz}_1] \, , \\
    [\hat{\calz}^1 , \hat{\calz}^A ; \hat{\bcalz}_B] &= 0 \; (A\neq B) \, , \\
    [\hat{\calz}^A , \hat{\calz}^B ; \hat{\bcalz}_C] &= 0 \, ,
\end{align}
\end{subequations}
which are supplemented by the Gauss's law constraint given in \cite{Lambert:2024uue}.

Then, the NR rescaling of the spacetime coordinates is\footnote{The rescaling parameter $\omega$ used in \cite{Lambert:2024uue} is related to our parameter via $\omega = 1/c$.}
\begin{equation} \label{eq: scaling 1}
    \hat{t} =  t \, , \qquad
    \hat{x}^i =  \frac{x^i}{c} \, . 
\end{equation}
In order for the BPS equations \eqref{eq: undeformed BPS equations} to remain invariant, one has to supplement the coordinate rescaling \eqref{eq: scaling 1} with a rescaling of the fields:
\begin{subequations} \label{eq: scaling 2}
\begin{align}
    \hat{\calz}^1(\hat{t},\hat{x}) &= c \calz^1(t,x) \, , 
    &\hat{\calz}^A(\hat{t},\hat{x}) &= \calz^A(t,x) \, , \\
    \hat{\cal{A}}_0(\hat{t}, \hat{x}) &= \cal{A}_0(t,x) \, , 
    &\hat{A}_i(\hat{t},\hat{x}) &= c A_i(t,x) \, ,
\end{align}
\end{subequations}
where the fields $A^{L/R}_0$ have been shifted accordingly to,  
\begin{equation} \label{eq: A0 shift}
    A^L_0 = \cal{A}^L_0 - \frac{2\pi}{k} \calz^1 \bcalz_1 \, , \qquad
    A^R_0 = \cal{A}^R_0 - \frac{2\pi}{k} \bcalz_1 \calz^1 \, .
\end{equation}
The reasoning behind this shift is to leave the Gauss's law invariant, resulting in a theory with non-trivial dynamics. 

After plugging the above coordinate and field rescalings inside the action \eqref{ABJM}, and expanding in large $c$, one finds that the action expands as 
\begin{equation}
    \hat{S}_{\text{ABJM}} = c^2 S^{(2)} + S^{(0)} + O(\omega^2) \, .
\end{equation}
The $S^{(2)}$ term is the sum of a squared quantity and a total derivative term. The total derivative term does not contribute to the dynamics, although it can contribute to generate an infinite ``rest energy''. Such rest energy term can be eliminated by adding a critical constant background $C_3$ field by coupling the ABJM action to the corresponding Wess-Zumino term, up to the non-abelian subtleties discussed in \cite{Lambert:2024uue}. After this, the squared quantity has been eliminated via a Hubbard-Stratonovich transformation. The final action in the limit $c\to \infty$ turns out to be, 
\begin{align} \label{NR_ABJM} \nonumber
    S_{\text{NR ABJM}} = \tr \int \dd^3x \Bigg[& D_0 \calz^1 D_0 \bcalz_1 + H D \bcalz_1 + \bar{D} \calz^1 \bar{H} - 2 D\calz^A \bar{D} \bcalz_A - 2 \bar{D} \calz^A D \bcalz_A \\ \nonumber
    &+ \frac{2\pi i }{k} D_0 \calz^A [\bcalz_1, \bcalz_A; \calz^1] + \frac{2\pi i}{k} [\calz^1, \calz^A; \bcalz_1] D_0 \bcalz_A \\ \nonumber
    &- \frac{4\pi^2}{3k^2} [\calz^A, \calz^1; \bcalz_A] [\bcalz_B, \bcalz_1; \calz^B] + \frac{16\pi^2}{3k^2} [\calz^A, \calz^1; \bcalz_B] [\bcalz_A, \bcalz_1, \calz^B] \\ \nonumber
    &+ \frac{8 \pi^2}{3k^2} [\calz^A, \calz^B; \bcalz_1] [\bcalz_A, \bcalz_B; \calz^1] - \frac{4\pi^2}{3k^2} [\calz^1, \calz^A; \bcalz_1] [\bcalz_B, \bcalz_A; \calz^B] \\ \nonumber
    &- \frac{4\pi^2}{3k^2} [\calz^B, \calz^A; \bcalz_B] [ \bcalz_1 , \bcalz_A; \calz^1] + \frac{ik}{2\pi} \bigg( A_0^L F_{z \bar{z}}^L + A_z^L F_{\bar{z}0}^L + A_{\bar{z}}^L F_{0z}^L \\
    &+ i A_0^L [A_z^L, A_{\bar{z}}^L] - (L \to R)
    \bigg) \Bigg] \, .
\end{align}
The Lagrange multiplier $H$ introduced via a Hubbard-Stratonovich transformation is a complex auxiliary field in the bifundamental of $U(N)\times U(N)$. Its equations of motion impose the constraint, 
\begin{equation} \label{eq: Z constraint}
    \bar{D} \calz^1 = 0 \, ,
\end{equation}
which is nothing else than the BPS equation \eqref{eq: holomorphic BPS equation}. 
It is interesting to point out a difference between this discussion and the NR limit of the DBI action discussed in section \ref{sec:NR_gauge_side}. In that case, setting the B-field to the critical value was enough to completely cancel the divergent term, and there was no Lagrange multiplier that imposes a constraint with a BPS equation interpretation. Instead, the action \eqref{NR_ABJM} has an interpretation as a quantum mechanics on Hitchin moduli space \cite{Lambert:2024uue, Lambert:2024yjk}.

\paragraph{NR limit of the black M2-brane.} The specific NR rescaling taken in the ABJM theory suggests that the NR limit to take in the gravity theory corresponds to a \emph{Membrane Newton-Cartan} (MNC) limit, where the NR longitudinal and M2-brane world-volume directions are chosen as,   
\begin{align} 
\begin{array}{rrrrrr}
M2: & 0 & 1 &2&& \\
MNC:& 0 &  & &3&4 \, . \\
\end{array}
\end{align}
The metric for a stack of M2-branes spanned by the spacetime coordinates $(t,z,\bar z,u,\bar u,  \vec v)$ is:
\begin{subequations}\label{M2_metric}
\begin{align}
    \dd s^2_{\text{M2-brane}} &= \hat{\ch}^{-\frac{2}{3}} \brac{-\dd t^2 + \dd z \dd\bar z} + \hat{\ch}^{\frac{1}{3}} \brac{\dd u  \dd\bar u  + \dd \vec v\cdot  \dd \vec v} \, , \\
    \hat{\ch} &= 1 + \frac{\hat{R}^6}{\brac{u\bar u +  \vec v\cdot  \vec v}^3} \, .
\end{align}
\end{subequations}
where $(t,z,\bar z)$ are the longitudinal M2-brane coordinates, $(u,\vec v)$ are the transverse coordinates and $\hat{R}$ is a parameter. Given the fact that, loosely speaking, $u$ corresponds to $\calz^1$ and $ \vec v$ to the real and imaginary parts of $\calz^A$ in the field theory, the NR rescaling of the coordinates was taken in \cite{Lambert:2024uue} as:
\begin{equation} \label{eq: scaling grav}
    \brac{t,z,  u  ,  \vec v} \to \brac{c t, c^{-\frac{1}{2}} z  , c u,  c^{-\inv{2}}  \vec v} \, , \qquad
     \hat{R} = c R \, .
\end{equation}
By taking the limit $c\to \infty$, the metric becomes a Membrane Newton-Cartan (MNC) geometry,\footnote{Notice that the relative factor between $\tau$ and $h$ in the MNC geometry is $c^3$, in contrast to the SNC geometry that requires $c^2$.}
\begin{subequations}\label{MNC_M2_metric}
\begin{align}
    \dd s^2_{\text{MNC M2-brane}} &= (c^2 \tau_{\mu\nu} + \frac{1}{c} h_{\mu\nu})\dd X^{\mu} \dd X^{\nu}  \, , \\
    \tau_{\mu\nu}\dd X^{\mu} \dd X^{\nu} &= - \ch^{-\frac{2}{3}} \dd t^2 + \ch^{\frac{1}{3}} \dd u \dd\bar u \, ,\\
    h_{\mu\nu}\dd X^{\mu} \dd X^{\nu} &= \ch^{-\frac{2}{3}} \dd z\dd\bar z + \ch^{\inv{3}} \dd\vec v \cdot \dd \vec v \, , \\
    \ch  &= 1 + \frac{R^6}{\brac{u \bar u }^3} \, .
\end{align}
\end{subequations}
The reasoning for rescaling the parameter $\hat{R}$ is to guarantee that the geometry \eqref{MNC_M2_metric} retains a horizon (a throat shape), as demanded by the \ref{Horizon condition} given at the beginning of section \ref{sec:consistency_conditions}. Moreover, rescaling $\hat{R}$ could potentially be viewed as sending the number of M2-branes sourcing the geometry to infinity. 

The next step is to take the near-horizon limit, corresponding to $R\to \infty$. This gives the following geometry,
\begin{subequations}\label{MNC_near_horizon}
\begin{align}
    \dd s^2_{\text{MNC AdS$_4\times$S$^7$}} &= (c^2 \tau_{\mu\nu} + \frac{1}{c} h_{\mu\nu})\dd X^{\mu} \dd X^{\nu}  \, , \\
    \tau_{\mu\nu}\dd X^{\mu} \dd X^{\nu} &= - \frac{\brac{u\bar{u}}^2 }{R^4} \dd t^2 + \frac{R^2 }{u\bar{u}} \dd u \dd\bar{u}  \, ,\\
    h_{\mu\nu}\dd X^{\mu} \dd X^{\nu} &= \frac{\brac{u \bar u }^{2}}{R^{4}} \dd z\dd\bar z + \frac{R^{2}}{u \bar u } \dd\vec v \cdot \dd \vec v \, .
\end{align}
\end{subequations}
The longitudinal $\tau$ metric describes AdS$_2\times$S$^1$, whereas $h$ describes $w_1(u) \mathbb{R}^2 \times w_2 (u) \mathbb{R}^6$, i.e. the product of two Euclidean spaces that grow and shrink accordingly to their warped factors $w_1$ and $w_2$. This MNC AdS$_4\times$S$^7$ geometry is expected to also appear by reversing the order of limits, i.e. first the near-horizon limit of \eqref{M2_metric} followed by the NR limit, although this has not been discussed in \cite{Lambert:2024uue}. 

The relativistic stack of M2-brane solution also has a 4-form RR field strength $\hat{F}_4$. In the NR and near-horizon limits, and up to irrelevant subleading terms of order $c^{-6}$, the solution is given by 
\begin{subequations}\label{NR_4_form}
\begin{align}
F_4 &=  F_4^{(0)} + \frac{1}{c^3} \tilde{F}^{(-3)}_4 \, ,  \\
F_4^{(0)} &= \frac{3i \brac{u\bar{u}}^2}{2R^6} \brac{u \dd\bar{u} + \bar{u} \dd u} \wedge \dd t\wedge \dd z\wedge \dd\bar z \, , \\
F_4^{(-3)} &= \frac{3i\brac{u \bar{u}}^2}{R^6} \brac{ \vec v\cdot  \vec v \brac{\frac{\dd u}{u} + \frac{\dd\bar{u}}{\bar{u}}} +  \vec v\cdot  \dd  \vec v} \wedge \dd t \wedge \dd z \wedge \dd\bar z \, . 
\end{align}
\end{subequations}
The NR limit of 11-dimensional supergravity was first taken for the bosonic sector in \cite{Blair:2021waq}. The MNC metric structure and RR form presented above via a limit procedure, in order to satisfy the equations of motion, need to be supplemented by a 4-form Lagrange multiplier field $G_{4}$, which nevertheless turns out to vanish on this solution, i.e.
\begin{eqnarray}
    G_{4} = 0 \, . 
\end{eqnarray}
The above MNC metric structure, RR form and $G_{4}$ was found to belong to a class of supersymmetric solutions of the NR 11-dimensional supergravity theory \cite{Bergshoeff:2024nin}. 

A different MNC limit was proposed in \cite{Lambert:2024ncn,3096754}, where $\hat{\ch}$ is taken as a ``smeared'' solution (in contrast with the ``localised'' solution considered above),
\begin{eqnarray}\label{ch_smeared}
   \hat{\ch} = 1 + \frac{\hat{R}^4}{\brac{ \vec v\cdot  \vec v}^2}
\end{eqnarray}
The steps we have to follow in this case are similar to the \emph{localised} solution, except that we have to rescale the radius as $\hat{R} = c^{-\inv{2}} R $. The metric we obtain at the end of the process is the same as \eqref{MNC_M2_metric} but with $\ch$ replaced by \eqref{ch_smeared}.
We notice that, in this solution, the longitudinal metric $\tau_{\mu\nu}$ depends on a transverse coordinate, which nevertheless might not be a concern as this is not a standard SNC limit.

\paragraph{Matching the symmetries.} The set of symmetries of the NR ABJM theory are discussed in \cite{Lambert:2024uue}. They were found to be infinite dimensional. However, only a subset of them has a non-vanishing Noether charge. The symmetries with vanishing Noether charge were interpreted as gauge redundancies, and therefore \emph{unphysical}. They consist of: 
\begin{subequations}
    \begin{align}
        H &= \partial_t \, , \qquad \qquad
        D = t \partial_t + z \partial_z \, , 
        &K &= t^2 \partial_t + 2 t z \partial_z \, , \\
        \hat{M}^{(m,n)} &= t^m z^n \partial_z \, , & \hat{N}^{(m,n)} &= i t^m z^n \partial_z \, , \\
        U &= i \calz_1 \frac{\partial \phantom{\calz_1}}{\partial {\calz_1}} +  i H \frac{\partial \phantom{H}}{\partial {H}} \, , \\
        V &= i \calz_1 \frac{\partial \phantom{\calz_1}}{\partial {\calz_1}} + i H \frac{\partial \phantom{H}}{\partial {H}} + i \calz_A \frac{\partial \phantom{\calz_A}}{\partial {\calz_A}}  \, , \\
        R_{AB} &= \calz_A \frac{\partial \phantom{\calz_B}}{\partial {\calz_B}} - \calz_B \frac{\partial \phantom{\calz_A}}{\partial {\calz_A}} \,  ,
    \end{align}
\end{subequations}
where hatted generators indicate unphysical symmetries.

In summary, the set of physical symmetries, namely the ones with non-vanishing Noether charges, are:
\begin{eqnarray}\label{G_gauge_11d}
    G_{gauge} = \mathfrak{sl}(2,\mathbb{R}) \oplus \mathfrak{u}(1)_{U}\oplus \mathfrak{u}(1)_{V} \oplus \mathfrak{su}(3) \, . 
\end{eqnarray}
However, it was pointed out in \cite{Lambert:2024ncn,3096754} that the non-trivial dynamics occurs when a scalar field possesses a VEV. Classically, this may be interpreted as a boundary condition of the scalar fields at infinity. One can ask whether the above symmetries preserve the boundary conditions. These are
\begin{eqnarray}
  G_{\text{VEV}} = \mathfrak{schr}(2)\oplus \mathfrak{u}(1)_{U} \oplus \mathfrak{su}(3)  \, . 
\end{eqnarray}

On the gravity theory, the symmetries of the full NR 11d supergravity are given by solving the system of equations:
\begin{eqnarray}\label{KV_NR_11d_sugra}
     \mathsterling_{\xi} \, \Theta = \delta \Theta \, , 
\end{eqnarray}
where $\Theta \in \{\tau_{\mu\nu}, h_{\mu\nu}, F_4, G_4\}$, and $\delta \Theta$ corresponds to the local transformations of the field $\Theta$ that leave the NR 11-dimensional supergravity action invariant. It appears that there is an infinite dimensional set of Killing vectors that fulfils \eqref{KV_NR_11d_sugra} for the metric structure, but not for the $p$-form sector. Demanding that these Killing vectors preserve the full solution reduces the infinite symmetries down to a finite subalgebra. For the \emph{localised} solution, the symmetries are 
\begin{eqnarray}\label{G_bulk_localised}
    G_{bulk -  localised} = \mathfrak{sl}(2,\mathbb{R})\oplus \mathfrak{iso}(2) \oplus \mathfrak{u}(1)\oplus \mathfrak{u}(1) \oplus \mathfrak{su}(3) \, . 
\end{eqnarray}
whereas for the \emph{smeared} solution, they are 
\begin{eqnarray}\label{G_bulk_smeared}
    G_{bulk -  smeared} = \mathfrak{schr}(2)\oplus \mathfrak{u}(1)\oplus \mathfrak{u}(1) \oplus \mathfrak{su}(4) \, . 
\end{eqnarray}
For the localised solution, $G_{gauge}$ matches $G_{bulk -  localised}$ only up to the $\mathfrak{iso}(2)$ factor, which was suggested in \cite{Lambert:2024uue} it might have a vanishing Noether charge. On the other hand, for the smeared solution, $G_{\text{VEV}}$ matches $G_{bulk -  smeared}$ only up to the mismatch between an $\mathfrak{su}(3)$ and $\mathfrak{su}(4)$ factor. This happens because the supergravity solution has not been orbifolded, in such case, the $\mathfrak{su}(4)$ algebra breaks into $\mathfrak{su}(3) \oplus \mathfrak{u}(1)$. The two $\mathfrak{u}(1)$ in excess can be attributed to an implicit compactification of the solution into a two-dimensional torus along the $(u,\bar{u})$ directions (for more detail, see the discussion around formula \eqref{torus_smearing}).

To establish holography, one expects to match $G_{asym}$ with $G_{gauge}$. Formally, this requires a near-boundary expansion of all fields, together with appropriate boundary conditions, which has not been done yet. As a first analysis, one could try to evaluate $G_{bulk}$ at the boundary. However, the bulk Killing vectors in \cite{Lambert:2024uue} are given in a set of coordinates that does not highlight the boundary structure and, for this reason, it is difficult to say how the bulk Killing vectors will act at the boundary. Assuming that the ``physical'' part of $G_{bulk}$ continues to be a symmetry at the boundary, one would still need to discuss whether there is any symmetry enhancement. This enhancement may come from symmetries of the metric structure which are not symmetries of the $p$-forms in the bulk, but they become at the boundary. In a heuristic way, in \cite{Lambert:2024uue} it is argued that this will not be the case. Another possibility of such enhancement is given by novel transformations of the near-boundary metric and $p$-forms. This also might not be the case, as the dual theory is defined in three dimensions, whereas an enhancement is typically expected from a boundary theory in two dimensions. 

Finally, we point out that in this M-theory holography example the symmetry matching between $G_{asym}$ and $G_{gauge}$ involves only the \emph{physical} symmetries, i.e. the ones with non-vanishing Noether charge, and the one preserving a scalar field VEV. Instead, the symmetry matching in the GGK/GYM holography discussed in section \ref{sec:consistency_conditions} holds without these extra restrictions. A difference between the two cases is also on the presence of Lagrange multipliers, which might be responsible for the presence of unphysical symmetries.    

\subsection{The aligned 3-NC limit of $D3$-branes: the AdS$_5$/CFT$_4$}  \label{sec:NRasNH}

Similarly to the Membrane Newton-Cartan limit, one can define a $D3$-brane Newton-Cartan ($3$-NC) limit as a generalisation of the $p=1$ non-relativistic limit discussed in section \ref{subsec:limit} by considering four longitudinal directions. Let us consider the black D$3$-brane of type IIB supergravity \eqref{D3_metric_z}, but this time in Cartesian coordinates
\begin{subequations}\label{D3-brane_sugra}
    \begin{align}
           \dd s^2_{\text{D}3} &= H^{-\frac12} \left[ -(\dd X^{0})^2 + \sum_{a=1}^3 (\dd X^{a})^2  \right] + H^{\frac12}  \sum_{\hat{I}=4}^9(\dd X^{\hat{I}})^2  \, ,  \\
    C_{4} &=  (H)^{-1} \dd X^0 \wedge \dots \wedge \dd X^4 + \text{self-dual} \, ,  \\    
    H &= 1 + \left( \frac{R}{\sqrt{\sum_{\hat{I}=4}^9 (X^{\hat{I}})^2}}\right)^{4} \, ,  \\
    R^4 &= 4\pi N G_s \alpha^{\prime 2}\, , 
    \end{align}
\end{subequations}
to which we perform the following coordinate rescaling
\begin{subequations}\label{D3-NR}
\begin{align} 
    (X^0 , \dots , X^3) &= \sqrt{c} (t, x^1, \dots, x^3) \, , \\
    (X^{4} , \dots , X^9) &= \frac{1}{\sqrt{c}} (x^{4}, \dots, x^9) \, ,\\
    G_s &=g_s
\end{align}
\end{subequations}
followed by the $c \rightarrow \infty$ limit. In this way, the non-relativistic and brane longitudinal directions are aligned.

The interest in this limit stems from a surprising observation in \cite{Guijosa:2025mwh}: this $3$-NC limit of a $D3$-brane coincides with Maldacena's near horizon limit, reviewed in section~\ref{MaldacenaGravity}. Indeed, the final result of the above limit is
\begin{subequations}
    \begin{align}
    \dd s^2_{3\text{NC-D}3} &= \frac{r^2}{R^2} \left[ -(\dd X^{0})^2 + \sum_{a=1}^3 (\dd X^{a})^2  \right] + \frac{R^2}{r^2}  \sum_{\hat{I}=4}^9(\dd X^{\hat{I}})^2  \, , \label{AdS5xS5_metric_G} \\
    C_{4} &=  \frac{r^4}{R^4} \dd X^0 \wedge \dots \wedge \dd X^4 + \text{self-dual} \, ,  \label{RR_5_form_G} \\
    r^2&=\sum_{\hat{I}=4}^9 (X^{\hat{I}})^2 \, ,
    \end{align}
\end{subequations}
which is the usual AdS$_5\times$S$^5$ geometry. We should keep in mind that, for the black brane geometry to be an accurate description, we need to work in the regime where $G_s N \gg 1$.

\subsubsection{The gauge theory side}

On the gauge theory side, applying the limit \eqref{D3-NR} to the ten-dimensional flat target spacetime of the non-abelian DBI action \cite{Tseytlin:1997csa,Myers:1999ps}, gives us
\begin{equation}
    S=-T_p \int{\dd^{4} \sigma \, g_s^{-1} \text{Tr} \left(\frac{1}{4} F_{\alpha\beta}F^{\alpha\beta} + \frac{1}{2} \partial_\alpha S^{\hat{I}} \partial^\alpha S^{\hat{I}} - \frac{1}{4} [X^{\hat{I}}, X^{\hat{J}}]^2\right)} \, ,
\end{equation}
which is the bosonic part of the compactification down to four dimensions of ten-dimensional $\mathcal{N}=1$ SYM, which corresponds to the usual 4d $\mathcal{N}=4$ SYM. We should keep in mind that, for the DBI action to be an accurate description, we need to work in the regime where $G_s N \ll 1$. 

As a direct consequence of these results, the original AdS/CFT correspondence can be understood instead as a gauge/gravity duality in the aligned $3$-NC limit of a stack of $N$ $D3$-branes.

%On the gravity theory side, when $G_s N \gg 1$, the branes backreact strongly, creating the black brane geometry \eqref{D3-brane_sugra}. When we perform the $3$-NC limit \eqref{D3-NR}, we obtain exactly the same AdS$_5\times$S$^5$ geometry as Maldacena's near-horizon limit. This is seen more clearly if we write the longitudinal metric in polar coordinates, with $z^2=\sum_{i=4}^9 (X^i)^2$.

\subsubsection{AdS$_5$ as a box and $D3$-branes as the lightest excitation}

An implication of this new perspective on the AdS$_5$/CFT$_4$ correspondence proposed in \cite{Guijosa:2025mwh}, is that it allows to reinterpret the idea of AdS space as a box.

Let us consider a specific D3-brane in the original stack of $N$ $D3$-branes and move it along one of the transverse direction. After integrating out the degrees of freedom of the remaining $N-1$ D3-branes in the stack, the effective action for the probe D3-brane takes the form of the abelian DBI action~\eqref{abelian_DBI} on a target space given by AdS$_5 \times$S$^5$ \eqref{AdS5xS5_metric_G} with Wess-Zumino term given by the RR 4-form generating the flux \eqref{RR_5_form_G}. Setting $F=0$ (although we can turn it on to describe the bound state of D3-D1 branes, see section 4.4 in \cite{Guijosa:2025mwh}), the action takes the form
\begin{equation}
    S_{D3}= \frac{1}{(2 \pi)^3 \alpha'^2 g_s} \int \dd\sigma^4 \left( \frac{r^2}{R^4}- \frac{r^2}{R^4} \sqrt{r^4 - R^4 \sum_{\hat{I},\hat{J}=4}^9( \partial_\alpha X^{\hat{I}} \partial_\beta X^{\hat{J}} \delta_{\hat{I}\hat{J}} \eta^{\alpha \beta} )} \right) \,.
\end{equation}
If we separate the probe brane enough, which means taking $r \gg R$, the action becomes
\begin{equation}
    S_{D3} \approx \frac{1}{(2 \pi)^3 \alpha'^2 g_s} \int \dd\sigma^4 \sum_{\hat{I},\hat{J}=4}^9\left( \frac12  \partial_\alpha X^{\hat{I}} \partial_\beta X^{\hat{J}} \delta_{\hat{I}\hat{J}} \eta^{\alpha \beta} \right)  \,,
\end{equation}
which means that the probe brane can move arbitrarily far from the other branes and behave like a free object.

This might seem surprising because we are used to thinking of AdS space as a box due to the fact that massive particles and light rays cannot escape to arbitrarily large radius in finite proper time. However, because AdS$_5 \times$S$^5$ is an asymptotically flat $D3$-brane Newton-Cartan geometry, objects carrying positive D3 charge can escape into the flat region. A simple explanation for it is that they are not subject to the same restrictions as massive particles because charged objects do not follow geodesics. In fact, from the above computation, we see that the D3-branes are the lightest excitations in the $D3$-brane Newton-Cartan theory.

\subsection{The misaligned $3$-NC limit of the AdS$_5$/CFT$_4$}
\label{sec:misaligned_D3NC}

After the discussion from the previous section, an immediate question to ask is what happens when the longitudinal directions of the 3-brane NC limit and the longitudinal directions of the $D3$-brane are not aligned. Following \cite{Lambert:2024yjk,Lambert:2024ncn,3096754}, let us consider the following setting
\begin{align}
\begin{array}{rrrrrrr}
D3: & 0 & 1 &2& 3 && \\
D3NC:& 0 & 1 & & &4&5 \, ,  \\
\end{array}
\end{align}
which has a natural interpretation as an intersecting D3-D3$'$ brane geometry, with the D3NC limit capturing its non-relativistic dynamics.   

\paragraph{The gauge theory.} The starting point considered in \cite{Lambert:2024yjk} is the $\mathcal{N}=4$ SYM action. Fermions are included in their discussion, but for simplicity we will not treat them here. To implement the D3NC limit, we shall first decompose the indices in the following way: $\sigma^{\alpha} = (x^A, y^a)$, with $A\in\{0,1\}$, $a\in\{2,3\}$; $S^{\hat{I}} = (S^I, \zeta^i)$, with $I \in\{ 1, ...., 4\}$, $i\in\{5,6\}$.  
Then, the D3NC limit is defined by the following rescaling of spacetime coordinates 
\begin{eqnarray}
    x^A \to \sqrt{c} x^A \, , \qquad\qquad
    y^a \to \frac{y^a}{\sqrt{c}} \, , 
\end{eqnarray}
together with the rescaling of adjoint scalar and 1-form fields
\begin{equation}
    \zeta^i \to \sqrt{c} \,\zeta^i \, , \qquad 
    S^I \to \frac{S^I}{\sqrt{c}}\, , \qquad 
    A_A \to \frac{A_A}{\sqrt{c}} \, , \qquad 
    A_a \to \sqrt{c} \,A_a \, , 
\end{equation}
and with no rescaling of $g_s$. 

By inserting the above rescalings in the bosonic $\mathcal{N}=4$ SYM \eqref{N=4_SYM}, one finds that the action expands at large $c$ as
\begin{eqnarray}
    S^{\text{bos.}}_{\text{SYM}} = c^2 S^{(2)} + S^{(0)} + \mathcal{O}(c^{-2})\, .
\end{eqnarray}
The divergent term of order $c^2$ is:
\begin{eqnarray}
    S^{(2)} = - \frac{1}{4\pi g_s} \int \dd x^2 \dd y^2 \left[ F^\mathtt{a} F^{\mathtt{a}} +  D_a \zeta^{i\, \mathtt{a}}D^a \zeta^{i\, \mathtt{a}} - \frac{1}{2} (f_{\mathtt{bc}}{}^\mathtt{a} \zeta^{i\, \mathtt{b}} \zeta^{i\, \mathtt{c}})^2 \right] \, . 
\end{eqnarray}
where $F^\mathtt{a}\equiv F_{23}^\mathtt{a}$.
By introducing the following redefinitions
\begin{eqnarray}
   \calz \equiv \zeta^5 + i \zeta^6 \, , \qquad
   z \equiv y^2 + i y^3 \, , \qquad
   x^{\pm} \equiv x^0 \pm x^1 \, , 
\end{eqnarray}
and by using the identity 
\begin{eqnarray}
   F^\mathtt{a} f_{\mathtt{bc}}{}^\mathtt{a} \calz^{\mathtt{b}} \bcalz^{\mathtt{c}} = i \bcalz^{\mathtt{a}} (D_2 D_3 - D_3 D_2) \calz^{\mathtt{a}} \, , 
\end{eqnarray}
the divergent term $S^{(2)}$ can be rewritten as
\begin{equation}
\label{S4_D3NC}
    S^{(2)} = - \frac{1}{4\pi g_s} \int \dd x^2 \dd y^2 \left[ \left(F^\mathtt{a} + \inv{2} f_{\mathtt{bc}}{}^\mathtt{a} \calz^{\mathtt{b}} \bcalz^{\mathtt{c}}  \right)^2 + \Bar{D} \calz^\mathtt{a} D \bcalz^\mathtt{a} \right] \, .
\end{equation}
In this rewriting process, we omitted a total derivative. Although not affecting the dynamics, such term can alter the value of global charges. However, as the gravitational dual background has a RR 4-form $C_4$, it was pointed out in \cite{Lambert:2024yjk} that the total derivative appearing in $S^{(2)}$ is precisely cancelled out by the Wess-Zumino term associated with the $C_4$ field.

Since \eqref{S4_D3NC} is the sum of a regular square and a Lorentz square, one can introduce two Hubbard-Stratonovich fields $B$ and $H$, to rewrite $c^2 S^{(2)}$ as a term of order $c^0$. After that, we can safely take the $c\to \infty$ limit, and arrive at the D3NC $\mathcal{N}=4$ super Yang-Mills action,
\begin{align}\label{D3NC_gauge}
\nonumber
    S_{\text{D3NC}} = - \frac{1}{4\pi g_s} \int \dd x^2 \dd y^2 
     & \bigg[ 8 \brac{F_{-z}^\mathtt{a} F_{+\bar{z}}^\mathtt{a} + F_{+z}^\mathtt{a} F_{- \bar{z}}^\mathtt{a}} + 2  \big( D_+ \calz^\mathtt{a} D_- \bcalz^\mathtt{a} + D_- \calz^\mathtt{a} D_+ \bcalz^\mathtt{a} \big)  \\ 
\nonumber
    &  - 4 D S^{I\, \mathtt{a}} \bar{D} S^{I\, \mathtt{a}} + (f_{\mathtt{bc}}{}^\mathtt{a}\calz^\mathtt{b} S^{I\, \mathtt{c}}) (f_{\mathtt{de}}{}^\mathtt{a}\bcalz^\mathtt{d} S^{I\, \mathtt{e}})  \\ 
    &  - H^\mathtt{a} \Bar{D} \calz^\mathtt{a} - \bar{H}^\mathtt{a} D \bcalz^\mathtt{a} 
    - B^\mathtt{a} \brac{F^\mathtt{a} + \inv{2} f_{\mathtt{bc}}{}^\mathtt{a} \calz^{\mathtt{b}} \bcalz^{\mathtt{c}}}
    \bigg] \, .
\end{align}
This action describes a 2d sigma model on Hitchin moduli space \cite{Lambert:2024yjk}.
It is important to note that the action \eqref{D3NC_gauge} has been derived directly from the D3NC limit of the $\mathcal{N}=4$ SYM action. It would be interesting to check that the same result appears by taking first the D3NC limit of the DBI action with Wess-Zumino term, followed by the low-energy limit $\alpha'\to 0$, as required by the \ref{Uniqueness} condition of section \ref{sec:consistency_conditions}. As for the case of the D1NC limit discussed in section \ref{sec:D1NC_limit}, also for the D3NC limit there are higher-order divergent terms that appear when first taking the D3NC limit of the DBI action with Wess-Zumino term. In order for the D3NC limit to be a consistent limit, these higher-order divergences should cancel once the constraint is imposed.

\vspace{3mm}
\noindent\emph{Symmetries.} The symmetries of the action \eqref{D3NC_gauge}, and their associated Noether currents, have been computed in \cite{Lambert:2024yjk}. These symmetries have been distinguished between \emph{unphysical} and \emph{physical}, depending on whether their Noether charges are vanishing or not, respectively. They are the following:
\begin{subequations}
    \begin{align}
        V_\pm^{(n)} &=(\sigma^\pm)^n \partial_\pm \,, &
        \hat{P}^{(n,m,\ell)}&=z^n (\sigma^+)^m (\sigma^-)^\ell \partial_z \,, &
        \hat{Q}^{(n,m,\ell)}&=i z^n (\sigma^+)^m (\sigma^-)^\ell \partial_z \,,\\
        Z&=i \mathcal{Z} \frac{\partial \phantom{\mathcal{Z}}}{\partial \mathcal{Z}} \,, &&&
        L_{IJ}&= S^I \frac{\partial \phantom{S^I}}{\partial S^J} -S^J \frac{\partial \phantom{S^I}}{\partial S^I} \,, 
    \end{align}
\end{subequations}
together with additional unphysical internal symmetries involving more complicated expressions, detailed in \cite{Lambert:2024yjk}. Here, as before, hatted generators indicate unphysical symmetries.

\paragraph{The gravity theory.} The gauge theory in the D3NC limit was suggested in \cite{Lambert:2024yjk,Lambert:2024ncn,3096754} to be dual to a gravitational system of intersecting D3-D3$'$ branes, where the D3$'$-brane is smeared. The relativistic geometry of this setting is ($A\in\{0,1\}$, $a\in\{2,3\}$, $I\in\{1,..., 4\}$, $i\in\{5,6\}$), 
\begin{subequations}\label{D3_D3'_branes}
\begin{align} \nonumber
    \dd s^2_{\text{D3-D3}'} &= - H_3^{-1/2} H_{3'}^{-1/2} \eta_{AB} \dd x^A \dd x^B  + H_3^{-1/2} H_{3'}^{1/2} \delta_{ab} \dd y^a\dd y^b  \\
    & \hspace{1.0cm} + H_3^{1/2} H_{3'}^{-1/2} \delta_{ij} \dd z^i \dd z^j + H_3^{1/2} H_{3'}^{1/2} \delta_{IJ} \dd w^I\dd w^J \, , \\
    C_4 &=  (H_3^{-1}-1) \dd x^0 \wedge \dd x^1 \wedge \dd y^2 \wedge \dd y^3 \, , \\
    C'_4 &= H_{3'}^{-1} \dd x^0 \wedge \dd x^1 \wedge \dd z^5 \wedge \dd z^6 \, , \\
    e^{\Phi} &= g_s \, ,
\end{align}
\end{subequations}
where $H_3$ and $H_{3'}$ satisfy the equations
\begin{subequations}
\begin{align}
    0 &= \partial^I \partial_I H_{3'} \, , \label{H3'_D3D3'} \\
    0 &= H_{3'} \partial^i \partial_i H_3 + \partial^I \partial_I H_3 \, .\label{H3_D3D3'}
\end{align}
\end{subequations}
The next step is to take the D3NC limit. Following \cite{Lambert:2024yjk}, the D3NC limit is engineered in such a way to create a codimension four foliation. This has the following consequence for a probing string: in the Polyakov action, it would generate a divergent term that is not cancelled via the usual Gomis-Ooguri critical B-field. For this reason, this limit is exotic, and it still deserves investigation. 

That said, the D3NC limit is achieved by taking $H_{3'}$ completely ``smeared'' over the $y^a$ directions, and $H_3$ as a solution of \eqref{H3_D3D3'}, 
\begin{eqnarray}
    H_{3'} = c^{-2} \, , \qquad\qquad
    H_3 = 1 + \frac{R^4}{\brac{z^iz^i + c^{-2} w^I w^I}^2} \, , 
\end{eqnarray}
where $R$ is an integration constant. By plugging this choice in \eqref{D3_D3'_branes}, and by taking $c\to \infty$, one arrives at the D3NC limit of the intersecting D3-D3$'$ brane system,  
\begin{subequations} \label{eq: metric c expansion D3}
\begin{align}
    \dd s^2_{\text{D3NC D3-D3}'} &= c \tau_{\mn} \dd X^{\mu} \dd X^{\nu} + c^{-1} h_{\mn} \dd X^{\mu} \dd X^{\nu} \, , \\
    \tau_{\mn} \dd X^{\mu} \dd X^{\nu} &= H^{-1/2} \eta_{AB} \dd x^A \dd x^B  + H^{1/2} \delta_{ij} \dd z^i \dd z^j  \, , \\
    h_{\mn} \dd X^{\mu} \dd X^{\nu} &= H^{-1/2} \delta_{ab} \dd y^a\dd y^b + H^{1/2} \delta_{IJ} \dd w^I\dd w^J   \, , \\
    C_4 &=  (H^{-1}-1) \dd x^0 \wedge \dd x^1 \wedge \dd y^2 \wedge \dd y^3 \, , \\
    C'_4 &= c^2 \dd x^0 \wedge \dd x^1 \wedge \dd z^5 \wedge \dd z^6 \, , \\
    e^{\Phi} &= g_s \, ,
\end{align}
\end{subequations}
where $H$ is the leading order of $H_3$, 
\begin{eqnarray}
     H = 1 + \frac{R^4}{(z^iz^i)^2} \, .
\end{eqnarray}
Finally, by taking the near-horizon limit $R\to \infty$, one arrives at a spacetime geometry identical to that in \eqref{eq: metric c expansion D3}, except that $H$ is replaced by the dominant term $R^4/(z^iz^i)^2$. Similarly to the D1NC limit, to arrive at this geometry one does not rescale the coordinates. Instead, the factors of $c$ enter purely by assuming that $H_{3'}$ is $c^{-2}$, which then propagate to $H_3$ upon solving \eqref{H3_D3D3'}.

A different solution to \eqref{H3_D3D3'} was proposed in \cite{Lambert:2024ncn, 3096754}, where also the $H_3$ is smeared along the $z^i$ directions
\begin{eqnarray}\label{D3_harmonic_smeared}
    H_{3'} = c^{-2} \, , \qquad\qquad
    H_3 = 1 + \frac{R^4}{\brac{w^I w^I}^2} \, , 
\end{eqnarray}
where $R$ is again an integration constant. The steps we have to follow in this case are same as for the \emph{localised} solution above mutatis mutandis. The metric we obtain at the end of the process is the same as \eqref{eq: metric c expansion D3} but with $H$ replaced by $1 + \frac{R^4}{\brac{w^I w^I}^2}$.

The \emph{smeared} limit is usually preferred because it does not require the rescaling of the number of branes implicitly required by the non-relativistic limit. We notice that, in this solution, the longitudinal metric $\tau_{\mu\nu}$ depends on a transverse coordinate, which nevertheless might not be a concern as this is not a standard SNC limit.

\vspace{3mm}
\noindent\emph{Symmetries.} 
As mentioned before, the D3NC is an exotic limit, since the spacetime acquires a codimension four foliation. The underlying supergravity theory has not yet been found, and therefore neither the Killing vectors preserving the full supergravity solution. The \ref{Symmetry matching} condition is still missing for this proposed holography, although some progress has been made on this topic \cite{3096754}.

\subsection{More generic configurations: $p$-NC limits of D$q$-branes}

The various examples of non-relativistic holography that we presented so far can be framed into a more general picture of $p$-NC limits of D$q$-branes, where the longitudinal/transverse directions of the $p$-NC limit have different overlaps with the longitudinal/transverse directions of the D$q$-brane, as proposed in \cite{Blair:2024aqz}. In the below discussion, we will only present the gravitational side of a putative $p$-NC holography. The identification of their dual gauge theory has not been done yet, although some discussions at the level of the DBI can be found in \cite{Blair:2024aqz, Blair:2025prd}.  

It is also important to note that in the $p$-NC limits that we shall present, the spacetime is a codimension $p+1$ singular foliation. Except for the case $p=1$, corresponding to the string foliation, it remains unclear how to define a consistent string theory for a generic $p$. The reason is that only for a codimension two foliation is it possible to turn on a closed $B$-field that cancels the $\tau$ divergence appearing in the Polyakov action, thereby leading to the Gomis–Ooguri non-relativistic action.

The starting point is the black D$q$-brane as a solution of type II supergravity, relevant for the gravity part of holography where $G_s N \gg 1$. The full supergravity solution is given by, 
\begin{subequations} \label{Dq-brane_sugra}
\begin{align}
    \dd s^2_{\text{D}q} &= H^{-\frac12} \dd X^{\parallel} \cdot \dd X^{\parallel} + H^{\frac12} \dd X^{\perp} \cdot \dd X^{\perp}\, , \\
    C_{q+1} &=  (G_s H)^{-1} \dd X^0 \wedge \dots \wedge \dd X^q \, , \\
    e^{\Phi} &= G_s  H^{\frac{3-q}{4}}\, , \\
    H &= 1 + \left( \frac{R}{r}\right)^{7-q} \, , 
\end{align}
\end{subequations}
where $X^{\parallel} \equiv (X^0, ..., X^q)$ and $X^{\perp} \equiv (X^{q+1}, ..., X^{9})$ are, respectively, the longitudinal and transverse coordinates to the D$q$-brane, and $r$ is defined as $r\equiv \sqrt{X^{\perp}\cdot X^{\perp}}$. 

The next step is to take a $p$-NC limit of \eqref{Dq-brane_sugra} with different alignments. First, we split $p$ into $(p^{\parallel}, p^{\perp})$, under the conditions that $p^{\parallel} + p^{\perp} = p$ and $p^{\parallel} \leq q$, $p^{\perp}\leq 9-q-1$. The numbers $(p^{\parallel}, p^{\perp})$ represent a splitting of the spatial longitudinal directions involved in the $p$-NC limit between the spatial longitudinal and the transverse directions of the D$q$-brane. More explicitly, the splitting is the following,
\begin{subequations} \label{index_splitting}
\begin{align}
\tag{\theparentequation}
    X^{\parallel} &= (X^A, X^I) \, , \qquad\text{with} \quad A\in\{0, ..., p^{\parallel}\} \, , \ I\in \{p^{\parallel}+1, ..., q\}\, , \\
    \notag
    X^{\perp} &= (X^a, X^i) \, , \qquad\ \text{with} \quad a\in\{q+1, ..., q+1 + p^{\perp}\} \, , \ i\in \{q+2 + p^{\perp}, ..., 9\}\, .
\end{align}
\end{subequations}
The case of $p^{\perp}=0$ and $p^{\parallel} = p = q$ corresponds to the near-horizon limit, e.g. the aligned 3\nobreakdash-NC limit of the D3-brane discussed in section \ref{sec:NRasNH}. The case of $p^{\parallel} = 1$ and $p^{\perp}=2$ corresponds to the misaligned 3-NC limit of the D3-brane discussed in section \ref{sec:misaligned_D3NC}, i.e. the so-called D3NC limit. 

To derive the $p$-NC limit of a D$q$-brane, we first need to distinguish the cases where $p^{\perp} = 0$ and $p^{\perp} \geq 1$. In addition, there are different procedures, which can either be a limit, a smearing or a large $N$ limit. Below, we shall review these cases separately. 

%At first sight, the D3NC limit and the setting considered here might look different. The reason is that the initial relativistic supergravity solution considered in section \ref{sec:misaligned_D3NC} describes a system of intersecting D3-D3$'$ branes, where the 3-NC limit physically plays the role of an additional D3$'$-brane that intersects the given D3-brane. On the other hand, the setting used in this section assumes the initial relativistic metric to be \eqref{Dq-brane_sugra}, which does not describe a system of intersecting D$q$-D$p$ branes. However, as we shall see, the smearing of the D3$'$-brane used to take the D3NC limit in section \ref{sec:misaligned_D3NC} can equivalently be realised as a pure coordinate rescaling, supplemented by a rescaling of the parameter $L$, in the D$q$-brane metric \eqref{Dq-brane_sugra}.    

\paragraph{The limit procedure for $p^{\perp} = 0$.}

In this case, we have that $p=p^{\parallel} \leq q$. The $p$-NC limit is given by rescaling the coordinates and parameters as follows, 
\begin{subequations} \label{rescaling_coords_pNC_Dq_brane_1}
\begin{align}
X^A &= \sqrt{c} \, x^A \, , 
&X^I &= \frac{x^I}{\sqrt{c}}  \, ,  &X^i &= \frac{x^i}{\sqrt{c}}  \, , \\
G_s &= c^{\frac{p-3}{2}} g_s \, ,  &R^{7-q} &= c^{\frac{p-3}{2}} \ell^{7-q} \, . && 
\end{align}
\end{subequations}
Due to this rescaling, $r$ scales as $r = c^{-\frac12} \sqrt{x^i x^i}$. Moreover, the rescaling of $R$ is not independent, but implied by the rescaling of $G_s$, via the defining relation between $R$, $G_s$, $\alpha'$ and the number of D$q$-branes $N$, as follows, 
\begin{eqnarray}
    R^{7-p} \sim N G_s \alpha'^{\frac{7-p}{2}} \, .
\end{eqnarray}

We first consider the case $p < q$. By implementing the rescaling \eqref{rescaling_coords_pNC_Dq_brane_1} inside \eqref{Dq-brane_sugra}, and by taking the $c\to \infty$ limit, we obtain
\begin{subequations}
\begin{align}
    \dd s^2_{p\text{NC-D}q} &= c \,\tau_{\mn} \dd X^{\mu} \dd X^{\nu} + \frac{1}{c}h_{\mn} \dd X^{\mu} \dd X^{\nu}\, , \\
    \tau_{\mn} \dd X^{\mu} \dd X^{\nu} &= H^{-\frac12} \dd x^A \dd x^B \eta_{AB}  \, , \\
    h_{\mn} \dd X^{\mu} \dd X^{\nu} &=H^{-\frac12}\dd x^I \dd x^J \delta_{IJ} 
    + H^{\frac12} \dd x^i \dd x^j \delta_{ij} \, , \\
    C_{q+1} &=  c^{2+p-q}(g_s H)^{-1} \dd x^0 \wedge \dots \wedge \dd x^q \, , \\
    e^{\Phi} &= c^{\frac{p-3}{2}} g_s  H^{\frac{3-q}{4}}\, , \\
    H &= 1 + c^{\frac{p-q+4}{2}}\left( \frac{\ell}{\sqrt{x^i x^i}}\right)^{7-q} \, . 
\end{align}
\end{subequations}
In this formalism, this limit has not been fully studied in the literature, in particular whether a critical RR potential is required for consistency with the supergravity equations. However, in the formalism where the $p$-NC limit of a D$q$-brane is seen as an intersecting D$p$-D$q$ branes, an example of this case was proposed in \cite{Lambert:2024ncn} as a D$0$-D$4$ system.

The case $p=q$ is special, as it defines the \emph{near-horizon} geometry of the D$q$-brane. In this case, the metric is Lorentzian, and given by, 
\begin{subequations} \label{NH_Dq-brane_sugra}
\begin{align}
    \dd s^2_{\text{NH-D}q} &= \chi^{-\frac12} \dd x^A \dd x^B \eta_{AB} + \chi^{\frac12} \dd x^i \dd x^j \delta_{ij}\, , \\
    C_{q+1} &=  (g_s \chi)^{-1} \dd x^0 \wedge \dots \wedge \dd x^q \, , \\
    e^{\Phi} &= g_s  \chi^{\frac{3-q}{4}}\, , \\
    \chi &= \left( \frac{\ell}{\sqrt{x^i x^i}}\right)^{7-q} \, . 
\end{align}
\end{subequations}
The aligned 3-NC limit of the D3-brane discussed in section \ref{sec:NRasNH} falls into this class of limit.

\paragraph{The limit procedure for $p^{\perp} \geq 1$.}
In this case, to take the $p$-NC limit of \eqref{Dq-brane_sugra}, one needs to rescale the coordinates and parameters as follows, 
\begin{subequations} \label{rescaling_coords_pNC_Dq_brane}
\begin{align}
X^A &= \sqrt{c} \, x^A \, , \qquad &X^a &=\sqrt{c} \, x^a \, , \\
X^I &= \frac{x^I}{\sqrt{c}}  \, , \qquad &X^i &= \frac{x^i}{\sqrt{c}}  \, , \\
G_s &= c^{\frac{p-3}{2}} g_s \, , \qquad &R^{7-q} &= c^{\frac{p-3}{2}} \ell^{7-q} \, . 
\end{align}
\end{subequations}
Due to this rescaling, $r$ will scale as $r = c^{\frac12} \sqrt{x^a x^a+ c^{-2} x^i x^i}$, where now $x^ax^a$ is the dominant term. By taking the $c\to \infty$ limit, the supergravity solution \eqref{Dq-brane_sugra} becomes,
\begin{subequations} \label{pNC_Dq-brane_sugra}
\begin{align}
    \dd s^2_{p\text{NC-D}q} &= c \,\tau_{\mn} \dd X^{\mu} \dd X^{\nu} + \frac{1}{c}h_{\mn} \dd X^{\mu} \dd X^{\nu}\, , \\
    \tau_{\mn} \dd X^{\mu} \dd X^{\nu} &= H^{-\frac12} \dd x^A \dd x^B \eta_{AB} 
    + H^{\frac12} \dd x^a \dd x^b \delta_{ab} \, , \\
    h_{\mn} \dd X^{\mu} \dd X^{\nu} &=H^{-\frac12}\dd x^I \dd x^J \delta_{IJ} 
    + H^{\frac12} \dd x^i \dd x^j \delta_{ij} \, , \\
    C_{q+1} &=  c^{2+p^{\parallel}-q}(g_s H)^{-1} \dd x^0 \wedge \dots \wedge \dd x^q \, , \\
    e^{\Phi} &= c^{\frac{p-3}{2}} g_s  H^{\frac{3-q}{4}}\, , \\
    H &= 1 + c^{\frac{p+q-10}{2}}\left( \frac{\ell}{\sqrt{x^a x^a+ c^{-2} x^i x^i}}\right)^{7-q} \, . 
\end{align}
\end{subequations}
The $p$-NC limit also requires a critical RR potential $C_{p+1}$, in order to be consistent with the $p$-brane Newton Cartan type II supergravity equations of motion. From the perspective of the dual gauge theory, this corresponds to turning on a Wess-Zumino term in the DBI action that cancels the corresponding divergence from taking the $p$-NC limit, see e.g. the D1NC and D3NC limits discussed in sections \ref{sec:D1NC_limit} and \ref{sec:misaligned_D3NC} for concrete examples. The critical RR potential is,
\begin{eqnarray}
    C_{p+1} = c^2 g_s^{-1} \dd x^0 \wedge \cdots \wedge \dd x^{p^{\parallel}} \wedge  \dd x^{q+1} \wedge\cdots \wedge \dd x^{q+1+p^{\perp}} \, .   
\end{eqnarray}
In order for the metric \eqref{pNC_Dq-brane_sugra} to preserve a horizon, as required by the \ref{Horizon condition} at the beginning of section \ref{sec:consistency_conditions}, one needs to demand that $H$ is finite and does not lose the coordinate dependent term. Moreover, for consistency with the supergravity equations of motion, one requires that $C_{q+1}$ remains finite. This restricts the parameters $p$ and $q$ as follows,  
\begin{eqnarray}\label{dim_constraint}
    p+q = 10 \, , \qquad\qquad
    p^{\parallel} = q -2 \, . 
\end{eqnarray}
The D3NC limit discussed in section \ref{sec:misaligned_D3NC} and characterised by $(p,p^{\parallel},q) = (3,1,3)$ does not fulfil the above conditions, and for this reason, it cannot be obtained via a limit procedure. Indeed, such D3NC geometry is obtained via a smearing procedure which we are going to discuss next.   

Finally, assuming that the dimension constraint \eqref{dim_constraint} is satisfied, the near-horizon limit of \eqref{pNC_Dq-brane_sugra} is given by $\ell \to \infty$, such that $H$ simply becomes $(\ell /\sqrt{x^a x^a})^{7-q}$. Equivalently, the near-horizon limit can be implemented as a $q$-NC limit as discussed around equation \eqref{NH_Dq-brane_sugra}.

\paragraph{The smearing procedure for $p^{\perp} \geq 1$.}

A way to remove the constraint $p+q = 10$ arising in the limiting procedure is to introduce smearing.\footnote{The limit and smearing procedure can also be rephrase in the language of intersecting D$p$- and D$q$-branes. Written in this way, the smearing procedure seems more natural than the limit procedure, as it does not require us to rescale the number of branes (at least, when $p+q=4$ and $q+p^\perp=2$). For more detail, we refer to section 2.1 of \cite{Lambert:2024ncn}.} Physically, this consists in compactifying the $x^a$ directions over a $p^{\perp}$-torus. Under the compactification, the supergravity solution remains unchanged except of the harmonic solution $H$, which needs to be periodic in $x^a$. To make $H$ periodic, it is enough to place an infinite number of branes at regular intervals along the periodic directions $x^a$. By implementing this procedure, and by taking the same rescaling as the one given in \eqref{rescaling_coords_pNC_Dq_brane}, $H$ becomes\footnote{This sum converges only if $7 - q > 1$. This is always the case if one assumes that $7 - q - p^{\perp} > 0$, see \cite{Blair:2024aqz}, which means that the smeared brane has effective codimension greater than two.}
\begin{eqnarray}\label{torus_smearing}
    H = 1 + \sum_{k_a = -\infty}^{+\infty} \frac{c^{\frac{p-3}{2}} \ell^{7-q}}{\left[ c (x^a + 2\pi k_a R_a)^2 + c^{-1} x^i x^i \right]^{\frac{7-q}{2}}} \, , 
\end{eqnarray}
where $R_a$ are the compactification radii of the $p^{\perp}$-torus, and $k_a \in \mathbb{Z}$. Smearing consists in replacing the sum for an integral, i.e. $\sum_{k_a} \to \int \dd^{p^{\perp}} \mathbf{k}$. Then, after performing the integral, $H$ becomes
\begin{eqnarray}
    H = 1 + \left(\frac{\boldsymbol{\ell}}{\sqrt{x^i x^i}} \right)^{7-q-p^{\perp}} \, , \qquad\qquad
    \boldsymbol{\ell}^{7-q-p^{\perp}} = \frac{\Gamma(\frac{7-q-p^{\perp}}{2})}{(4\pi)^{\frac{p^{\perp}}{2}} \Gamma (\frac{7-q}{2})} 
    \frac{\ell^{7-q}}{R_{q+1} \cdots R_{q+ p^{\perp}}} \, . 
\end{eqnarray}
This procedure makes $H$ finite while keeping a non-trivial dependence on the transverse coordinates, and without the need of assuming $p+q = 10$. The previously discussed D1NC and D3NC limits of the D3-brane fall into this class of procedure.

\paragraph{The large $N$ limit for $p^{\perp} \geq 1$.}

Another possibility to generate a harmonic function that is finite while keeping a non-trivial dependence on the transverse coordinates is by allowing  the number of D$q$-branes to scale as well. Indeed, if we rescale $N$ as 
\begin{eqnarray}
    N \to c^{\frac{10-p-q}{2}} N \, , 
\end{eqnarray}
then $\ell$ will scale as $\ell^{7-q} \to c^{\frac{10-p-q}{2}}\ell^{7-q}$, which eliminates the $c$ dependency in the harmonic function $H$ in \eqref{pNC_Dq-brane_sugra}, and therefore the constraint $p+q = 10$ is no longer required. To be of physical meaning, $N$ should scale with a positive power of $c$, namely $N$ is large. This is guaranteed when $p+q < 10$. Otherwise, $N$ goes to zero, meaning there is no backreaction that generates a non-trivial bulk.

\paragraph{Web of dualities.}

So far we have seen how to take the $p$-NC limit of a D$q$-brane, but not each of these limits produces a novel configuration.
As pointed out in \cite{Lambert:2024ncn}, if the $p$-NC limit of a D$q$-brane gives the non-relativistic limit of an intersecting D$p$-D$q$ brane system, then the same result is obtained by inverting the order of $p$ and $q$, i.e. by taking the $q$-NC limit of a D$p$-brane. This implies that the two limits in the two configurations lead to the same result.

Furthermore, the known string dualities, namely T-, S-, and U-duality, can also be applied to non-relativistic string theory. These dualities map different types of non-relativistic limits into one another; the interested reader is referred to \cite{Blair:2023noj, Blair:2024aqz, Blair:2025prd}. While this mapping is defined at the level of the limits, it is reasonable to expect that it also extends to the corresponding theories. Consequently, not all the holographic dualities presented so far are independent: some may be related to each other through string dualities. We have already encountered an example of this phenomenon: the SNC holography discussed in section \ref{sec:consistency_conditions} is expected to be S-dual to the D1NC holography of section \ref{sec:D1NC_limit}.  

A more general framework comes from uplifting these geometries to M-theory. It was proposed in \cite{Blair:2024aqz} that each BPS decoupling limit (i.e. each $p$-NC limit) corresponds to a ``Discrete Light Cone Quantisation'' (DLCQ). The conjectured holography is between DLCQ$^n$/DLCQ$^m$, with $n<m$. For example, the near-horizon limit can be seen as a $p$-NC limit of a D$p$-brane. At the level of the dual theory, this corresponds to applying DLCQ once, i.e. DLCQ$^1$. In the bulk, instead, the near-horizon leaves the geometry Lorenztian, so the limit acts as a DLCQ$^0$. In this logic, the relativistic AdS/CFT is seen as a DLCQ$^0$/DLCQ$^1$ correspondence. By applying a further BPS decoupling limit, one obtains a non-relativistic holography denoted by DLCQ$^1$/DLCQ$^2$, and so on. In this framework, one can also find the non-relativistic gravitational solitons discussed in \cite{Harmark:2025ikv}.

\section{Beyond the non-relativistic limit}
\label{sec:Holo_other_NR_limits}

So far, we discussed a methodology to incorporate various types of non-relativistic limits inside the brane construction of the AdS/CFT correspondence, by demanding a series of consistency conditions outlined at the beginning of section \ref{sec:consistency_conditions}. The same method can be applied to other types of limits, not necessarily non-relativistic. The purpose of this section is to review the state of the art concerning the application of the \emph{Carroll} and \emph{flat space} limits to the brane construction of the AdS/CFT correspondence.

\subsection{The Carroll limit}

The Carroll limit is the opposite of the non-relativistic limit, as the parameter that plays the role of the speed of light $c$ is sent to zero. This limit is less well understood than the non-relativistic one, and several questions remain open - see \cite{Bagchi:2025vri, Ruzziconi:2026bix, Nguyen:2025zhg} for recent reviews on aspects of the Carroll limit. A method to incorporate the Carroll limit into the brane construction of AdS/CFT was proposed in \cite{Fontanella:2025tbs}, which we shall review here. The logic parallels that of the non-relativistic limit: the Carroll limit must satisfy the consistency conditions outlined at the beginning of section~\ref{sec:consistency_conditions}. There are two inequivalent types of Carroll limits, typically called \emph{electric} and \emph{magnetic} limits \cite{Henneaux:2021yzg}.  
As we shall see, only theories in the magnetic Carroll limit will appear from this construction. A very legitimate question is why theories in the electric Carroll limit are not appearing. Although this question is still open, a possible suggestion may come from \cite{Bagchi:2024rje}, where they noticed that in order to obtain an electric Carroll limit of the Polyakov action one needs to use the phase space action formalism. It may be that the electric Carroll $\mathcal{N}=4$ SYM only appears after rewriting the non-abelian DBI action in the phase space formalism.

\paragraph{Carroll limit in the gravity side.}

The Carroll limit of the Polyakov string action is somehow bizarre: there is no mechanism, such as fine-tuning of a critical field, that allows to eliminate the divergence coming from the transverse metric $h_{\mu\nu}$. Instead, the divergence has to be eliminated via a Hubbard-Stratonivich transformation. Because of that, it was shown in \cite{Cardona:2016ytk} that a Carroll limit with string foliation, i.e. of codimension two, gives the same type of dynamics of a Carroll limit with particle foliation, i.e. of codimension one. For this reason, it is enough to consider the particle Carroll limit that aligns with the D3-brane directions as follows, 
\begin{align}
\begin{array}{rrrrrrr}
D3: & 0 & 1 &2& 3 && \\
\text{particle Carroll}:& 0 &  & & & &  \, .  \\
\end{array}
\end{align}

Following the construction of \cite{Fontanella:2025tbs}, which mimics the logic applied to the non-relativistic limit, one starts from the metric of a stack of D3-branes \eqref{D3_metric_z} and then take the particle Carroll limit,   
\begin{eqnarray}
\label{Carroll_coords_resc_1_2NH}
    \alpha' \to \frac{\alpha'}{c^2}\, , \qquad
    z\to \frac{z}{c} \, , \qquad
    x^i \to \frac{x^i}{c} \, , \qquad
    R \to \frac{R}{c} \, . 
\end{eqnarray}
By plugging the rescaling \eqref{Carroll_coords_resc_1_2NH} inside the stack of D3-branes \eqref{D3_metric_z}, and after taking the $c\to 0$ limit, one arrives at the Carroll D3-brane metric,
\begin{subequations}\label{Carroll_NC_D3_metric}
    \begin{align}
        \dd s^2_{\text{Carroll D3-brane}} &=  \left(\frac{1}{c^2} h_{\mu\nu} + \tau_{\mu\nu}\right) \dd X^{\mu} \dd X^{\nu}\, , \\
        h_{\mu\nu} \dd X^{\mu} \dd X^{\nu} &= \frac{R^4}{\alpha^{\prime 2}\sqrt{f(z)}} \dd x^i \dd x_i + \frac{\alpha^{\prime 2} \sqrt{f(z)}}{z^4} \dd z^2 + \frac{\alpha^{\prime 2} \sqrt{f(z)}}{z^2}  \dd \Omega_5^2\, , \\
        \tau_{\mu\nu} \dd X^{\mu} \dd X^{\nu} &= -\frac{R^4}{\alpha^{\prime 2}\sqrt{f(z)}} \dd t^2  \, .
    \end{align}
\end{subequations}
The next step is to take the near-horizon limit $\alpha'\to 0$, arriving at the Carroll AdS$_5\times$S$^5$ geometry,\footnote{Here there is an order of limits to resolve. Similarly to the argument provided for the non-relativistic limit, we demand that the ``old'' $\alpha'$ goes to zero, which implies that the ``new'' $\alpha'$ appearing after taking the rescaling \eqref{Carroll_coords_resc_1_2NH} goes to zero faster than $c^2$.}
\begin{subequations}\label{Carroll_NC_AdS5xS5_metric}
\begin{align}
    \dd s^2_{\text{Carroll AdS}_5\times\text{S}^5} &= \left(\frac{1}{c^2} h_{\mu\nu} + \tau_{\mu\nu}\right) \dd X^{\mu} \dd X^{\nu}\,  , \\
    h_{\mu\nu} \dd X^{\mu} \dd X^{\nu} &= \frac{R^2}{z^2} \left( \dd z^2 + \dd x^i \dd x_i \right) + R^2  \dd \Omega_5^2 \, , \\
    \tau_{\mu\nu} \dd X^{\mu} \dd X^{\nu} &= -\frac{R^2}{z^2} \dd t^2 \, , 
\end{align}
\end{subequations}
where $h_{\mu\nu}$ and $\tau_{\mu\nu}$ describe $\mathbb{H}^4 \times S^5$ and warped $\mathbb{R}$, respectively. 
The Penrose boundary of the geometry \eqref{Carroll_NC_AdS5xS5_metric} is given by the Carroll Mink$_4$ spacetime \cite{Fontanella:2025tbs}, characterised by $\tilde{h} = R^2 \dd x^i \dd x_i$ and $\tilde{\tau} = - R^2 \dd t^2$.  
Moreover, the limits can also be taken in the opposite order and the final metric will be again \eqref{Carroll_NC_AdS5xS5_metric}, therefore the \ref{Uniqueness} consistency condition is fulfilled.

The dynamics of a string propagating on the background \eqref{Carroll_NC_AdS5xS5_metric} is described by the magnetic Carroll string action discussed in \cite{Harksen:2024bnh}, where the worldsheet is relativistic. 
The worldsheet itself cannot be embedded inside the spacetime spanned by the $\tau$ metric tensor, as it is one-dimensional. Furthermore, the $\mathfrak{so}(1,1)$ Lorentz invariance of the worldsheet prevents us from embedding the worldsheet in a mixed way between $\tau$ and $h$.
Instead, the magnetic Carroll string will propagate in the space spanned by $h$. As $h$ is purely spatial, the only possibility is to embed the Lorentzian worldsheet as a tachyon. This is somehow analogue to the tachyonic particle that propagates outside the light-cone - something expected when taking the magnetic Carroll limit.

From the rescaling of $\alpha'$ and $R$ in \eqref{Carroll_coords_resc_1_2NH}, and from the definition of $R$ given below \eqref{AdS5xS5_metric}, one deduces that $g_s N$ needs to remain finite. Since $g_s$ has to rescale as in the gauge theory, namely as $g_s \to c g_s$ as we shall see in a moment, then $N$ has to rescale as $N\to N/c$. Then this implies that the self-dual 5-form RR field \eqref{RR_5_form}, after taking the Carroll limit, will diverge as     
\begin{eqnarray}
   F^{(5)}_{\text{Carroll}} &=& \frac{1}{c^4} \frac{1}{g_s R}  \text{dvol}(\text{AdS}_5)  + \frac{1}{c^5} \frac{1}{g_s R}  \text{dvol}(\text{S}^5) \, .
\end{eqnarray}
At this point, it would be important to check that the above expressions for the metric structure and fluxes obtained via a limit procedure represent a proper solution of the Carroll type IIB supergravity. Unfortunately, this is still a missing point in literature because such supergravity theory has not been found yet. Connected to this, the Killing vectors preserving the full solution have not been found yet, due to the lack of understanding of the local symmetries of all supergravity fields. Therefore, a symmetry matching with the dual gauge theory is still an open question.

\paragraph{Carroll limit in the gauge theory side.}

In the weak coupling regime, we start by considering the non-abelian DBI action in 10d flat spacetime. 
Following \cite{Fontanella:2025tbs}, the particle Carroll limit is taken at the level of the target space coordinates. There are two possibilities, one is 
\begin{eqnarray}\label{Carroll_DBI_1}
    X^0 \to c X^0 \, , \qquad
    X^a \to X^a \, , 
\end{eqnarray}
and the other one is 
\begin{eqnarray}\label{Carroll_DBI_2}
    X^0 \to  X^0 \, , \qquad
    X^a \to \frac{X^a}{c}  \, .
\end{eqnarray}
Both of them  create a 4d flat Carroll geometry, once the six transverse directions to the D3-brane have been decoupled by the low-energy limit.
However, as discussed in \cite{Fontanella:2025tbs}, only the rescaling \eqref{Carroll_DBI_1} commutes with the low-energy limit, i.e. it fulfils the \ref{Uniqueness} condition given at the beginning of section \ref{sec:consistency_conditions}. For this reason, the rescaling \eqref{Carroll_DBI_2} has to be discarded. In addition to the rescaling \eqref{Carroll_DBI_1}, $g_s$ also needs to rescale as 
\begin{eqnarray}\label{dilaton_resc_gauge}
    g_s \to c g_s \, .
\end{eqnarray}
By plugging the rescaling \eqref{Carroll_DBI_1} and \eqref{dilaton_resc_gauge} inside the non-abelian DBI action, together with the limit $c\to 0$, and then by taking the low-energy limit $\alpha'\to 0$ one arrives at the magnetic Carroll limit of the bosonic sector of $\mathcal{N}=4$ super Yang-Mills, 
\begin{eqnarray}\label{Carroll_N=4_SYM}
    &&S_{\text{CSYM}} = \frac{1}{2\pi g_s}\int \dd t\, \dd^{3}x \left[-\frac{1}{4} (F^{ij\, \mathtt{a}})^2 
    - \frac{1}{2} (D_i S^{\hat{I}\, \mathtt{a}} )^2  + \frac{1}{4} \left( \mathfrak{f}_{\mathtt{b}\mathtt{c}}{}^{\mathtt{a}} S^{\hat{I}\, \mathtt{b}} S^{\hat{J}\, \mathtt{c}} \right)^2 \right. \\
    \notag
    && \hspace{4.8cm} \left. -\frac{1}{4} \kappa^{i \, \mathtt{a}} F^{\mathtt{a}}_{t i} + \frac{1}{2} \lambda^{\mathtt{a}}_{\hat{I}} D_t S^{\hat{I}\, \mathtt{a}}   \right]\, ,
\end{eqnarray} 
The Lagrange multipliers $\lambda^{\mathtt{a}}_{\hat{I}}$ and $\kappa^{i \, \mathtt{a}}$ emerge from the elimination of the divergent terms via a Hubbard-Stratonovich transformation, similarly to the gravity picture. Inverting the order in which we perform the limits gives us exactly the same action, fulfilling the \ref{Uniqueness} condition.

Completing the action \eqref{Carroll_N=4_SYM} with fermions is a delicate matter. Carroll fermions have been studied in the literature \cite{Bergshoeff:2023vfd, Bergshoeff:2024ytq, Grumiller:2025rtm, Banerjee:2022ocj, Bagchi:2022eui, Koutrolikos:2023evq}, however it remains an open question how to couple them to a bosonic action in a supersymmetric manner. While a solution is known for the \emph{electric} Carroll fermion, no such construction exists for the \emph{magnetic} case. Since the action \eqref{Carroll_N=4_SYM} arises as the magnetic Carroll limit of the bosonic sector of $\mathcal{N}=4$ super Yang–Mills theory, it is therefore necessary to first understand magnetic Carroll supersymmetry.

\vspace{3mm}
\noindent\emph{Symmetries.} The symmetries of the action \eqref{Carroll_N=4_SYM} can be divided into spacetime and internal symmetries \cite{Fontanella:2025tbs}. The spacetime symmetries consist in the following generators 
\begin{subequations}
\begin{align}
   D&=t\partial_t + x^i\partial_i \, , &  P_i&=\partial_i \, , & &K_i = 2x_i t \partial_t - x_jx^j\partial_i+ 2x_i x^j\partial_j \, , \\
   L_{ij}&= x_j\partial_i-x_i\partial_j \, , & & &
    &M^{(\ell,m,n)}=x_1^{\ell}x_2^m x_3^n \partial_t \, .
\end{align}
\end{subequations}
which they form the so-called \emph{Conformal Carroll Algebra} (CCA) in $d=4$, isomorphic to the Bondi, van der Burg, Metzner and Sachs (BMS) algebra in $d=5$ \cite{Duval:2014uva}. These symmetries, in addition of acting on the spacetime coordinates, also need to act on the fields, see \cite{Fontanella:2025tbs} for the detail.

The internal symmetries, which act on the fields but not on the coordinates, are quite involved. We therefore refrain from presenting them in full detail here and refer the interested reader to \cite{Fontanella:2025tbs}. 
However, it is important to comment on a peculiarity of the symmetry algebra CCA$\oplus$Internal. This algebra does not close under the ordinary Lie bracket. Indeed, while
\begin{eqnarray}
    [\mathrm{CCA},\mathrm{CCA}] \subset \mathrm{CCA}, \qquad
[\mathrm{Internal},\mathrm{Internal}] \subset \mathrm{Internal} \, , 
\end{eqnarray}
the mixed bracket $[\mathrm{Internal},\mathrm{CCA}]$ does not close within either the CCA or the internal algebra. It is still an open problem to understand the underlying Lie algebra structure of these symmetries. 
Moreover, the corresponding Noether charges have not been computed yet, and it could well be that many of them are vanishing, similarly to the non-relativistic examples discussed in \cite{Lambert:2024uue, Lambert:2024yjk}. This would make such symmetries unphysical, and therefore not expected to be matched in the dual gravity theory. These are open questions for the future.

An application of this \emph{Carroll bulk} holography to flat space holography was suggested in \cite{Fontanella:2025tbs}, to which we refer the interested reader.

\subsection{The Flat Space limit}

Another limit of physical interest is the flat space limit. It consists of taking the common radius of AdS$_5$ and S$^5$ to infinity, which flattens the AdS$_5\times$S$^5$ geometry and results in 10d Minkowski spacetime. As we shall explain below, the flat space limit cannot be consistently incorporated inside the brane construction of the AdS/CFT, for the reason that it does not fulfil the \ref{Horizon condition} and the \ref{Holographic realisation} condition given at the beginning of section \ref{sec:consistency_conditions}.

Let us start by considering, as usual, the metric of a stack of $N$ D3-branes \eqref{D3_metric_z}.  After applying the near-horizon limit, it gives the AdS$_5\times$S$^5$ metric \eqref{AdS5xS5_metric}, which we write here for convenience, 
\begin{eqnarray}
\dd s^2_{\text{AdS}_5\times\text{S}^5} &=& R^2 \left[\frac{-\dd t^2 +  \dd z^2 + \dd x^i \dd x_i}{z^2} + \left(\frac{4-y^2}{4+y^2} \right)\dd \phi^2 + \frac{16\ \dd y^2 }{\left( 4+y^2\right)^2} \right]\, , 
\end{eqnarray}
To take the flat space limit, we need to perform the following coordinate rescaling, 
\begin{eqnarray}\label{flat_space_limit}
    z\to z + \omega \, , \qquad
    \phi \to \frac{\phi}{\omega} \, , \qquad
    y^m \to \frac{y^m}{\omega} \, , \qquad
    R \to \omega R \, ,  
\end{eqnarray}
where $\omega$ is a dimensionless parameter. After taking the limit $\omega \to \infty$, we obtain the geometry of 10d Minkowski, 
\begin{eqnarray}\label{Mink_10_metric}
\dd s^2_{\text{Mink}_{10}} &=& R^2 \left(-\dd t^2 +  \dd z^2 + \dd x^i \dd x_i + \dd \phi^2 + \dd y^2 \right)\, . 
\end{eqnarray}

Let us now reverse the order of limits. By applying first the flat space limit \eqref{flat_space_limit} to the stack of D3-branes \eqref{D3_metric_z}, we arrive directly to the flat spacetime metric \eqref{Mink_10_metric}, therefore losing the horizon region. Moreover, allowing for a rescaling of the parameter $\alpha'$ does not make any change, as in the limit $\omega\to \infty$ one always get the flat metric \eqref{Mink_10_metric}.

Since the flat space limit of the stack of D3-branes metric loses the horizon region, we would be led to conclude that the flat space limit cannot be consistently taken at the level of the brane setting of the AdS/CFT, just because the \ref{Horizon condition} is not fulfilled. Although this is in part correct, one may also argue that the \ref{Horizon condition} is not required for the flat space limit. The reason is that the flat space limit of AdS$_5\times$S$^5$ is 10d Minkowski spacetime, which matches precisely the asymptotic geometry from which AdS$_5\times$S$^5$ decouples in the near-horizon limit. From this perspective, and after taking the flat space limit, the near-horizon and asymptotic geometries coincide, so effectively there is just a single string theory to consider. Because of that, one may argue that a stack of D3-branes that loses the horizon and produces the flat metric is perfectly acceptable, as it matches the result obtained in the reversed order of taking the limits.   
The issue comes when trying to holographically match the theory appearing in this scenario with the ones in the weak coupling regime: in this case the near-horizon limit produces an ``asymptotic'' theory, that is supergravity in 10d, and a ``near-horizon'' theory, that is $\mathcal{N}=4$ super Yang-Mills. To holographically match the result obtained in the bulk, one expects that also in the boundary the two theories boil down to a single one in the flat space limit. However, that sounds highly improbable, because of the different nature of the two theories: one involves gauge interactions in 4d, the other one gravity in 10d. This leads us to regard the flat space limit as inconsistent at the level of the brane setting of the AdS/CFT.

A second issue is the violation of the \ref{Holographic realisation} condition at the beginning of section \ref{sec:consistency_conditions}. The flat space limit of AdS$_5\times$S$^5$ gives us Mink$_{10}$, whose null boundary is a nine-dimensional Carroll flat spacetime. On the other hand, the flat space limit of four-dimensional $\mathcal{N}=4$ super Yang-Mills can at most produce a Carrollian gauge theory in a 3+1 dimensional flat spacetime, $\mathfrak{Carroll}_{4}$. From here, it is very clear that 
\begin{eqnarray}
    \partial \text{Mink}_{10} \neq \mathfrak{Carroll}_{4} \, .
\end{eqnarray}
This is the boundary miss-match problem that occurs when taking the flat space limit of the AdS$_5$/CFT$_4$. Possible solutions have been presented in \cite{Fontanella:2025tbs}, either in terms of a suitable compactification of Mink$_{10}$, which would be desirable to occur via a dynamical mechanism, or in terms of holography involving a Carroll-bulk gravity theory. For a recent analysis of this issue from the wave function perspective, see \cite{Berenstein:2025tts, Berenstein:2025qhb}.

%%%%%%%%%%%%%%%%%%%%%%%%%%%%%%%%%%%%%%%%%%%%%%%%%%%%%%%%%%%%%%%%%%%%%%%%%%%%%%%%%%%%%%%%%%%%%%%%%%%%%%%%%%%%%%%%%%%%%%%%%%%%%%%%%%%%%%%%%%%%%%%%%%%%%%%%%%%%%%%%%%%%%%%%%%%%%%%%%%%%%%%%%%%%%%%%%%%%%%%%%%%%%%%%%%%%%%%%%%%%%%%%%%%%%%%%%%%%%%%%%%%%%%%%%%%%%%%%%%%%%%%%%%%%%%%%%%%%%%%%%%%%%%%%%%%%%%%%%%%%
\chapter{Classical strings and Quantisation}
\label{chap:quantisation}

Quantisation of relativistic strings in flat spacetime is simple enough that it is explained in the first chapters of any string theory book, e.g. \cite{Polchinski_book1}. Its simplicity boils down to the fact that the Polyakov action in conformal gauge is quadratic in fields, making its quantisation equivalent to the quantisation of a set of free bosons with the additional constraint of having to cancel the worldsheet stress-energy tensor. 

Quantisation of strings in curved spacetime is more difficult to handle, although the logic is quite straightforward. The first step is to find a suitable classical solution that can be used as vacuum for the quantum field theory. This means solving the classical equations of motion. After that, one expands the embedding coordinate fields around the chosen solution. The action in terms of deviations from the classical solution can be brought to the form of a quadratic action plus interaction terms, which can be quantised via the usual perturbative methods, like canonical quantisation, or series expansion of the path integral.

In this chapter, we discuss the whole process for NR strings in SNC AdS$_5\times$S$^5$ (GGK theory): first, we present the construction of classical strings around which we can expand the string theory action, and, after that, we perform canonical quantisation around the simplest of these solutions.

\section{Static gauge, conformal gauge and light-cone gauge}

For convenience, let us write once again the Polyakov action of non-relativistic strings, equation (\ref{NR_action_Eric}), where we set $b^{\text{NR}}_{\mu\nu}=0$,
\begin{equation}\label{NR_class_sol}
S^{NR} = - \frac{T}{2} \int \dd^2 \sigma \, \bigg[ \gamma^{\alpha\beta}\partial_{\alpha} X^{\mu} \partial_{\beta} X^{\nu} H_{\mu\nu} + \varepsilon^{\alpha\beta} (\lambda_+ \theta_{\alpha}{}^+ \tau_{\mu}{}^+ + \lambda_- \theta_{\alpha}{}^- \tau_{\mu}{}^- )\partial_{\beta}X^{\mu}  \bigg] \, ,
\end{equation}
where $\theta_{\alpha}{}^{i}$ is the zweibein for the worldsheet metric $\mathsf{h}_{\alpha\beta} = \theta_{\alpha}{}^{i}\theta_{\beta}{}^{j} \eta_{ij}$, and $\gamma^{\alpha\beta} \equiv \sqrt{-\mathsf{h}} \mathsf{h}^{\alpha\beta}$. The indices $\pm$ are defined as $\theta_\alpha{}^\pm=\theta_\alpha{}^0 \pm \theta_\alpha{}^1$ and similarly for $\tau$.

As discussed in section~\ref{subsec:weyl}, this action is invariant under Weyl transformations of the worldsheet metric, and this invariance is not broken at the quantum level. Therefore, as we are keeping the worldsheet relativistic, we do not have to deviate from the standard philosophy of using the redundancy associated with the Weyl symmetry to our advantage and simplify the equations of motion by fixing a convenient gauge. Notice that, apart from the Weyl gauge freedom inherited from $\mathsf{h}_{\alpha \beta}$, the zweibein has an additional SO$(1,1)$ freedom we can use to simplify computations.

There are three gauge choices that prove to be specially convenient. These are:
\begin{itemize}
    \item \textbf{Static gauge:} it is defined by fixing the two longitudinal coordinates to be equal to the appropriate worldsheet directions. If we call $t$ the temporal and $z$ the spatial longitudinal directions, this amounts to fixing $z=\sigma$ and $t=\tau$. In addition, if we eliminate the Lagrange multipliers $\lambda_\pm$, as in eq.~\eqref{PolytoNG}, the worldsheet metric inherits the structure of the transverse metric $\tau_{\mu \nu}$. In particular, for the SNC AdS$_5\times$S$^5$ background, it inherits an AdS$_2$ structure. 

    This choice is particularly useful, as the equation of motion for all transverse embedding coordinates take the form of free bosons in AdS$_2$.

    \item \textbf{Conformal gauge:} it is defined by fixing $\mathsf{h}_{\alpha \beta}=\text{diag}(-1,1)$, making it particularly useful for solving the equations of motion. A possible choice of zweibein compatible with this fixing takes the form $\theta_{\alpha}{}^0=(1,0)$ and $\theta_{\alpha}{}^1=(0,1)$.
    As usual, the equations of motion have to be supplemented by the Virasoro conditions, which are nothing but the equations of motion of the worldsheet metric.

    Similarly to the relativistic case, the conformal gauge is not enough to fix all the gauge degrees of freedom of the worldsheet metric. There are residual diffeomorphisms that transform $\sigma + \tau$ and $\sigma - \tau$ separately \cite{Bergshoeff:2018yvt,Fontanella:2021btt}. This residual symmetry is usually eliminated by additionally fixing $t\propto \tau$.

    \item \textbf{Light-cone gauge:} this gauge works especially well for point-like strings moving on an isometric direction, but can be used on extended strings as long as they have a non-vanishing angular momentum. Given an angular direction $\phi$ in the S$^5$, the light-cone gauge fixing consists in setting $a\phi + (1-a) t=\tau$ and $(1-a)p_\phi - a p_t=1$ for any $0\leq a \leq 1$, where $p_i$ is the conjugate momentum to $X^i$.

    This choice is very convenient not only for computing quadratic fluctuations (and thus, quantum corrections) around the BMN string for relativistic strings, but it is also convenient for non-relativistic strings, as we will discuss in section~\ref{LCG spectrum}.

\end{itemize}

\section{Classical closed string solutions in conformal gauge}

The first step to quantisation is to find appropriate vacua around which we can expand our action. Furthermore, classical string solutions have been widely studied in relativistic string theory, among other reasons, due to their relevance in the context of the AdS/CFT correspondence. Naturally, we expect them to play an equally important role in the non-relativistic case.

For relativistic strings, the dispersion relation of classical states described by solitonic closed string solutions can be matched with the dimensions of their corresponding gauge theory operator in the limit of large quantum numbers, even beyond operators protected from quantum corrections by supersymmetry \cite{Beisert:2003xu, Frolov:2003xy, Kruczenski:2003gt}. Prime examples of classical string solutions are the BMN string \cite{Berenstein:2002jq} and the GKP string \cite{Gubser:2002tv,Frolov:2002av}. Both strings move along a great circle of the S$^5$ space but, while the BMN string is a point-like string, the GKP string is an extended string in the AdS$_5$ space. We will describe here how the BMN string does not have a non-relativistic counterpart, while the non-relativistic limit of an infinitely extended GKP string becomes one of the simplest classical non-relativistic string. As the topic of classical strings in AdS$_5\times$S$^5$ is very extensive, we refer the reader to \cite{Plefka:2005bk,Tseytlin:2010jv} for a review on the subject.

In this section, we review the construction of the classical strings propagating in SNC AdS$_5\times$S$^5$, both in Cartesian coordinates \cite{Fontanella:2021btt} and in polar coordinates \cite{Fontanella:2023men} defined in section~\ref{subsec:SNCAdscoords}. In this review, we do not  discuss classical strings propagating in SNC flat spacetime. The interested reader can refer to \cite{Gomis:2004ht} and \cite{Banerjee:2025gyh}.\footnote{Although the authors start from relativistic $\mathbb{R}\times$ S$^2\subset\,$AdS$_5\times$S$^5$, the non-relativistic limit considered, involving the time direction and one angle of S$^2$, leads to SNC flat spacetime instead of GGK theory.} In addition, we do not discuss the case of classical solutions with non-relativistic worldsheet. For this topic, we refer the reader to \cite{Roychowdhury:2021wte, Roychowdhury:2020yun, Roychowdhury:2020dke, Roychowdhury:2020cnj, Roychowdhury:2020kma, Roychowdhury:2019sfo, Roychowdhury:2019olt}.\footnote{Despite our reference, we encourage the reader to approach these articles with some caution, as we have encountered some mistakes.}

\subsection{Lagrange multipliers}

It is instrumental to start by analysing the equations of motion coming from the Lagrange multipliers. The following analysis holds as long as the longitudinal metric $\tau_{\mu \nu}$ is diagonal and either of the longitudinal directions is isometric, so the metric cannot depend on it. This is the case of all the coordinate systems considered in section~\ref{subsec:SNCAdscoords}.

Let us assume, for concreteness, that the longitudinal vielbein $\tau_\mu{}^A$ only depends on $z$ (the steps are the same \textit{mutatis mutandis} for a longitudinal vielbein that only depends on  $t$). In addition, we assume that the longitudinal metric is diagonal, implying that we can take $\tau_t{}^1 =\tau_z{}^0 =0$. Under these two assumptions, a linear combination of the equations of motion for the Lagrange multipliers $\lambda_\pm$ in conformal gauge take the form
\begin{subequations} \label{LagrangeMultipliersEquation}
\begin{align}
    \varepsilon^{\alpha\beta} \left(\theta_{\alpha}{}^+ \tau_{\mu}{}^+ \partial_{\beta}X^{\mu} +\theta_{\alpha}{}^- \tau_{\mu}{}^- \partial_{\beta}X^{\mu} \right)
    = 2\tau_t{}^0  \partial_\sigma t - 2\tau_z{}^1 \partial_\tau z
    =0 \, , \\
    \varepsilon^{\alpha\beta} \left(\theta_{\alpha}{}^+ \tau_{\mu}{}^+ \partial_{\beta}X^{\mu} -\theta_{\alpha}{}^- \tau_{\mu}{}^- \partial_{\beta}X^{\mu}\right) 
    = 2\tau_z{}^1  \partial_\sigma z - 2\tau_t{}^0 \partial_\tau t
    =0 \, .
\end{align}
\end{subequations}
To simplify these two equations, we introduce the coordinate
\begin{equation}
    Z=-\int{ \frac{\tau_z{}^1 (z)}{\tau_t{}^0(z)} dz} \, .
\end{equation}
After performing a change of variable from $z$ to $Z$, these two equations become
\begin{eqnarray}
    \dot{t} + Z' = t' + \dot{Z} = 0 \, ,  
\end{eqnarray}
where dot and prime denotes derivatives with respect to $\tau$ and $\sigma$, respectively. 
As a consequence, the most general solution is of the form,
\begin{equation}
    t= f_-(\tau - \sigma) - f_+(\tau + \sigma)  \, ,  \qquad \qquad Z=f_+(\tau + \sigma) + f_-(\tau - \sigma) \, ,
\end{equation}
for two generic functions $f_{\pm}$. The fact that these two functions are unfixed is a consequence of conformal gauge being a partial gauge fixing. It reduces full diffeomorphisms on the worldsheet to independent diffeomorphisms of the light‑cone coordinates $\tau \pm \sigma$. Thus, it has to be supplemented with the condition $t=\kappa \tau$, which in turn implies $Z=-\kappa (\sigma - \sigma_0)$, where $\kappa$ and $\sigma_0$ are constants.\footnote{More exotic solutions can be found by choosing a different residual gauge choice than $t=\kappa \tau$, see appendix A of \cite{Fontanella:2021btt}.}

Let us consider now some particular coordinate systems to illustrate the process. For the Cartesian coordinates (\ref{Tau_cartesian}), setting $R=1$, we have that
\begin{equation}
    Z=\int{ \frac{1}{1+\left( \frac{z_1}{2} \right)^2} dz}=2 \arctan\left( \frac{z_1}{2} \right) \quad \Longrightarrow\quad  z_1=2 \tan \left[ -\frac{\kappa}{2} (\sigma - \sigma_0) \right] \, . \label{z1forCartesian}
\end{equation}
For polar coordinates (\ref{Tau_polar}), we have instead,
\begin{equation}
    Z=\int{ \frac{-1}{\cosh \rho} d\rho}=-2\arctan \left[\tanh \left(\frac{\rho}{2}\right) \right]=-\text{gd}(\rho)\  \Longrightarrow\  \rho=\text{gd}^{-1} [\kappa (\sigma-\sigma_0)] \, ,
\end{equation}
where gd and gd$^{-1}$ are known as the Gudermannian and inverse Gudermannian functions. The case of GGK coordinates (\ref{Tau_GGK}) is a little different, as the vielbein depends on $t$ instead of $z$, but the procedure is the same. We define the coordinate
\begin{equation}
    T=-\int{ \frac{\tau_t{}^0(t)}{\tau_z{}^1 (t)} dt}=\int{ \sec (t) dt}=\text{gd}^{-1}(t) \ \Longrightarrow \ t=\text{gd} (\kappa \tau) \, , \quad z=-\kappa (\sigma-\sigma_0) \, .
\end{equation}

In all these examples we observe that any non-trivial solution in $t$, which has $\kappa \neq 0$, must have winding on the spatial longitudinal coordinate (in the first two cases, the winding comes from the periodicity of the tangent and the Gudermannian function, in the third case it is a consequence of the $z$ coordinate being compact). As a consequence, we cannot construct point-like strings in this theory. This peculiarity was first observed in the context of SNC flat spacetime solutions \cite{Gomis:2000bd},\footnote{Strings without winding may nevertheless survive off-shell. However, the non-relativistic limit of the worldsheet action does not seem to capture these states. See sections 2.1 and 2.3 of \cite{Danielsson:2000mu} for a more detailed discussion in SNC flat spacetime.} and seems to hold also for SNC AdS$_5\times$S$^5$.

In this review, we will only discuss closed strings, meaning that all embedding coordinates have to satisfy
\begin{equation}
    X^\mu (\tau, \sigma+2\pi)=X^\mu (\tau , \sigma) \, .
\end{equation}
As a consequence, in all the above solutions $\kappa$ is restricted to be either an integer or half-integer. In what follows, we shall describe classical string solutions in both Cartesian and polar coordinates.

\subsection{Cartesian coordinates}

We begin by substituting the longitudinal vielbein~\eqref{NRdatacartesian} and boost-invariant metric~\eqref{Hcartesian} into the action~\eqref{NR_class_sol}, obtaining
\begin{align}
    S^{NR}_{\text{Cartesian}}&=-\frac{T}{2} \int \dd^2 \sigma \, \Bigg[ - \frac{(1+ (\frac{z_1}{2})^2)\zeta^2}{(1- (\frac{z_1}{2})^2)^3} \, (-\dot{t}^2+t^{\prime 2}) + \frac{\zeta^2}{2(1- (\frac{z_1}{2})^2)^3}\,  (-\dot{z}_1^2+z_1^{\prime 2}) \notag \\
    &+ \frac{1}{(1- (\frac{z_1}{2})^2)^2} \, (-\dot{\zeta}^2+\zeta^{\prime 2}) + (-\dot{\phi}^2+\phi^{\prime 2}) + (-\dot{y}_i^2+y_i^{\prime 2})  \notag\\
    &-\lambda_0 \frac{1+ (\frac{z_1}{2})^2}{1- (\frac{z_1}{2})^2} \left( \dot{t} + \frac{z'_1}{1 + (\frac{z_1}{2})^2} \right) + \lambda_1 \frac{1+ (\frac{z_1}{2})^2}{1- (\frac{z_1}{2})^2} \left( t' + \frac{\dot{z}_1}{1 + (\frac{z_1}{2})^2} \right)  \Bigg] \, , \label{Action_cartesian}
\end{align}
where we recall that the transverse coordinates in AdS$_5$, namely $z_m$ with $m\in\{2,3,4\}$, are also collectively indicated as $\zeta \equiv (z_2, z_3, z_4)$.

This action is invariant under shifts of $t$ and the directions formerly associated to S$^5$, $\phi$ and $y_i$, so we can define the following conserved quantities
\begin{subequations}
\begin{align}
	E&\equiv \int_0^{2\pi} \dd\sigma \frac{\dd\mathcal{L}}{\dd\dot{t}} =  \frac{T}{2} \int_0^{2\pi} \dd\sigma \frac{1+ (\frac{z_1}{2})^2}{1- (\frac{z_1}{2})^2} \left( \lambda_0 - \frac{2 \zeta^2 \dot{t} }{\left( 1- (\frac{z_1}{2})^2 \right) ^2} \right) \, ,\label{classicalE} \\
	p_{i} &\equiv \int_0^{2\pi} \dd\sigma \frac{\dd\mathcal{L}}{\dd\dot{y}_i}= T \int_0^{2\pi} \dd\sigma \ \dot{y}_i \, , \\
    J &\equiv \int_0^{2\pi} \dd\sigma \frac{\dd\mathcal{L}}{\dd\dot{\phi}}= T \int_0^{2\pi} \dd\sigma \ \dot{\phi} \, .
\end{align}
\end{subequations}

In addition to solving the equations of motion coming from this action, fixing conformal gauge forces us to impose Virasoro constraints, which take the form
\begin{subequations}\label{VC}
\begin{align}
	V_1&=H_{\mu \nu} \dot{X}^\mu X^{\prime \nu}+\frac{1}{2} \lambda_A \tau^A_\nu X^{\prime \nu}= 0 \, , \label{VC1} \\
	V_2&=H_{\mu \nu} \dot{X}^\mu \dot{X}^\nu + H_{\mu \nu}  X^{\prime \mu} X^{\prime \nu} + \lambda_0 \tau^1_\mu X^{\prime \mu} +\lambda_1 \tau^0_\mu X^{\prime \mu}=0 \label{VC2} \, .
\end{align}
\end{subequations}
Notice that the terms involving the Lagrange multipliers are never accompanied by a $\dot{X}$ factor. In the case of relativistic strings, one may consider point-like strings that only depend on $\tau$, making those terms irrelevant. In contrast, all non-relativistic classical string solution are extended in $\sigma$, and these terms will always give a contribution to the Virasoro constraints.

As it is the usual situation with systems of partial differential equations, solving them in full generality is an insurmountable task. These equations are usually solved by considering some particular ansätze that simplify the problem while being interesting from a physical point of view. This is the approach used in \cite{Fontanella:2021btt, Fontanella:2023men} to construct classical solutions. 

\subsubsection{Folded BMN-like string} \label{subsec:bmnfolded}

The simplest solution to the equations of motion that we can write, found in \cite{Fontanella:2021btt}, has the following form
\begin{subequations}\label{BMNfoldedcartesian}
\begin{align}
	t&=\kappa \tau \, , & z_1 &=2 \tan \left[ -\frac{\kappa}{2} (\sigma - \sigma_0) \right] \, , &
    z_m &=0 \, , & y_i &=0 \, , \\
    \phi &=w\tau  \, , & \lambda_1 \pm \lambda_0 &= \pm\frac{w^2}{\kappa} \cos \left[ \kappa (\sigma-\sigma_0) \right]\, , &
    \kappa &\in \mathbb{Z}\,, & w &\in \mathbb{R} \, . 
\end{align}
\end{subequations}
The non-vanishing conserved quantities associated to this solution are
\begin{equation}
	J=2\pi w T \, , \qquad E=\frac{T}{2} \, \frac{2\pi w^2}{\kappa}=\frac{J^2}{4\pi \kappa T} \, .
\end{equation}

There are several points regarding the construction and properties of this solution that should be discussed.
\begin{itemize}
    \item Fixing $\phi =w\tau$ may seem restrictive, but it is not restrictive at all. $\phi$ was an angular coordinate on the sphere, so naively we would have had to impose periodicity in $\sigma$ up to winding factors, i.e. $\phi (\sigma+2\pi)=\phi(\sigma)+2\pi m$ for $m\in \mathbb{Z}$. However, the non-relativistic limit has transformed the S$^5$ space into $\mathbb{R}^5$, and winding is no longer justified. Combining this with the fact that $\phi$ satisfies a wave equation makes $\phi=w\tau$ the only physically relevant solution. 
    \item The Lagrange multipliers $\lambda^\pm$ can be computed either through the equations of motion for $z_1$ and $t$ or through the Virasoro constraints. Both methods give the same result, but it is simpler to use the Virasoro constraints, as they are algebraic constraints instead of differential equations.
    \item Although this solution can be considered the equivalent in terms of simplicity to the BMN string in relativistic AdS$_5\times$S$^5$, the relation $E= 2\pi \kappa T= J$, characteristic of relativistic BMN string, no longer holds here. We have instead a dispersion relation $E\propto J^2$ more reminiscent of a non-relativistic system.
    \item  Furthermore, the constants $w$ and $\kappa$ are no longer related. In the relativistic case, the Virasoro constraints forced them to be equal, but in this setting the Virasoro constraints are used to fix the Lagrange multipliers.
    \item  The two points above are associated with the fact that the $H_{tt}$ element of the boost-invariant metric vanishes when we turn off all the other fields instead of becoming $-1$. This behaviour hinders the direct applications of some of the methods developed for relativistic string theory.
\end{itemize}

\subsubsection{Generalised folded BMN-like string}

The previous solution was generalised in \cite{Fontanella:2021btt} by including a non-trivial dependence on $\sigma$ for the remaining directions formerly associated to AdS$_5$
\begin{subequations}
\begin{align}
	t&=\kappa \tau \, , & z_1 &=2\tan \left[ -\frac{\kappa}{2} (\sigma-\sigma_0) \right] \, , \\
    y_i &=0 \, , & z_m &=C_m^1 \tan \left[ -\frac{\kappa}{2} (\sigma-\sigma_0) \right] \, , \\
     \phi &=w \tau \, ,  & \lambda_1 \pm \lambda_0 &= \pm\frac{w^2}{\kappa} \cos \left[ \kappa (\sigma-\sigma_0) \right]\, , \\
     \kappa &\in \mathbb{Z} \,, & w&\in \mathbb{R} \,,
\end{align}
\end{subequations}
where $C_m^1$ are constants. The angular momentum and energy of this solution are the same as the ones for the previous solution.

It is worth commenting that the equation of motion for the $z_m$ coordinates hints the existence of integrability in this action, a topic we will discuss in the next chapter. Assuming $z_m=z_m(\sigma)$ and $z_1=z_1(\sigma)$, if we write the equation of motion in terms of $x=z_1/2$ instead of $\sigma$, it takes the form
\begin{equation}
    \frac{d^2 z_m}{d x^2}+ \left( \frac{1}{x-i}+\frac{1}{x+i} -\frac{2}{x-1}-\frac{2}{x+1} \right) \frac{d z_m}{d x}+ \frac{2(3 +x^2)}{x^4-1}z_m=0 \, .
\end{equation}
This differential equation falls into the category of \emph{generalized Lamé equations}, and it can be solved in terms of simple functions if the numerator in the $z_m$ coefficient has the appropriate form.\footnote{Generalized Lamé equations play a central role in the context of the ODE/IM correspondence \cite{Dorey:2007zx}. In particular, a generalized Lamé equation has polynomial solutions, called ``Heine–Stieltjes polynomials'', if the numerator in front of the $z_m$ coefficient is a Van Vleck polynomial. The zeroes of the Heine–Stieltjes polynomials fulfil an equation that highly resembles the Bethe Ansatz equations for the Gaudin Model, while the zero of the Van Vleck polynomials fulfil an equation that resembles an auxiliary Bethe equation (as it depends also on the positions of the zeros of the Heine–Stieltjes polynomials). We refer the reader to \cite{VanVleck} for more information.} In this particular case
\begin{equation}
	z_m=C^1_m x +C^2_m \left[ 1+4 x \arctan (x) -x^2 \right] \, , 
\end{equation}
where $C^1_m$ and $C^2_m$ are integrations constants. Periodicity of $z_m$ forces us to set $C^2_m=0$.

\subsection{Polar coordinates}

From the point of view of relativistic string theory, studying the Polyakov action both in Cartesian and in polar coordinates is redundant, as the invariance under target space diffeomorphisms makes them equivalent. In the non-relativistic case, the equivalence between two coordinate systems can be proven if the transformation that relates the two coordinate systems is analytic in the $1/c$. However, the map between polar and Cartesian coordinates has singularities at points where angles are not well-defined. Although we believe that they should be equivalent, a proper check would require us to use the global SNC gauge transformations, which are not known - see the discussion in section 1.1. of \cite{Fontanella:2021hcb}. Therefore, we are forced to study the two actions separately.

Similarly to the Cartesian case, we substitute the longitudinal vielbein~\eqref{NRdatapolar} and boost-invariant metric~\eqref{Hpolar} into the action~\eqref{NR_action}, obtaining in this case
\begin{align}
    S^{NR}_{\text{Polar}}&=-\frac{T}{2} \int \dd^2 \sigma \, \bigg[ \sinh^2 \rho (-\dot{\beta}_1^2+\beta_1^{\prime 2}) + \sinh^2 \rho (-\dot{\beta}_2^2+\beta_2^{\prime 2}) \notag\\
    &+ \sinh^2 \rho \, (\beta_2 + \pi/2)^2 (-\dot{\beta}_3^2+\beta_3^{\prime 2}) +(-\dot{\varphi}_1^2+\varphi_1^{\prime 2}) +(-\dot{\varphi}_2^2+\varphi_2^{\prime 2})\notag\\
    &+ (\varphi_2 - \pi/2)^2 (-\dot{\varphi}_3^2+\varphi_3^{\prime 2})+ (\varphi_2 - \pi/2)^2 \, \cos^2 \varphi_3 (-\dot{\varphi}_4^2+\varphi_4^{\prime 2}) \notag \\
    &+ (\varphi_2 - \pi/2)^2 \, \cos^2 \varphi_3 \, \cos^2 \varphi_4 (-\dot{\varphi}_5^2+\varphi_5^{\prime 2})\notag  \\
    &+\lambda_0  \left( \cosh \rho \,\dot{t} - \rho' \right) - \lambda_1 \left( \cosh \rho \,t' - \dot{\rho} \right)  \bigg] \, , \label{actionpolar}
\end{align}
which has to be supplemented with the Virasoro constraints \eqref{VC}.

As the action \eqref{actionpolar} is invariant under shifts of $t$, $\beta_1$ and $\varphi_1$, we can define the following conserved quantities
\begin{subequations}
\begin{align}
    E&\equiv \int_0^{2\pi} \dd\sigma \frac{\dd\mathcal{L}}{\dd\dot{t}}= -\frac{T}{2} \int_0^{2\pi} \dd\sigma \lambda_0 \cosh \rho \, , \label{energysimplest}\\
    S&\equiv \int_0^{2\pi} \dd\sigma \frac{\dd\mathcal{L}}{\dd\dot{\beta}_1}=T\int_0^{2\pi} \dd\sigma \, \dot{\beta}_1 \, , \\
    J&\equiv \int_0^{2\pi} \dd\sigma \frac{\dd\mathcal{L}}{\dd\dot{\varphi}_1}=T\int_0^{2\pi} \dd\sigma \, \dot{\varphi}_1 \, .
\end{align}
\end{subequations}

\subsubsection{Folded BMN-like string}

The folded BMN-like string in polar coordinates, found in \cite{Fontanella:2023men}, takes the form
\begin{subequations} \label{polarBMNlike}
\begin{align}
    t &=\kappa \tau \, , & \rho &= \text{gd}^{-1} (\kappa \sigma) \, , &
    \lambda_0\pm \lambda_1 &= - \frac{\nu^2 |\cos (\kappa \sigma)|}{2 \kappa} \, , \\
    \varphi_1&=\nu \tau \, , & \nu &\in \mathbb{R} \,,  & \kappa &\in \mathbb{Z} \,, 
\end{align}
\end{subequations}
with all the remaining coordinates set to zero. Due to the original angular nature of the coordinate $\rho$, which may be periodic up to winding, and the fact that it appears in the metric only inside hyperbolic sines and cosines, we do not demand periodicity of $\rho$ but periodicity of $\cosh \rho$ and $\sinh \rho$. This imposes $\kappa$ to be an integer.

Similarly to $\phi$ in the case of Cartesian coordinates, $\varphi_1$ was an angular coordinate before taking the non-relativistic limit, but now it resembles a radial coordinate. As it fulfils a wave equation but winding is forbidden, $\varphi_1=\nu \tau$ is the only physically relevant solution.

The only non-vanishing Noether charges of this solution are the energy $E$ and a linear momentum $J$, which take the form
\begin{equation}
    E= \frac{\pi\nu^2 T}{\kappa} \, ,  \qquad  J=2\pi \nu T \, , \qquad \Longrightarrow \qquad E = \frac{J^2}{4\pi \kappa T} \, . \label{dispersion1}
\end{equation}

\subsubsection{Generalised folded BMN-like string}

A generalisation of the folded BMN-like string, found in \cite{Fontanella:2023men}, can be obtained if we introduce a spin on the $\mathbb{R}^3$ directions associated to the AdS part of the metric. It takes the form
\begin{subequations}\label{eq:more_involved}
\begin{align}
    t&=\kappa \tau \, , & \rho &= \text{gd}^{-1} (\kappa \sigma) \, , &
    \
    \lambda_0\pm \lambda_1 &= - \frac{\nu^2 + \omega^2 \tan^2 (\kappa \sigma)}{2 \kappa  |\sec (\kappa \sigma)|} \, , \\
    \beta_1&=\omega \tau \, , & \varphi_1&=\nu \tau \, ,  & \nu \in \mathbb{R} \,,  &\qquad \kappa \in \mathbb{Z} \,,
\end{align}
\end{subequations}
with all the remaining coordinates set to zero.

Similarly to $\varphi_1$, $\beta_1$ is also an angular coordinate that becomes a radial coordinate after performing the non-relativistic limit. As they both fulfil a wave equation but winding is forbidden, they can only be linear in $\tau$.

The Noether charges associated to this solution are
\begin{subequations}
\begin{align}
    E&= T  \left( \frac{\pi\nu^2}{\kappa} + \frac{\pi \omega^2}{\kappa}  \int_0^{2\pi} \frac{\dd\sigma}{2\pi}  \tan^2 (\kappa \sigma) \right)  \, , \label{dispersion2} \\
    S&= \omega T \int_0^{2\pi} \dd\sigma \tan^2 (\kappa \sigma) \, , \label{SpinInvolved}\\
    J&= 2\pi \nu T \, . \label{JInvolved}
\end{align}
\end{subequations}
Notice that the integral that appears in both $E$ and $S$ diverges, as it has a second order pole in the integration path. Thus, the only solution with finite energy is the one with $\omega=0$, namely, \eqref{polarBMNlike}. However, there exist another interesting case, which corresponds to fine-tuning $\omega=\pm \kappa$. In this case, energy and spin are both divergent, but their linear combination is finite
\begin{equation}
   E \mp\frac{S}{2}=\frac{J^2}{4 \pi \kappa T} \, . \label{dispersion3}
\end{equation}
This is reminiscent of the Hofman-Maldacena giant magnon \cite{Hofman:2006xt}.\footnote{This behaviour is not unique to string theory. Analogous situations also arise in gravitational systems. For example, for a supermassive Kerr black hole close to extremality, both the mass and angular momentum become arbitrarily large, while the difference $M^2 - J$ remains finite. } 

\section{Flowing from relativistic to non-relativistic classical strings}
\label{flowing}

Up to this point, the classical closed string solutions we have considered were obtained in \cite{Fontanella:2021btt,Fontanella:2023men} from solving the equations of motion of the non-relativistic limit of the action. Nevertheless, \cite{Fontanella:2023men} also approaches the problem with the same \ref{Uniqueness} idea we used for the gauge/gravity correspondence. The idea is to check if the same classical non-relativistic string solutions can be obtained by performing the non-relativistic limit on the classical relativistic string solutions.

In this section we will review \cite{Fontanella:2023men}, where the non-relativistic limit of the BMN string and the folded string were considered. To perform the non-relativistic limit of these solutions, they have to be re-derived from the Polyakov action, including the critical B-field \eqref{critical_B} and the rescaling of the embedding fields as in \eqref{rescaling_polar_AdSxS}.\footnote{Another method to obtain these solutions would be to use the fact that the B-field is closed (see the discussion around equation \eqref{torsionless_constraint} for more details), which means that the equations of motion are not modified, and directly apply the rescaling \eqref{rescaling_polar_AdSxS} to the solutions presented in \cite{Berenstein:2002jq} and \cite{Frolov:2002av}. This leads to equivalent results up to a redefinition of some parameters. It is worth commenting that, although the B-field does not modify the equations of motions, this does not mean that the B-field plays no role. It will be necessary to obtain the correct dispersion relations.} This effectively introduces $c$ as a deformation parameter in the relativistic action. 
We then take the $c\rightarrow \infty$ limit of these classical solutions and compare them with those described in the previous sections.

\subsection{BMN string: an inconsistent non-relativistic solution}

Using the polar coordinates (\ref{coordpolar}) with the rescaling~\eqref{rescaling_polar_AdSxS}, the BMN string \cite{Berenstein:2002jq} takes the form 
\begin{equation}
    t= \kappa \tau \, , \qquad \varphi_1=c\kappa \tau \, . \label{RelBMN}
\end{equation}
with all the other coordinates set to zero.

To perform the large $c$ limit, we first have to know how $\kappa$ scales with $c$. In order to have a non-trivial and non-divergent solution in $t$, we want $\kappa$ be constant for large values of $c$. However, this means that $\varphi_1$ would diverge. We may argue that this is not a problem because $\varphi_1$ is an angular variable defined modulo $2\pi$, but trigonometric functions acting on $\varphi_1$  will be ill-defined. Therefore, the BMN string does not have a well-defined non-relativistic limit.

We can also see the issue from the dispersion relation. With the factor of speed of light, and taking into account the B-field contribution, the BMN dispersion relation is given by
\begin{align}
    E= 2\pi c^2 \kappa T \, , \quad
    J= 2\pi c \kappa T \ \quad \Longrightarrow  \quad
    E= cJ \, ,
\end{align}
meaning that either $E$ diverges, or we have to set $J=0$ to get a finite $E$, which trivialises the dynamics and actually sets $E=0$.

\subsection{Folded string}

For the polar coordinates (\ref{coordpolar}) with the rescaling~\eqref{rescaling_polar_AdSxS}, the counterpart of the folded string \cite{Gubser:2002tv,Frolov:2002av, Frolov:2006qe, Beccaria:2008dq} with factors of $c$ takes the form\footnote{A different choice would be to have factors of $c$ appearing in $\beta_1$ and $\varphi_1$ and not in $\rho$. This choice can be discarded because trigonometric functions acting on these angles would be ill-defined for large $c$.}
\begin{equation}
    t=\kappa \tau \, , \quad \beta_1=\omega \tau \, , \quad \varphi_1=\nu \tau \, , \quad \rho=- i \text{ am} \left( \sqrt{ \frac{\nu^2}{c^2} - \kappa^2 } \sigma , \frac{c^2 \kappa^2 - \omega^2}{c^2 \kappa^2 - \nu^2} \right)\, , \label{foldedansatz}
\end{equation}
with all the other coordinates set to zero. The function am is the elliptic amplitude function.\footnote{Here we follow the convention of \cite{Abramowitz} regarding elliptic functions, where dn$(x,m) + m$ sn$(x,m)=1$ is fulfilled.} For this solution to have finite energy, it is necessary to have $\nu > c \kappa$. This condition also makes $\sinh \rho$ a compact function and prevents the metric from diverging.

To recover the non-relativistic solution we are interested in, we need to focus on the opposite regime, $\nu < c \kappa$, where $\sinh \rho$ is a non-compact function given by
\begin{align}
\label{noncompactsol}
    \sinh \rho &= -\sqrt{\frac{c^2\kappa^2 - \nu^2}{c^2\kappa^2 - \omega^2 }}\text{sc} \left(\frac{\sqrt{c^2\kappa^2 - \omega^2 }}{c}  \sigma , \frac{\nu^2 -\omega^2}{c^2\kappa^2 - \omega^2 }\right) \, .
\end{align}
where we have also assumed that $\nu^2 >\omega^2$. Here sc$(x,k)=\tan($am$(x,k))$.

As $\rho$ behaves in a way akin to an angle, it is more convenient to impose the periodicity condition $\sinh \rho(\sigma + 2\pi) = \sinh \rho (\sigma)$. This condition gives us
\begin{eqnarray}
\label{periodicity}
    \frac{\pi}{2}  \sqrt{\frac{c^2\kappa^2 - \omega^2 }{c^2}} = n \, \text{K}\left( \frac{\nu^2 -\omega^2}{c^2\kappa^2 - \omega^2 } \right)
\end{eqnarray}
where $n$ is an integer and K is the complete elliptic integral of first kind. The large $c$ limit of (\ref{foldedansatz}) then gives 
\begin{equation}
    t=\kappa \tau \, , \quad \beta_1=\omega \tau \, , \quad \varphi_1=\nu \tau \, , \quad \rho=-i\, \text{gd} ( i\kappa \sigma)=\text{gd}^{-1} ( \kappa \sigma) \, , \label{foldedansatz_large_c}
\end{equation}
furthermore, the periodicity condition becomes
\begin{equation}
    \frac{\pi}{2} |\kappa|=n \, \text{K}(0) \ \ \Longrightarrow \ \ \kappa\in \mathbb{Z}
\end{equation}
which matches perfectly (\ref{eq:more_involved}). Thus, the non-compact folded string is the relativistic uplift of the folded BMN-like solution.

We have checked that the limit of the relativistic solution matches the non-relativistic solution, but that is not enough. We should also check that the dispersion relation of the folded string reduces to the correct dispersion relation in this limit, especially because the energy of a folded string with $\nu < c \kappa$ diverges. At this point, the critical B-field that must be included to take the non-relativistic limit becomes relevant, because it modifies the energy of the string. Therefore, there are two contributions to the energy of the relativistic string, one coming from the metric and another coming from the B-field. In particular
\begin{subequations}
\begin{align}
    E_{\text{metric}} &= c^2 T \int_0^{2\pi} \dd\sigma \cosh^2\rho \, \dot{t} \, , \\
    E_B  &= - c^2 T \int_0^{2\pi} \dd\sigma \cosh\rho \, \rho' \, , \\
    E_{\text{total}}  &= E_{\text{metric}} + E_B=  c^2 T \int_0^{2\pi} \dd\sigma\cosh\rho \, (\cosh\rho \, \dot{t}-\rho') \, . \label{renormalisedenergy}
\end{align}
\end{subequations}
When we substitute the solution~\eqref{foldedansatz} into these integrals, the contributions from the metric and the B-field diverge in opposite ways, making the total energy finite. 

Let us consider the case $\omega =0$. This solution has not been studied in the literature due to its infinite ``metric'' energy. However, its total energy turns out to be finite,\footnote{Sadly, it is not possible to write a closed form for the relativistic dispersion relation because it requires us to solve the periodicity condition, which cannot be solved analytically due to the elliptic integral.}
\begin{eqnarray}
\label{energy_omega_0}
    E_{\text{total}} = 2\pi \kappa c^2 T \left[ 1 - \frac{\text{E} \left( \frac{\nu^2}{c^2 \kappa^2} \right)}{\text{K} \left( \frac{\nu^2}{c^2 \kappa^2} \right)} \right] \qquad
    \xrightarrow[\quad c\to\infty \quad]{}
    \qquad \frac{\pi\nu^2 T}{\kappa}\, ,
\end{eqnarray}
where E is the complete elliptic integral of second kind. Since the total energy is finite, we can interpret the contribution from the closed B-field as playing the same role as subtracting the divergent term $mc^2$ that appears when taking the non-relativistic limit of the energy for a relativistic free particle. In the $c\to \infty$ limit, this perfectly matches (\ref{dispersion1}).

The angular momentum of this solution is simpler to compute, giving us
\begin{equation}
\label{J_omega_0}
    J =2\pi \nu T \, ,
\end{equation}
which remains the same in the large $c$ limit. The dispersion relation we obtain combining \eqref{energy_omega_0} and \eqref{J_omega_0} matches (\ref{dispersion1}).

\section{Non-relativistic string spectrum}

Quantisation of relativistic strings in AdS$_5\times$S$^5$ is usually performed around the BMN string using light-cone gauge,\footnote{An expansion around the GKP string has also been studied, see e.g. \cite{Frolov:2002av, Basso:2010in, Dorey:2011gr}.} chapter 2 of \cite{Arutyunov:2009ga} contains a very complete and detailed explanation of the procedure. We may be tempted to take the relativistic result, apply some sort of non-relativistic limit, and extract from that a spectrum and S-matrix of excitations for non-relativistic strings, but this is not possible. To begin with, we have just shown that the BMN string does not have a well-defined non-relativistic limit, and therefore we cannot simply start from the usual relativistic result. We have to perform a similar derivation starting from the simplest non-relativistic classical solution: the folded BMN-string. 

In this section, we are going to consider quantisation of the non-relativistic string theory action using two different gauges: static gauge and light-cone gauge.

Quantisation in static gauge is the simplest of the two, as the action immediately becomes quadratic in fields. In contrast to the flat space case, where the action being quadratic in field implies that we have free fields in Mink$_2$, here we get free fields propagating in AdS$_2$. The problem with the static gauge quantisation is the fact that the vacuum we are expanding around is not interesting for the computations needed to check the holographic correspondence described in chapter~\ref{chap:holography}.

The second option we will discuss is quantisation of the action in the light-cone gauge. This procedure is adapted for solutions more relevant for holography, like the folded BMN-like string described in~\ref{subsec:bmnfolded}, but it leads to very computationally demanding calculations.

\subsection{Static gauge}

To start the quantisation, the first step is to pick a suitable classical solution that can be used as the vacuum. If we use GGK coordinates \eqref{all_GGK}, the simplest solution compatible with the static gauge is\footnote{Notice that this is not a solution for conformal gauge, as it requires a non-trivial worldsheet metric.}
\begin{equation}
    x^0=\tau \,, \qquad x^1=\sigma \,, \qquad x^a=0 \,, \qquad x^{a'}=0 \,,
\end{equation}
which has vanishing energy.

Now that we have a classical solution, we have to go back to our action and expand the embedding coordinate fields around it. As this solution has a non-trivial worldsheet metric, it is more convenient to work with the Nambu-Goto action than with the Polyakov action. There is no problem with that because, as discussed at the end of section~\ref{subsec:limit}, solving the equations of motion from the Lagrange multipliers and substituting them back into the action~\eqref{NR_action_Eric} gives us exactly the same result as performing the limiting procedure on the Nambu-Goto action. Following \cite{Sakaguchi:2007ba}, we can start from the non-relativistic Nambu-Goto action~\eqref{NR_NG}, substitute the GGK coordinates \eqref{all_GGK} and impose static gauge $x^0=\tau$ and $x^1=\sigma$. As the classical solution vanishes in the transverse directions, we can directly identify the transverse embedding coordinate fields with the fluctuation fields, giving us the following action for the fluctuations\footnote{In \cite{Sakaguchi:2007ba}, the authors use a modified version of GGK coordinates, but the same result can be obtained using regular GGK coordinates. This is also true for the Cartesian coordinates \eqref{All_cartesian} after performing the field redefinition $x_a=\frac{4}{4-\sigma^2}z_a$, and for the polar coordinates \eqref{All_polar} after performing the field redefinition $x^2=(\beta_2+\pi/2) \sin (\beta_3)\sinh (\sigma)$, $x^3=(\beta_2+\pi/2) \cos (\beta_3)\sinh (\sigma)$ $x^4=\beta_1 \sinh (\sigma)$.}
\begin{equation}
    S = - \frac{T}{2} \int \dd^2 \sigma \,  \sqrt{-\det (\tau_{\gamma \delta})}\left( \tau^{\alpha\beta}\partial_{\alpha} x^{a} \partial_{\beta} x^{a} +\tau^{\alpha\beta}\partial_{\alpha} x^{a'} \partial_{\beta} x^{a'} +2 x^{a} x^{a} \right)\, , \label{gaugefixedKentaroh}
\end{equation}
where $\tau_{\alpha \beta}=(-1,\cos^2 \tau)$ is the AdS$_2$ metric in the Friedmann–Robertson–Walker open slicing coordinates. Therefore, what we have to quantise is the action of five massless and three massive free bosons in AdS$_2$ space.

Although these are non-interacting fields, quantisation is not an easy task because the AdS space is not globally hyperbolic. There are two main problems that make the AdS$_n$ space non globally hyperbolic: the presence of closed time-like curves and the existence of a time-like boundary. The first problem can be circumvented by considering the universal covering of the AdS space, where the compact time coordinate is unravelled into the topology of the real line. The second problem is more important and more difficult to solve. The presence of a time-like boundary implies that information can be lost or gained from it. In this situation, the Cauchy problem is ill-defined.

The solution proposed in \cite{PhysRevD.18.3565} is to realise that the universal covering of the AdS space is conformally equivalent to half of the Einstein static universe. As a consequence, a solution to the Klein-Gordon equation in the universal covering of AdS differs only by a conformal factor from the solution to the Klein-Gordon equation in the Einstein static universe. Therefore, the idea is to solve the equations of motion in the Einstein static universe (where the problem is well posed because the space is globally hyperbolic), impose that there is no probability flux though the boundary, and undo the conformal transformation.

After applying the above argument, we find that the field operator $\psi$ that fulfils the differential equation
\begin{equation}
    (-\partial_\tau^2 + \partial_\sigma^2) \psi = \frac{m^2 \kappa^2}{\cos ^2 \kappa \sigma} \psi \, , \label{EoMfield}
\end{equation}
takes the form\footnote{There are actually four sets of solutions \cite{SAKAI1985661}, but this is the only one relevant for our analysis. Notice also that the solution is normalisable only if the corresponding $\Delta$ is real and larger than $-1/2$, which is equivalent to the Breitenlohner-Freedman bound $m^2 \geq -1/4$.}
\begin{align}
    \psi^\pm=\sum_{n=0}^\infty N_n &\left[ e^{i (n+\Delta^\pm) \kappa \tau} (\cos \kappa \sigma )^{\Delta^\pm} C^{\Delta^\pm}_n (\sin \kappa \sigma)\, \hat{a}_n^{\phantom{\dagger}}  \right. \notag\\
    +& \left. e^{-i (n+\Delta^\pm) \kappa \tau} (\cos \kappa \sigma )^{\Delta^\pm} C^{\Delta^\pm}_n (\sin \kappa \sigma)\, \hat{a}_n^\dagger \right] \, , \label{field}
\end{align}
where $C^p_q(x)$ are the Gegenbauer polynomials, $\hat{a}_n $ and $\hat{a}_n^\dagger$ are operators that satisfy canonical bosonic commutation relations, and the coefficients $N_n$ and $\Delta$ are defined as follows
\begin{equation}
    N_n = \frac{ \Gamma(\Delta) 2^{\Delta -1}}{\sqrt{\pi}} \sqrt{\frac{n!}{\Gamma (n+2\Delta)}} \, , \qquad \Delta^{\pm}= \frac{1 \pm \sqrt{1+4 m^2}}{2} \, .
\end{equation}
At this point, the construction of the Fock space proceeds as in flat space. Computing the Hamiltonian associated with \eqref{gaugefixedKentaroh} and substituting the field \eqref{field}, we find the same spectrum as for a regular harmonic oscillator with $n+\Delta^+$ energy. Thus, as we have 3 bosons with mass $m^2=2$ and 5 massless bosons, it takes the form
\begin{equation}
    H=\sum_{j=1}^3\sum_{n=0}^\infty (n+2) a_{j,n}^\dagger a_{j,n} + \sum_{k=1}^5\sum_{n=0}^\infty (n+1) b_{k,n}^\dagger b_{k,n} \, .
\end{equation}
that is
\begin{equation}
    E^{a,j}_n=(n+2) \,, \qquad E^{b,k}_n=(n+1) \, .
\end{equation}
As these fields behave like harmonic oscillators, we can write the 1-loop quantum correction to the energy as the sum of the zero-point energy of these oscillators, namely as 
\begin{equation}
    E=E_{\text{clas}}+E_{\text{1-loop}}=0+\frac{1}{2}\sum_{n=-\infty}^\infty \left[ 3(n+2)+5(n+1) -(\text{fermionic contributions})\right] \,.
\end{equation}
where the term `fermionic contributions' includes the energy coming from fermionic modes, which has not yet been computed.  
If we assume that the 1-loop correction is finite, convergence of the sum demands the fermionic contributions to cancel the orders $n$ and $1$ of the sum. From supersymmetry arguments, we should only expect contributions of order $n$ and $1$ from fermions. Therefore, we conclude that the energy still vanishes after taking into account 1-loop corrections.

\subsection{Light-cone gauge}
\label{LCG spectrum}

Let us move now to quantisation in light-cone gauge, following \cite{Fontanella:2021hcb,deLeeuw:2024uaq}. In this case, the folded BMN-like string \eqref{BMNfoldedcartesian} can be used as a classical solution around which we can expand, with some caveat that we will address in a moment.

First and foremost, we have to appropriately define light-cone gauge fixing. We begin by writing the action~\eqref{NR_action} in first-order formalism
\begin{equation}
\label{first_NR_action}
S = \int \dd \tau \int_0^{2 \pi} \dd \sigma \, ( p_{\mu} \dot{X}^{\mu} - \mathcal{H} ) = \int \dd \tau \int_0^{2 \pi} \dd \sigma \, ( p_{\mu} \dot{X}^{\mu} +\frac{\gamma^{01}}{\gamma^{00}} C_1 + \frac{1}{2 T \gamma^{00}} C_2 )  \, , 
\end{equation}
where $\mathcal{H}$ is the Hamiltonian, and 
\begin{subequations}
\begin{align}
C_1 &= p_{\mu} X^{\prime \mu} \, , \\
\notag
C_2 &= H^{\mu\nu} p_{\mu}p_{\nu} + T^2 H_{\mu\nu} X^{\prime \mu}X^{\prime\nu} - T^2 \lambda_A \varepsilon^{AB} \eta_{BC} \tau_{\mu}{}^C X^{\prime\mu}  \\
&+ T \lambda_A \tau_{\mu}{}^A p_{\nu} H^{\mu\nu} 
+ \frac{T^2}{4} \lambda_A \lambda_B \tau_{\mu}{}^A \tau_{\nu}{}^B H^{\mu\nu}
 \, ,
\end{align}
\end{subequations}
are combinations of the Virasoro constraints. Here $H^{\mu\nu}$ is the usual inverse of $H_{\mu\nu}$, i.e. $H^{\mu\rho}H_{\rho \nu}=\delta^\mu_\nu$. In addition, we introduce the following one-parameter family of light-cone coordinates, 
\begin{eqnarray}
\notag
X_+ &=& (1-a) t + a \phi \, , \qquad\ \ 
X_- = \phi - t \, , \qquad \\
p_+ &=& (1-a) p_{\phi} - a p_t \, , \qquad
p_- = p_{\phi} + p_t \, , 
\end{eqnarray}
where $0 \leq a \leq 1$ parameterises a gauge freedom. Fixing the light-cone gauge is done in four steps:
\begin{enumerate}
    \item Fix $X_+ = \tau$ and $p_+ = 1$.
    \item Eliminate the Lagrange multiplier fields. This is done by imposing $\dot{p}_{\lambda_A} = \{ p_{\lambda_A}, \mathcal{H}\} \approx 0$, which enforces that the initial condition $p_{\lambda_A} = 0$ is fulfilled at all times.
    \item Eliminate $X'_-$ using the Virasoro condition $C_1 \approx 0$.
    \item Eliminate $p_-$ using the Virasoro condition $C_2 \approx 0$.
\end{enumerate}
After these four steps, we get an action of the form
\begin{eqnarray}
S_{\text{g.f.}} &=& \int \dd^2 \sigma \, (p_I \dot{X}^I - \mathcal{H}_{\text{red}} ) \, ,
\end{eqnarray}
where the reduced Hamiltonian $\mathcal{H}_{\text{red}} \equiv - p_- (X^I, X^{\prime I}, p_I)$ only depends on the light-cone transverse\footnote{Here, and only here, we are using the word transverse with respect to the light-cone gauge coordinates, i.e. all the coordinates except $t$ and $\phi$. This is in contrast to the non-relativistic transverse directions, i.e. all the coordinates except $t$ and $z_1$.} coordinate and momenta.

For the case of Cartesian coordinates \eqref{Hcartesian} and $a=0$, where the action heavily simplifies, the gauge fixed action takes the following form
\begin{eqnarray}
\label{gauge_fixed_action}
    S^{a=0}_{\text{g.f.}} &=& \int \dd^2 \sigma \, \frac{1}{8T (z_1^2 - 4)^3 (z_1^2 + 4) z_1^{\prime}} \bigg[ z_1^{10} (1 + T^2 y_j^{\prime 2}) - 4 z_1^8 (1 + T^2 y_j^{\prime 2} - 2 T z_1^{\prime}) \\
\notag
    &-& 16 z_1^6 \bigg(2 + 4 T z_1^{\prime} - T^2 z_m^{\prime 2} + T^2 z_1^{\prime 2} \dot{y}_j^2 - 2 T^2 y_j^{\prime} \dot{y}_j z_1^{\prime}  \dot{z}_1 + \dot{z}_1^2 + T^2 y_j^{\prime 2} (2 + \dot{z}_1^2)\bigg) \\
\notag
    &+& 32 z_1^4 \bigg( 4 - T^2 z_m^2 z_1^{\prime 2} + 2 T^2 z_m^{\prime 2} + 6 T^2 z_1^{\prime 2} \dot{y}_j^2  - 12 T^2 y_j^{\prime} \dot{y}_j z_1^{\prime}  \dot{z}_1 + 6 \dot{z}_1^2 + 2 T^2 y_j^{\prime 2} (2 + 3 \dot{z}_1^2) \bigg) \\
\notag
    &-& 256 z_1^2 \bigg(-1 + T^2 z_m^{\prime 2} - 6 T^2 y_j^{\prime} \dot{y}_j z_1^{\prime}  \dot{z}_1 + 3 \dot{z}_1^2 + T^2 y_j^{\prime 2} (-1 + 3 \dot{z}_1^2) - 2 T z_1^{\prime} (2 + T z_m^{\prime} \dot{z}_m \dot{z}_1 ) \\
\notag
    &+& T^2 z_m^{\prime 2} \dot{z}_1^2 + T^2 z_1^{\prime 2} (2 z_m^2 + 3 \dot{y}_j^2 + \dot{z}_m^2 ) \bigg) - 512 \bigg( 4T^2 y_j^{\prime} \dot{y}_j z_1^{\prime}  \dot{z}_1 - 2 T^2 y_j^{\prime 2} (-1 + \dot{z}_1^2) \\
\notag
    &-& 2(1 + T^2 z_m^{\prime 2})(-1 + \dot{z}_1^2) + 4T z_1^{\prime} (1 + T z_m^{\prime} \dot{z}_m \dot{z}_1 ) + T^2 z_1^{\prime 2} \big(3 z_m^2 - 2 (\dot{y}_j^2 + \dot{z}_m^2) \big) \bigg) \bigg]\, ,  
\end{eqnarray}
where sums over $m \in \{2, 3, 4\}$ and $j\in\{1,..., 4\}$ are to be understood.

Now we can proceed with the quantisation of the action. We start by choosing a classical solution of the equations of motion around which we expand. As the folded BMN-like string \eqref{BMNfoldedcartesian} is the simplest non-relativistic solution we can write, we would be interested in expanding around it. However, this solution was computed on the conformal gauge, and it is not guaranteed that it would solve the equations of motion in the light-cone gauge. It turns out that a solution with the same profile and energy solves the equation of motion for the light-cone gauge at $a=0$\footnote{The parameter $\nu$ is not a free parameter, as it is fixed by $\gamma^{00}\dot{\phi}\ T=p_+\overset{!}{=}1$.}
\begin{eqnarray}
\label{BMNlike_vacuum}
    t= \tau \, , \qquad
    z_1 = 2 \tan \left( -\frac{\kappa}{2} \sigma \right) \, ,  \qquad \kappa \in \mathbb{Z} \,, \qquad
    \phi = \nu \tau \,,\qquad E=\frac{J^2}{4\pi \kappa T}  \, ,
\end{eqnarray}
where, from the action \eqref{first_NR_action} and $a=0$, $E$ and $J$ are defined as
\begin{eqnarray}
   E=-\int_0^{2\pi} \dd\sigma \, p_t = \int_0^{2\pi} \dd\sigma \, (\mathcal{H}_{\text{red}} + p_+)  \, , \qquad 
   J=\int_0^{2\pi} \dd\sigma \, p_{\phi}= \int_0^{2\pi} \dd\sigma \, p_{+} \, . 
\end{eqnarray}

We should also comment that, as the classical theory corresponds to the large tension limit, the inverse of the tension is a convenient expansion parameter for the quantisation
\begin{equation}
    \hbar\equiv \frac{1}{\sqrt{T}} \, .
\end{equation}
Now we are ready to expand the transverse fields $X^I$ around the folded BMN-like vacuum,
\begin{eqnarray}\label{exapnsion_BMN_vacuum}
    z_1 = 2 \tan \left( -\frac{\kappa}{2} \sigma \right) + \hbar \, \delta z_1 \, , \qquad     z_m =  \hbar \, \delta z_m \, , \qquad     y_j = \hbar \, \delta y_j \, .
\end{eqnarray}
where $\delta z_1, \delta z_m, \delta y_j$ are the fluctuation fields, and they must satisfy periodic boundary conditions. The gauge fixed action expands in small $\hbar$ as follows\footnote{To facilitate the computations, it is useful to eliminate the $\hbar$ dependence that appears in several places inside the gauge fixed action (\ref{gauge_fixed_action}), before taking the expansion \eqref{exapnsion_BMN_vacuum}. This is achieved in the same way as it was done for the relativistic action, that is, by rescaling $\sigma \to T\sigma$.% This also has the effect to decompactifying the worldsheet, which is important to have a well-defined S-matrix for the quantum fields.
}
\begin{eqnarray}
\label{action_expansion}
\mathcal{S}^{a=0}_{\text{g.f.}} =  \int \dd^2 \sigma \,  \sum_{j=0}^\infty \hbar^{j-2} \mathcal{L}_j  \, , 
\end{eqnarray}
where $\mathcal{L}_i$ are homogeneous polynomials of degree $i$ in the fluctuation fields and their derivatives. The lowest order, $\mathcal{L}_0$, is a constant, and thus it can be disregarded; the second lowest order, $\mathcal{L}_1$, is a total derivative because we are expanding around a solution of the classical equations of motion. The first non-trivial contribution, i.e. $\mathcal{L}_2$, appears to be a rather complicated quadratic Lagrangian that depends on $\sigma$ in a cumbersome way. After a field redefinition, it takes the form
\begin{eqnarray}
\label{quadratic_free_AdS2}
\mathcal{L}_2 = \sqrt{-g} \bigg[ \frac{1}{2} g^{\alpha\beta}(\partial_{\alpha} \delta z_1 \partial_{\beta} \delta z_1 + \partial_{\alpha} \delta z_m \partial_{\beta} \delta z_m
    + \partial_{\alpha} \delta y_j \partial_{\beta} \delta y_j) + \kappa^2 \delta z_m^2 \bigg] \, ,
\end{eqnarray}
where $g_{\alpha\beta} = \sec (\kappa \sigma)^2 \, \eta_{\alpha\beta}$ is the AdS$_2$ metric. The quadratic action describes the dynamics of 3 massive fields $\delta z_m$ and 5 massless fields $(\delta z_1, \delta y_j)$.

Higher orders can be simplified after we perform field redefinitions perturbatively, following the method described in Appendix A of \cite{Kruczenski:2004kw}. After some complicated field redefinitions (see Appendix B in \cite{deLeeuw:2024uaq} for the explicit expressions), one finds that
\begin{equation}
    \mathcal{L}_3 = \mathcal{L}_4 = \mathcal{L}_5 = \mathcal{L}_6 = 0 \, , 
\end{equation}
up to total derivatives, which can be disregarded due to the closed string boundary conditions. This result gives strong evidence that the non-relativistic action in light-cone gauge expanded around the folded BMN-like vacuum is equivalent to free fields in AdS$_2$.\footnote{Although the final result is exactly the same as for static gauge, its origin and interpretation is completely different. In particular, we want to highlight two key differences. First, that we are computing excitations around a different vacuum (for example, the static solution always has $E=0$ while the folded BMN-string has $E\propto J^2/\kappa$). Second, that we are exciting different coordinates. For the static gauge all the fields are transverse with respect to the non-relativistic limit, while for the light-cone gauge one of the fields is longitudinal.} This result may be extended at all loops, although this task requires solving a complicated PDE \cite{deLeeuw:2024uaq}.

As we have already discussed quantisation of free fields in AdS$_2$ in the previous section, here we do not repeat the discussion and directly present the final form of the Hamiltonian
\begin{align}
    H=\sum_{n=0}^\infty &\left[ (n+1) \left(\hat{a}_{n}^\dagger \hat{a}_{n}^{\phantom{\dagger}} +\sum_j \hat{c}_{j,n}^\dagger \hat{c}_{j,n}^{\phantom{\dagger}} \right) +\left(  n+2 \right) \sum_m \hat{b}_{m,n}^\dagger \hat{b}_{m,n}^{\phantom{\dagger}} \right] \, ,
\end{align}
where the oscillators $a$, $b$ and $c$ correspond to the fluctuations $\delta z_1$, $\delta z_m$ and $\delta y_i$ respectively. As these fields behave like harmonic oscillators, we can write the quantum correction to the energy at 1-loop  as follows
\begin{equation}
    E=E_{\text{clas}}+E_{\text{1-loop}}= \frac{J^2}{4\pi \kappa T}+\frac{1}{2T}\sum_{n=-\infty}^\infty \left[ 3(n+2)+5(n+1) -(\text{fermionic contributions})\right] \,.
\end{equation}
where the term `fermionic contributions' includes the energy coming from fermionic modes, which has not yet been computed. 
If we assume that the 1-loop correction is finite, convergence of the sum demands the fermionic contributions to cancel the orders $n$ and $1$ of the sum. From supersymmetry arguments, we should only expect contributions of order $n$ and $1$ from fermions. Therefore, we conclude that the one-loop correction to the energy vanishes.

%%%%%%%%%%%%%%%%%%%%%%%%%%%%%%%%%%%%%%%%%%%%%%%%%%%%%%%%%%%%%%%%%%%%%%%%%%%%%%%%%%%%%%%%%%%%%%%%%%%%%%%%%%%%%%%%%%%%%%%%%%%%%%%%%%%%%%%%%%%%%%%%%%%%%%%%%%%%%%%%%%%%%%%%%%%%%%%%%%%%%%%%%%%%%%%%%%%%%%%%%%%%%%%%%%%%%%%%%%%%%%%%%%%%%%%%%%%%%%%%%%%%%%%%%%%%%%%%%%%%%%%%%%%%%%%%%%%%%%%%%%%%%%%%%%%%%%%%%%%%%%%%%%%%%%%%%%%%%%%%%%%%%%%%%%%%%%%%%%%%%%%%%%%%%%%%%%%%%%%%%%%%%%%%%%%%%%%%%%%%%%%%%%%%%%%%%%%%%%%%%%%%%%%%%%%%%%%%%%%%%%%%%%%%%%%%%%%%%%%%%%%%%%%%%%%%%%%%%%%%%%%%%%%%%%%%%%%%%%%%%%%%%%%%%%%%%%%%%%%%%%%%%%%%%%%%%%%%%%%%%%%%%%%%%%%%%%%%%%%%%%%%%%%%%%%%%%%%%%%%%%%%%%%%%%%%%%%%%%%%%%%%%%%%%%%%%%%%%%%%%%%%%%%%%%%%%%%%%%%%%%%%%%%%%%%%%%%%%%%%%%%%%%%%%%%%%%%%%%%%%%%%%%%%%%%%%%%%%%%%%%%%%%%%%%%%%%%%%%%%%%%%%%%%%%%%%%%%%%%%%%%%%%%%%%%%%%%%%%%%%%%%%%%%%%%%%%%%%%%%%%%%%%%%%%%%%%%%%%%%%%%%%%%%%%%%%%%%%%%%%%%%%%%%%%%%%%%%%%%%%%%%%%%%%%%%%%%%%%%%%%%%%%%%
\chapter{Integrability in non-relativistic string theory} \label{chap:integrability}

One of the reasons behind the huge interest in the AdS$_5$/CFT$_4$ correspondence is the fact that the two theories involved are integrable. For classical systems, integrability means that we have at least as many independent conserved quantities as degrees of freedom, allowing us to exactly determine the dynamics of the systems (at least, formally). At the level of quantising the theory, integrability allows us to bypass perturbation theory in some computations and obtain results that are valid for all values to the coupling constant. Consequently, it was used in the original correspondence to check the matching between computations on $\mathcal{N}=4$ SYM and computations in string theory in AdS$_5\times$S$^5$. 
These observables obtained from exact integrability methods can then be compared with a perturbative analysis on both sides, and show that they holographically match. In this chapter we discuss integrability only in the context of the GGK/GYM holography described in  section~\ref{sec:NR_limit_gravity}.

On the string theory side of the holography, classical integrability was first described in \cite{Bena:2003wd} and later explored in \cite{Alday:2005gi}. One of the consequences of integrability is the relationship between some special string configurations and simpler integrable mechanical models, e.g. the action associated to a general class of rotating closed string can be mapped to the Neumann-Rosochatius system \cite{Arutyunov:2003uj,Arutyunov:2003za}. In addition, there are computational techniques that only work for integrable modes. Among those, we want to emphasise the classical spectral curve, which has been used to describe classical string solutions \cite{Kazakov:2004qf, Kazakov:2004nh, Beisert:2004ag, Beisert:2005bm}, and to compute quantum corrections to them \cite{Gromov:2007aq, Schafer-Nameki:2010qho}. As integrability plays a prominent role in relativistic string theory and holography, it is natural to ask whether the non-relativistic SNC AdS$_5 \times$S$^5$ string action is integrable. Since integrability of the AdS$_5 \times$S$^5$ string action becomes manifest when it is written in terms of a coset space, we follow the same approach and rewrite the non-relativistic SNC AdS$_5 \times$S$^5$ string action in this form. 
Once that is done, classical integrability, understood as the existence of a Lax connection, can be demonstrated. However, we will also show that integrability is not realised in the standard form.

On the field theory side, integrability is more manifest after mapping the problem of computing conformal dimension of operators to the problem of finding the spectrum of a quantum integrable spin chain \cite{Minahan:2002ve,Sieg:2010jt,Beisert:2005fw}. Integrability of the GYM dual theory is currently under research, so we will not discuss it in this review.

\section{The relativistic string action as a coset model} \label{sec:symmetricspaces}

Although there are several methods to construct a string theory with fermions, each of them have different problems. The Ramond–Neveu–Schwarz (RNS) formalism cannot be used because we do not know how to include the self-dual Ramond-Ramond five-form flux supporting the AdS$_5\times$S$^5$, the pure spinor formalism is somewhat obscure and, while the Green-Schwarz formalism can be used, it cannot be covariantly quantised.

Although the Green-Schwarz formalism is difficult to implement in practice, luckily it has been shown for AdS$_5\times$S$^5$ to be equivalent to a Wess-Zumino-Novikov-Witten (WZNW) non-linear sigma model on a coset superspace \cite{Metsaev:1998it}. Here we will review how this is done for the bosonic sector and, in the following sections, how it generalises to the non-relativistic case. For more in-depth review on coset construction and integrability in the relativistic case, we refer to \cite{Arutyunov:2009ga,Zarembo:2017muf}.

Let us start with the following definition: we say that a manifold $\mathcal{M}$ is an \emph{homogeneous space} with respect to the group $G$ if any two points can be connected by the action of an element of $G$. Mathematically, we demand that for any two points $x,y\in \mathcal{M}$ there exists $g\in G$ such that $T_g(x)=y$, where $T_g$ is a representation of $g$.
Let us also define the \emph{stability group} of a point $x_0\in \mathcal{M}$, denoted by $H_{x_0}$, as the set of elements that leave the point invariant, that is, $h\in H_{x_0}\subset G$ if $T_h(x_0)=x_0$

We are interested in homogeneous spaces because they can be written as the coset space $G/H_{x_0}$ for any point $x_0\in \mathcal{M}$.\footnote{From the definition of homogeneous space, it is immediate to see that the stability groups of any two given points of $\mathcal{M}$ are related by conjugation, making them isomorphic. Therefore, from here on we drop the point label of the stability group because the point we pick is inconsequential for constructing the quotient.}  The construction of the metric associated with $\mathcal{M}$ described in section~\ref{sec:coordinates} can always be applied to homogeneous spaces.

The goal is to construct the Polyakov action in $\mathcal{M}$ directly from the coset $G/H$. We consider a map from the worldsheet, parameterised by coordinates $\tau$ and $\sigma$, into the group $G$, denoted $g(\tau,\sigma) \in G$, and define the Maurer-Cartan (MC) 1-form $A_{\alpha} (\tau,\sigma) \equiv g^{-1} (\tau,\sigma) \partial_{\alpha} g(\tau,\sigma)$, with $\alpha\in\{\tau , \sigma\}$.\footnote{From here on, we will not explicitly write the dependence on $\tau$ and $\sigma$ to not clutter our notation, unless it is necessary.} From its definition, $A$ is then an element of the Lie algebra $\mathfrak{g}=\text{Lie}(G)$, and it can be split into the Lie algebra $\mathfrak{h}$ associated with the stability group $H$ and its complement in $\mathfrak{g}$, denoted as $\mathfrak{p}$. In particular\footnote{The reader should be careful, as some authors (see, e.g. \cite{Arutyunov:2009ga}) define the MC 1-form with an additional minus sign, $A=-g^{-1} \dd g$}
\begin{equation} \label{MCdef}
    A_\alpha=g^{-1} \partial_\alpha g=A_\alpha^{(\mathfrak{h})}+A_\alpha^{(\mathfrak{p})} \, , \qquad A_\alpha^{(\mathfrak{h})}\in \mathfrak{h} , \, A_\alpha^{(\mathfrak{p})}\in \mathfrak{p} \, .
\end{equation}
If we can define a non-degenerate inner product in the group $G$, such split is always possible. 

Let us look deeper into the algebra structure of $\mathfrak{g}$. The fact that $H$ is a stability group, and thus a subgroup, means that $[\mathfrak{h}, \mathfrak{h}]\subset \mathfrak{h}$. In addition, if the algebra fulfils $[\mathfrak{h}, \mathfrak{p}]\subset \mathfrak{p}$, the space associated with it is called \emph{homogeneous reductive space}. It will later prove useful to also demand the condition $[\mathfrak{p}, \mathfrak{p}]\subset \mathfrak{h}$. Spaces that fulfil this additional condition are called \emph{symmetric spaces}. A geometrical origin for this additional condition on the Lie bracket is that the group of isometries of $\mathcal{M}$ contains an inversion symmetry about every point, that is, for every $y\in \mathcal{M}$ there is a discrete isometry $\Sigma$ that leaves $y$ invariant and acts on the tangent space $TM_{y}$ as minus the identity. This isometry $\Sigma$ appears as a $\mathbb{Z}_2$ outer automorphism of $\mathfrak{g}$, with $\mathfrak{h}$ being the subspace with eigenvalue $+1$ and $\mathfrak{p}$ being the subspace with eigenvalue $-1$,
\begin{equation}
    \Sigma (\mathfrak{h})=\mathfrak{h} \, , \qquad \Sigma (\mathfrak{p})=-\mathfrak{p} \, . \label{Z2outer}
\end{equation}

As the information regarding the manifold $\mathcal{M}$ is fully contained in $\mathfrak{p}$, the simplest action we can write using only the $\mathfrak{p}$-valued components of the MC takes the form
\begin{equation}
    S=-\frac{T}{2} \int{\dd^2 \sigma \, \gamma^{\alpha \beta} \langle A_\alpha^{(\mathfrak{p})} , A_\beta^{(\mathfrak{p})} \rangle} \, , \label{PCMaction}
\end{equation}
where $\langle \cdot , \cdot \rangle$ is an inner product in $\mathfrak{g}$ that is invariant under adjoint action of the group $G$, $\text{Ad}_k(A)=k A k^{-1}$
\begin{equation}
    \langle \text{Ad}_k(A),\text{Ad}_k(B) \rangle= \langle A, B \rangle \, , \qquad\ \forall A,B\in \mathfrak{g} \, , \ \forall k\in G \,. \label{adinvariant}
\end{equation}
If the algebra $\mathfrak{g}$ admits a matrix representation, the simplest inner product we can write is the trace of the product $\langle A , B \rangle=$Tr$(AB)$. Comparing \eqref{PCMaction} with the Polyakov action \eqref{rel_Pol_action}, we get the identification \eqref{metric_from_MC} between $\langle A^{(\mathfrak{p})} , A^{(\mathfrak{p})} \rangle$ and the metric we discussed in section \ref{sec:coordinates}.

\paragraph{Equations of motion.}
The equations of motion associated to the variation of the group element $g$ used to construct the MC take the form
\begin{equation}
    \partial_\alpha (\gamma^{\alpha \beta} A^{(\mathfrak{p})}_\beta) + [A^{(\mathfrak{h})}_\alpha , A^{(\mathfrak{p})}_\beta] \gamma^{\alpha \beta}=D_\alpha (A^{(\mathfrak{p})})^\alpha=0 \, , \label{eomsigmarelativista}
\end{equation}
where $D_\alpha \equiv \partial_\alpha + [A^{(\mathfrak{h})}_\alpha , \cdot]$ behaves as a covariant derivative. The equations of motion obtained from the variation of the worldsheet metric $\mathsf{h}_{\alpha\beta}$ (Virasoro constraints) take the form
\begin{equation}
    \langle A^{(\mathfrak{p})}_\alpha, A^{(\mathfrak{p})}_\beta \rangle - \frac12 \gamma_{\alpha \beta} \gamma^{\rho \sigma} \langle A^{(\mathfrak{p})}_\rho, A^{(\mathfrak{p})}_\sigma \rangle=0 \, . \label{VirasoroCoset}
\end{equation}
These equations have to be supplemented with the flatness condition of $A$, as it is an exact form,
\begin{equation}
    \left(\partial_\alpha A^{(\mathfrak{h})}_\beta - \partial_\beta A^{(\mathfrak{h})}_\alpha + [A^{(\mathfrak{h})}_\alpha , A^{(\mathfrak{h})}_\beta]\right) + D_\alpha A^{(\mathfrak{p})}_\beta - D_\beta A^{(\mathfrak{p})}_\alpha + [A^{(\mathfrak{p})}_\alpha , A^{(\mathfrak{p})}_\beta]=0 \, , 
\end{equation}
where the term in parentheses can be written as a field strength $F_{\alpha \beta}$ for $A^{(\mathfrak{h})}$. The flatness condition simplifies for symmetric spaces, where it becomes two separate equations
\begin{equation}
    \left(\partial_\alpha A^{(\mathfrak{h})}_\beta - \partial_\beta A^{(\mathfrak{h})}_\alpha + [A^{(\mathfrak{h})}_\alpha , A^{(\mathfrak{h})}_\beta]\right)  + [A^{(\mathfrak{p})}_\alpha , A^{(\mathfrak{p})}_\beta]=D_\alpha A^{(\mathfrak{p})}_\beta - D_\beta A^{(\mathfrak{p})}_\alpha=0 \, . \label{flatness}
\end{equation}

\paragraph{Degrees of freedom and gauge symmetry.} Although the action \eqref{PCMaction} depends on the $\mathfrak{p}$-components of the Maurer-Cartan, the equations of motion \eqref{eomsigmarelativista} and the flatness condition \eqref{flatness} depend also on the $\mathfrak{h}$-components. However, the $\mathfrak{h}$-components should not be physical, as all the physical information is contained in the $\mathfrak{p}$-components.\footnote{This may seem in contradiction with the fact that we used elements of both $\mathfrak{h}$ and $\mathfrak{p}$ to construct the polar coordinates in \eqref{PolarCoset}. However, a gauge transformation lets us express a coset representative entirely in $\mathfrak{p}$, at the expense of introducing the coordinates via non-linear functions.} To check that the action contains the appropriate degrees of freedom, we can perform the following transformation of the group element $g(\sigma,\tau)\in G$ entering in the action
\begin{equation}
    g(\tau,\sigma)\rightarrow g(\tau,\sigma)\cdot h(\tau,\sigma)\,, \qquad h(\tau,\sigma)\in H \,.
\end{equation}
From \eqref{MCdef} it is immediate to see that the MC transforms as
\begin{align}
    A_\alpha^{(\mathfrak{h})} &\rightarrow h^{-1} A_\alpha^{(\mathfrak{h})} h + h^{-1} \partial_\alpha h \,,\notag \\
    A_\alpha^{(\mathfrak{p})}&\rightarrow h^{-1} A_\alpha^{(\mathfrak{p})} h  \,.
\end{align}
As the inner product is adjoint-invariant \eqref{adinvariant}, the action \eqref{PCMaction} does not change under this transformation. Therefore, it only depends on the degrees of freedom of $\mathfrak{p}$, and the degrees of freedom associated to $\mathfrak{h}$ are gauge degrees of freedom that can be fixed by selecting an appropriate parameterisation of $g(\sigma,\tau)$.

\paragraph{Symmetries of the action.} Let us consider the adjoint action of a constant element $k\in G$ on $g(\tau,\sigma)\in G$
\begin{equation}
    g(\tau,\sigma)\rightarrow \text{Ad}_k[g(\tau,\sigma)]=k\cdot g(\tau,\sigma)\cdot k^{-1}\,.
\end{equation}
This transformation leaves the action invariant by construction, as the inner product used to define it is invariant under the adjoint action of the group. It is clear that $kgk^{-1}$ is still an element of $G$, but it might not be parameterised in $G/H$ in the same fashion as $g$. Nevertheless, we can always introduce a compensatory field $h'(\tau,\sigma)\in H$ to bring it to the right parameterisation
\begin{equation}
    k\cdot g(\tau,\sigma)\cdot k^{-1}=g'(\tau,\sigma)\, h'(\tau,\sigma)\,.
\end{equation}
Consequently, the action is invariant under the adjoint action of a constant element of $G$, as the coset space $G/H$ is. The conserved current associated to this symmetry is 
\begin{equation}
    J_\alpha=g A_\alpha^{(\mathfrak{p})} g^{-1} \,. \label{RelConserved}
\end{equation}
In fact, the continuity equation $\partial_\alpha J^\alpha=0$ can be derived from the equation of motion \eqref{eomsigmarelativista}.

\paragraph{Integrability.} The equation of motion for $g$ and the flatness condition can be combined into the flatness conditions of the current
\begin{equation}
    L_\alpha=A^{(\mathfrak{h})}_\alpha +\eta_1 A^{(\mathfrak{p})}_\alpha + \eta_2 \, \gamma_{\alpha \beta} \varepsilon^{\beta \rho} A^{(\mathfrak{p})}_\rho \, , 
\end{equation}
with the condition $\eta_1^2-\eta_2^2=1$. We can indeed check that the projection on $\mathfrak{h}$ of the curvature of $L$ gives us the $\mathfrak{h}$ component of the curvature of $A$
\begin{equation} \label{laxcurvatureh}
    \left(\partial_\alpha L_\beta - \partial_\beta L_\alpha + [L_\alpha , L_\beta]\right)^{(\mathfrak{h})}=\left(\partial_\alpha A^{(\mathfrak{h})}_\beta - \partial_\beta A^{(\mathfrak{h})}_\alpha + [A^{(\mathfrak{h})}_\alpha , A^{(\mathfrak{h})}_\beta]\right)  + (\eta_1^2-\eta_2^2)[A^{(\mathfrak{p})}_\alpha , A^{(\mathfrak{p})}_\beta] \,,
\end{equation}
while the projection on $\mathfrak{p}$ gives us the $\mathfrak{p}$ component of the curvature of $A$ plus the equations of motion
\begin{equation} \label{laxcurvaturep}
    \left(\partial_\alpha L_\beta - \partial_\beta L_\alpha + [L_\alpha , L_\beta]\right)^{(\mathfrak{p})}=\eta_1 \left( D_\alpha A^{(\mathfrak{p})}_\beta - D_\beta A^{(\mathfrak{p})}_\alpha \right) + 2 \eta_2 \epsilon_{\alpha \beta} D_\rho (A^{(\mathfrak{p})})^\rho \,.
\end{equation}

Although we can write $\eta_2=\sqrt{\eta_1^2-1}$, the square root introduces an unnecessary complexity and it is usually avoided through ingenious parameterisations, like $\eta_1=\frac12(1\pm\cosh \xi)$ (see e.g. \cite{Bena:2003wd}), $2\eta_1=\xi^2+\frac{1}{\xi^2}$ (see e.g. \cite{Arutyunov:2009ga}), and $\eta_1=\pm \frac{1+\xi^2}{1-\xi^2}$ (see e.g. \cite{Alday:2005gi,Zarembo:2017muf}). The parameter $\xi$ is usually called \emph{spectral parameter}. For simplicity, from now on we will use the last parameterisation, where the current takes the form
\begin{equation}
    L_\alpha(\xi;\tau,\sigma)=A^{(\mathfrak{h})}_\alpha +\frac{\xi^2+1}{\xi^2-1} A^{(\mathfrak{p})}_\alpha - \frac{2\xi}{\xi^2-1} \gamma_{\alpha \beta} \varepsilon^{\beta \rho} A^{(\mathfrak{p})}_\rho \, . \label{RelativisticLax}
\end{equation}

The existence of this flat one-parameter-dependent current, called \emph{Lax connection}, implies the existence of an infinite tower of conserved charges. If we can show that these conserved charges Poisson-commute with each other and are linearly independent, the model is classically integrable, see e.g. \cite{Babelon:2003qtg,Torrielli:2016ufi,Arutyunov:2019}.\footnote{Quantum integrability of \eqref{PCMaction} is more involved to prove because not all its degrees of freedom are physical. Therefore, first we have to fix the degrees of freedom associated to $H$ and solve the Virasoro constraints \eqref{VirasoroCoset}. See \cite{Alday:2005gi} for a discussion on the topic.} To see that, first we define a \emph{monodromy matrix} as the path-ordered exponential of the Lax connection
\begin{equation}
    T(\xi,\tau)=\mathcal{P}\exp \oint_\Gamma \Big[ \dd\tau(s) \, L_\tau (\xi; \tau,\sigma) + \dd\sigma(s) \, L_\sigma (\xi; \tau,\sigma) \Big] \, ,
\end{equation}
where $\Gamma$ is any path wrapping the worldsheet cylinder once, parameterised by the variable $s$. Notice that flatness of the Lax connection guarantees that the monodromy matrix is independent of the path chosen. If we choose to perform the integral along a constant $\tau$ path, that is
\begin{equation}
        T(\xi,\tau)=\mathcal{P}\exp \int_0^{2\pi} \dd\sigma \, L_\sigma (\xi; \tau,\sigma) \, , \label{monodromy}
\end{equation}
it is immediate to show that the monodromy matrix fulfills the equation of motion
\begin{equation}
    \partial_\tau T(\xi,\tau)= L_\sigma (\xi; \tau, 2\pi) T(\xi,\tau) - T(\xi,\tau) L_\sigma (\xi; \tau, 0 ) \, .
\end{equation}
If we are interested in closed strings and demand periodicity in $\sigma$, this equation implies that the eigenvalues of $T(\xi,\tau)$ are independent of $\tau$. Thus, said eigenvalues are related to the conserved quantities of the system. As this statement is true for any value of the parameter $\xi$, we can consider the formal expansion of the trace of $T(\xi,\tau)$ in powers of $\xi$ as a generating function for the conserved quantities of the theory. This generates an infinite tower of conserved charges, implying that the action \eqref{PCMaction} is integrable in the sense of Liouville.\footnote{Liouville theorem says that an action is integrable if it has as many linearly independent conserved charges in involution as degrees of freedom. The construction of conserved quantities using the monodromy matrix guarantees that they are indeed conserved charges. Requiring that they are in involution, that is, that we can use any linear combination of them as the generator of the time evolution and all of these quantities will still be conserved quantities, would require more effort but can be done, see for example \cite{Maillet:1985ec,MAILLET198654}. Linear independence, in contrast, has to be checked in a case-by-case basis.}

This construction can be generalised to include fermionic degrees of freedom as long as we use a supergroup that fulfills a generalisation of the symmetric space condition, where instead of a $\mathbb{Z}_2$ outer automorphism, the associated superalgebra has a $\mathbb{Z}_4$ outer automorphism. Spaces constructed using these superalgebras are called semi-symmetric. We will not discuss them in this review, but we encourage the interested reader to look at chapter 1 of \cite{Arutyunov:2009ga}.

\section{The non-relativistic string action as a coset model}
\label{sec:NR_action_as_coset}
A natural way to construct the non-relativistic limit of the coset action is to follow the spirit of the expansion approach discussed in section \ref{sec.expansionapproach} and use the Lie algebra expansion to determine the structure of the resulting non-relativistic algebra.\footnote{A different approach based on null-reduction was developed in \cite{Figueroa-OFarrill:2025nmo} which, at the moment, requires the coset to be of the form $(G\times G)/H$.} The question is at what point of the construction should we implement it, at the level of the MC current or at the level of the coset construction? Here we examine the two options, developed in \cite{Fontanella:2020eje} and \cite{Fontanella:2022fjd,Fontanella:2022pbm} respectively. For preliminary studies in the literature of non-relativistic coset spaces, see \cite{Grosvenor:2017dfs, Brugues:2006yd}.

\subsection{The Lie Algebra Expansion}

The Lie Algebra Expansion (LAE) is a technique to generate new Lie algebras from a given one \cite{deAzcarraga:2002xi,deAzcarraga:2007et, Hatsuda:2001pp}. The main idea is to decompose the original algebra into subspaces $\mathfrak{g}=\bigoplus V_n$ in such a way that the structure is compatible with a series expansion in a parameter $\epsilon$. For our purposes, we only need two subspaces $V_0$ and $V_1$ with a symmetric space structure, that is
\begin{equation}
    [V_0 , V_0] \subseteq V_0 \, , \qquad [V_0 , V_1] \subseteq V_1 \, , \qquad [V_1 , V_1] \subseteq V_0 \, .
\end{equation}
Compatibility of the series expansion with this structure implies that generators in $V_0$ expand in series in even powers of $\epsilon$, while generators in $V_1$ expand in series in odd powers of $\epsilon$
\begin{equation} \label{LAExpansionDefinition}
    V_0 \ni T_{A_0}=\sum_{n=0}^{N_0/2} T^{(2n)}_{A_0} \epsilon^{2n} \,,\qquad V_1 \ni T_{A_1}=\sum_{n=0}^{(N_1-1)/2} T^{(2n+1)}_{A_1} \epsilon^{2n+1} \,,
\end{equation}
where the truncation has to satisfy $N_0=N_1 \pm 1$ for the expansion to be consistent with the flatness of the MC. Furthermore, as the structure constants do not depend on the expansion parameter $\epsilon$, the expanded algebra inherits the following graded relations
\begin{equation}
    [T_A^{(m)},T_B^{(n)}]=f_{AB}\null^C T_C^{(m+n)} \, ,
\end{equation}
where $f_{AB}\null^C$ are the structure constants of the original algebra $\mathfrak{g}$.

For the $\mathfrak{so}(2,4)\oplus \mathfrak{so}(6)$ algebra that is relevant for us, see equation \eqref{so(4,2)+so(6)} and above for notation, we choose the following splitting\footnote{Here we only discuss the bosonic part of the AdS$_5\times$S$^5$ background, but this construction can be generalised to include the fermionic generators \cite{Fontanella:2020eje}.}
\begin{equation}
    V_0=\text{span}\{ J_{AB}, J_{ab}, J_{a',b'} , P_A \} \, , \qquad V_1=\text{span}\{ J_{Ab}, P_a , P_{a'} \} \, . 
\end{equation}
The reason for this splitting is that we want the leading orders in $V_0$ and $V_1$ to be the \.In\"on\"u-Wigner contraction of the $\mathfrak{so}(2,4)\oplus \mathfrak{so}(6)$ algebra to the string Newton-Hooke plus Euclidean algebra \eqref{Newton-Hooke-algebra}. Notice also that the splitting from the Lie algebra expansion is completely independent of the $(\mathfrak{h},\mathfrak{p})$ splitting of the coset construction.

With this split, the truncation \eqref{LAExpansionDefinition} for $N_0=N_1+1=2$ takes the form
\begin{subequations}\label{LieAlgebraExpansion}
\begin{align}
J_{AB} &\rightarrow \varepsilon_{AB} ( M + \epsilon^2 Z ) \, , &
J_{Aa} &\rightarrow \epsilon G_{Aa}\, , \\
J_{ab} &\rightarrow J_{ab} + \epsilon^2 Z_{ab} \, , &
J_{a'b'}&\rightarrow J_{a'b'} + \epsilon^2 Z_{a'b'}\, ,   \\
P_{A}&\rightarrow H_A + \epsilon^2 Z_A \, , &
P_{a} &\rightarrow \epsilon P_a \, ,  \\ 
P_{a'}&\rightarrow \epsilon P_{a'} \, .
\end{align}
\end{subequations}
This is nothing but an extension of the string Newton-Hooke algebra plus an extension of the Euclidean algebra. Notice the similarity of these extensions with the non-central extensions of the String Newton-Cartan algebra \eqref{eq:generatorslistbulk}.
%This is nothing but the string Newton-Cartan algebra \eqref{eq:generatorslistbulk} under the constraint $Z_{AB}=-Z_{BA}$, which allows us to write $Z_{AB}$ as $\varepsilon_{AB}Z$, and extended by the generators $Z_{ab}$ and $Z_{a'b'}$. The SNC algebra \eqref{eq:generatorslistbulk} with the constraint $Z_{AB}=-Z_{BA}$ is usually called string Bargmann algebra, see the discussion bellow equation (3.39) in \cite{Bergshoeff:2019pij}.

At the level of commutation relations, this algebra inherits them from the original $\mathfrak{so}(2,4)\oplus \mathfrak{so}(6)$, with the appropriate restrictions to ensure that the series is defined consistently
\begin{subequations}
 \begin{align} 
	[J_{ab},P_{c}] & =  \delta_{bc} P_{a} - \delta_{ac} P_{b}\,, &
    [P_{a'}, J_{b'c'}] &= \delta_{a'b'} P_{c'} - \delta_{a'c'} P_{b'} \, ,\\
    [P_a, P_b] &= Z_{ab} \, , &
	[P_{a'}, P_{b'}] &= - Z_{a'b'} \, , \\ 
    [J_{ab},G_{Ac}] & = \delta_{bc} G_{Aa} - \delta_{ac} G_{Ab} \,, &
	[M,G_{Aa}]  & = -\varepsilon_A{}^B G_{Ba}\,, \\
	[G_{Aa}, P_{b}] & = \delta_{ab} Z_A\,, &
	[G_{Aa},H_B]  & = -\eta^{}_{AB} P_{a}\,, \\ 
    [H_A,Z_B] & = -\varepsilon_{AB} Z \,, &    [M,Z_A] & = -\varepsilon_A{}^B Z_B\,, \\ 
	[Z,H_A]  & = -\varepsilon_A{}^B Z_B\,, &	[M,H_A] & = -\varepsilon_A{}^B H_B\,, \\
	[H_A, H_B]& = -\varepsilon_{AB} M \, , &
	[H_A, P_b] &= G_{Ab} \, .
\end{align}
\begin{gather} 
	{}[J_{ab}, J_{cd}]  = \delta_{bc} J_{ad} - \delta_{ac} J_{bd} + \delta_{ad} J_{bc} - \delta_{bd} J_{ac} \,, \\
 [J_{ab}, Z_{cd}]  = \delta_{bc} Z_{ad} - \delta_{ac} Z_{bd} + \delta_{ad} Z_{bc} - \delta_{bd}
 Z_{ac} \,,  \\
[J_{a'b'}, J_{c'd'}] = \delta_{b'c'} J_{a'd'} - \delta_{a'c'} J_{b'd'} + \delta_{a'd'} J_{b'c'} - \delta_{b'd'} J_{a'c'} \, ,\\
 [J_{a'b'}, Z_{c'd'}] = \delta_{b'c'} Z_{a'd'} - \delta_{a'c'} Z_{b'd'} + \delta_{a'd'} Z_{b'c'} - \delta_{b'd'} Z_{a'c'} \, ,\\
	[G_{Aa}, G_{Bb}]  = \delta_{ab} \varepsilon_{AB} Z - \eta_{AB} Z_{ab} \,,
 \end{gather}
\end{subequations}
It can be checked that this algebra is not semisimple, as the Killing form is degenerate. For example, we can take any of the $P_a$ and check that
\begin{equation}
    \text{Tr}[\hat{\rho}_{\text{ad}}(P_a)\,\hat{\rho}_{\text{ad}}(T)]=0 \,,
\end{equation}
for any generator $T$, where $\hat{\rho}_{\text{ad}}$ means the adjoint representation of the generator. Nevertheless, despite it being a non-semisimple algebra, it has an adjoint-invariant bilinear inner product\footnote{As we do not have the full group structure, we impose invariance under the adjoint action only at the level of the algebra instead of at the level of a group, like \eqref{adinvariant}. We say that a bilinear inner product $\langle \cdot , \cdot \rangle$ is \emph{adjoint-invariant} if for any three generators of the algebra, it fulfils $\langle [T_3,T_1] , T_2 \rangle+\langle T_1 , [T_3,T_2] \rangle=0$.} inherited from the fact that the parent algebra is semisimple\footnote{Notice that the generators $Z_{ab}$ and $Z_{a'b'}$ form an ideal subalgebra, so we might be tempted to eliminate them.  However, as we later will need $\phi_1$ and $\phi_5$ to be non-vanishing to have $\langle \mathfrak{p}, \mathfrak{p} \rangle \neq 0$,  their presence is necessary for the existence of an ad-invariant bilinear inner product \cite{Fontanella:2022pbm}. See also the discussion at the end of section \ref{sec:cosetoftheLAE}.}
	\begin{align}\label{inner_product_extended_algebra}
	\langle M, Z \rangle &= \phi_1\, , &
	\langle M, M \rangle &= \phi_2 \, , &
	\langle H_+, H_- \rangle &= - \phi_2 /2 \, , \notag \\
	\langle H_{\pm}, Z_{\mp} \rangle &=  -\phi_1/2 \, , &
	\langle G_{\pm a}, G_{\mp b} \rangle &=   \delta_{ab}  \phi_1/2 \, , \\
	\langle J_{ab}, J_{cd} \rangle &=  \delta_{[a[c} \delta_{b]d]}  \phi_3 \, , &
	\langle P_a, P_b \rangle &=    \delta_{ab} \phi_1 \, , &
	\langle J_{ab}, Z_{cd}  \rangle &= -2 \delta_{[a[c} \delta_{b]d]}  \phi_1  \, ,  \notag \\
	\langle J_{a'b'}, J_{c'd'} \rangle &=  \delta_{[a'[c'} \delta_{b']d']} \phi_4 \, , &
	\langle P_{a'}, P_{b'} \rangle &=   \delta_{a'b'} \phi_5  \, , &
	\langle J_{a'b'}, Z_{c'd'} \rangle &= 2 \delta_{[a'[c'} \delta_{b']d']} \phi_5 \, ,  \notag
\end{align}
where $\phi_i$ are arbitrary constants.

The Lie algebra expansion also gives us a method to construct matrix representations for the truncated algebra. As the commutation relations are inherited from the parent algebra, for any representation $\rho$ of the parent algebra we can construct an induced representation of the truncated algebra, $\hat{\rho}$. The procedure goes as follows
\begin{equation}
    \hat{\rho}(T_A^{(k)})=\rho(T_A) \otimes \omega^k \,, \label{inducedrepresentation}
\end{equation}
where $\omega$ is matrix such that $\omega^{1+\text{Max}(N_0,N_1)}=0$, the zero matrix. In this way, $\omega$ plays at the level of the representation the role that $\epsilon$ is playing at the level of the expansion. For the $\mathfrak{so}(2,4)\oplus \mathfrak{so}(6)$ algebra truncated at $N_0=N_1+1=2$ we are interested in, we can pick \cite{Fontanella:2022pbm}
\begin{equation}
    \omega=\begin{pmatrix}
        0 & 1 & 0 \\
        0 & 0 & 1 \\
        0 & 0 & 0
    \end{pmatrix} \, ,
\end{equation}
while for $\rho$ we can pick, e.g., the spinorial representation of $\mathfrak{so}(2,4)\oplus \mathfrak{so}(6)$, see e.g. \cite{Arutyunov:2009ga}.

\subsection{The Lie Algebra Expansion of the coset}

Let us now apply the idea of the Lie Algebra Expansion to the coset construction of the action. As we already have the form of the action after the coset construction, \eqref{PCMaction}, we can just take this final result and apply the Lie Algebra Expansion to the MC. To do so, we consider as starting algebra the Lie Algebra Expansion of $\mathfrak{g}$, which we will denote as $\text{LAE}_{N_0,N_1}(\mathfrak{g})$ from now on, and we split it into $\text{LAE}_{N_0,N_1}(\mathfrak{h})$ and $\text{LAE}_{N_0,N_1}(\mathfrak{p})$. With the notation of \eqref{LieAlgebraExpansion}, the algebras relevant for AdS$_5\times$S$^5$ up to the truncation $N_0=N_1+1=2$ are
\begin{align}
    \mathfrak{p} & =\text{Span}\{P_A, P_a,P_{a'}\}  \,, & \mathfrak{h} & =\text{Span}\{J_{AB}, J_{Aa}, J_{ab}, J_{a'b'}\} \,, \notag\\
    \text{LAE}_{0,-1}(\mathfrak{p}) & =\text{Span}\{H_A\} \,, & \text{LAE}_{0,-1}(\mathfrak{h}) & =\text{Span}\{M,J_{ab},J_{a'b'}\} \,, \\
    \text{LAE}_{2,1}(\mathfrak{p}) & =\text{Span}\{H_A, P_a, P_{a'}, Z_A\} \,, & \text{LAE}_{2,1}(\mathfrak{h}) & =\text{Span}\{M,Z,J_{ab},Z_{ab},G_{Aa}, J_{a'b'}, Z_{a'b'}\} \,. \notag
\end{align}

If we substitute the expansion \eqref{LieAlgebraExpansion} into the action~\eqref{PCMaction}, we get an expansion of the form
\begin{equation}
    S=S^{(0)}+\epsilon^2 S^{(2)} \,,
\end{equation}
as the action has to be truncated at order Max$(N_0, N_1)$, otherwise we would miss terms. The leading order of this expansion is based on $\text{LAE}_{0,-1}(\mathfrak{g})$, and given by
\begin{equation} \label{leadingLAEC}
    S^{(0)}=-\frac{T}{2} \int{\dd^2 \sigma \ \gamma^{\alpha \beta} \langle A_\alpha^{(\mathfrak{p},0)} , A_\beta^{(\mathfrak{p},0)} \rangle} \, ,
\end{equation}
where $A^{(\mathfrak{p},n)}$ represent the components in $\mathfrak{p}$ of the MC 1-form at the $n$-th order in the Lie algebra expansion. Consequently, $A^{(\mathfrak{p},0)}$ contains only the contribution from $H_A$. Similarly, $A^{(\mathfrak{p},1)}$ contains the contributions from $P_a$ and $P_{a'}$, while $A^{(\mathfrak{p},2)}$ contains only the contribution from $Z_A$. The inner product we use here is the induced inner product inherited from the original one in $\mathfrak{so}(2,4)\oplus \mathfrak{so}(6)$, i.e. \eqref{inner_product_extended_algebra} with $\phi_i=1$ for all coefficients. To compare with the expressions we wrote in section~\ref{subsec:limit}, we write the components of the MC as
\begin{equation} \label{MCLAE}
    A_\mu^{(\mathfrak{p},0)}=\tau_\mu \null^A H_A \, , \qquad A_\mu^{(\mathfrak{p},1)}=e_\mu \null^a P_a +e_\mu \null^{a'} P_{a'} \, , \qquad A_\mu^{(\mathfrak{p},2)}=m_\mu \null^A Z_A \, .
\end{equation}
Substituting into the above action, we get
\begin{eqnarray}
\notag
      S^{(0)}&=&-\frac{T}{2} \int{\dd^2 \sigma\gamma^{\alpha \beta} \langle A^{(\mathfrak{p},0)}_\alpha , A^{(\mathfrak{p},0)}_{\beta} \rangle} \\
      &=&  -\frac{T}{2} \int{\dd^2 \sigma (-\phi_2) \gamma^{\alpha \beta} \tau_\mu \null^+ \tau_\nu \null^- \partial_\alpha X^\mu \partial_\beta X^\nu} = -\frac{T}{2} \int{\dd^2 \sigma \gamma^{\alpha \beta} \tau_{\alpha \beta}} \,,
\end{eqnarray}
where we have used that $\phi_2=1$. This corresponds exactly to the divergent part of the Polyakov action before introducing the B-field, i.e. the divergent part of \eqref{rel_Pol_action} after substituting the expansion \eqref{gscaling}. We can also see that \eqref{leadingLAEC} matches equation (4.3) in \cite{Gomis:2005pg}.

The next-to-leading order action is based on $\text{LAE}_{2,1}(\mathfrak{g})$, and takes the form
\begin{equation} \label{subleadingLAEC}
    S^{(2)}=-\frac{T}{2} \int{\dd^2 \sigma \ \gamma^{\alpha \beta} \left( 2 \langle A^{(\mathfrak{p},2)}_\alpha , A^{(\mathfrak{p},0)}_\beta \rangle + \langle A^{(\mathfrak{p},1)}_\alpha , A^{(\mathfrak{p},1)}_\beta \rangle \right) } \, .
\end{equation}
which, in terms of the components of the MC, takes the form
\begin{align}
    &2\langle A^{(\mathfrak{p},2)}_\alpha , A^{(\mathfrak{p},0)}_\beta \rangle=-\frac{\phi_1}{2} \left( \tau^+_\alpha m^-_\beta + \tau^-_\alpha m^+_\beta \right) \,, \notag \\
   & \langle A^{(\mathfrak{p},1)}_\alpha , A^{(\mathfrak{p},1)}_\beta \rangle=\phi_1 e^a_\alpha e^a_\beta + \phi_5 e^{a'}_\alpha e^{a'}_\beta \,, \notag \\
   & S^{(2)}=-\frac{T}{2} \int{\dd^2 \sigma \ \gamma^{\alpha \beta} H_{\alpha \beta}} \,,
\end{align}
where we have used that $\phi_1=\phi_5=1$. This corresponds exactly to the first term in \eqref{NR_action}. This also matches equation (4.2) in \cite{Gomis:2005pg}.

However, we should have in mind that $S^{(0)}$ and $S^{(2)}$ are not tied in any way and have to be treated as two independent actions.

\paragraph{Degrees of freedom and symmetries of the action.} As the actions \eqref{leadingLAEC} and \eqref{subleadingLAEC} have the same structure of ``inner product of the $\mathfrak{p}$-components of the MC'' as the relativistic action, we can immediately borrow the arguments we used in the relativistic case to state that these actions are invariant under the adjoint action by a constant element of the group $G$ and under right multiplication by a local element of the group $H$, the only difference being that $G$ and $H$ are substituted by their LAEs. Therefore, \eqref{leadingLAEC} and \eqref{subleadingLAEC} are invariant under transformations constructed using the extended string Newton-Hooke plus Euclidean algebra \eqref{LieAlgebraExpansion}. In fact, for a truncation to the $(N_0,N_0-1)$ order, the action will be invariant under a global transformation constructed from the $(N_0,N_0-1)$ truncation of the $\mathfrak{so}(2,4)\oplus \mathfrak{so}(6)$ algebra.

Regarding the degrees of freedom, the dimension of $\text{LAE}_{2,1}(\mathfrak{p})=\text{LAE}_{2,1}(\mathfrak{g})\setminus\text{LAE}_{2,1}(\mathfrak{h})$ indicates that we have 12 dynamical degrees of freedom. This is a problem because the action \eqref{NR_action} has only 10 dynamical degrees of freedom. It was proposed in \cite{Fontanella:2020eje} that this mismatch can be healed by truncating the theory to a subsector, e.g. by identifying the degrees of freedom associated to $H_A$ and $Z_A$. In the next section we will discuss a different approach that solves this issue.

\paragraph{Lax connection.} Finding the expanded Lax connection equivalent to the expanded equations of motion is a bit more subtle. This is both because the Lax connection does not start at order $\epsilon^0$ and because we have to keep the truncation consistent. As these are very technical details, we refer the reader to chapter 3 of \cite{Fontanella:2020eje} and only provide the final result. The Lax connection associated to the leading order of the action can be obtained from substituting the Lie algebra expansion $\text{LAE}_{0,-1}[\mathfrak{so}(2,4)\oplus \mathfrak{so}(6)]$ (i.e., equation \eqref{LieAlgebraExpansion} with $\epsilon=0$) into \eqref{RelativisticLax}
\begin{equation}
    L^{(0)}_\alpha=A^{(\mathfrak{h},0)}_\alpha +\frac{\xi^2+1}{\xi^2-1} A^{(\mathfrak{p},0)}_\alpha - \frac{2\xi}{\xi^2-1} \gamma_{\alpha \beta} \varepsilon^{\beta \rho} A^{(\mathfrak{p},0)}_\rho  \,,
\end{equation}
where, similarly to $A^{(\mathfrak{p},n)}$, $A^{(\mathfrak{h},n)}$ represent the components in $\mathfrak{h}$ of the MC 1-form at the $n$-th order in the Lie algebra expansion. The Lax connection for the next-to-leading order of the action is obtained by substituting \eqref{LieAlgebraExpansion} into \eqref{RelativisticLax}
\begin{equation}
    L^{(2)}_\alpha=\tilde{L}^{(2,0)}_\alpha+\tilde{L}^{(2,1)}_\alpha+\tilde{L}^{(2,2)}_\alpha \,, 
\end{equation}
where
\begin{equation}
    \tilde{L}^{(2,n)}_\alpha=A^{(\mathfrak{h},n)}_\alpha +\frac{\xi^2+1}{\xi^2-1} A^{(\mathfrak{p},n)}_\alpha - \frac{2\xi}{\xi^2-1} \gamma_{\alpha \beta} \varepsilon^{\beta \rho} A^{(\mathfrak{p},n)}_\rho  \,.
\end{equation}

\subsection{The coset of the Lie Algebra Expansion} \label{sec:cosetoftheLAE}

Although applying the Lie algebra expansion at the level of the action already gives us a coset action for the SNC AdS$_5\times$S$^5$, we have seen that it has too many degrees of freedom. We can instead take a step back and look for a way to write the SNC AdS$_5\times$S$^5$ background as a homogeneous space by crafting a split of the algebra $\text{LAE}_{2,1}[\mathfrak{g}]$ that gives us the correct number of degrees of freedom.\footnote{Although in this review we discuss only the SNC AdS$_5\times$S$^5$ space, the case of SNC flat space is constructed similarly \cite{Fontanella:2022fjd}.}

In \cite{Fontanella:2022fjd}, and later refined in \cite{Fontanella:2022pbm}, it was found that the algorithm for the construction of the coset model works if the following split is used
\begin{equation}
    \text{LAE}_{2,1}[\mathfrak{so}(2,4)\oplus \mathfrak{so}(6)]=\left( \text{LAE}_{2,1} (\mathfrak{p}) \setminus \{Z_A\} \right) \oplus \left( \text{LAE}_{2,1} (\mathfrak{h}) \oplus \{Z_A\} \right) \,,
\end{equation}
that is, compared to the previous split of the algebra, we have moved $Z_A$ from the expanded $\mathfrak{p}$ to the expanded $\mathfrak{h}$. Written explicitly,
\begin{equation}
    \tilde{\mathfrak{h}}=\text{Span}\{ M, J_{ab}, J_{a'b'}, G_{Aa} , Z_A, Z, Z_{ab}, Z_{a'b'}\} \, , \qquad 
    \tilde{\mathfrak{p}}=\text{Span}\{ H_A, P_a , P_{a'}\} \, . \label{splitNR}
\end{equation}
With a clear split of the algebra and an ad-invariant bilinear form \eqref{inner_product_extended_algebra}, the construction of a coset action follows similar steps to the relativistic one.

In addition to the MC 1-form, defined in the usual form $A=g^{-1} \dd g$, we define a current containing the information of what will be the Lagrange multipliers
\begin{equation}
    \Lambda_\alpha=\Lambda^A_\alpha Z_A=\lambda_+ \theta^+_\alpha Z_- + \lambda_- \theta^-_\alpha Z_+ \, , \label{LambdaCurrent}
\end{equation}
and a projector 
\begin{equation}
    \mathcal{P}:\text{LAE}_{2,1}(\mathfrak{g})\to \text{LAE}_{2,1}(\mathfrak{p})=\text{Span}\{H_A, P_a, P_{a'}, Z_A\} \, . \label{ProjectorNR}
\end{equation}
Under the $\mathbb{Z}_2$ automorphism, $\text{LAE}_{2,1}(\mathfrak{p})$ and $\text{LAE}_{2,1}(\mathfrak{h})$ have grading -1 and +1 respectively. However, $\tilde{\mathfrak{p}}$ and $\tilde{\mathfrak{h}}$ do not have a well-defined grading. This is because $Z_A$ is an element of $\tilde{\mathfrak{h}}$, which nevertheless will play a role in the action. This situation is different from the relativistic case, where $\mathfrak{p}$ and $\mathfrak{h}$ coincide with the subspaces with $\mathbb{Z}_2$ grading $\pm 1$.    

With these ingredients, and fixing $\phi_2=0$ and $\phi_1=\phi_5=1$ in the inner product \eqref{inner_product_extended_algebra}, the action
\begin{equation}
    S=-\frac{T}{2} \int{\dd^2 \sigma \ (\gamma^{\alpha \beta} \langle \mathcal{P} A_\alpha , \mathcal{P} A _\beta\rangle + 2\varepsilon^{\alpha \beta} \langle \mathcal{P} A_\alpha, \mathcal{P} \Lambda_\beta \rangle) } \, , \label{NRPCMaction}
\end{equation}
is equivalent to \eqref{NR_action_Eric} with $b^{\text{NR}}_{\mu\nu}=0$. To see that, we substitute \eqref{MCLAE} and \eqref{LambdaCurrent} into the above action, giving us
\begin{subequations}
    \begin{align}
        \gamma^{\alpha \beta}\langle \mathcal{P} A_\alpha , \mathcal{P} A _\beta\rangle &= \gamma^{\alpha \beta}\left( e^a_\alpha e^a_\beta +e^{a'}_\alpha e^{a'}_\beta -\frac{\tau^+_\alpha m^-_\beta + \tau^-_\alpha m^+_\beta}{2} \right)=\gamma^{\alpha \beta} H_{\alpha \beta} \,,\\
        2\varepsilon^{\alpha \beta}\langle \mathcal{P} A_\alpha, \mathcal{P} \Lambda_\beta \rangle &= -\varepsilon^{\alpha \beta} ( \lambda_+ \theta^+_\beta \tau^+_\alpha +\lambda_- \theta^-_\beta \tau^-_\alpha )= \varepsilon^{\alpha \beta} ( \lambda_+ \theta^+_\alpha \tau^+_\beta +\lambda_- \theta^-_\alpha \tau^-_\beta ) \,.
    \end{align}
\end{subequations}

Furthermore, if we define the generalised current
\begin{equation}
    J_\alpha \equiv A_\alpha + \gamma_{\alpha \beta} \varepsilon^{\beta \rho} \Lambda_\rho \, ,
\end{equation}
the action can be written in a more compact form
\begin{equation}
    S=-\frac{T}{2} \int{\dd^2 \sigma \, \gamma^{\alpha \beta} \langle \mathcal{P} J_{\alpha} , \mathcal{P} J_{\beta} \rangle} \, , \label{NRPCMactionJ}
\end{equation}

\paragraph{Equations of motion.}
The equations of motion for $g$ and the worldsheet metric $\mathsf{h}_{\alpha\beta}$ (Virasoro constaints) are:
\begin{subequations}
\begin{gather}
    \partial_\alpha (\gamma^{\alpha \beta} \mathcal{P} J_\beta )+ \gamma^{\alpha \beta}[J_\alpha , \mathcal{P} J_\beta] =0 \, . \\
    \langle \mathcal{P} J_\alpha, \mathcal{P} J_\beta \rangle - \frac12 \gamma_{\alpha \beta} \gamma^{\rho \sigma} \langle \mathcal{P} J_\rho, \mathcal{P} J_\sigma \rangle=0 \, . 
\end{gather}
\end{subequations}
There is also an additional equation of motion imposed by the Lagrange multipliers,
\begin{eqnarray}\label{Lambda_eom_MC}
    \varepsilon^{\alpha \beta}  \langle Z_{\mp}, A_\alpha \rangle \theta^{\pm}_\beta= 0 \, .
\end{eqnarray}

The flatness condition has exactly the same form as the relativistic one, if we project into $\text{LAE}_{2,1} (\mathfrak{p})$ and $\text{LAE}_{2,1} (\mathfrak{h})$ instead. Thus, it we denote $(A - \mathcal{P} A)=\mathcal{H}A$, the flatness condition takes the form
\begin{subequations}
\begin{align}
    \Big(\partial_\alpha (\mathcal{H}A_\beta) - \partial_\beta (\mathcal{H} A_\alpha) + [(\mathcal{H} A_\alpha ) , (\mathcal{H} A_\beta] )\Big)  + [(\mathcal{P} A_\alpha ) , (\mathcal{P} A_\beta )] &=0 \,, \\
    D_\alpha (\mathcal{P} A_\beta ) - D_\beta (\mathcal{P} A_\alpha ) &=0 \, ,
\end{align}
\end{subequations}
where $D_\alpha\equiv \partial_\alpha + [\mathcal{H} A_\alpha , \cdot]$. If we want to write it in terms of projections onto $\tilde{\mathfrak{h}}$ and $\tilde{\mathfrak{p}}$, we arrive to a similar expression except for the fact that $[H_A,Z_B]\in \tilde{\mathfrak{h}}$ while for a reductive homogeneous space we expected it to contribute to $\tilde{\mathfrak{p}}$
\begin{subequations}
\begin{align}
    \left(\partial_\alpha A^{(\tilde{\mathfrak{h}})}_\beta - \partial_\beta A^{(\tilde{\mathfrak{h}})}_\alpha + [A^{(\tilde{\mathfrak{h}})}_\alpha , A^{(\tilde{\mathfrak{h}})}_\beta]\right)  + [A^{(\tilde{\mathfrak{p}})}_\alpha , A^{(\tilde{\mathfrak{p}})}_\beta]-\frac{1}{2}\Big(\tau_{[\alpha}^+ m^-_{\beta]} +\tau_{[\alpha}^- m^+_{\beta]} \Big))&=0  \,,\\
    \tilde{D}_\alpha A^{(\tilde{\mathfrak{p}})}_\beta - \tilde{D}_\beta A^{(\tilde{\mathfrak{p}})}_\alpha+\frac{1}{2}\Big(\tau_{[\alpha}^+ m^-_{\beta]} +\tau_{[\alpha}^- m^+_{\beta]} \Big)&=0 \, .
\end{align}
\end{subequations}
where $\tilde{D}_\alpha\equiv \partial_\alpha + [A^{(\tilde{\mathfrak{h}})}_\alpha , \cdot]$.

Notice that the degrees of freedom associated to $\{Z_{ab},Z_{a'b'}\}$ appear neither on the equations of motion nor on the flatness condition associated to the other degrees of freedom, making them inconsequential for the dynamics (although they are needed to construct an inner product with good properties). This is a direct consequence of $\{Z_{ab},Z_{a'b'}\}$ being an ideal subalgebra of the numerator and denominator of our coset.

\paragraph{Integrability.}On the one hand, similarly to the relativistic case, we can combine the equations of motion with the flatness condition of $A$ into the flatness condition of a Lax connection \cite{Fontanella:2022fjd}
\begin{equation}
    L_\alpha=(A_\alpha -\mathcal{P} A_\alpha) +\frac{\xi^2+1}{\xi^2-1}  \mathcal{P} A_\alpha - \frac{2\xi}{\xi^2-1} \, \gamma_{\alpha \beta} \varepsilon^{\beta \rho} \mathcal{P} J_\rho \,. \label{NRLax}
\end{equation}
This can be checked by repeating the computation \eqref{laxcurvatureh} and \eqref{laxcurvaturep}.\footnote{There is a caveat to this. Here there is again an unwanted contribution to the flatness equation that appears because the split $\mathfrak{h} \oplus \mathfrak{p}$ is not reductive. This is not a problem here because such term is proportional to $\varepsilon^{\alpha \beta}\langle \mathcal{P} A_\alpha, \mathcal{P} \Lambda_\beta \rangle$, which vanishes due to the conditions imposed by the Lagrange multipliers \eqref{Lambda_eom_MC}.} On the other hand, in contrast to the relativistic case, where flatness of $L$ contains all the equations of motion, flatness of $L$ does not contain the constraints associated to the Lagrange multipliers and they have to be imposed separately.

The existence of this Lax connection implies classical integrability of the non-relativistic coset action (assuming linear independence and involution are inherited from the relativistic theory).\footnote{A similar construction works for SNC flat space, showing that it is also integrable \cite{Fontanella:2022fjd}. It is nevertheless worth highlighting that integrability on flat SNC was first shown in \cite{Kluson:2017ufb} without using the coset construction. Although it is claimed in \cite{Roychowdhury:2019vzh} that nonrelativistic strings propagating over any SNC geometry are integrable, the proof explicitly uses extended string Galilei invariance, rendering the result only valid for flat SNC.} Consequently, the next step would be the construction of the monodromy matrix and the extraction of the conserved quantities of the theory.

\paragraph{Degrees of freedom.} It might look like the action \eqref{NRPCMaction} also has the structure of ``inner product of the $\mathfrak{p}$-components of the MC'' and we can directly borrow the arguments from the relativistic case, but this is not the case. The projector $\mathcal{P}$ keeps the degrees of freedom associated to the generators $Z_A$, which are now included in the subalgebra $\tilde{\mathfrak{h}}$. As a consequence, it seems like we have too many equations of motion compared to the ten degrees of freedom and two Lagrange multipliers. However, thanks to gauge invariance for the generators $Z_A$, there exist Noether identities between the equations of motion that appropriately reduce their number. For example, the equations of motion associated to $Z_A$ are proportional to the equations of motion of the Lagrange multipliers \cite{Fontanella:2022fjd}.

Another point to discuss is the enhanced gauge symmetry coming from $\{Z_{ab},Z_{a'b'}\}$. They are not expected in an SNC geometry and, therefore, they should not contribute to the gauge symmetries. This is indeed the case, as they only contribute to the gauge transformation of themselves, making the remaining fields of the SNC geometry blind to them. Although they can be considered spurious fields, they are necessary to get an adjoint invariant inner product.

\paragraph{Symmetries of the action.}Regarding the symmetries of the action, we can apply the same argument as in relativistic case because we have constructed the action \eqref{NRPCMaction} using an adjoint-invariant inner product. Thus, this action is invariant under the adjoint action of $\text{LAE}_{2,1}[\mathfrak{so}(2,4)\oplus \mathfrak{so}(6)]$, giving rise to conserved currents.

\section{Classical spectral curve}

\subsection{Relativistic strings in AdS$_5\times$S$^5$}

Let us start by reviewing how the classical spectral curve works for AdS$_5\times$S$^5$.\footnote{Regarding backgrounds relevant for non-relativistic holography, the classical spectral curve has also been applied to strings in five-dimensional Schr\"odinger target space\cite{Ouyang:2017yko}.} As we have commented at the end of section \ref{sec:symmetricspaces}, for classical integrable theories with a Lax connection one can define a monodromy matrix that contains all the conserved quantities of the theory. If the monodromy matrix has simple spectrum (i.e. the eigenvalues appear with multiplicity one), and because the physical information is only codified in its conjugacy class (in fact, changing the value of $\tau$ at which we construct the monodromy matrix amounts to a similarity transformation), we can diagonalise it by using an appropriate matrix $\Omega (\xi)$
\begin{equation}\label{T_diag}
   \Omega^{-1} (\xi) T(\xi, \tau) \Omega (\xi) = \text{diag} (\hat{\zeta}_1 , \hat{\zeta}_2 , \hat{\zeta}_3 , \hat{\zeta}_4 | \tilde{\zeta}_1 , \tilde{\zeta}_2 , \tilde{\zeta}_3 , \tilde{\zeta}_4) \, ,
\end{equation}
where we used the 4-dimensional matrix representations of $\mathfrak{so}(2,4)$ and $\mathfrak{so}(6)$, and we have indicated the eigenvalues associated to the $\mathfrak{so}(2,4)$ and $\mathfrak{so}(6)$ by a hat and a tilde respectively. Instead of the eigenvalues, it is usually more useful to consider the quasi-momenta, $\hat{p}_i$ and $\tilde{p}_i$, obtained from the logarithm of these eigenvalues
\begin{equation}
    \hat{\zeta}_i = e^{i \hat{p}_i} \, , \qquad \tilde{\zeta}_i = e^{i \tilde{p}_i} \, .
\end{equation}
We have to emphasise that neither the eigenvalues $\zeta$ nor the quasi-momenta $p$ are analytic in the spectral parameter $\xi$. In fact, $\xi$ and $\zeta$ define an algebraic surface in $\mathbb{C}^2$ given by the characteristic equation of the monodromy matrix
\begin{equation}
    \Gamma (\zeta,\xi)= \det (T(\xi,\tau) - \zeta \, \mathbb{I} ) \, . \label{CharacteristicEquation}
\end{equation}
This algebraic curve is usually called \emph{spectral curve}, as it codifies the spectral properties of the system. Here we want to highlight only five interesting analytic properties of these eigenvalues, while referring to \cite{Beisert:2004ag,Beisert:2005bm,Schafer-Nameki:2010qho} for more properties and detailed discussions,
\begin{itemize}
    \item \emph{Leading asymptotic structure}: from the definition of the Lax connection \eqref{RelativisticLax}, we see that $L_\alpha \to A_\alpha-\frac{2}{\xi} \gamma_{\alpha \beta} \varepsilon^{\beta \rho} A^{(\mathfrak{p})}_\rho$  for $\xi\to \infty$. Let us focus for the moment on the leading contribution. Integrating this contribution, we get
\begin{equation}\label{rep_free_asympt_monodromy}
    \lim_{\xi\rightarrow \infty} T (\xi,\tau)={\cal P}\!\exp \left[ \int_0^{2\pi} A_\sigma \, \dd\sigma \right] = {\cal P}\!\exp \left[ \int_0^{2\pi} g^{-1} \partial_\sigma g \, \dd\sigma \right] = g(\tau, 2\pi) g(\tau, 0 )^{-1} \, .
\end{equation}
For periodic solutions this imposes $\lim_{\xi\rightarrow \infty} \zeta=1$, but the quasi-momenta are able to catch the winding of angular variables, behaving as $\lim_{\xi\rightarrow \infty} p(\xi)= 2\pi m$.
    \item \emph{Subleading asymptotic structure}: let us focus now on the $\xi^{-1}$ order. As the monodromy matrix is defined as the path-ordered exponential of $L_\sigma$, a gauge transformation of the Lax connection produces a similarity transformation of $T(\xi)$. We can use this to our advantage and consider instead  $\tilde{L}_\alpha=g L_\alpha g^{-1} - (\partial_\alpha g)g^{-1} $, giving us $\tilde{L}_\sigma \to \frac{2}{\xi} g A^{(\mathfrak{p})}_\tau g^{-1}$ provided we work in the conformal gauge. As the $\sigma$ component of the Lax connection is proportional to the $\tau$ component of the conserved current \eqref{RelConserved}, the coefficient of order $\frac{1}{\xi}$ in the quasimomenta $p$ for large values of $\xi$ takes the form of linear combinations of the $SO(2,4)\times SO(6)$ conserved quantities. We refer to eq. (2.18) in \cite{Schafer-Nameki:2010qho} for the specific way they are encoded.
    \item \emph{Branch cuts}: from analysing how the characteristic equation \eqref{CharacteristicEquation} behaves around a point $\xi_0$ where two eigenvalues $\tilde{\zeta}_i$ and $\tilde{\zeta}_j$ degenerate, we find that these eigenvalues need to have a square-root singularity
    \begin{equation}
        \tilde{\zeta}_{i,j}(\xi)=\tilde{\zeta}_i (\xi_0) \left(1+ \tilde{\alpha}_0 \sqrt{\xi - \xi_0} + \dots \right) \, ,
    \end{equation}
    where $\tilde{\alpha}_0$ is a constant. Hatted eigenvalues have a similar behaviour.
    \item \emph{Singularities at $\pm 1$}: around $\xi=\pm 1$, the Lax connection develops a pole with residue proportional to $A^{(\mathfrak{p})}_\alpha \pm \gamma_{\alpha \beta} \varepsilon^{\beta \rho} A^{(\mathfrak{p})}_\rho$. The square of this quantity is proportional to the Virasoro constraint \eqref{VirasoroCoset}. This, together with the tracelessness of $A^{(\mathfrak{p})}$, implies that the residues of the poles appearing at $\xi=\pm 1$ on the hatted and tilded eigenvalues $p_i$ have to be correlated. We refer to eq. (2.17) in \cite{Schafer-Nameki:2010qho} for the specifics.
    \item \emph{Inversion symmetry}: the $\mathbb{Z}_2$ outer automorphism $\Sigma$ on a symmetric space reduces the number of independent quasi-momenta. Schematically, the quasi-momenta must satisfy $p_{2n}(\xi)=-p_{2n-1}(1/\xi) + (\text{winding})$, hence the name. We refer to eq. (2.20) in \cite{Schafer-Nameki:2010qho} for the specific form.
\end{itemize}
These properties, together with some mild assumptions, are enough to construct algebraic curves without the explicit computation of the path-ordered integral, which can become rather difficult even for simple stings solutions. For example, the algebraic curve for the folded string \eqref{foldedansatz} was found using these analyticity arguments in \cite{Gromov:2011de}.

To get a better understanding of how information is codified in these algebraic curves, let us have a look at two examples that can be obtained by brute-force computing the path-ordered exponential. For the BMN string \eqref{RelBMN} (with $c=1$), we find
\begin{equation}
    (\hat{p}_1 , \hat{p}_2 , \hat{p}_3 , \hat{p}_4 | \tilde{p}_1 , \tilde{p}_2 , \tilde{p}_3 , \tilde{p}_4)=\frac{2\pi \xi \kappa }{\xi^2-1} (1,1,-1,-1|1,1,-1,-1) \, .
\end{equation}
Energy and angular momentum can be read from here as $E\propto\lim_{\xi\to \infty} \xi (\hat{p}_1 + \hat{p}_2)$ and $J_1\propto\lim_{\xi\to \infty} \xi (\tilde{p}_1 + \tilde{p}_2)$ \cite{Schafer-Nameki:2010qho}, giving us the well known dispersion relation $E=J\propto \kappa$.

Another very simple solution is the one with two angular momenta and winding. If we write the $\mathbb{R}\times $S$^3\subset$AdS$_5\times$S$^5$ as
\begin{equation}
    \dd s^2=-\dd t^2 + \dd r^2 + r^2 \dd \varphi_1^2 + (1-r^2)\dd\varphi_2 \, ,
\end{equation}
this solution corresponds to
\begin{equation}
    \varphi_1=\omega \tau +m\sigma \, , \qquad \varphi_2=\omega \tau -m\sigma \, , \qquad r=\sqrt{\frac{1}{2}} \, , \qquad t=\sqrt{\omega^2 + m^2}\,\tau \, ,
\end{equation}
whose algebraic curve is given by
\begin{align}
    (\hat{p}_1 , \hat{p}_2 , \hat{p}_3 , \hat{p}_4 ) &=\frac{2\pi \xi \sqrt{\omega^2 + m^2}}{\xi^2-1} (1,1,-1,-1) \, , \label{curvespinningstring} \\
    (\tilde{p}_1 , \tilde{p}_2 , \tilde{p}_3 , \tilde{p}_4) &=\left(\frac{2\pi \xi K(1/\xi)}{\xi^2-1}  , \frac{2\pi \xi K(\xi)}{\xi^2-1}-m , -\frac{2\pi \xi K(\xi)}{\xi^2-1}+m , -\frac{2\pi \xi K(1/\xi)}{\xi^2-1} \right) \, , \notag
\end{align}
where $K(\xi)=\sqrt{m^2 \xi^2 + \omega^2}$. The properties we have discussed above are very clear in this example: inversion symmetry is fulfilled, residues at $\pm 1$ are correlated, the winding numbers appear at $\xi\to \infty$, we can read the energy of this string from $\lim_{\xi\to \infty} \xi(\hat{p}_1 + \hat{p}_2)=4\pi \sqrt{\omega^2 + m^2}$, and we can read the angular momentum from $\lim_{\xi\to \infty} \xi (\tilde{p}_1 + \tilde{p}_2+m)=4\pi \omega$.

\subsection{Strings in SNC AdS$_5\times$S$^5$}

Because the action of non-relativistic strings in SNC AdS$_5\times$S$^5$ can also be described in terms of a Lax connection, we can try to replicate the construction of an algebraic curve in this setting. Let us start by first discussing the analytical properties of the quasi-momenta before doing any brute-force computations. 

Similarly to relativistic AdS$_5 \times S^5$, the Lax connection \eqref{NRLax} becomes the Maurer-Cartan current in the $\xi\rightarrow \infty$ limit. This means that the quasi-momenta still encode the winding numbers of the classical solution we are considering. As some form of winding is always necessary for non-relativistic strings, these contributions will be always present.

In addition, the properties regarding the branch cuts and the inversion symmetry still hold for the non-relativistic string, because they are tied to the structure of the characteristic equations (which remains the same) and inversion symmetry (as the transformation of the Lax under the outer automorphism $\Sigma$ can be traded for a redefinition of the spectral parameter) respectively.

In contrast, the poles at $\pm 1$ are now proportional to $\mathcal{P}A_\alpha \pm \gamma_{\alpha \beta} \varepsilon^{\beta \rho} \mathcal{P}J_\rho$, which is not exactly the ``square root of the Virasoro'', meaning that the correlation of the poles is not as clear as in the relativistic case (although we retain the tracelessness of the residue, which gives us a partial correlation). Similarly, the matching with the symmetries at $\xi \to \infty$ is not completely clear for a similar reason.

As we now have an intuition of what to expect, we can review an explicit computation done in \cite{Fontanella:2022wfj}. Let us consider the group parametrisation that gives us the GGK coordinates \eqref{coset_GGK} with $R=1$
\begin{eqnarray}
    g = e^{x^1 H_1} e^{x^0 H_0} e^{x^a P_a} e^{x^{a'} P_{a'}}\, . 
\end{eqnarray}
Substituting this parameterisation into the Lax connection \eqref{NRLax}, we find that the eigenvalues of its $\sigma$ component associated with $\text{LAE}_{2,1}[\mathfrak{so}(2,4)]$, $\hat{\mu}$, have the following structure
\begin{equation}
    \hat{\mu}^2=a+b\pm 2\sqrt{ab} \, ,\qquad a\equiv\left( L_\sigma^M \right)^2 -L_\sigma^{H_+}L_\sigma^{H_-} \, , \qquad b\equiv -\left( L_\sigma^{J_{23}} \right)^2-\left( L_\sigma^{J_{24}} \right)^2-\left( L_\sigma^{J_{34}} \right)^2 \, ,
\end{equation}
where $L_\sigma^A$ is the component of $L_\sigma$ associated with generator $A$. On the other hand, the eigenvalues of $L_\sigma$ associated with $\text{LAE}_{2,1}[\mathfrak{so}(6)]$, $\tilde{\mu}$, are functions of $L_\sigma^{J_{a'b'}}$. If we now take into account the form of the Lax \eqref{NRLax} and of the projector \eqref{ProjectorNR}, we can see that only the components $L_\sigma^{H_\pm}$ can depend on the parameter $\xi$. Consequently, every $\tilde{\mu}$ is independent of $\xi$ and, in principle, only $\hat{\mu}$ may depend on $\xi$. To check if that is the case, let us examine the terms $L_\sigma^{H_\pm}$ in more detail. These two terms become
\begin{equation}
L^{H_{\pm}}_{\sigma} = \frac{\xi^2+1}{\xi^2-1} \left( x'^0 \pm \cos x^0 x'^1 \right) 
+ \frac{2 \xi}{\xi^2-1} \left( \dot{x}^0 \pm \cos x^0 \dot{x}^1 \right)  \, ,
\end{equation}
so the product takes the form
\begin{equation}
    L^{H_{+}}_{\sigma}L^{H_{-}}_{\sigma}=\frac{[(1+\xi^2)x^{\prime 0} +2\xi\,\dot{x}^0]^2- \cos^2(x^0) [(1+\xi^2)x^{\prime 1} +2\xi\,\dot{x}^1]^2 }{(\xi^2-1)^2} \, .
\end{equation}
The key observation now is that, if we use the equation of motion from the Lagrange multipliers \eqref{LagrangeMultipliersEquation}, this product takes the form
\begin{equation}
    L^{H_{+}}_{\sigma}L^{H_{-}}_{\sigma}=\cos^2 x^0 \left[ (\dot{x}^1)^2 - (x'^1)^2\right] \, ,
\end{equation}
consequently, the eigenvalues of the $\sigma$-component of the Lax connection are completely independent of the parameter $\xi$. Additionally, if $L_\sigma$ is independent of $\sigma$, the same holds for the eigenvalues of the monodromy matrix.

This is a mysterious result: in the relativistic case, the trace of the monodromy $T(\xi , \tau)$ can be used as a generating function for the conserved quantities of the action, but the eigenvalues of the monodromy are independent of $\xi$ in the non-relativistic case, so we cannot use them to generate an infinite tower of conserved charges. Therefore, Liouville integrability of the non-relativistic action is not straightforward. We should stress that $T(\xi , \tau)$ is not completely independent of the spectral parameter, but it appears in the non-diagonalisable contributions.

To get a better understanding of the situation, let us compute brute-force the algebraic curve associated to the folded BMN-like string \eqref{BMNfoldedcartesian}. This solution in GGK coordinates takes the form
\begin{equation}\label{BMN_like_sol}
x^0 = \text{gd}(\tau) \, , \qquad
x^1 = \sigma \, , \qquad
x^5 = J \tau \, , \qquad
\lambda_{\pm} = \pm \frac{J^2}{2} \cosh \tau \, ,
\end{equation}
with the other coordinates set to zero, giving us the following monodromy matrix
\begin{equation}\label{T_BMN_like}
T(\xi, \tau) = e^{2\pi (a_+ H_+ + a_- H_- + b_+ Z_+ + b_- Z_- + c M + d P_5)} \, , 
\end{equation}
where the coefficients $a_{\pm}, b_{\pm}, c, d$ are 
\begin{align}
a_{\pm} &= \pm \frac{\xi\pm 1}{\xi \mp 1} \text{sech}\, \tau \, , &
b_{\pm} &= \frac{J^2 \xi}{\xi^2-1} \cosh \tau \, , \notag \\
c &= - \tanh \tau \, , &
d &= \frac{2 J \xi}{\xi^2 - 1}\, . \label{coefficients_T_BMN_like}
\end{align}
Interestingly, if we substitute here either the adjoint representation or the induced representation \eqref{inducedrepresentation} obtained from the spinorial representation of $\mathfrak{so}(2,4)\oplus \mathfrak{so}(6)$, we find that the monodromy matrix is not diagonalisable. This means that some of the usual understanding of spectral curves cannot be applied.

Although the knowledgeable reader may be familiar with other non-diagonalisable monodromy matrices, e.g., the ones appearing in the context of Yang-Baxter deformations of AdS$_5\times$S$^5$ \cite{Ouyang:2017yko, Borsato:2021fuy, Borsato:2022drc, Driezen:2024mcn}, those monodromy matrices are non-diagonalisable only for isolated values of the spectral parameter.\footnote{Monodromy matrices that are non-diagonalisable for arbitrary values of the spectral parameter have been studied for quantum integrable systems \cite{Ahn:2020zly, Ahn:2021emp, NietoGarcia:2021kgh, NietoGarcia:2022kqi, NietoGarcia:2023jeb, deLeeuw:2025sfs}, although the quantum version of the spectral curve of these systems has not been constructed yet.} In contrast, the monodromy matrix associated to the BMN-like solution we are interested in is non-diagonalisable for nearly every value of $\xi$ and cannot be brought into a diagonal form.

The main reason why the monodromy matrix is non-diagonalisable is the fact that the extended string Newton-Hooke plus Euclidean algebra we used to construct the coset action \eqref{splitNR} is not a semisimple algebra. To better understand the implication, we have to define first what a Cartan subalgebra is. Given an algebra $\mathfrak{g}$, we define the Cartan subalgebra $\mathfrak{c}$ as a nilpotent and self-normalising subalgebra.\footnote{A \emph{nilpotent} algebra $\mathfrak{c}$ is the one that, if we construct the series of subalgebras
\begin{equation}
    \mathfrak{c}_0=\mathfrak{c} \,, \qquad \mathfrak{c}_1=[\mathfrak{c}_0,\mathfrak{c}_0] \,, \qquad \mathfrak{c}_2=[\mathfrak{c}_0,\mathfrak{c}_1] \,, \qquad \dots \qquad \mathfrak{c}_n=[\mathfrak{c}_0,\mathfrak{c}_{n-1}] \,,
\end{equation}
then, there exists an $n_0$ such that $\mathfrak{c}_n=\emptyset$ for all $n\geq n_0$.

Roughly speaking, we say that $\mathfrak{c}\subset \mathfrak{g}$ is \emph{self-normalised} if there is no element outside of $\mathfrak{c}$, i.e. in $\mathfrak{g}\setminus\mathfrak{c}$, that commutes with all the elements of $\mathfrak{c}$. More rigorously, we define first the normaliser as $\mathcal{N}(\mathfrak{c})=\{ x : x\in \mathfrak{g} \text{ such that } [x,c]=0 \ \forall c\in \mathfrak{c}\}$. Then, the self-normalising property can be written as $\mathcal{N}(\mathfrak{c})=\mathfrak{c}$.} For semisimple algebras, this definition is equivalent to saying that $\mathfrak{c}$ is the maximal abelian subalgebra consisting of elements whose adjoint representation is diagonalisable. However, as we commented above, the extended string Newton-Hooke plus Euclidean algebra we are working with is not semisimple. In fact, we can pick the following Cartan subalgebra\footnote{Here we are refining the concept of MAS (maximal abelian subalgebra) introduced in the paper \cite{Fontanella:2022wfj} by requiring nilpotency.}
\begin{equation}
    \mathfrak{c}=\{M,Z,P_2,J_{34},P_5,J_{67},J_{89}\} \,,
\end{equation}
as it is a nilpotent and self-normalising subalgebra, but $Z$, $P_2$ and $P_5$ have a non-diagonalisable adjoint representation.

We can write the monodromy matrix \eqref{T_BMN_like} in terms of the elements of the above Cartan subalgebra after we perform a similarity transformation
\begin{equation}
T(\xi, \tau) = S e^{2\pi \big(M + f Z + d P_5 \big) } S^{-1} \, , 
\end{equation}
where $d$ is the same as in \eqref{coefficients_T_BMN_like} while $f$ takes the form
\begin{equation}
f = - \left(\frac{\sqrt{2} J \xi}{\xi^2 - 1}\right)^2 \, .
\end{equation}
It is worth here to come back to the discussion on the independence of the eigenvalues of $T$ of the parameter $\xi$. Both in the adjoint representation and on the representation inherited from the spinorial representation of $\mathfrak{so}(2,4)\oplus \mathfrak{so}(6)$, the matrix associated to $M$ is diagonalisable but the matrices associated to both $Z$ and $P_5$ are not. Consequently, the eigenvalues of $T(\xi, \tau)$ are independent of $\xi$, but it also makes clear that the non-diagonalisable structure must carry non-trivial physical information.

\subsection{Non-diagonalisability in relativistic flat space}

So far, we have seen that the non-relativistic Lax connection is non-diagonalisable and its eigenvalues are independent of the spectral parameter $\xi$. Here, we argue that these peculiarities are not due to the non-relativistic nature of the theory we are considering, but related to the non-semisimple algebra we are working with. To show that this is the case, we want to consider the monodromy matrix for relativistic strings in Minkowski space. As Minkowski space is a symmetric space, as discussed at the beginning of section \ref{sec:SNCflatspacetime}, the construction from section \ref{sec:symmetricspaces} holds.

Let us consider the following coset representative
\begin{equation}
g = e^{P_{\hat{A}} x^{\hat{A}}}\, ,
\end{equation}
which generates Minkowski space in Cartesian coordinates. In these coordinates, let us consider the point-like string
\begin{equation}\label{point_like_flat}
t = \kappa \tau \, , \qquad
x_1 = \kappa \tau \, , 
\end{equation}
with the other coordinates to zero. The monodromy matrix associated to this solution is non-diagonalisable and is given by
\begin{equation}
 T(\xi, \tau) = e^{i q (P_0 + P_1)} \, , \qquad
q = \frac{i 4 \pi \kappa \xi}{1-\xi^2} \, . 
\end{equation}
If we take the adjoint representation of the $\mathfrak{iso} (1,d-1)$ algebra, the monodromy matrix is not diagonalisable and its eigenvalues are independent of the spectral parameter. This shows that relativistic theories can have the same issues we have found for the folded BMN-like solution in the non-relativistic theory.

%%%%%%%%%%%%%%%%%%%%%%%%%%%%%%%%%%%%%%%%%%%%%%%%%%%%%%%%%%%%%%%%%%%%%%%%%%%%%%%%%%%%%%%%%%%%%%%%%%%%%%%%%%%%%%%%%%%%%%%%%%%%%%%%%%%%%%%%%%%%%%%%%%%%%%%%%%%%%%%%%%%%%%%%%%%%%%%%%%%%%%%%%%%%%%%%%%%%%%%%%%%%%%%%%%%%%%%%%%%%%%%%%%%%%%%%%%%%%%%%%%%%%%%%%%%%%%%%%%%%%%%%%%%%%%%%%%%%%%%%%%%%%%%%%%%%%%%%%%%%
\chapter{Open questions and future directions}

In this review, we have collected recent results on string and 
$p$-brane Newton–Cartan holography and their integrability structure. 
Despite the important advances done in the last years, the field is far from closed and there are several open questions. 
We have collected these questions in different sections organised by topic. We hope that this review attracts new interest to the field and soon renders this chapter outdated.

\section{Formal aspects of holography}

\paragraph{$p$-brane limits of String Theory.} The $p$-NC limits of D$q$-branes have so far been proposed only at the level of the spacetime geometry. To be fully incorporated within the AdS/CFT correspondence, it is necessary to understand how string theory can be consistently defined on such geometries. If the probing object is the D$q$-brane itself, the non-relativistic limit of the DBI action is well-defined, provided that critical Ramond-Ramond $p+1$ and $q+1$ form fields are turned on in order to cancel the divergent term. However, if the probing object is the string - which is the natural situation when taking the limit of the AdS/CFT correspondence - the corresponding Polyakov action develops a divergent term arising from the longitudinal metric $\tau_{\mu\nu}$. If the limit is of stringy type, i.e.\ $p=1$, one may invoke the Gomis-Ooguri mechanism and turn on a critical B-field in order to cancel the divergence. However, if the limit involves a more general $p$, the critical B-field mechanism does not work. It is therefore an important question how to make sense of these $p$-NC limits of D$q$-branes when the probing object is a string.    

\paragraph{Asymptotic symmetries.} Matching symmetries is a very important test supporting the proposed holographic dualities. However, at the current stage, the computation of the symmetries of the bulk theory is still missing for several of the cases discussed in this review. The only two cases in which the symmetries have been computed are the GGK/GYM duality and the M2-brane limit of AdS$_4$/ABJM. To make progress in the other cases, it would be important to identify the underlying non-relativistic supergravity theories.

In the two cases where the symmetries have been computed, they correspond to Killing vectors of the bulk geometry. For the purpose of holography, however, one needs the asymptotic symmetries. The set of asymptotic symmetries consists of the bulk Killing vectors evaluated at the boundary (i.e.\ in the asymptotic region), up to a possible symmetry enhancement. We reviewed the arguments supporting why such an enhancement is not expected in either of these two cases. It would nevertheless be desirable to provide a more formal proof that no such enhancement occurs, for instance by determining the set of Killing vectors that act in the near-boundary region.

\paragraph{Holographic dual of GED.} The SNC limit of AdS$_5$/CFT$_4$ allows two different limits of the DBI action. The first, which we called limit 1, leads to GYM, while the second, called limit 2, leads to GED. Only limit 1 (GYM) is consistent with GGK theory at the level of matching the symmetries. Nevertheless, limit 2 is still a legitimate limit that fulfils the conditions \ref{Holographic realisation} and \ref{Uniqueness}. It is therefore natural to ask what the corresponding holographic string dual of GED theory with five scalar fields might be.
To obtain the GED theory from the DBI action it is necessary to turn on a critical B-field. Such theory appears from the DBI action as $N^2$ copies of the GED theory with free scalar fields. Therefore, it is a free gauge theory.

From this perspective, the dual SNC limit of AdS$_5\times$S$^5$ should also lead to a free non-relativistic string theory. One expects that a B-field also enters the derivation of the dual string theory, which should therefore arise from an SNC limit of the relativistic string in AdS$_5\times$S$^5$.
Since AdS$_5$ and S$^5$ share the same radius, the possibilities for an SNC limit appear to be rather restricted, essentially to the GGK one. A straightforward way to obtain a free non-relativistic string theory is to take an additional large radius limit, which yields the Gomis-Ooguri string. It would therefore be interesting to investigate whether the GED theory (supplemented by free scalars) admits a holographic string dual, and if so what its precise form is.

\paragraph{Application to other limits.} The method to derive a new holographic duality from AdS/CFT, based on imposing five axioms, can be applied to limits that are not necessarily non-relativistic. In this review we have already shown some examples, namely the Carroll and flat space limits. It would be interesting to apply the method to other types of limits. For instance, one may consider the limit that takes AdS$_5\times$S$^5$ to the pp-wave background. Such an analysis may help to elucidate the dual gauge theory of string theory in the pp-wave background. 

Another context in which this method could be applied is that of deformations of AdS/CFT. 
For example, one could consider the holographic  correspondence between the Lunin–Maldacena deformation of AdS$_5 \times$S$^5$ and the Leigh–Strassler $\mathcal{N}=4$ SYM. On the gauge theory side, a ``fishnet limit'' was proposed in \cite{Gurdogan:2015csr}, and its corresponding ``dual'' effect was conjectured in \cite{Gromov:2019jfh,Gromov:2019bsj}. If a brane setting of this deformed duality were known, then one could ask if the conjectured limit satisfies the set of five consistency conditions, or, if not, whether the method can be used to identify a consistent alternative.

\section{Quantitative tests for the GGK/GYM duality}

In this review, we have seen that the GGK/GYM duality is well-developed and ready to be tested at a quantitative level. Here we summarise some of the tests that could further probe its validity.

\paragraph{Field-operator map.} The relativistic AdS/CFT correspondence is characterised by a dictionary between bulk fields and gauge-invariant operators in the boundary theory, also known as the GKPW prescription \cite{Gubser:1998bc, Witten:1998qj}. This dictionary was crucial for establishing a holographic procedure to compute gauge theory correlation functions from gravity in AdS. It would be important to develop a similarly complete holographic dictionary for the GGK/GYM duality.

\paragraph{String spectrum = Scaling dimensions.} One of the most important tests of AdS/CFT is that the string spectrum matches the scaling dimensions of the dual gauge-invariant operators, $E = \Delta$. It would be important to see how this test applies to the GGK/GYM duality. In particular, the string spectrum in GGK theory, as discussed in this review, consists of three massive and five massless \emph{free} fields in AdS$_2$. This spectrum arises both around the static string and the folded BMN-like string.

A natural question is what the corresponding dual observables are in GYM. To argue that the dual observables of the string spectrum are the scaling dimensions of gauge-invariant operators, one needs to invoke the state-operator correspondence. This has not yet been studied for GYM, and it is possible that the familiar relation $E = \Delta$ may require corrections from additional quantum numbers (see e.g.\ \cite{Lambert:2021nol, Baiguera:2024vlj} in a different non-relativistic gauge theory context).

If the state-operator correspondence holds for GYM,\footnote{This looks like to be the case \cite{state_operator}.} it would then be useful to employ the conformal Milne algebra to constrain the most general two-point functions via Ward–Takahashi identities. One could then consider specific dual gauge-invariant operators, compute their two-point functions using Feynman diagrams, and extract their full scaling dimensions, including quantum corrections (provided that non-renormalisation theorems hold in GYM). The results should match the exact spectrum of non-relativistic strings, implying a cancellation of the quantum corrections beyond one-loop.

\paragraph{GYM at finite temperature and black holes.} In the GYM theory one could turn on a finite temperature. The expected effect on the GGK theory would be the appearance of a black hole with SNC AdS$_5\times$S$^5$ asymptotic geometry. So far, naive arguments suggest the non-existence of non-relativistic black holes, based on the observation that taking the limit $c \to \infty$ makes the horizon disappear, as can be seen from the formula for the Schwarzschild radius. However, this argument is too naive, since it assumes that the non-relativistic limit arises from a particle foliation. In the GGK/GYM duality one instead has to consider a string foliation, corresponding to a String Newton–Cartan geometry. In this case, there could exist a mechanism that retains the horizon in the SNC limit, similarly to what we have seen for the horizon of a stack of non-relativistic D3-branes. Indications of black holes with 3-brane NC foliations have already appeared in \cite{Blair:2025ewa}.

\paragraph{Hagedorn/confinement correspondence.} In the relativistic AdS$_5$/CFT$_4$ duality, there is a phase transition on the string theory side near the Hagedorn temperature, while on the gauge theory side there is a confinement–deconfinement transition. These phase transitions on the two sides of the duality exhibit matching behaviour \cite{Aharony:2003sx}. A similar Hagedorn/confinement correspondence also arises in the duality between non-relativistic strings with a non-relativistic worldsheet and Spin Matrix Theory \cite{Harmark:2006ta}. It would be interesting to investigate whether such a correspondence also holds in the GGK/GYM duality, both at zero and at finite temperature.

\paragraph{Generalise the above tests to $p$-NC holography.} The tests described above may also be applied to the other examples of $p$-NC holography. This would first require understanding how to consistently define non-relativistic strings in $p$-NC spacetimes, for $p\neq 1$, and then computing their energy spectrum.

\paragraph{Wilson loops.} Wilson loops are very interesting operators to study in gauge theories, as their non-local nature can be used to probe if the theory is confining or not. As GYM is still a gauge theory, we can define Wilson loops in the usual way. We can then ask what would be the holographic dual of a Wilson loop in GYM. For AdS/CFT, Maldacena-Wilson loops (a supersymmetric generalisation of the Wilson loop where we include scalars in the path-ordered exponential) over a closed curve $\Gamma$ are dual to minimal surfaces in AdS$_5\times$S$^5$ with the condition of ending on the curve $\Gamma$ at the boundary. Thus, we would expect that a Wilson loop in GYM (or a sensible modification of it) to be dual to minimal surfaces in the GGK theory. However, it is not clear what a minimal surface would be in a String Newton Cartan geometry, as we have to work with two metrics. To overcome this problem, first one could study paths that are non-trivial only on the longitudinal coordinates.

Maldacena-Wilson loops are of special interest because, for specific shapes, they are tightly constrained by supersymmetry and can be exactly computed by supersymmetric localisation \cite{Pestun:2007rz}. As a consequence, they are an ideal ground to test the correspondence. Thus, it would be interesting to see if the localisation machinery can be transferred to the GGK/GYM duality. This would require first a better understanding of the supersymmetric extension of the non-semisimple algebras involved in the duality.

\section{Classical strings and spectrum}

\paragraph{Analogy with the near pp-wave expansion.} 
In this review, we have seen that in the GGK theory the spectrum around the folded BMN-like string is exact. This result can be viewed as the analogue of what is encountered when expanding the relativistic string action in AdS$_5\times$S$^5$ around the BMN vacuum. In that case, the string energy $E$ receives a series of quantum corrections in an expansion in $1/T$. In the limit of large angular momentum $J\to \infty$, with $J/T$ kept fixed, only the classical dispersion relation $E_{\text{cl}} = J$ and the contribution from the quadratic action of harmonic oscillators, $E_2$, survive. All higher-loop contributions beyond one-loop are suppressed in this limit, i.e.
\begin{eqnarray}\label{BMN_spectrum}
E - J = E_2 + \frac{1}{J} E_4 + \ldots \quad\xrightarrow[\ J\to\infty \ ]{} \quad E_2 \, .
\end{eqnarray}
This result also admits a geometric interpretation. The exact spectrum \eqref{BMN_spectrum} coincides with the energy spectrum of strings in the pp-wave background. Geometrically, the pp-wave background can be obtained via a Penrose limit of the AdS$_5\times$S$^5$ geometry. Therefore, the $J\to\infty$ limit can be interpreted as zooming into a sector of strings in AdS$_5\times$S$^5$ that is effectively described by strings propagating in the pp-wave background (see \cite{Plefka:2005bk} for a review).

We expect a similar mechanism to be at work for the SNC limit of AdS$_5\times$S$^5$. One could attempt to compute the energy spectrum of relativistic strings in AdS$_5\times$S$^5$ around the relativistic uplift of the folded BMN-like string (i.e. equation \eqref{foldedansatz} with $\omega = 0$). In a large string tension expansion, the string energy $E$ should again receive quantum corrections in powers of $1/T$. In the SNC limit $c\to \infty$, with $c/T$ kept fixed, we expect to recover precisely the exact spectrum found around the folded BMN-like string. In particular, the energy is expected to behave analogously to the relativistic BMN case, namely
\begin{eqnarray}\label{SNC_spectrum}
E - E_{\text{cl}} = E_2 + \frac{1}{c} E_4 + \ldots \quad\xrightarrow[\ c\to\infty \ ]{} \quad E_2 \, .
\end{eqnarray}
Here, there is a geometric interpretation as well. The exact energy emerging in the $c\to \infty$ limit in \eqref{SNC_spectrum} coincides with the energy of non-relativistic strings that one would find in the SNC AdS$_5\times$S$^5$ background. It would be interesting to explicitly compute the spectrum of \emph{relativistic} strings in AdS$_5\times$S$^5$ around the relativistic uplift of the folded BMN-like string, and verify that the expected behaviour \eqref{SNC_spectrum} indeed arises.

\paragraph{More examples of classical non-relativistic string solutions.}
It would be interesting to further expand our catalogue of classical non-relativistic string solutions. For instance, one could try to generalise the folded BMN-like string to solutions with several non-vanishing spins, and find their relativistic origin. The catalogue of relativistic string solutions is incredibly broad: rigid strings, rotating strings, pulsating strings, spiky strings, giant magnons, etc. We can, in principle, follow the same steps as in section \ref{flowing} and construct the non-relativistic counterparts from the relativistic ones, once we have included the appropriate factors of $c$.

\paragraph{Exactness of the non-relativistic energy spectrum.}
Once a sufficiently large catalogue of non-relativistic string solutions, together with their relativistic uplifts, has been constructed, it will be interesting to compute the spectrum around these backgrounds, similarly to e.g. \cite{Drukker:2000ep, Beccaria:2010ry, Forini:2010ek, Forini:2014kza, Forini:2015mca}. There are at least two important questions to address.

The first is whether the non-relativistic string spectrum remains exact when expanded around different vacua. In particular, this would clarify whether the exact expansion of the GGK non-relativistic action in terms of three massive and five massless free fields in AdS$_2$ is a special feature of the static and folded BMN-like string vacua, or instead a universal property.

The second question is whether the expected behaviour in equation \eqref{SNC_spectrum} (i.e.  corrections beyond one-loop are suppressed by powers of $1/c$) also holds for vacua other than the folded BMN-like string. Establishing this would provide concrete settings in which the relativistic string action simplifies.

\section{Integrability in non-relativistic theories}

\paragraph{Conserved higher charges.} 
To explicitly demonstrate the integrability of the GGK theory, it is desirable to construct the infinite set of conserved charges. An immediate obstacle arises from the fact that the eigenvalues of the $\sigma$-component of the Lax pair, $L_{\sigma}$, become independent of the spectral parameter when evaluated on the constraint surface. Moreover, if $L_{\sigma}$ is independent of $\sigma$ the same holds for the monodromy matrix. This has a direct consequence: the transfer matrix is likewise independent of the spectral parameter and therefore cannot be expanded to generate the infinite tower of higher conserved charges required for integrability.

There are three possible directions one could explore to circumvent this issue. One is to modify the algebra representation, a second one is to modify the Lax pair, and a third one is to go ``beyond the trace'' of the monodromy matrix.

The first possibility is inspired by relativistic strings in flat spacetime. In that case, if one chooses, for example, the adjoint representation of the Poincaré generators, the $\sigma$-component of the Lax pair is again independent of the spectral parameter. However, one may instead consider a unitary infinite-dimensional representation in which the momentum generators are diagonal, $P_a \ket{p} = p_a \ket{p}$. In such a representation, the transfer matrix depends on the spectral parameter, thereby generating the conserved charges required for integrability. It would be interesting to investigate whether a representation with analogous properties can be constructed for the extended Newton–Hooke and Euclidean algebras.

A second possibility is to modify the Lax pair itself. At the moment, the zero-curvature condition of the Lax connection reproduces the equations of motion only upon imposing the constraint associated with the Lagrange multipliers. This constraint is also responsible for rendering the eigenvalues of $L_{\sigma}$ independent of the spectral parameter. Therefore it would be interesting to find a mechanism by which the zero-curvature condition automatically incorporates the constraint. In such a scenario, the constraint would not need to be imposed by hand, and the issue described above might be resolved. One possible idea would be to incorporate the extension generators $Z_A$ into the coset representative, interpreting them as generators associated with the Lagrange multipliers $\lambda_A$, and to reformulate the constraint term in the coset action as a Wess–Zumino term.

A third possibility is to support the trace of the monodromy and the trace of powers of the monodromy with more information. There is physical information codified in the non-diagonalisable part of the monodromy matrix, therefore traces of powers of the monodromy will always fail to retrieve this information. The key question here would how to extract information regarding conserved charges from the off-diagonal elements of the monodromy matrix.

\paragraph{A Non-Relativistic Classical Spectral Curve.}
The fact that the monodromy matrix has eigenvalues independent of the spectral parameter appears to preclude the existence of a non-trivial classical spectral curve. This issue seems closely related to the fact that the Lax pair is based on a non-semisimple Lie algebra, namely, an extended version of the Newton–Hooke and Euclidean algebras, and possibly to the related property that the momentum-type generators are non-diagonalisable in finite-dimensional representations.

Furthermore, it is clear that for non-diagonalisable monodromy matrices, the classical spectral curve cannot codify all the physical information. The monodromy matrices are invariant under conjugation, thus physical information has to be stored in their conjugacy class. For diagonalisable matrices (over $\mathbb{C}$), such information is stored in the eigenvalues, which in turns are found from the characteristic equation. However, this is not true for non-diagonalisable matrices. It is obvious that just knowing the characteristic equation of a non-diagonalisable matrix is not enough to know its Jordan block decomposition. From this point of view, it is natural that the classical spectral curve is unable to describe systems with non-diagonalisable monodromy matrices. Thus, we need to support the classical spectral curve with information regarding the non-diagonalisable structure of the monodromy matrix we want to study. 

Since finite-dimensional representations are required in the standard construction of the classical spectral curve, we are bounded to work with non-diagonalisable transfer matrices. Thus, it would be interesting to investigate how we can use the classical spectral curve to extract all the physical information encoded in the monodromy matrix.

The current notion of the spectral curve relies heavily on the properties of AdS$_5 \times$S$^5$, in particular on the fact that the Lax connection is based on a semisimple Lie algebra. A generalisation of the spectral curve to Lax pairs built from non-semisimple Lie algebras would therefore be beneficial for a broad class of systems.

\paragraph{Non-relativistic Yang-Baxter deformations.}
An interesting unexplored avenue is whether one can construct integrable deformations of non-relativistic string theory. This does not seem to be entirely straightforward. For example, the SNC limit of the $\eta$-deformation of AdS$_5 \times$S$^5$ yields directly the undeformed GGK theory, where the deformation parameter drops out completely from the action. It would be interesting to study the non-relativistic limit of more general types of Yang–Baxter deformations and determine whether the deformation parameters can be retained in this limit.

\paragraph{Further developments concerning the non-relativistic coset action.}
In its current formulation, the coset action for GGK theory does not include fermions. If we recall how fermions enter in the relativistic AdS$_5 \times$S$^5$ coset action, the $\mathbb{Z}_2$ grading of the bosonic algebra is promoted to a $\mathbb{Z}_4$ grading once fermions are introduced. However, there is an important difference compared to the relativistic case. In that setting, both bosonic and fermionic degrees of freedom align with the $\mathbb{Z}_4$ projections. In the non-relativistic case, we already know that this alignment fails for the bosonic sector. Therefore, we expect that a similar misalignment with the $\mathbb{Z}_4$ grading will also appear in the fermionic sector. Constructing this explicitly would be an interesting direction for future work.

At the level of the Polyakov action, the limit approach and the null reduction approach are equivalent. Is there a way for this equivalence to manifest itself at the level of coset actions? An example of null reduction implemented directly at the level of a coset action was given in \cite{Figueroa-OFarrill:2025nmo}. However, the starting point there is a relativistic Wess–Zumino–Witten action based on the coset $(G\times G)/H$. This setup does not allow for an immediate comparison with the GGK coset action \cite{Fontanella:2022fjd,Fontanella:2022pbm}, but it could instead be compared with the coset action arising from the SNC limit of an AdS$_3 \times$ S$^3$ target space. A natural future direction would therefore be to work out the SNC limit of string theory in AdS$_3 \times$ S$^3$ (by taking into account the non-trivial relativistic B-field), construct the corresponding coset action, and compare it with the null reduction approach.

An intriguing point concerns the structure of the coset itself. The construction is based on a Klein pair $(\mathfrak{g}, \mathfrak{h})$. From this coset, one can derive a metric, which is either non-degenerate in the Lorentzian case, or a pair of degenerate metric tensors in the non-Lorentzian case. One can then compute the associated Killing vectors, which generate an algebra $\mathfrak{g}_{\text{iso}}$. A striking feature is that $\mathfrak{g} = \mathfrak{g}_{\text{iso}}$ only in the Lorentzian case, but not in the non-Lorentzian ones. In fact, SNC spacetimes generically admit an infinite-dimensional isometry algebra, as illustrated by few examples in this review.
One possible way to reduce this infinite-dimensional symmetry is to introduce a connection. In the Lorentzian case, this has no effect, but in the non-Lorentzian setting it does. For instance, if one imposes a vanishing Christoffel symbol, $\Gamma^{\rho}_{\mu\nu} = 0$, and requires it to be preserved by Killing vectors, then $(\mathsterling_{\xi} \Gamma)^{\rho}_{\mu\nu} = \partial_{\mu} \partial_{\nu} \xi^{\rho} = 0$, which restricts the isometry algebra to be finite-dimensional. It would be interesting to explore the relation between these finite-dimensional isometries and the underlying Klein pair.

\paragraph{Integrable systems from classical non-relativistic string solutions.}
Integrability can also play an important role in the construction of the classical string solutions discussed above. In the relativistic case, the string dynamics can be reduced, via suitable ansätze, to simpler integrable mechanical models. For instance, a rotating string ansatz in AdS$_5 \times$S$^5$ reduces the problem to the Neumann–Rosochatius system \cite{Arutyunov:2003uj,Arutyunov:2003za}, which describes a set of oscillators constrained to move on a sphere or in AdS space.
It would be interesting to investigate whether a similar reduction can be achieved in the GKK theory, and whether it leads to Galilean integrable mechanical systems.

\paragraph{Non-relativistic spin-chains.}
 Another setting in which integrability might manifest is in the computation of scaling dimensions of gauge-invariant operators in GYM theory with five scalars. Provided that the state–operator correspondence works as in the relativistic AdS/CFT case, these scaling dimensions are expected to be exact and to receive corrections only up to one-loop in the planar limit. This suggests that the dilatation operator truncates at one-loop, and it would be particularly interesting to identify it.
In the relativistic AdS$_5$/CFT$_4$ correspondence, Minahan and Zarembo showed that the one-loop dilatation operator is described by the Heisenberg XXX spin chain. In that case, the spectrum is further modified by long-range corrections, which, however, are not expected to arise in the GYM theory.

\section*{Acknowledgments}
%%%%%%%%%%%%%%%%%%%%%%%%%%%%%%%%%%%%%%

This review benefits from conversations with scientists working in the topics of Non-Lorentzian geometries and Integrability. In particular, we are very happy to thank Eric Bergshoeff, Chris Blair, Marius de Leeuw, Jos\'e Miguel Figueroa-O'Farrill, Sergey Frolov, Jaume \& Quim Gomis, Daniel Grumiller, Troels Harmark, Jelle Hartong, Neil Lambert, Tristan McLoughlin, Niels Obers, Gerben Oling, Stefan Prohazka, Jan Rosseel, Joseph Smith, Alessandro Torrielli, Arkady Tseytlin, Pedro Vieira and Ziqi Yan. 
AF is supported by the SFI-Royal Society under the grant number RFF$\backslash$EREF$\backslash$210373. JMNG is supported by ERC-2022-CoG - FAIM 101088193.
AF thanks Lia for her permanent support.

%%%%%%%%%%%%%%%% BIBLIOGRAPHY %%%%%%%%%%%%%%%%

\bibliographystyle{nb}

\bibliography{Biblio.bib}

\end{document}